# The solution classical and quantum feedback optimal control problem without the Bellman Equation.
# Part I. The Solution classical feedback optimal control problem without the Bellman Equation. Dissipative differential game imbeded into a 'small' white noise.

Jaykov Foukzon

Israel Institute of Technology

jaykovfoukzon@list.ru

**Abstract:** A new approach, which is proposed in this paper allows one to construct the Bellman function $u(t,x)$ and optimal control $\alpha^*(t) = (\alpha_1^*(t),\ldots,\alpha_m^*(t))$ directly, i.e., without any reference to the Bellman equation, by way of using strong large deviations principle of Non-Freidlin-Wentzell type for Colombeau-Ito's **SDE**. An approach is proposed for solving problems of positional control of a dynamical system under conditions of incomplete information about its current phase states.

## Contents.

# I.Introduction.

Many optimal control problems essentially come down to constructing Colombeau generalized function $(u_\epsilon(t,x))_\epsilon, \epsilon \in (0,1]$ that has the properties:

$$(u_\epsilon(t,x))_\epsilon \triangleq (u_\epsilon(t,x;T))_\epsilon =$$
$$= \inf_{\alpha_\epsilon(t) \in U \subsetneq \mathbb{R}^n} \left[ \left( \bar{\mathbf{J}}_\epsilon \left( t, T, \{\mathbf{X}_{s,\epsilon}^{x,0}\}_{s \in [t,T]}, \{\alpha_\epsilon(t)\}_{s \in [t,T]} \right) \right)_\epsilon \right],$$

$$(u_\epsilon(t,x))_\epsilon =$$
$$= \left( \inf_{\{\alpha_\epsilon(s)\}_{s \in [t,T]} \in U, U \subsetneq \mathbb{R}^n} \left[ \bar{\mathbf{J}}_\epsilon \left( t, t', \{\mathbf{X}_{s,D,\epsilon}^{x,0}\}_{s \in [t,t']}, \{\alpha_\epsilon(s)\}_{s \in [t,t']} \right) + u_\epsilon(t', \mathbf{X}_{t',D,\epsilon}^{x,0}) \right] \right)_\epsilon$$

(1.1)

where $\bar{\mathbf{J}}_\epsilon(\circ,\circ,\circ,\circ)$ is the termination payoff functional, $\alpha_\epsilon(t) = (\alpha_{\epsilon,1}(t),\ldots,\alpha_{\epsilon,m}(t))$ is a control and $(\mathbf{X}_{t,\epsilon}^x)_{\epsilon, t \geq 0}$ is process governed by some ordinary differential equation:

$$\frac{d\mathbf{X}_{s,\epsilon}^{x,t'}}{ds} = g_\epsilon\left(s, \mathbf{X}_{s,\epsilon}^{x,t'}, \alpha_\epsilon(s)\right), \mathbf{X}_{t',\epsilon}^{x,t'} = x.$$

$$s \in [t',t].$$

(1.2)

$$x \in \mathbb{R}^n,$$

or integral equation:

$$\left(\mathbf{X}^{x,t'}_{t,\epsilon}\right)_\epsilon = x + \left(\int_{t'}^{t} g_\epsilon\left(s, \mathbf{X}^{x,t'}_{s,\epsilon}, \mathbf{\alpha}_\epsilon(s)\right) ds\right)_\epsilon$$

$$\left(\mathbf{X}^{x,t'}_{t',D,\epsilon}\right)_\epsilon = x \in \widetilde{\mathbb{R}}^n,$$

(1.2')

$$s \in [t',t].$$

Here $\epsilon \in (0,1]$ the singular perturbation parameter.
**Example.1.1.** Homing missile control:

$$\ddot{R} = R\dot{\sigma}^2 - a_\mathsf{T}\sin(\Psi_\mathsf{T} - \sigma) + a_\mathsf{M}\sin(\Psi_\mathsf{M} - \sigma),$$

$$\ddot{\sigma} = -\frac{2R\dot{R}\dot{\sigma}}{R_\epsilon^2 + \epsilon} + \frac{a_\mathsf{T} R\cos(\Psi_\mathsf{T} - \sigma)}{R^2 + \epsilon} - \frac{a_\mathsf{M} R\cos(\Psi_\mathsf{M} - \sigma)}{R^2 + \epsilon},$$

$$\dot{\Psi}_\mathsf{M} = \frac{a_\mathsf{M} V_\mathsf{M}}{V_\mathsf{M}^2 + \epsilon},$$

(1.3)

$$\dot{\Psi}_\mathsf{T} = \frac{a_\mathsf{T} V_\mathsf{T}}{V_\mathsf{T}^2 + \epsilon},$$

$$\epsilon \in (0,1].$$

In this paper we will be considered a solutions to the equations of the above singular form in the sense of Colombeau algebra $\mathcal{G}(\mathbb{R}_+)$, i.e.:

$$(\ddot{R}_\epsilon)_\epsilon = (R_\epsilon)_\epsilon (\dot{\sigma}_\epsilon^2)_\epsilon - a_T \sin((\Psi_{T,\epsilon})_\epsilon - (\sigma_\epsilon)_\epsilon) + a_M \sin((\Psi_{M,\epsilon})_\epsilon - (\sigma_\epsilon)_\epsilon),$$

$$(\ddot{\sigma}_\epsilon)_\epsilon = -\frac{2(R_\epsilon)_\epsilon (\dot{R}_\epsilon)_\epsilon (\dot{\sigma}_\epsilon)_\epsilon}{(R_\epsilon^2)_\epsilon + \epsilon} + \frac{a_T (R_\epsilon)_\epsilon \cos((\Psi_{T,\epsilon})_\epsilon - (\sigma_\epsilon)_\epsilon)}{(R_\epsilon^2)_\epsilon + \epsilon} -$$

$$-\frac{a_M (R_\epsilon)_\epsilon \cos((\Psi_{M,\epsilon})_\epsilon - (\sigma_\epsilon)_\epsilon)}{(R_\epsilon^2)_\epsilon + \epsilon},$$

$$(\dot{\Psi}_{M,\epsilon})_\epsilon = \left(\frac{a_M V_M}{V_M^2 + \epsilon}\right)_\epsilon,$$

$$(\dot{\Psi}_{T,\epsilon})_\epsilon = \left(\frac{a_T V_T}{V_T^2 + \epsilon}\right)_\epsilon,$$

$$\epsilon \in (0,1].$$

(1.4)

Thus in general case we will be considered a solutions to the equations Eq.(1.2) in the sense of Colombeau algebra $\mathcal{G}(\mathbb{R}_+)$, i.e.:

$$\frac{d(\mathbf{X}_{t,\epsilon}^x)_\epsilon}{dt} \triangleq \left(\frac{d\mathbf{X}_{t,\epsilon}^x}{dt}\right)_\epsilon = (g_\epsilon(t, \mathbf{X}_{s,\epsilon}^x, \alpha(t)))_\epsilon, \mathbf{X}_{0,\epsilon}^x = x.$$

$$x \in \widetilde{\mathbb{R}}^n,$$

$$\epsilon \in (0,1].$$

(1.4′)

Many stochastic optimal control problems essentially come down to constructing Colombeau generalized function $(u_\epsilon(t,x))_\epsilon \in \mathcal{G}(\mathbb{R}_+ \times \mathbb{R})$ that has the properties:

$$(u_\epsilon(t,x))_\epsilon \triangleq (u_\epsilon(t,T,x))_\epsilon =$$

$$= \left( \inf_{\{\alpha(s)\}_{s\in[t,T]} \in U, U \subsetneq \mathbb{R}^n} \mathbf{E}\left[ \bar{\mathbf{J}}_\epsilon\left(t, T, \{\mathbf{X}^{x,0}_{s,D,\epsilon}(\omega)\}_{s\in[t,T]}, \{\boldsymbol{\alpha}_\epsilon(s)\}_{s\in[t,T]}\right)\right]\right)_\epsilon,$$

$$(u_\epsilon(t,x))_\epsilon =$$

(1.5)

$$\left( \inf_{\{\alpha_\epsilon(s)\}_{s\in[t,T]} \in U, U \subsetneq \mathbb{R}^n} \mathbf{E}\left[ \bar{\mathbf{J}}_\epsilon\left(t, t', \{\mathbf{X}^{x,0}_{s,D,\epsilon}(\omega)\}_{s\in[t,t']}, \{\boldsymbol{\alpha}_\epsilon(s)\}_{s\in[t,t']}\right) + u_\epsilon(t', \mathbf{X}^{x,0}_{t',D,\epsilon}(\omega))\right]\right)_\epsilon,$$

$$t \leq t' \leq T,$$

$$t' - t \approx 0,$$

where $(\bar{\mathbf{J}}_\epsilon(\circ,\circ,\circ,\circ))_\epsilon$ is the Colombeau generalized termination payoff functional, such the next conditions is satisfied:

**Condition 1.**

(1.1) $\forall s, t' \leq s \leq t : (\bar{\mathbf{J}}_\epsilon(t',t,\circ,\circ))_\epsilon = (\bar{\mathbf{J}}_\epsilon(t',s,\circ,\circ))_\epsilon + (\bar{\mathbf{J}}_\epsilon(s,t,\circ,\circ))_\epsilon,$

(1.2) $\forall t \in [0,\infty), \forall \epsilon \in (0,1] : \exists \left( \lim_{t \to t'} \frac{\bar{\mathbf{J}}_\epsilon(t',t,\circ,\circ)}{t-t'} \right) \triangleq (\overline{\Delta\mathbf{J}}_\epsilon(t'\circ,\circ))_\epsilon,$

(1.3) $(\overline{\Delta\mathbf{J}}_\epsilon(t'\circ,\circ))_\epsilon \in \mathcal{G}(\mathbb{R}_+ \times \mathbb{R} \times \mathbb{R}).$

$(\alpha_\epsilon(t))_\epsilon = ((\alpha_{\epsilon,1}(t))_\epsilon, \ldots, (\alpha_{\epsilon,m}(t))_\epsilon)$ is a generalized control and $\left((\mathbf{X}^{x,t'}_{t,D,\epsilon})\right)_{\epsilon, t \geq 0}$ is some Colombeau-Markov generalized process governed by some stochastic equation driven by white noise, of the form:

$$\left(\mathbf{X}_{t,D,\epsilon}^{x,t'}(\omega)\right)_{\epsilon} = x + \left(\int_{t'}^{t} g_{\epsilon}\left(s, \mathbf{X}_{s,D,\epsilon}^{x,t'}(\omega), \alpha_{\epsilon}(s)\right) ds\right)_{\epsilon} +$$

$$+ \sqrt{D} \left(\int_{t'}^{t} \sigma_{\epsilon}\left(s, \mathbf{X}_{s,D,\epsilon}^{x,t'}(\omega), \alpha(s)\right) d\mathbf{W}(s,\omega)\right)_{\epsilon}, \quad (1.6)$$

$$\left(\mathbf{X}_{t',D,\epsilon}^{x,t'}(\omega)\right)_{\epsilon} = x \in \widetilde{\mathbb{R}}^{n},$$

$$s \in [t', t].$$

Using Ito formula we obtain:

$$\left(\mathbf{E}\left[u_{\epsilon}(t', X_{t,D,\epsilon}^{x,0}(\omega))\right]\right)_{\epsilon} - (u_{\epsilon}(t,x))_{\epsilon} =$$

$$= \left(\mathbf{E}\left[\int_{t'}^{t} L_{\epsilon}^{\alpha_{\epsilon}(s)}\left(s, X_{s,D}^{x,0}(\omega)\right) u_{\epsilon}(s, X_{s,D,\epsilon}^{x,0}(\omega)) ds\right]\right)_{\epsilon}, \quad (1.7)$$

Where

$$L_{\epsilon}^{\alpha_{\epsilon}} = L_{\epsilon}^{\alpha_{\epsilon}}(s,x) = \frac{\partial}{\partial s} + \sum_{i=1}^{n} g_{\epsilon}^{i}(s,x,\alpha_{\epsilon}) \frac{\partial}{\partial x_{i}} + D^{2} \sum_{i,j=1}^{n} \sigma_{\epsilon}^{ij}(s,x,\alpha_{\epsilon}) \frac{\partial^{2}}{\partial x_{i} \partial x_{j}}. \quad (1.8)$$

Thus from Eqs.(1.5),(1.7)-(1.8) we obtain:

$$0 =$$

$$\left( \inf_{\alpha_\epsilon} \left\{ \mathbf{E}\left[ \bar{\mathbf{J}}_\epsilon\left(t',t, \{\mathbf{X}^{x,0}_{s,D,\epsilon}(\omega)\}_{s\in[t',t]}, \{\boldsymbol{\alpha}_\epsilon(s)\}_{s\in[t',t]}\right) + u(t', \mathbf{X}^{x,0}_{t',D,\epsilon}(\omega)) \right] - u_\epsilon(t,x) \right\} \right)_\epsilon =$$

$$= \left( \inf_{\alpha_\epsilon} \mathbf{E}\left[ \bar{\mathbf{J}}_\epsilon\left(t',t, \{\mathbf{X}^{x,t'}_{s,D,\epsilon}(\omega)\}_{s\in[t',t]}, \{\boldsymbol{\alpha}_\epsilon(s)\}_{s\in[t',t]}\right) + \int_{t'}^{t} L^{\alpha(s)} u_\epsilon(s, X^{x,0}_{s,D,\epsilon}(\omega)) ds \right] \right)_\epsilon.$$

(1.9)

Thus we obtain:

$$\left( \inf_{\alpha_\epsilon} \mathbf{E}\left[ \bar{\mathbf{J}}_\epsilon\left(t,t', \{\mathbf{X}^{x,0}_{s,D,\epsilon}(\omega)\}_{s\in[t,t']}, \{\boldsymbol{\alpha}_\epsilon(s)\}_{s\in[t,t']}\right) + \int_{t'}^{t} L^{\alpha(s)} u_\epsilon(s, \mathbf{X}^{x,0}_{s,D,\epsilon}(\omega)) ds \right] \right)_\epsilon$$

$$= 0.$$

(1.10)

By multiplying the equation Eq.(1.10) on the quantity $(t' - t)^{-1}$ and "proceeding formally" we obtain the Colombeau-Bellman equation:

$$\left( \inf_{\alpha_\epsilon} [\overline{\Delta\mathbf{J}}_\epsilon(t,x,\alpha_\epsilon) + L^{\alpha_\epsilon}_\epsilon(t,x)u_\epsilon(t,x)] \right)_\epsilon = 0,$$

$$\left( \inf_{\alpha_\epsilon} \left[ \inf_{\alpha_\epsilon} \overline{\Delta\mathbf{J}}_\epsilon(t,x,\alpha_\epsilon) + \frac{\partial u_\epsilon(t,x)}{\partial t} + \sum_{i=1}^{n} g^i_\epsilon(t,x,\alpha_\epsilon) \frac{\partial u_\epsilon(t,x)}{\partial x_i} + \right.\right.$$

$$\left.\left. +D^2 \sum_{i,j=1}^{n} \sigma^{ij}_\epsilon(t,x,\alpha_\epsilon) \frac{\partial^2 u_\epsilon(t,x)}{\partial x_i \partial x_j} \right] \right)_\epsilon = 0.$$

(1.11)

Many "quasi stochastic" optimal control problems essentially come down to constructing Colombeau generalized function $(u_\epsilon(t,x))_\varepsilon$ that has the properties:

$$(u_\epsilon(t,x))_\epsilon =$$

$$= \left( \inf_{\{\alpha_\epsilon(s)\}_{s\in[0,t]}\in U, U\subsetneq \mathbb{R}^n} \mathbf{E}\left[\bar{\mathbf{J}}_\epsilon\left(t, T, \{\mathbf{X}^{x,0}_{s,\varepsilon,\epsilon}(\omega)\}_{s\in[t,T]}, \{\alpha_\epsilon(s)\}_{s\in[t,T]}\right)\right]\right)_\epsilon, \quad (1.12)$$

$$u(t,x) =$$

$$= \inf_{\{\alpha(s)\}_{s\in[t,t']}\in U, U\subsetneq \mathbb{R}^n} \mathbf{E}\left[\bar{\mathbf{J}}_\epsilon\left(t, t'\{\mathbf{X}^{x,0}_{S,\varepsilon}(\omega)\}_{s\in[t,t']}, \{\boldsymbol{\alpha}(s)\}_{s\in[t,t']}\right) + u(t', \mathbf{X}^{x,0}_{t',\varepsilon,\epsilon}(\omega))\right],$$

where $(\bar{\mathbf{J}}_\varepsilon(\circ,\circ,\circ,\circ))_\epsilon$ is the termination payoff functional, $(\alpha_\epsilon(t))_\epsilon = ((\alpha_{\epsilon,1}(t))_\epsilon, \ldots, (\alpha_{\epsilon,m}(t))_\epsilon)$ is a control and $((X^x_{t,\varepsilon,\epsilon}))_{\epsilon,t\geq 0}$ is some Colombeau-Markov process governed by some stochastic equation driven by a "small" white noise, of the form:

$$(\mathbf{X}^x_{t,\varepsilon,\epsilon}(\omega))_\epsilon = (x_\epsilon)_\epsilon + \left(\int_0^t g_\epsilon(\mathbf{X}^x_{s,\varepsilon,\epsilon}(\omega), \alpha_\epsilon(s))ds\right)_\epsilon + \sqrt{\varepsilon}\mathbf{W}(t,\omega), \quad (1.13)$$

$$\varepsilon \ll 1.$$

Traditionally, the function $u(t,x)$ has been computed by way of solving the associated *Bellman equation*, for which various numerical techniques – mostly variations of the finite difference scheme – have been developed.

**A new approach, which is proposed in this paper allows one to construct the Bellman function $u(t,x)$ and optimal control $\alpha^*(t) = (\alpha_1^*(t), \ldots, \alpha_m^*(t))$ directly, i.e., without any reference to the Bellman equation, by way of using strong large deviations principle [19] for infinitesimal stochastic differential game is formulated in section III, Theorem 3.2.1.**

# I.1. Description of the classical deterministic problem.

We consider in the interval $[0, T]$ a dynamic system which evolves according to the ordinary differential equation (ODE)

$$\frac{dy(s)}{ds} = g(s, y(s), \alpha(s)), \qquad (1.1.1)$$

$$0 \leq t \leq s \leq T.$$

with initial condition

$$y(t) \in \mathbb{R}^n. \qquad (1.1.2)$$

The set of controls is denoted genetically by $\mathbf{A}(t,s)$

$$\mathbf{A}(t,s) = \{\alpha : [t,s] \to \mathbf{A} \subset \mathbb{R}^m | \alpha(\cdot) \text{ measurable}\}. \qquad (1.1.3)$$

The optimal control problem consists in minimizing the functional

$$\mathbf{J}(t, x, \alpha(\cdot)),$$

where

$$\mathbf{J}(\cdot, \cdot, \cdot) : [0, T] \times \mathbb{R}^m \times \mathbf{A}(t, T) \to \mathbb{R},$$

$$\mathbf{J}(t, x, \alpha(\cdot)) = \operatorname*{ess\,sup}_{s \in [t,T]} f(s, y(s), \alpha(s)) \qquad (1.1.4)$$

Let us define the optimal cost :

$$\mathbf{V} : [0, T] \times \mathbb{R}^m \to \mathbb{R},$$

(1.1.5)

$$\mathbf{V}(t,x) = \inf\{\mathbf{J}(t,x,\alpha(\cdot)) | \alpha(\cdot) \in \mathbf{A}(t,T)\}$$

The value function $V(t,x)$ satisfies the following dynamic programming principle:

$$\forall t(t \in [0,T]), x \in \mathbb{R}^n :$$

$$V(t,x) = \inf_{\alpha(\cdot) \in \mathbf{A}[t,s]} \left\{ \max\left( \left\{ \operatorname*{ess\,sup}_{\tau \in [t,s]} f(\tau, y(\tau), \alpha(\tau)) \right\}; \{V(s, y(s))\} \right) \right\},$$

(1.1.6)

$$V(T,x) = \min_{\alpha \in A} f(T,x,\alpha).$$

## I.1.1. The Hamilton-Jacobi-Bellman equation.

The aim of this section is to arrive to an equation of HJB type associated to the optimal cost of the original minimax problem.

**Definition** *1.1.1. To simplify notation, we will define*

$$\mathbf{J}(t,s,x,\alpha(\cdot)) \triangleq \operatorname{ess\,sup}\{f((\tau, y(\tau), \alpha(\tau))) | \tau \in (t,s)\}.$$

(1.1.7)

We start from the basic dynamical programming equation (1.1.6)

$$V(t,x) =$$

(1.1.8)

$$= \min_{\alpha(\cdot) \in \mathbf{A}(t,t+\delta)} \{\max(\{\mathbf{J}(t,t+\delta, x, \alpha(\cdot))\}; \{V(t+\delta, y(t+\delta))\})\}$$

and "proceeding formally" by standard way we get the **HJB** equation:

$$\min_{\alpha \in A} \left\{ \max\left[ f(t,x,\alpha) - V(t,x)\frac{\partial V(t,x)}{\partial x} + \frac{\partial V(t,x)}{\partial x} g(t,x,\alpha) \right] \right\} = 0. \quad (1.1.9)$$

## I.2. Notation and basic notions from Colombeau theory.

We use [43],[44] as standard references for the foundations and various applications of standard Colombeau theory. We briefly recall the basic Colombeau construction.

Throughout the paper $\Omega$ will denote an open subset of $\mathbb{R}^n$. Stanfard Colombeau generalized functions on $\Omega$ are defined as equivalence classes $u = [(u_\varepsilon)_\varepsilon]$ of nets of smooth functions $u_\varepsilon \in C^\infty(\Omega)$ (regularizations) subjected to asymptotic norm conditions with respect to $\varepsilon \in (0,1]$ for their derivatives on compact sets. The basic idea of *classical Colombeau's theory of nonlinear generalized functions* [43],[44] is regularization by sequences (nets) of smooth functions and the use of asymptotic estimates in terms of a regularization parameter $\varepsilon$.

Let $(u_\varepsilon)_{\varepsilon \in (0,1]}$ with $(u_\varepsilon)_\varepsilon \in C^\infty(M)$ for all $\varepsilon \in \mathbb{R}_+$, where $M$ a separable, smooth orientable Hausdorff manifold of dimension $n$.

**Definition** *1.2.1. The classical Colombeau's algebra of generalized functions on $M$ is defined as the quotient:*

$$\mathcal{G}(M) \triangleq \mathcal{E}_M(M)/\mathcal{N}(M) \quad (1.2.1)$$

*of the space $\mathcal{E}_M(M)$ of sequences of moderate growth modulo the space $\mathcal{N}(M)$ of negligible sequences. More precisely the notions of moderateness resp. negligibility are defined by the following asymptotic estimates (where $\mathfrak{X}(M)$ denoting the space of smooth vector fields on $M$):*

$$\mathcal{E}_M(M) \triangleq \{(u_\varepsilon)_\varepsilon |\ \forall K(K \subsetneq M)\, \forall k(k \in \mathbb{N})\, \exists N(N \in \mathbb{N})$$

$$\forall \xi_1,\ldots,\xi_k(\xi_1,\ldots,\xi_k \in \mathfrak{X}(M))\left[\sup_{p \in K} |L_{\xi_1}\ldots L_{\xi_k} u_\varepsilon(p)| = O(\varepsilon^{-N}) \text{ as } \varepsilon \to 0\right]\}, \quad (1.2.2)$$

$$\mathcal{N}(M) \triangleq \{(u_\varepsilon)_\varepsilon |\ \forall K(K \subsetneq M),\, \forall k(k \in \mathbb{N}_0)\, \forall q(q \in \mathbb{N})$$

$$\forall \xi_1,\ldots,\xi_k(\xi_1,\ldots,\xi_k \in \mathfrak{X}(M))\left[\sup_{p \in K} |L_{\xi_1}\ldots L_{\xi_k} u_\varepsilon(p)| = O(\varepsilon^q) \text{ as } \varepsilon \to 0\right]\}. \quad (1.2.3)$$

In particular the classical Colombeau's algebra of generalized functions on $M = \mathbb{R}^d$ is defined as the quotient:

$$\mathcal{G}(\mathbb{R}^d) \triangleq \mathcal{E}_{\mathbb{R}^d}(\mathbb{R}^d)/\mathcal{N}(\mathbb{R}^d) \quad (1.2.1')$$

of the space $\mathcal{E}_{\mathbb{R}^d}(\mathbb{R}^d)$ of sequences of moderate growth modulo the space $\mathcal{N}(\mathbb{R}^d)$ of negligible sequences. More precisely the notions of moderateness resp. negligibility are defined by the following asymptotic estimates (where $\mathfrak{X}(M)$ denoting the space of smooth vector fields on $\mathbb{R}^d$):

$$\mathcal{E}_{\mathbb{R}^d}(\mathbb{R}^d) \triangleq \{(u_\varepsilon)_\varepsilon |\ \forall K(K \subsetneq \mathbb{R}^d)\, \forall k(k \in \mathbb{N})\, \exists N(N \in \mathbb{N})$$

$$\forall \xi_1,\ldots,\xi_k(\xi_1,\ldots,\xi_k \in \mathfrak{X}(\mathbb{R}^d))\left[\sup_{p \in K} |L_{\xi_1}\ldots L_{\xi_k} u_\varepsilon(p)| = O(\varepsilon^{-N}) \text{ as } \varepsilon \to 0\right]\}, \quad (1.2.2')$$

$$\mathcal{N}(M) \triangleq \{(u_\varepsilon)_\varepsilon | \ \forall K(K \subsetneq \mathbb{R}^d), \forall k(k \in \mathbb{N}_0) \forall q(q \in N)$$

$$\forall \xi_1,\ldots,\xi_k(\xi_1,\ldots,\xi_k \in \mathfrak{X}(\mathbb{R}^d)) \left[ \sup_{p \in K} |L_{\xi_1}\ldots L_{\xi_k} u_\varepsilon(p)| = O(\varepsilon^q) \text{ as } \varepsilon \to 0 \right] \}. \quad (1.2.3')$$

**Definition** *1.2.2. In the definition above the Landau symbol $a_\varepsilon = O(\psi(\varepsilon))$ appears, having the following meaning:*

$$\exists C(C > 0) \exists \varepsilon_0(\varepsilon_0 \in (0,1]) \forall \varepsilon(\varepsilon < \varepsilon_0)[a_\varepsilon \leq C\psi(\varepsilon)].$$

**Definition** *1.2.3. Elements of $\mathcal{G}(M)$ are denoted by:*

$$u = \mathbf{cl}[(u_\varepsilon)_\varepsilon] \triangleq (u_\varepsilon)_\varepsilon + \mathcal{N}(M). \quad (1.2.4)$$

**Remark** *1.2.1. In particular lements of $\mathcal{G}(\mathbb{R}^d)$ are denoted by:*

$$u = \mathbf{cl}[(u_\varepsilon)_\varepsilon] \triangleq (u_\varepsilon)_\varepsilon + \mathcal{N}(M). \quad (1.2.4')$$

**Remark** *1.2.2. With componentwise operations $(\cdot, \pm)\mathcal{G}(M)$ is a fine sheaf of differential algebras with respect to the Lie derivative defined by*

$$L_\xi u \triangleq \mathbf{cl}[(L_\xi u_\varepsilon)_\varepsilon].$$

The spaces of moderate resp.negligible sequences and hence the algebra itself may be characterized locally, i.e., $u \in \mathcal{G}(M)$ iff $u \circ \psi_\alpha \in \mathcal{G}(\psi_\alpha(V_\alpha))$ for all charts $(V_\alpha, \psi_\alpha)$, where on the open set $\psi_\alpha(V_\alpha) \subset \mathbb{R}^n$ in the respective estimates Lie derivatives are replaced by partial derivatives.

**Remark** *1.2.3. Smooth functions $f \in C^\infty(M)$ are embedded into $\mathcal{G}(M)$ simply by the "constant" embedding $\sigma$, i.e., $\sigma(f) = cl[(f)_\varepsilon]$, hence $C^\infty(M)$ is a faithful subalgebra of $G(M)$.*

# I.2.1. Point Values of a Generalized Functions on $M$. Colombeau's Generalized Numbers.

Within the classical distribution theory, distributions cannot be characterized by their point values in any way similar to classical functions. On the other hand, there is a very natural and direct way of obtaining the point values of the elements of Colombeau's algebra: points are simply inserted into representatives.

The objects so obtained are sequences of numbers, and as such are not the elements in the field $\mathbb{R}$ or $\mathbb{C}$. Instead, they are the representatives of *Colombeau's generalized numbers.*

We give the exact definition of these "numbers".

**Definition** *1.2.4. Inserting $p \in M$ into $u \in \mathcal{G}(M)$ yields a well defined element of the ring of constants (also called generalized numbers) $K$ (corresponding to $K = R$ resp. $C$), defined as the set of moderate nets of numbers $((r_\varepsilon)_\varepsilon \in K^{(0,1]}$ with $|r_\varepsilon| = O(\varepsilon^{-N})$ for some $N$) modulo negligible nets ($|r_\varepsilon| = O(\varepsilon^m)$ for each $m$); componentwise insertion of points of $M$ into elements of $G(M)$*

*yields well-defined generalized numbers, i.e., elements of the ring of constants:*

$$\mathcal{K} = \mathcal{E}_{\mathbf{c}}(M)/\mathcal{N}_{\mathbf{c}}(M) \tag{1.2.5}$$

*(with $\mathcal{K} = \widetilde{\mathbb{R}}$ or $\mathcal{K} = \widetilde{\mathbb{C}}$ for $\mathbf{K} = \mathbb{R}$ or $\mathbf{K} = \mathbb{C}$), where*

$$\mathcal{E}_{\mathbf{c}}(M) = \left\{(r_\varepsilon)_\varepsilon \in \mathbf{K}^I | \exists n(n \in \mathbb{N})\bigl[|r_\varepsilon| = O(\varepsilon^{-n}) \text{ as } \varepsilon \to 0\bigr]\right\},$$

$$\mathcal{N}_{\mathbf{c}}(M) = \left\{(r_\varepsilon)_\varepsilon \in \mathbf{K}^I | \forall m(m \in \mathbb{N})\bigl[|r_\varepsilon| = O(\varepsilon^{m}) \text{ as } \varepsilon \to 0\bigr]\right\}, \qquad (1.2.6)$$

$$I = (0,1].$$

Generalized functions on $M$ are characterized by their generalized point values, i.e., by their values on points in $\widetilde{M}_c$, the space of equivalence classes of compactly supported nets $(p_\varepsilon)_\varepsilon \in M^{(0,1]}$ with respect to the relation $p_\varepsilon \sim p'_\varepsilon :\Leftrightarrow d_h(p_\varepsilon, p'_\varepsilon) = O(\varepsilon^m)$ for all $m$, where $d_h$ denotes the distance on $M$ induced by any Riemannian metric.

**Definition** 1.2.5. *For $u \in \mathcal{G}(M)$ and $x_0 \in M$, the point value of $u$ at the point $x_0, u(x_0)$, is defined as the class of $(u_\varepsilon(x_0))_\varepsilon$ in $\mathcal{K}$.*

**Definition** 1.2.6. *We say that an element $r \in \mathcal{K}$ is strictly nonzero if there exists a representative $(r_\varepsilon)_\varepsilon$ and a $q \in \mathbb{N}$ such that $|r_\varepsilon| \geq \varepsilon^q$ for $\varepsilon$ sufficiently small. If $r$ is strictly nonzero, then it is also invertible with the inverse $[(1/r_\varepsilon)_\varepsilon]$. The converse is true as well.*

# I.3. Colombeau Stochastic Differential Equations (CSDE)

One of the fundamental concepts in stochastic differential equations is the white noise process standing for the noisy term in an equation. Here we are interested in nonlinear SDE's with a additive generalized stochastic process. We are interested in considering equations with such a noise but now viewed as a Colombeau generalized process.

We consider as Brownian motion **B** is the $d$-dimensional coordinate process on the classical Wiener space $(\Omega, \mathcal{F}, \mathbf{P})$ i.e., is the set of continuous functions from $[0,T]$ to $\mathbb{R}^d$ starting from 0 ($\Omega = C_0([0,T]; \mathbb{R}^d)$), $\mathcal{F}$ the completed Borel $\sigma$-algebra over $\Omega$, **P** the Wiener measure and **B** the canonical process:

$$B_s(\omega) = \omega_s, s \in [0,T], \omega \in \Omega.$$

By $\{\mathcal{F}_s | 0 \leq s \leq T\}$ we denote the natural filtration generated by $\{B_s\}_{0 \leq s \leq T}$ and augmented by all **P**-null sets, i.e.,

$$\mathcal{F}_s = \{B_r | r \leq s\} \vee N_P, s \in [0, T], \quad (1.3.1)$$

where $N_P$ is the set of all **P**-null subsets, and $T > 0$ a fixed real time horizon.

For any $n \geq 1$, $\|z\|$ denotes the Euclidean norm of $z \in \mathbb{R}^n$.

**Definition** 1.3.1. *Let $(\Omega, \Sigma, \mu)$ be a probability space. A generalized stochastic process on $\mathbb{R}^d$ is a weakly measurable mapping $X(\omega) : \Omega \to \mathcal{D}(\mathbb{R}^d)$. We denote by $\mathcal{D}_\Omega'(\mathbb{R}^d)$ the space of generalized stochastic processes.*
*For each fixed function $\varphi \in \mathcal{D}(\mathbb{R}^d)$, the mapping $\Omega \to \mathbb{R}$ defined by $\omega \mapsto \langle X(\omega), \varphi \rangle$ is a random variable.*

**Definition** 1.3.2. *White noise $\dot{W}$ on $\mathbb{R}^d$ can be constructed as follows. We take as probability space the space of tempered distributions $\Omega = S'(\mathbb{R}^d)$ with $\Sigma$ the Borel $\sigma$-algebra generated by the weak topology. By the Bochner-Minlos theorem, there is a unique probability measure $\mu$ on $\Omega$ such that:*

$$\int exp[i\langle \omega, \varphi \rangle] d\mu(\omega) = exp\left(-\frac{1}{2} \|\varphi\|^2_{L^2(\mathbb{R}^d)}\right) \quad (1.3.2)$$

*for $\varphi \in S(\mathbb{R}^d)$. The white noise process $\dot{W}$ is defined as the identity mapping*

$$\dot{W}(\omega) : \Omega \to S'(\mathbb{R}^d), \langle \dot{W}(\omega), \varphi \rangle = \langle \omega, \varphi \rangle \quad (1.3.3)$$

*for $\varphi \in \mathcal{D}(\mathbb{R}^d)$. $\dot{W}(\omega)$ is a generalized Gaussian process with mean zero and variance*

$$V(\dot{W}(\varphi)) = \mathbf{E}\left[\dot{W}^2(\varphi)\right] = \|\varphi\|^2_{L^2(\mathbb{R}^d)} \quad (1.3.4)$$

*Its covariance is the bilinear functional*

$$\mathbf{E}[\dot{W}(\varphi_1)\dot{W}(\varphi_2)] = \int_{\mathbb{R}^d} \varphi_1(x)\varphi_2(x)dx \qquad (1.3.5)$$

*represented by the Dirac measure on the diagonal $\mathbb{R}^d \times \mathbb{R}^d$, showing the singular nature of white noise. A net $\varphi_\varepsilon$ of mollifiers given by*

$$\varphi_\varepsilon(y) = \frac{1}{\varepsilon^d}\varphi\left(\frac{y}{\varepsilon}\right),$$

$$\varphi \in \mathcal{D}(\mathbb{R}^d), \varphi \geq 0, \qquad (1.3.6)$$

$$\int_{\mathbb{R}^d} \varphi(y)dy = 1,$$

*is called a nonnegative model delta net. Smoothed white noise process on $\mathbb{R}^d$ is defined as*

$$\dot{W}_\varepsilon(x) = \langle \dot{W}_\varepsilon(y), \varphi_\varepsilon(x-y)\rangle, \qquad (1.3.7)$$

*where $\dot{W}$ is white noise on $\mathbb{R}^d$ and $\varphi_\varepsilon$ is a nonnegative model delta net. It follows from (1.3.5) that the covariance of smoothed white noise is*

$$\mathbf{E}[\dot{W}_\varepsilon(x)\dot{W}_\varepsilon(y)] = \int_{\mathbb{R}^d} \varphi_\varepsilon(x-z)\varphi_\varepsilon(y-z)dz = [\varphi_\varepsilon * \tilde{\varphi}_\varepsilon](x-y) \qquad (1.3.8)$$

*where $\tilde{\varphi}(z) = -\varphi(z)$.*

We now introduce by standard way Colombeau generalized stochastic processes in the $d$-dimensional case.

**Notation** 1.3.1. Denote by $\mathcal{E}(\mathbb{R}^d)$ the space of nets $(X_\varepsilon(t,\omega))_\varepsilon, \varepsilon \in (0,1]$ of processes $X_\varepsilon(t,\omega)$ with almost surely continuous paths, i.e., the space of nets of processes

$$X_\varepsilon(t,\omega) : (0,1] \times \mathbb{R}^d \times \Omega \to \mathbb{R} \tag{1.3.9}$$

such that

$(t,x) \mapsto X_\varepsilon(t,\omega)$ is jointly measurable, for all $\varepsilon \in (0,1]$,

$t \mapsto X_\varepsilon(t,\omega)$ belongs to $C^\infty(\mathbb{R}^d)$ for all $\varepsilon \in (0,1]$ and  (1.3.10)

almost all $\omega \in \Omega$.

**Definition** 1.3.3. $\mathcal{E}_M^\Omega(\mathbb{R}^d)$ is the space of nets of processes $(X_\varepsilon(t,\omega))_\varepsilon$ belonging to $\mathcal{E}(\mathbb{R}^d), \varepsilon \in (0,1]$, with the property that for almost all $\omega \in \Omega$, for all $T > 0$ and $\alpha \in \mathbb{N}_0$, there exist constants $N, C, > 0$ and $\varepsilon \in (0,1]$ such that:

$$\sup_{t \in [0,T]} |\partial^\alpha X_\varepsilon(t,\omega)| < C\varepsilon^{-N}, \varepsilon \leq \varepsilon_0. \tag{1.3.11}$$

$\mathcal{N}^\Omega(\mathbb{R}^d)$ the is the space of nets of processes $(X_\varepsilon(t,\omega))_\varepsilon \in \mathcal{E}(\mathbb{R}^d), \varepsilon \in (0,1]$ with the property that for almost all $\omega \in \Omega$, for all $T > 0$ and $\alpha \in \mathbb{N}_0$ and all $b \in \mathbb{R}$, there exist constants $C > 0$ and $\varepsilon \in (0,1]$ such that:

$$\sup_{t \in [0,T]} |\partial^\alpha X_\varepsilon(t,\omega)| < C\varepsilon^b, \varepsilon \leq \varepsilon_0. \tag{1.3.12}$$

**Definition** 1.3.4. The differential algebra of Colombeau generalized stochastic processes is the factor algebra

$$\mathcal{G}^{\Omega}(\mathbb{R}^d) = \mathcal{E}_M^{\Omega}(\mathbb{R}^d)/\mathcal{N}^{\Omega}(\mathbb{R}^d) \qquad (1.3.13)$$

The elements of $\mathcal{G}^{\Omega}(\mathbb{R}^d)$ will be denoted by $\mathbf{X} = \mathbf{cl}[(X_\varepsilon)_\varepsilon]$, where $(X_\varepsilon)_\varepsilon$ is a representative of the class.

White noise can be viewed as a Colombeau generalized stochastic processes having a representative given by (1.3.7). This follows from the usual imbedding arguments of Colombeau theory, since its paths are distributions. For evaluation of generalized stochastic process at fixed points of time, we introduce the concept of a Colombeau generalized random variable as follows. Let $\mathcal{E}[\mathbb{R}^d]$ be the space of nets of measurable functions on $\Omega$.

**Definition** 1.3.5. $\mathcal{E}[\mathbb{R}_M^d]$ *is the space of nets* $(X_\varepsilon)_\varepsilon, \varepsilon \in (0,1]$ *with the property that for almost all* $\omega \in \Omega$ *there exist constants* $N, C > 0$, *and* $\varepsilon_0 \in (0,1]$ *such that*

$$|X_\varepsilon(t,\omega)| < C\varepsilon^{-N}, \varepsilon \leq \varepsilon_0. \qquad (1.3.14)$$

**Definition** 1.3.6. $\mathcal{N}[\mathbb{R}^d]$ *is the space of nets* $(X_\varepsilon)_\varepsilon, \varepsilon \in (0,1]$, *with the property that for almost all* $\omega \in \Omega$ *and all* $b \in \mathbb{R}$, *there exist constants* $C > 0$ *and* $\varepsilon_0 \in (0,1]$ *such that*

$$|X_\varepsilon(t,\omega)| < C\varepsilon^b, \varepsilon \leq \varepsilon_0. \qquad (1.3.15)$$

**Definition** 1.3.7. *The differential algebra* $\mathcal{G}[\mathbb{R}^d]$ *of Colombeau generalized random variables is the factor algebra*

$$\mathcal{E}[\mathbb{R}_M^d]/\mathcal{N}[\mathbb{R}^d].$$

We also shall introduce the following both spaces of Colombeau generalized random processes which will be used frequently in the sequel:

**Definition** 1.3.8.

$$\mathbf{S}^2(0,T;\widetilde{\mathbb{R}}) \triangleq$$

$$\triangleq \left\{ (\psi_{s,\varepsilon})_{\varepsilon;0\leq s\leq T} \text{ is a } \widetilde{\mathbb{R}}\text{-valued adapted cádlág generalized process}: \right. \quad (1.3.16)$$

$$\left. \forall \varepsilon(0<\varepsilon) \mathbf{E}\left[\sup_{0\leq s\leq T}|\psi_{s,\varepsilon}|^2\right] < +\infty \right\}, \varepsilon \in (0,1]$$

$$\mathbf{H}^2(0,T;\widetilde{\mathbb{R}}^n) \triangleq$$

$$\left\{ (\psi_{s,\varepsilon})_{\varepsilon;0\leq s\leq T} \text{ is a } \widetilde{\mathbb{R}}^n\text{-valued progressively measurable generalized process}: \right.$$

$$\|(\psi_{s,\varepsilon})_\varepsilon\|_2^2 = \left(\mathbf{E}\left[\int_0^T \|\psi_{s,\varepsilon}\|^2\right]\right)_\varepsilon \wedge \quad (1.3.17)$$

$$\left. \wedge \forall \varepsilon(0<\varepsilon)\left[\left(\mathbf{E}\left[\int_0^T \|\psi_{s,\varepsilon}\|^2\right] < +\infty\right) \wedge (\varepsilon \in (0,1])\right]\right\}.$$

## I.4. Description of the Colombeau deterministic problem.

Let us consider in the interval $[0,T]$ a dynamic system which evolves according to the Colombeau ordinary differential equation (CODE)

$$\frac{d(y_\varepsilon(s))_\varepsilon}{ds} = (g_\varepsilon(s,y_\varepsilon(s),\alpha_\varepsilon(s)))_\varepsilon, 0 \leq t \leq s \leq T,$$
(1.4.1)

$$(g_\varepsilon(s,y_\varepsilon(s),\alpha_\varepsilon(s)))_\varepsilon \in \mathcal{G}(\mathbb{R}^d)$$

with initial condition

$$(y_\varepsilon(t))_\varepsilon \in \mathbb{R}^n. \quad (1.4.2)$$

The set of controls is denoted genetically by $\mathbf{A}_\varepsilon(t,s), \varepsilon \in (0,1]$

$$\mathbf{A}_\varepsilon(t,s) = \{\alpha_\varepsilon : [t,s] \to \mathbf{A} \subset \mathbb{R}^m | \alpha_\varepsilon(\cdot) \text{ measurable}\}. \quad (1.4.3)$$

The optimal control problem consists in minimizing the functional
$(\mathbf{J}_\varepsilon(t,x,\alpha_\varepsilon(\cdot)))_\varepsilon,$ where

$$(\mathbf{J}_\varepsilon(\cdot,\cdot,\cdot))_\varepsilon : [0,T] \times \widetilde{\mathbb{R}}^m \times (\mathbf{A}_\varepsilon(t,T))_\varepsilon \to \widetilde{\mathbb{R}},$$

$$(\mathbf{J}_\varepsilon(t,x,\alpha_\varepsilon(\cdot)))_\varepsilon = \left(\operatorname*{ess\,sup}_{s \in [t,T]} f_\varepsilon(s, y_\varepsilon(s), \alpha_\varepsilon(s))\right)_\varepsilon \quad (1.4.4)$$

Let us define the Colombeau optimal cost :

$$(\mathbf{V}_\varepsilon(t,x))_\varepsilon : [0,T] \times \widetilde{\mathbb{R}}^m \to \widetilde{\mathbb{R}},$$

$$(\mathbf{V}_\varepsilon(t,x))_\varepsilon = (\inf\{\mathbf{J}_\varepsilon(t,x,\alpha_\varepsilon(\cdot)) | \alpha_\varepsilon(\cdot) \in \mathbf{A}_\varepsilon(t,T)\})_\varepsilon \quad (1.4.5)$$

The Colombeau value function $(\mathbf{V}_\varepsilon(t,x))_\varepsilon$ satisfies the following dynamic programming principle:

$$\forall t(t \in [0,T]), x \in \mathbb{R}^n :$$

$$(\mathbf{V}_\varepsilon(t,x))_\varepsilon =$$

$$\left( \inf_{\alpha_\varepsilon(\cdot) \in \mathbf{A}_\varepsilon[t,s]} \left\{ \max\left( \left\{ \mathbf{ess}\sup_{\tau \in [t,s]} f_\varepsilon(\tau, y_\varepsilon(\tau), \alpha_\varepsilon(\tau)) \right\}; \{\mathbf{V}_\varepsilon(s, y(s))\} \right) \right\} \right)_\varepsilon, \quad (1.4.6)$$

$$(\mathbf{V}_\varepsilon(T,x))_\varepsilon = \left( \min_{\alpha_\varepsilon \in A_\varepsilon} f_\varepsilon(T, x, \alpha) \right)_\varepsilon.$$

## I.4.1. The Colombeau Hamilton-Jacobi-Bellman equation.

The aim of this section is to arrive to an equation of CHJB type associated to the optimal cost of the generalized minimax problem.
**Definition** *1.4.1. To simplify notation, we will define*

$$(\mathbf{J}_\varepsilon(t,s,x,\alpha_\varepsilon(\cdot)))_\varepsilon \triangleq (\mathbf{ess}\sup\{f_\varepsilon((\tau, y_\varepsilon(\tau), \alpha_\varepsilon(\tau)))| \tau \in (t,s)\})_\varepsilon. \quad (1.4.7)$$

We start from the basic dynamical programming equation (1.4.6)

$$(\mathbf{V}_\varepsilon(t,x))_\varepsilon =$$

$$= \left( \min_{\alpha_\varepsilon(\cdot) \in \mathbf{A}_\varepsilon(t,t+\delta)} \{\max(\{\mathbf{J}_\varepsilon(t, t+\delta, x, \alpha_\varepsilon(\cdot))\}; \{\mathbf{V}_\varepsilon(t+\delta, y_\varepsilon(t+\delta))\})\} \right)_\varepsilon \quad (1.4.8)$$

and "proceeding formally" by standard way we get the **CHJB** equation:

$$\left(\min_{\alpha_\varepsilon \in A_\varepsilon} \left\{\max\left[f_\varepsilon(t,x,\alpha) - \mathbf{V}_\varepsilon(t,x)\frac{\partial \mathbf{V}_\varepsilon(t,x)}{\partial x} + \frac{\partial \mathbf{V}_\varepsilon(t,x)}{\partial x} g_\varepsilon(t,x,\alpha)\right]\right\}\right)_\varepsilon = 0 \quad (1.4.9)$$

## II. m-Persons antagonistic dissipative differential game imbeded into a 'small' white noise.

Let us consider an *m*-persons differential game $DG_{m;T}(f,g,y)$, with nonlinear dynamics

$$\frac{dx_i(t)}{dt} = f_i(x_1,\ldots,x_m;\alpha_1,\ldots,\alpha_n),$$

$$x(0) = x_0 \in \mathbb{R}^n,$$

$$\alpha_i(t) \in U_i, i = 1,\ldots,n.$$

$$t \in [0,T].$$

(2.1)

Here $t \to \alpha_i(t)$ is the control chosen by the *i*-th player, within a set of admissible control values $U_i$, and the payoff for the *i*-th player:

$$\mathbf{J}_i = \int_0^T g_i(x_1,\ldots,x_m;\alpha_1,\ldots,\alpha_n)dt + \sum_{i=1}^m [x_i(T) - y_{1,i}]^2. \quad (2.2)$$

where $t \mapsto x(t)$ is the trajectory of (2.1). Optimal control problem for the *i*-th player:

$$\min_{\alpha_i(t)} \left(\max_{\alpha_j(t), j \neq i} \mathbf{J}_i\right). \quad (2.3)$$

**Definition** *2.1. Let be a scalar function $V : \mathbb{R}^n \to \mathbb{R}$. V is a Lyapunov-candidate-function if it is a locally positive-definite* function, i.e. (i) $V(0) = 0$, (ii) $V(x) > 0, \forall x, x \in U\setminus\{0\}$, with $U$ being a neighborhood region around $x = 0$.

**Definition** 2.2. Let $\dot{x}_i = f_i(x_1,\ldots,x_m; \alpha_1,\ldots,\alpha_n), i = 1,\ldots,m.$

$$\dot{V}(x;f) \triangleq \sum_{i=1}^{m} \frac{\partial V(x)}{\partial x_i} f(x). \quad (2.4)$$

**Definition** 2.3. Differential game $DG_m(\mathbf{f}, \mathbf{0}, \mathbf{y})$ is dissipative iff exist Lyapunov-candidate-function $V(x)$ and constants $C > 0, R \geq 0$ such that

$$\dot{V}(x;f) \leq -CV(x), \|x\| \geq R,$$

$$V_r(x) = \lim_{r\to\infty}\left(\inf_{\|x\|>r} V(x)\right) = \infty. \quad (2.5)$$

**Definition** 2.4. Let us consider an $m$-persons stochastic differential game $SDG_{m;T}(\mathbf{f}, \mathbf{g}, \mathbf{y})$, with nonlinear dynamics:

$$\frac{dx_i(t)}{dt} = f_i(x_1,\ldots,x_m; \alpha_1,\ldots,\alpha_m) + \sqrt{\varepsilon}\,\dot{W}(t),$$

$$x(0) = x_0 \in \mathbb{R}^n,$$

$$\alpha_i(t) \in U_i, i = 1,\ldots,m.$$

$$t \in [0,T], \varepsilon \ll 1. \quad (2.6)$$

Here $t \to \alpha_i(t)$ is the determined control chosen by the $i$-th player, within a set of admissible control values $U_i$, and the payoff for the $i$-th player:

$$J_i = \mathbf{E}\left[\int_0^T g_i(x_1(t,\omega),\ldots,x_m(t,\omega);\alpha_1(t),\ldots,\alpha_m(t))dt\right] +$$

(2.7)

$$+\mathbf{E}\left[\sum_{i=1}^m [x_i(T,\omega) - y_i]^2\right]$$

where $t \mapsto x(t,\omega)$ is the trajectory of (2.6).

**Definition** 2.5. Stochastic differential game $SDG_{m;T}(f,g,y)$ is the determined differential game $DG_{m;T}(f,g,y)$ imbeding into a 'small' white noise.

# III.1. Infinitesimal Reduction. Strong large deviations principle of Non-Freidlin-Wentzell type for infinitesimal stochastic differential game. "Step by step" strategy.

Let us consider now a family $X_t^\varepsilon$ of the solutions SDE:

$$d\mathbf{X}_t^\varepsilon = \mathbf{b}(\mathbf{X}_t^\varepsilon, t)dt + \sqrt{\varepsilon}\, d\mathbf{W}(t),$$

(3.1.1)

$$\mathbf{X}_0^\varepsilon = x_0 \in \mathbb{R}^n, t \in [0,T],$$

where $\mathbf{W}(t)$ is $n$-dimensional Brownian motion, $\mathbf{b}(\circ, t) : \mathbb{R}^n \times \mathbb{R} \to \mathbb{R}^n$ is a polinomial transform, i.e.

$$b_i(x,t) = \sum_{|\alpha|} b_\alpha^i(x,t) x^\alpha,$$

$$\alpha = (i_1, \ldots, i_k), |\alpha| = \sum_{j=0}^k i_j,$$

(3.1.2)

$$i = 1, \ldots, n$$

**Definition** *3.1.1. SDE (3.1.1) is dissipative if exist Lyapunov-candidate-function $V(x)$ and constants $C > 0, R \geq 0$ such that*

$$\dot{V}(x; \mathbf{b}) \leq -CV(x), \|x\| \geq R,$$

$$\tilde{V}(x) = \lim_{r \to \infty} \left( \inf_{\|x\| > r} V(x) \right) = \infty. \tag{3.1.3}$$

*Let us consider now a family $X_t^\varepsilon$ of the solutions dissipative SDE (3.1.1).*

**Theorem** *3.1.1.( **Strong large deviations principle**).[19] For the all solutions $\mathbf{X}_t^\varepsilon = (X_{1,t}^\varepsilon, \ldots, X_{n,t}^\varepsilon)$ dissipative SDE (3.1.1) and $\mathbb{R}$ valued parameters $\lambda_1, \ldots, \lambda_n, \lambda = (\lambda_1, \ldots, \lambda_n) \in \mathbb{R}^n$, there exists Colombeau constant $(C_\varepsilon)_\varepsilon \in \widetilde{\mathbb{R}}, (C_\varepsilon)_\varepsilon \geq 0$, such that:*

$$\liminf_{\varepsilon \to 0} \mathbf{E}\left[ \|\mathbf{X}_t^\varepsilon - \lambda\|^2 \right] \leq (C_\varepsilon)_\varepsilon \|\mathbf{U}(t, \lambda)\|^2$$

$$\lambda = (\lambda_1, \ldots, \lambda_n) \in \mathbb{R}^n \tag{3.1.4}$$

*where $\mathbf{U}(t, \lambda) = (U_1(t, \lambda), \ldots, U_n(t, \lambda))$ the solution of the linear differential master equation:*

$$\frac{d\mathbf{U}(t, \lambda)}{dt} = \mathbf{J}[\mathbf{b}(\lambda, t)]\mathbf{U} + \mathbf{b}(\lambda, t),$$

$$\mathbf{U}(0, \lambda) = x_0 - \lambda, \tag{3.1.5}$$

*where $\mathbf{J} = \mathbf{J}[\mathbf{b}(\lambda, t)]$ the Jacobian, i.e. $\mathbf{J}$ is a $n \times n$-matrix:*

$$\mathbf{J}[\mathbf{b}(\lambda,t)] = \mathbf{J}[\mathbf{b}(x,t)]|_{x=\lambda} =$$

$$= \begin{bmatrix} \dfrac{\partial b_1(x,t)}{\partial x_1} & \cdots & \dfrac{\partial b_1(x,t)}{\partial x_n} \\ \cdot & \cdots & \cdot \\ \cdot & \cdots & \cdot \\ \cdot & \cdots & \cdot \\ \dfrac{\partial b_n(x,t)}{\partial x_1} & \cdots & \dfrac{\partial b_n(x,t)}{\partial x_n} \end{bmatrix}_{x=\lambda} \quad (3.1.6)$$

**Corollary** 3.1.1. Assume the conditions of the Theorem 3.1 for any

$$\lambda = (\lambda_1,\ldots,\lambda_n) \in \mathbb{R}^n, t \in [0,T] :$$

$$\|\mathbf{U}(t,\lambda)\| = 0 \Rightarrow \liminf_{\varepsilon \to 0} \mathbf{E}\left[\|\mathbf{X}_t^\varepsilon - \lambda\|^2\right] = 0 \quad (3.1.7)$$

More precisely, for any $t \in [0,T]$ and
$\lambda = \lambda(t) = (\lambda_1(t),\ldots,\lambda_n(t)) \in \mathbb{R}^n$ sutch that

$$U_1(t,\lambda_1(t),\ldots,\lambda_n(t)) = 0,$$

$$\ldots\ldots\ldots\ldots\ldots\ldots\ldots \quad (3.1.8)$$

$$U_n(t,\lambda_1(t),\ldots,\lambda_n(t)) = 0,$$

the equalities is satisfaed

$$\liminf_{\varepsilon \to 0} \mathbf{E}\left[\|X_{1,t}^\varepsilon - \lambda_1(t)\|^2\right] = 0,$$

$$\ldots\ldots\ldots\ldots\ldots\ldots \quad (3.1.9)$$

$$\liminf_{\varepsilon \to 0} \mathbf{E}\left[\|X_{n,t}^\varepsilon - \lambda_n(t)\|^2\right] = 0.$$

**Definition**

# III.1.1. The solution master equation in comparison with a coresponding solution nonlinear ODE.

**Example** 3.1.

$$\dot{x} = -ax^3 - bx^2 - \sigma t^n - \varkappa \cdot t^m \sin(\Omega \cdot t) + \sqrt{\varepsilon}\, \dot{W}(t),$$

$$0 < a, \qquad (E.3.1.1)$$

$$x(0) = x_0.$$

**Corollary** *From general master equation 3.1.5 one obtain the next differential linear master equation*

$$\dot{u}(t) = -(3a\lambda^2 + 2b\lambda)u(t) - (a\lambda^3 + b\lambda^2) -$$
$$- \sigma \cdot t^n - \varkappa \cdot t^m \sin(\Omega \cdot t), \qquad (E.3.1.2)$$

$$u(0) = x_0 - \lambda.$$

*From linear master equation (E.3.2.3) one obtain the next transcendental master equation*

$$(x_0 - \lambda(t)) \exp[-(3a\lambda^2(t) + 2b\lambda(t)) \cdot t] -$$

$$-(a\lambda^3(t) + b\lambda^2(t)) \int_0^t \exp[-(3a\lambda^2(t) + 2b\lambda(t))(t - \tau)]d\tau -$$

$$-\sigma \int_0^t \tau^n \exp[-(3a\lambda^2(t) + 2b\lambda(t))(t - \tau)]d\tau - \qquad (E.3.2.3)$$

$$-\varkappa \int_0^t \tau^m \sin(\Omega \cdot \tau) \exp[-(3a\lambda^2(t) + 2b\lambda(t))(t - \tau)]d\tau = 0.$$

*From master equation (E.3.2.3) one obtain the next transcendental master equation:*

$$(x_0 - \lambda(t))e^{t \cdot v(t)} + e^{t \cdot v(t)}(-\lambda(t) + 2\sigma v^{-3}(t) + \phi(t)v^{-1}(t)) -$$

$$-\sigma(t^2 v^{-1}(t) + 2t v^{-2}(t) + 2v^{-3}(t)) - \phi(t)v^{-1}(t) = 0,$$

$$v(t) \triangleq -(3a\lambda^2(t) + 2b\lambda(t)), \qquad (E.3.2.4)$$

$$\phi(t) \triangleq -(a\lambda^3(t) + b\lambda^2(t)).$$

$$n = 1, \varkappa = 0.$$

**Example**  3.1.1.

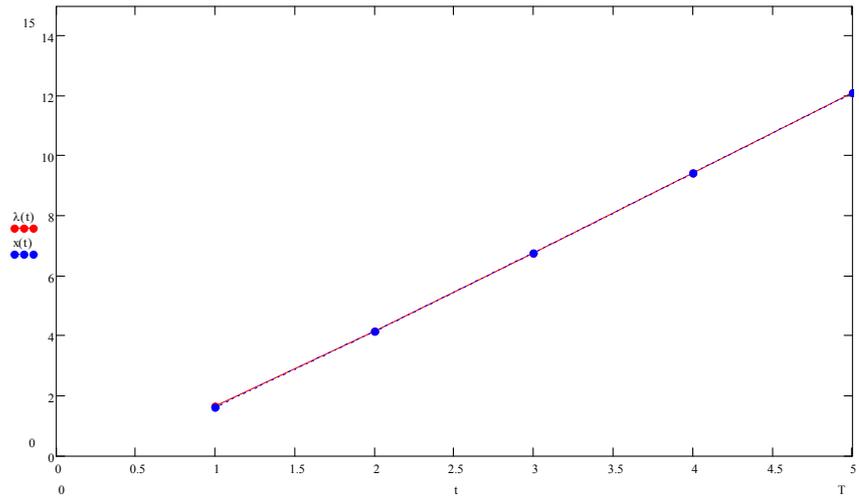

The solution master equation (3.1.8) in comparison with a coresponding solution x(t) nonlinear ODE.

*Pic.3.1.1.1.* $a = 1, b = 1, \sigma = -20, \varkappa = 0, n = 3, x_0 = 2.$

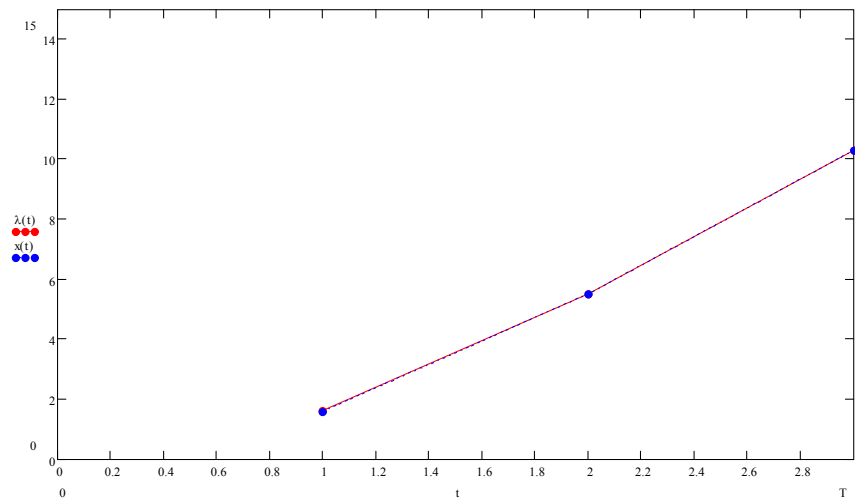

The solution master equation (3.1.8) in comparison with a coresponding solution x(t) nonlinear ODE.

*Pic.3.1.1.2.* $a = 1, b = 1, \sigma = -20, \varkappa = 0, n = 4, x_0 = 2.$

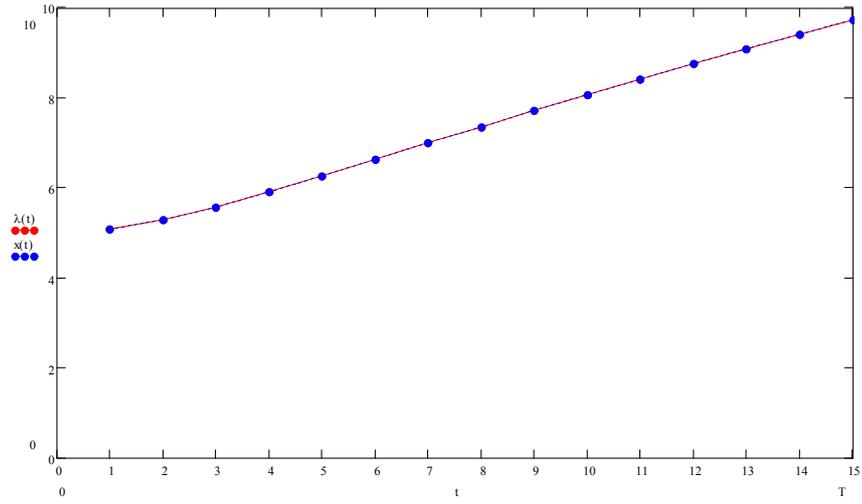

The solution master equation (3.1.8) in comparison with a coresponding solution x(t) nonlinear ODE.

*Pic.3.1.1.3.* $a = 1, b = -5, \sigma = -2, \varkappa = 0, n = 2, x_0 = 2.$

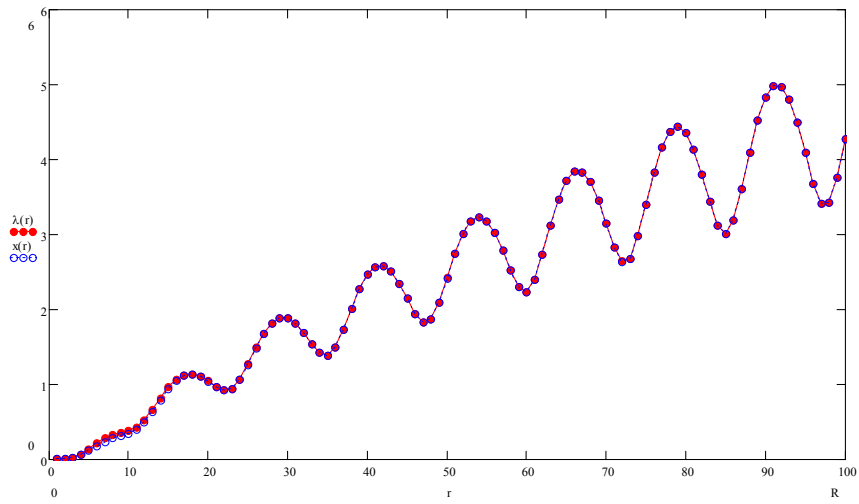

The solution master equation (3.1.8) in comparison with a coresponding solution x(t) nonlinear ODE.

*Pic.3.1.1.4.* $a = 1, b = 5, \sigma = -2, \varkappa = -1, \Omega = 5, n = 2,$

$$m = 2, x_0 = 0,$$

$$R = T/0.1, T = 10.$$

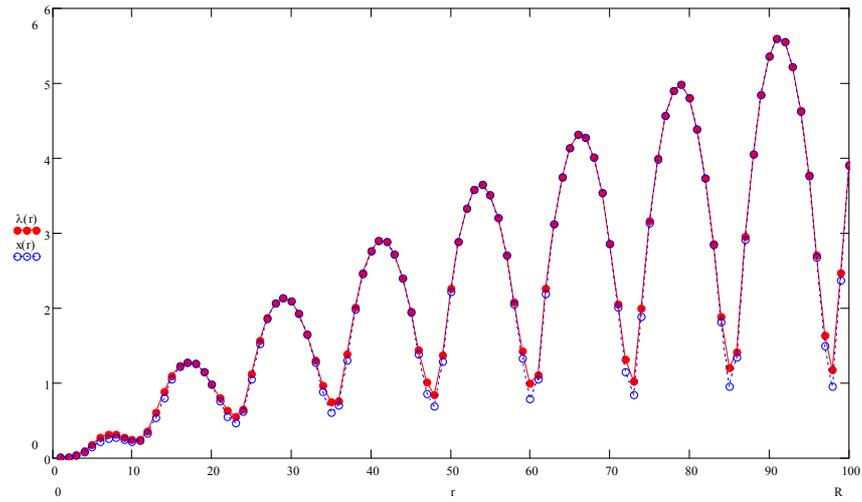

The solution master equation (3.1.8) in comparison with a coresponding solution x(t) nonlinear ODE.

**Pic.3.1.1.5.** $a = 1, b = 5, \sigma = -2, \varkappa = -2, \Omega = 5, n = 2,$

$m = 2, x_0 = 0,$

$R = T/0.1, T = 10.$

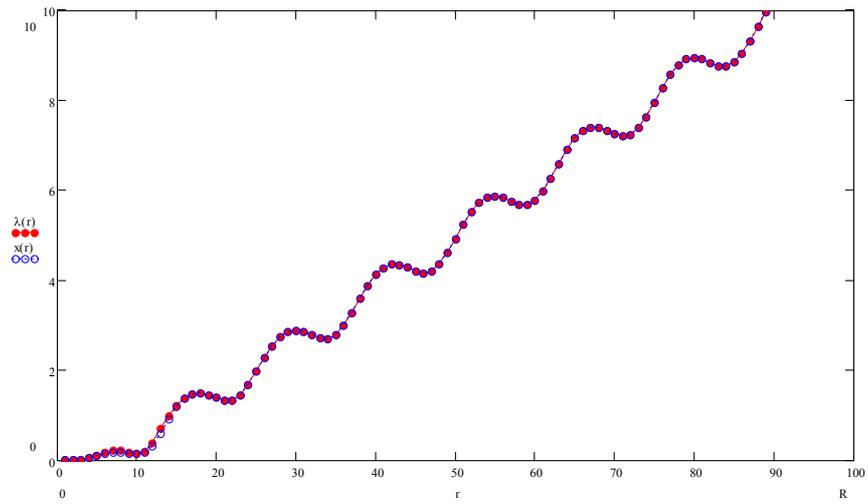

The solution master equation (3.1.8) in comparison with a coresponding solution x(t) nonlinear ODE.

**Pic.3.1.1.6.** $a = 1, b = 5, \sigma = -2, \varkappa = -2, \Omega = 5, n = 3,$

$m = 2, x_0 = 0,$

$R = T/0.1, T = 10.$

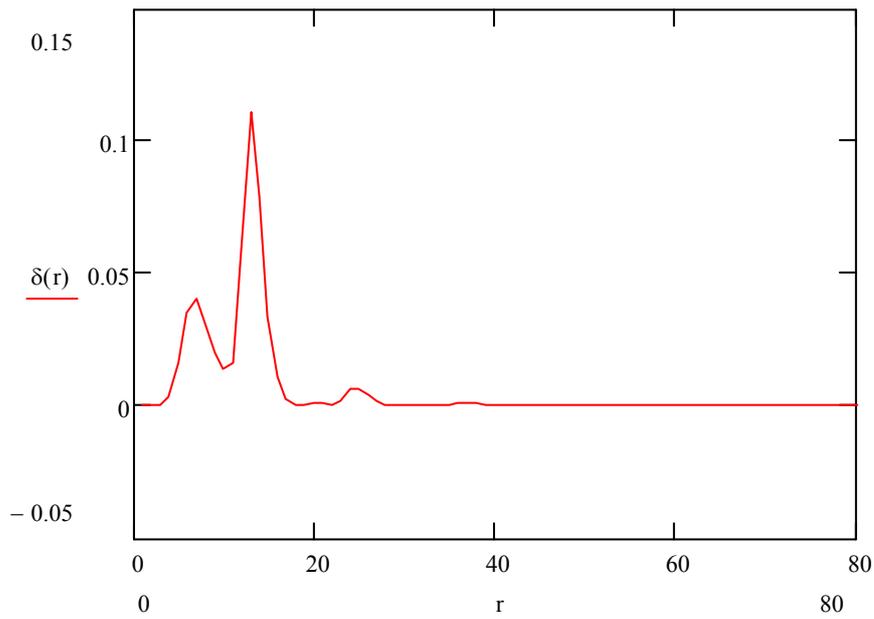

*Pic.3.1.1.7.*$\delta(r) = |x(r) - \lambda(r)|.$

$$a = 1, b = 5, \sigma = -2, \varkappa = -2, \Omega = 5, n = 3,$$

$$m = 2, x_0 = 0,$$

**Example** 3.2.

$$\dot{x} = -ax^3 - bx^2 - cx - \sigma \cdot t^n - \varkappa \cdot t^m \cdot sin(\Omega t^k) + \sqrt{\varepsilon}\,\dot{W}(t),$$

$$\varepsilon \ll 1,$$

$$0 < a,$$

$$x(0) = x_0.$$

(E.3.2.1)

*From general master equation 3.1.5 one obtain the next differential linear master equation*

$$\dot{u}(t) = -(3a\lambda^2 - 2b\lambda - c)u(t) - (a\lambda^3 - b\lambda^2 - c\lambda) - \sigma \cdot t^n - x \cdot t^m \cdot sin(\Omega t^k),$$

$$(E.3.2.2)$$

$$u(0) = x_0 - \lambda.$$

*From the differential linear master equation (E.3.2.2) one obtain the next transcendental master equation:*

$$(x_0 - \lambda(t))exp[-(3a\lambda^2(t) + 2b\lambda(t) + c) \cdot t] -$$

$$-(a\lambda^3(t) + b\lambda^2(t) + c\lambda)\int_0^t exp[-(3a\lambda^2(t) + 2b\lambda(t) + c)(t-\tau)]d\tau -$$

$$(E.3.2.3)$$

$$-\sigma \int_0^t \tau^n exp[-(3a\lambda^2(t) + 2b\lambda(t) + c)(t-\tau)]d\tau -$$

$$-x \int_0^t \tau^m sin(\Omega \tau^k) exp[-(3a\lambda^2(t) + 2b\lambda(t) + c)(t-\tau)]d\tau = 0.$$

**Example**   3.2.1.

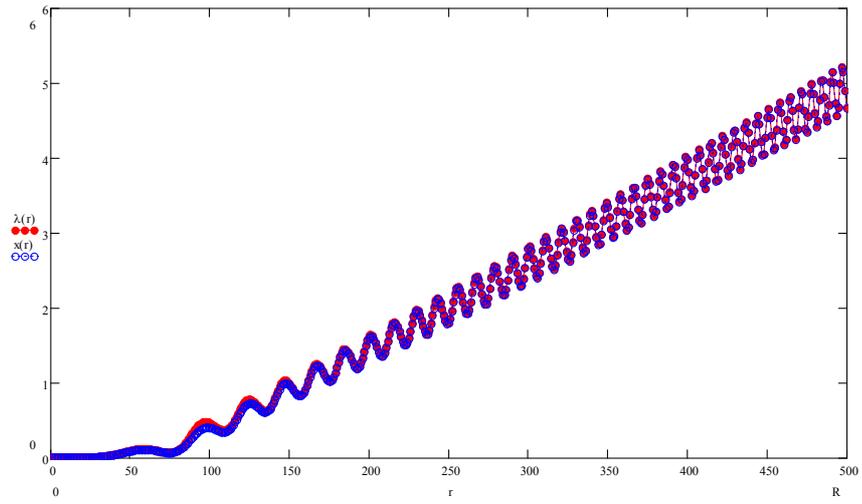

The solution master equation (3.1.8) in comparison with a coresponding solution x(t) nonlinear ODE.

*Pic.3.2.1.1.a = 1, b = 5, c = 1, σ = −2, ϰ = −2, Ω = 10, n = 3, m = 2,*

$$k = 2, x_0 = 0,$$

$$R = T/0.01, T = 5.$$

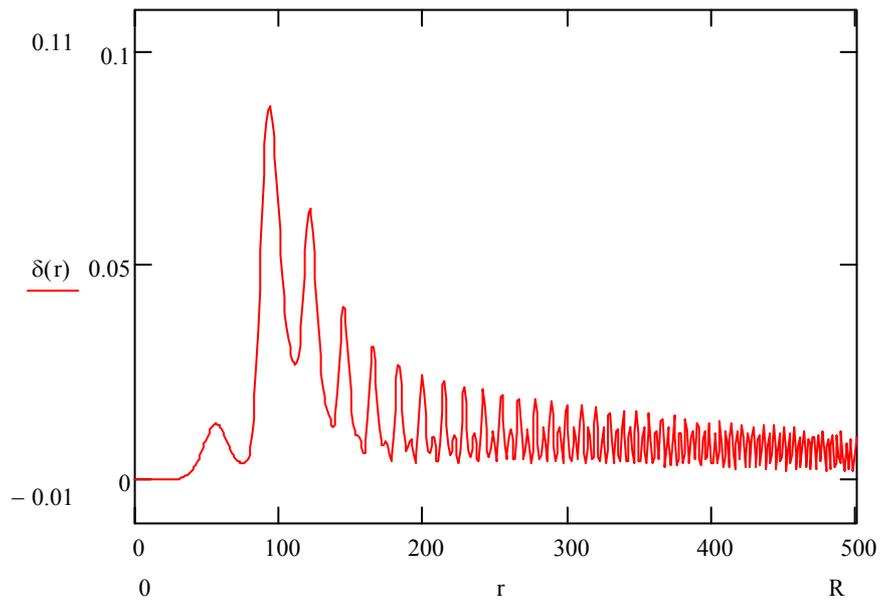

*Pic.3.2.1.2.* $\delta(r) = |x(r) - \lambda(r)|$.

$a = 1, b = 5, c = 1, \sigma = -2, \varkappa = -2, \Omega = 10, n = 3, m = 2,$

$k = 2, x_0 = 0,$

$R = T/0.01, T = 5.$

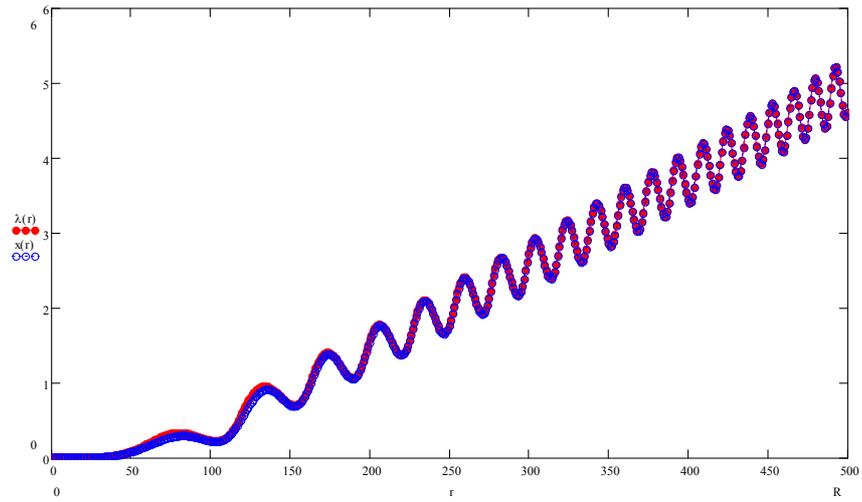

The solution master equation (3.1.8) in comparison with a coresponding solution x(t) nonlinear ODE.

**Pic.3.2.1.3.** $a = 1, b = 5, c = 1, \sigma = -2, \varkappa = -2, \Omega = 5, n = 3, m = 2,$

$$k = 2, x_0 = 0,$$

$$R = T/0.01, T = 5.$$

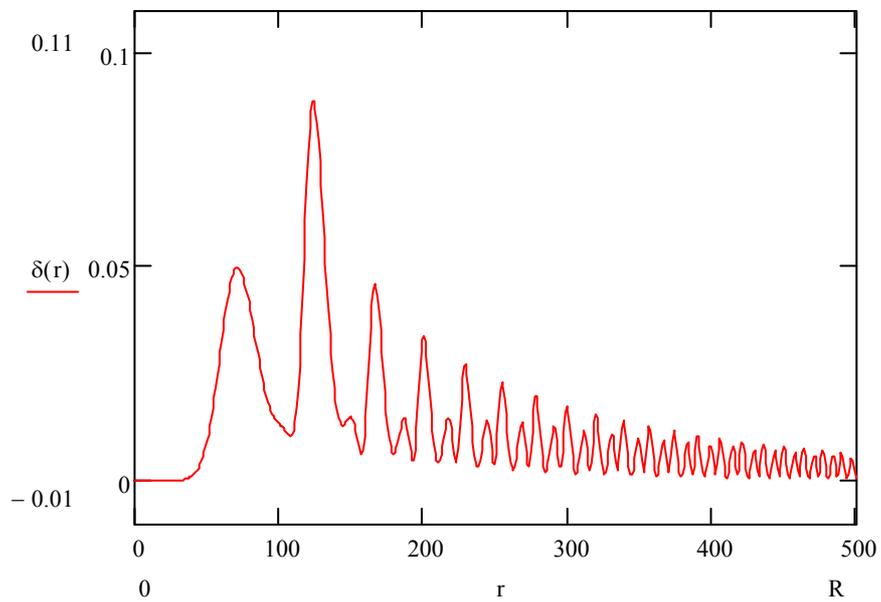

**Pic.3.2.1.4.** $\delta(r) = |x(r) - \lambda(r)|.$

$a = 1, b = 5, c = 1, \sigma = -2, \varkappa = -2, \Omega = 5, n = 3, m = 2,$

$$k = 2, x_0 = 0,$$

$$R = T/0.01, T = 5.$$

Let us consider an $m$-persons stochastic differential game $SDG^\delta_{m;\bar{t}}(f,0,y_1)$, with nonlinear dynamics:

$$\frac{dx_i(t)}{dt} = f_i(x_1,\ldots,x_m;\alpha_1,\ldots,\alpha_m) + \sqrt{\varepsilon}\,\dot{W}(t),$$

$$x(\bar{t}) = x_{\bar{t}} \in \mathbb{R}^n,$$

$$\alpha_i(t) \in U_i, i = 1,\ldots,m. \qquad (3.1.10)$$

$$t \in [\bar{t},\bar{t}+\delta],$$

$$\delta,\varepsilon \ll 1.$$

Here

$$t \to \alpha(t) = (\alpha_1[x_1(t),\ldots,x_m(t)],\ldots,\alpha_n[x_1(t),\ldots,x_m(t)]), \qquad (3.1.11)$$

$\alpha_i(t) = \alpha_i[x_1(t),\ldots,x_m(t)]$ is the determined feedback control chosen by the $i$-th player, within a set of admissible control values $U_i$, and the payoff of the $i$-th player:

$$\bar{J}_i = \mathbf{E}\left[\sum_{i=1}^m [x_i(\bar{t}+\delta,\omega) - y_i]^2\right], \qquad (3.1.12)$$

where $t \mapsto x(t,\omega)$ is the trajectory of (3.1.10).

**Definition** 3.1.2. Stochastic differential game $SDG^\delta_{m;\bar{t}}(f,0,y_1,y_2)$, is the infinitesimal stochastic differential game.

Suppose that: $\check{\alpha}(t,\lambda;x_1(t),\ldots,x_m(t)) = \mathbf{Q}(t) \times [x_1(t),\ldots,x_m(t)]^\mathsf{T}$, $t \in [\bar{t},\bar{t}+\delta]$, where $\mathbf{Q}(t)$ is a $m \times m$ matrix.
Thus:

$$\frac{dx_i(t)}{dt} = f_i(x_1,\ldots,x_m; \check{\alpha}_1(t,\lambda),\ldots,\check{\alpha}_m(t,\lambda)) + \sqrt{\varepsilon}\,\dot{W}(t),$$

$$x(\bar{t}) = x_{\bar{t}} \in \mathbb{R}^n,$$

$$\check{\alpha}_i(t) \in U_i, i = 1,\ldots,m.$$

$$t \in [\bar{t}, \bar{t} + \delta],$$

(3.1.13)

$$\delta, \varepsilon \ll 1.$$

$$\bar{J}_i = \mathbf{E}\left[\sum_{i=1}^{m}[x_i(\bar{t}+\delta,\omega) - \lambda_i]^2\right].$$

## III.1.2. Strong large deviations principle of Non-Freidlin-Wentzell type for infinitesimal stochastic differential game.

**Theorem.** *3.1.2. For the all solutions*
$$\{\mathbf{X}_{\bar{t}}^{\varepsilon}, \check{\alpha}(t,\lambda)\} = (X_{1,t}^{\varepsilon},\ldots,X_{m,t}^{\varepsilon}), (\check{\alpha}_1(t,\lambda),\ldots,\check{\alpha}_m(t,\lambda))$$

*dissipative infinitesimal SDG (3.13) and $\mathbb{R}$ valued parameters $\lambda = (\lambda_1,\ldots,\lambda_m) \in \mathbb{R}^m$, there exists a Colombeau constant $(C_\varepsilon)_\varepsilon \in \widetilde{\mathbb{R}}, (C_\varepsilon)_\varepsilon \geq 0$, such that:*

$$\liminf_{\varepsilon \to 0} \mathbf{E}\left[\|\mathbf{X}_t^\varepsilon - \lambda\|^2\right] \leq (C_\varepsilon)_\varepsilon \|\mathbf{U}(t,\lambda)\|^2,$$

$$\lambda = (\lambda_1, \ldots, \lambda_n) \in \mathbb{R}^n, \qquad (3.1.14)$$

$$t \in [\bar{t}, \bar{t} + \delta],$$

where $\mathbf{U}(t,\lambda) = (U_1(t,\lambda), \ldots, U_n(t,\lambda))$ *is the trajectory of the differential master game:*

$$\frac{d\mathbf{U}(t,\lambda)}{dt} = \mathbf{J}[\mathbf{f}(\lambda, \check{\alpha}(t,\lambda))]\mathbf{U} + \mathbf{f}(\lambda, \check{\alpha}(t,\lambda)),$$

$$\mathbf{U}(\bar{t}, \lambda) = x_{\bar{t}}, \qquad (3.1.15)$$

$$\mathbf{J}_i = \|\mathbf{U}(\bar{t}+\delta)\|^2 = \sum_{i=1}^m [U_i(\bar{t}+\delta)]^2.$$

Where $\mathbf{J}[\mathbf{f}(\lambda, \check{\alpha}(t,\lambda))]$ the Jacobian, i.e. $\mathbf{J}$ is a $n \times n$-matrix:

$$\mathbf{J}[\mathbf{f}(\lambda, \check{\alpha}(t,\lambda))] = \mathbf{J}[\mathbf{f}(x, \check{\alpha}(t,x))]\big|_{x=\lambda} =$$

$$= \begin{bmatrix} \dfrac{\partial f_1(x,\check{\alpha}(t,x))}{\partial x_1} & \cdots & \dfrac{\partial f_1(x,\check{\alpha}(t,x))}{\partial x_n} \\ \cdot & \cdots & \cdot \\ \cdot & \cdots & \cdot \\ \cdot & \cdots & \cdot \\ \dfrac{\partial f_n(x,\check{\alpha}(t,x))}{\partial x_1} & \cdots & \dfrac{\partial f_n(x,\check{\alpha}(t,x))}{\partial x_n} \end{bmatrix}\Bigg|_{x=\lambda} \qquad (3.1.16)$$

**Corollary** *3.1.2. Assume the conditions of the Theorem 3.2 for any $\lambda = (\lambda_1, \ldots, \lambda_n) \in \mathbb{R}^n, t \in [\bar{t}, \bar{t} + \delta]$ :*

$$\|\mathbf{U}(t,\lambda)\| = 0 \Rightarrow$$

$$\liminf_{\varepsilon \to 0} \mathbf{E}\left[\|\mathbf{X}_t^\varepsilon - \lambda\|^2\right] = 0. \tag{3.1.17}$$

$$\liminf_{\varepsilon \to 0} \left[\min_{\alpha_i(t)} \left(\max_{\alpha_j(t), j \neq i} \mathbf{J}_i\right)\right] = 0.$$

*More precisely, for any $t \in [\bar{t}, \bar{t}+\delta]$ and $\lambda = \lambda(t) = (\lambda_1(t), \ldots, \lambda_n(t)) \in \mathbb{R}^n$ sutch that*

$$U_1(t, \lambda_1(t), \ldots, \lambda_n(t)) = 0,$$

$$\ldots\ldots\ldots\ldots\ldots\ldots\ldots \tag{3.1.18}$$

$$U_n(t, \lambda_1(t), \ldots, \lambda_n(t)) = 0,$$

the equalities is satisfaed

$$\liminf_{\varepsilon \to 0} \mathbf{E}\left[\|X_{1,t}^\varepsilon - \lambda_1(t)\|^2\right] = 0,$$

$$\ldots\ldots\ldots\ldots\ldots\ldots\ldots$$

$$\liminf_{\varepsilon \to 0} \mathbf{E}\left[\|X_{n,t}^\varepsilon - \lambda_n(t)\|^2\right] = 0, \tag{3.1.19}$$

$$\liminf_{\varepsilon \to 0} \left[\min_{\alpha_i(t)} \left(\max_{\alpha_j(t), j \neq i} \bar{\mathbf{J}}_i\right)\right] = 0.$$

# III.1.3. Strong large deviations principle of Non-Freidlin-Wentzell type for Colombeau-Ito's SDE.

Let us consider now a family $(\mathbf{X}_{t,\epsilon}^{\varepsilon})_{\epsilon}$ of the solutions **CSDE:**

$$(d\mathbf{X}_{t,\epsilon}^{\varepsilon})_{\epsilon} = (\mathbf{b}_{\epsilon}(\mathbf{X}_{t,\epsilon}^{\varepsilon},t))_{\epsilon}dt + \sqrt{\varepsilon}\,d\mathbf{W}(t),$$

$$\epsilon \in (0,1], \tag{3.1.20}$$

$$(\mathbf{X}_{0,\epsilon}^{\varepsilon})_{\epsilon} = x_0 \in \widetilde{\mathbb{R}}^n, t \in [0,T],$$

where $\mathbf{W}(t)$ is $n$-dimensional Brownian motion, $(\mathbf{b}_{\epsilon}(\circ,t))_{\epsilon} : \widetilde{\mathbb{R}}^n \times \mathbb{R}_+ \to \widetilde{\mathbb{R}}^n$ is a polinomial transform, i.e. $b_{\epsilon,i}(x,t) = \sum_{|\alpha|} b_{\epsilon,\alpha}^i(x,t)x^{\alpha}, \epsilon \in (0,1]$, $\alpha = (i_1,\ldots,i_k), |\alpha| = \sum_{j=0}^{k} i_j, i = 1,\ldots,n.$

**Definition** 3.1.3. CSDE (3.1.20) is $\widetilde{\mathbb{R}}$-dissipative if exist generalized Lyapunov-candidate-function $(V_{\epsilon}(x,t))_{\epsilon}$ and Colombeau constants $\widetilde{\mathbb{R}} \ni (C_{\epsilon})_{\epsilon} = \widetilde{C} > 0, \widetilde{\mathbb{R}} \ni (R_{\epsilon})_{\epsilon} = \widetilde{R} > 0,$ such that:
(1) $\forall x, \|x\| \geq \widetilde{R} : (\dot{V}_{\epsilon}(x,t;\mathbf{b}_{\epsilon}))_{\epsilon} \leq -\widetilde{C} \cdot (V_{\epsilon}(x,t))_{\epsilon}$ where

$$(\dot{V}_{\epsilon}(x,t;f_{\epsilon}))_{\epsilon} \triangleq \left(\frac{\partial V_{\epsilon}(x,t)}{\partial t}\right)_{\epsilon} + \sum_{i=1}^{n}\left(\frac{\partial V_{\epsilon}(x,t)}{\partial x_i}f_{\epsilon}(x)\right)_{\epsilon}, \epsilon \in (0,1];$$

(2) $\left(\widetilde{V}_{\epsilon}(x)\right)_{\epsilon} = \left(\lim_{r\to\infty}\left(\inf_{\|x\|>r} V_{\epsilon}(x)\right)\right)_{\epsilon} = \infty.$

**Theorem** 3.1.3. ( "*Strong large deviations principle of Non-Freidlin-Wentzell type for Colombeau-Ito's SDE.*"). For any solution $(\mathbf{X}_t^{\varepsilon,\epsilon})_{\epsilon} = \left((X_{1,t}^{\varepsilon,\epsilon})_{\epsilon},\ldots,(X_{n,t}^{\varepsilon,\epsilon})_{\epsilon}\right)$ dissipative CSDE (3.1.20) and $\widetilde{\mathbb{R}}^n$ valued vector-parameters $(\lambda_{\epsilon})_{\epsilon} = ((\lambda_{\epsilon,1})_{\epsilon},\ldots,(\lambda_{\epsilon,n})_{\epsilon}),$ there exists Colombeau constant $\widetilde{C'} = \left(C'_{\varepsilon(\epsilon)}\right)_{\varepsilon(\epsilon)} \in \widetilde{\mathbb{R}}, \left(C'_{\varepsilon(\epsilon)}\right)_{\varepsilon(\epsilon)} \geq 0$ such that:

$$\lim_{\varepsilon \to 0} \inf \mathbf{E}\left[\|(\mathbf{X}_t^{\varepsilon,\epsilon})_\epsilon - (\lambda_\epsilon)_\epsilon\|^2\right] \leq \widetilde{C'}\|(\mathbf{U}_\epsilon(t,\lambda_\epsilon))_\epsilon\|^2. \qquad (3.1.21)$$

Where $(\mathbf{U}_\epsilon(t,\lambda_\epsilon))_\epsilon = ((U_{\epsilon,1}(t,\lambda_\epsilon))_\epsilon,\ldots,(U_{\epsilon,n}(t,\lambda))_\epsilon)$ the solution of the master equation:

$$\left(\dot{\mathbf{U}}_\epsilon(t,\lambda)\right)_\epsilon = (\mathbf{J}_\epsilon[\mathbf{b}_\epsilon(\lambda_\epsilon,t)])_\epsilon(\mathbf{U}_\epsilon)_\epsilon + (\mathbf{b}_\epsilon(\lambda,t))_\epsilon, (\mathbf{U}_\epsilon(0,\lambda_\epsilon))_\epsilon = x_0 - \widetilde{\lambda},$$
$$(3.1.22)$$

$$\widetilde{\lambda} \triangleq (\lambda_\epsilon)_\epsilon$$

where $(\mathbf{J}_\epsilon)_\epsilon = (\mathbf{J}_\epsilon[(\mathbf{b}_\epsilon(\lambda_\epsilon,t))])_\epsilon$ the Jacobian, i.e. $(\mathbf{J}_\epsilon)_\epsilon$ is $n \times n$-matrix:

$$(\mathbf{J}_\epsilon[\mathbf{b}_\epsilon(\lambda_\epsilon,t)])_\epsilon = \left(\mathbf{J}_\epsilon[\mathbf{b}_\epsilon(x,t)]|_{x=\lambda_\epsilon}\right)_\epsilon =$$

$$= \begin{bmatrix} \left(\dfrac{\partial b_{\epsilon,1}(x,t)}{\partial x_1}\right)_\epsilon & \cdots & \left(\dfrac{\partial b_{\epsilon,1}(x,t)}{\partial x_n}\right)_\epsilon \\ \cdot & \cdots & \cdot \\ \cdot & \cdots & \cdot \\ \cdot & \cdots & \cdot \\ \dfrac{\partial b_{\epsilon,n}(x,t)}{\partial x_1} & \cdots & \dfrac{\partial b_{\epsilon,n}(x,t)}{\partial x_n} \end{bmatrix}_{x=\lambda_\epsilon} \qquad (3.1.23)$$

**Corollary** 3.1.3. Assume the conditions of the Theorem 3.1.1 for any

$$\widetilde{\lambda} = \left(\lambda_\epsilon\right)_\epsilon = (\lambda_{1,\epsilon},\ldots,\lambda_{n,\epsilon})_\epsilon \in \widetilde{\mathbb{R}}^n, t \in [0,T]:$$

$$(\|\mathbf{U}_\epsilon(t,\lambda_\epsilon)\|)_\epsilon = 0 \Rightarrow \left(\lim_{\varepsilon \to 0}\inf \mathbf{E}\left[\|\mathbf{X}_t^{\varepsilon,\epsilon} - \lambda_\epsilon\|^2\right]\right)_\epsilon = 0 \qquad (3.1.24)$$

More precisely, for any $t \in [0,T]$ and
$\widetilde{\lambda}(t) = (\lambda_\epsilon(t))_\epsilon = ((\lambda_{1,\epsilon}(t))_\epsilon,\ldots,(\lambda_{n,\epsilon}(t))_\epsilon) \in \widetilde{\mathbb{R}}^n$ sutch that

$$(U_{1,\epsilon}(t,\lambda_{1,\epsilon}(t),\ldots,\lambda_{n,\epsilon}(t)))_\epsilon = 0,$$

$$\cdots\cdots\cdots\cdots\cdots\cdots \quad (3.1.25)$$

$$(U_{n,\epsilon}(t,\lambda_{1,\epsilon}(t),\ldots,\lambda_{n,\epsilon}(t)))_\epsilon = 0,$$

*the equalities is satisfaed*

$$\left(\liminf_{\varepsilon\to 0} \mathbf{E}\left[\|X_{1,t}^{\varepsilon,\epsilon} - \lambda_{1,\epsilon}(t)\|^2\right]\right)_\epsilon = 0,$$

$$\cdots\cdots\cdots\cdots\cdots\cdots \quad (3.1.26)$$

$$\left(\liminf_{\varepsilon\to 0} \mathbf{E}\left[\|X_{n,t}^{\varepsilon,\epsilon} - \lambda_{n,\epsilon}(t)\|^2\right]\right)_\epsilon = 0.$$

**Remark.** Note that

$$(\delta_\epsilon(t))_\epsilon \triangleq \left(\liminf_{\varepsilon\to 0} \mathbf{E}\left\|(\mathbf{X}_t^{\varepsilon,\epsilon})_\epsilon - (\mathbf{X}_t^{0,\epsilon})_\epsilon\right\|\right)_\epsilon \neq 0. \quad (3.1.26')$$

## III.1.4. Generalized strong large deviations principle of Non-Freidlin-Wentzell type for Ito's type SDE.

Let us consider now a family $\left(X_t^{\varepsilon,\epsilon}(\omega,\omega')\right)_\epsilon$ of stochastic srocesses (where pair $(\omega,\omega') \in \Omega \times \Omega', \Omega \cap \Omega' = \varnothing$) of the solutions Ito's type SDE:

$$(d\mathbf{X}_t^{\varepsilon,\epsilon}(\omega,\omega'))_\epsilon = \mathbf{b}((\mathbf{X}_t^{\varepsilon,\epsilon}(\omega,\omega'),t;\omega))_\epsilon dt + \sqrt{\varepsilon}\,d\mathbf{W}(t,\omega'),$$

$$\mathbf{b}_\epsilon(\circ,t;\omega)dt = \bar{\mathbf{b}}(\circ,t;\omega)dt + \sqrt{D}\,\dot{\mathbf{W}}_\epsilon(t,\omega) \quad (3.1.27)$$

$$\mathbf{X}_0^{\varepsilon,\epsilon}(\omega,\omega') = x_0 \in \mathbb{R}^n, t \in [0,T],$$

where $\mathbf{W}(t,\omega')$ is $n$-dimensional Brownian motion, $(\mathbf{W}_\epsilon(t,\omega))_\epsilon$

is *n*-dimensional Colombeau Brownian motion, $\mathbf{b}_\epsilon(\circ, t; \omega) : \mathbb{R}^n \times \mathbb{R}_+ \times \Omega \to \widetilde{\mathbb{R}}^n$ is a stochastic polinomial transform, i.e.

$$b_{\epsilon,i}(x,t;\omega) = \sum_{|\alpha|} b^i_{\epsilon,\alpha}(x,t;\omega) x^\alpha,$$

$$\alpha = (i_1, \ldots, i_k), |\alpha| = \sum_{j=0}^{k} i_j, \qquad (3.1.28)$$

$$i = 1, \ldots, n.$$

**Definition** 3.1.4. SDE (3.1.27) is dissipative if exist Lyapunov-candidate-function $V_\epsilon(x,t,\omega; f_\epsilon)$ and constants $C > 0, R \geq 0$ such that

$$\dot{V}_\epsilon(x,t,\omega; \mathbf{b}) \leq -C V_\epsilon(x,t,\omega), \|x\| \geq R,$$

$$\dot{V}_\epsilon(x,t; f_\epsilon)_\epsilon \triangleq \frac{\partial V_\epsilon(x,t)}{\partial t} + \sum_{i=1}^{n} \frac{\partial V_\epsilon(x,t,\omega)}{\partial x_i} f_\epsilon(x)$$

$$\widetilde{V}(x,t,\omega,t,\epsilon) = \lim_{r \to \infty} \left( \inf_{\|x\| > r} V_\epsilon(x,t,\omega) \right) = \infty. \qquad (3.1.29)$$

Let us consider now a family $X_t^{\varepsilon,\epsilon}$ of the solutions dissipative SDE (3.1.27).

**Theorem** 3.1.4.( **Generalized Strong large deviations principle**).[19] For the all solutions $\mathbf{X}_t^{\varepsilon,\epsilon} = (X_{1,t}^{\varepsilon,\epsilon}, \ldots, X_{n,t}^{\varepsilon,\epsilon})$ dissipative SDE (3.1.27) and $\mathbb{R}$ valued parameters $\lambda_1, \ldots, \lambda_n$, $\lambda = (\lambda_1, \ldots, \lambda_n) \in \mathbb{R}^n$, there exists Colombeau constant $(C_\varepsilon)_\varepsilon \in \widetilde{\mathbb{R}}, (C_\varepsilon)_\varepsilon \geq 0,$ such that:

$$\left( \liminf_{\varepsilon \to 0} \mathbf{E}_{\Omega'} \left[ \|\mathbf{X}_t^{\varepsilon,\epsilon} - \lambda\|^2 | \Omega \right] \right)_\epsilon \leq (C_\varepsilon)_\varepsilon \|\mathbf{U}_\epsilon(t,\lambda,\omega)\|^2 \qquad (3.1.30)$$

$$\lambda = (\lambda_1, \ldots, \lambda_n) \in \mathbb{R}^n$$

where $\mathbf{U}_\epsilon(t,\lambda,\omega) = (U_{\epsilon,1}(t,\lambda,\omega), \ldots, U_{\epsilon,n}(t,\lambda,\omega))$ the solution of the linear differential master equation:

$$\frac{d\mathbf{U}_\epsilon(t,\lambda,\omega)}{dt} = \mathbf{J}[\mathbf{b}_\epsilon(\lambda,t,\omega)]\mathbf{U}_\epsilon + \mathbf{b}_\epsilon(\lambda,t,\omega),$$

(3.1.31)

$$\mathbf{U}_\epsilon(0,\lambda,\omega) = x_0 - \lambda,$$

where $\mathbf{J}_\epsilon = \mathbf{J}_\epsilon[\mathbf{b}_\epsilon(\lambda,t,\omega)]$ the Jacobian, i.e. $\mathbf{J}_\epsilon$ is a $n \times n$-matrix:

$$\mathbf{J}_\epsilon[\mathbf{b}_\epsilon(\lambda,t,\omega)] = \mathbf{J}_\epsilon[\mathbf{b}_\epsilon(x,t,\omega)]|_{x=\lambda} =$$

$$= \begin{bmatrix} \frac{\partial b_{\epsilon,1}(x,t,\omega)}{\partial x_1} & \cdots & \frac{\partial b_{\epsilon,1}(x,t,\omega)}{\partial x_n} \\ \cdot & \cdots & \cdot \\ \cdot & \cdots & \cdot \\ \cdot & \cdots & \cdot \\ \frac{\partial b_{\epsilon,n}(x,t,\omega)}{\partial x_1} & \cdots & \frac{\partial b_{\epsilon,n}(x,t,\omega)}{\partial x_n} \end{bmatrix}_{x=\lambda}$$

(3.1.32)

**Corollary** 3.1.4. Assume the conditions of the Theorem 3.1.4 for any

$$\lambda = (\lambda_1,\ldots,\lambda_n) \in \mathbb{R}^n, t \in [0,T]:$$

$$\|\mathbf{U}_\epsilon(t,\lambda,\omega)\| = 0 \Rightarrow \liminf_{\varepsilon \to 0} \mathbf{E}_{\Omega'}\left[\|\mathbf{X}_t^{\varepsilon,\epsilon} - \lambda\|^2 | \Omega\right] = 0$$

(3.1.33)

*More precisely, for any $t \in [0,T]$ and $\lambda = \lambda(t) = (\lambda_1(t),\ldots,\lambda_n(t)) \in \widetilde{\mathbb{R}}^n$ sutch that*

$$U_{\epsilon,1}(t,\lambda_{\epsilon,1}(t,\omega),\ldots,\lambda_{\epsilon,n}(t,\omega),\omega) = 0,$$

$$\cdots\cdots\cdots\cdots\cdots\cdots\cdots \qquad (3.1.34)$$

$$U_{\epsilon,n}(t,\lambda_{\epsilon,1}(t,\omega),\ldots,\lambda_{\epsilon,n}(t,\omega),\omega) = 0,$$

*the equalities is satisfaed*

$$\liminf_{\varepsilon\to 0} \mathbf{E}_{\Omega'}\left[\|X_{1,t}^{\varepsilon,\epsilon}(\omega,\omega') - \lambda_{\epsilon,1}(t,\omega)\|^2 \big|\Omega\right] = 0,$$

$$\cdots\cdots\cdots\cdots\cdots\cdots\cdots \qquad (3.1.35)$$

$$\liminf_{\varepsilon\to 0} \mathbf{E}_{\Omega'}\left[\|X_{n,t}^{\varepsilon,\epsilon}(\omega,\omega') - \lambda_n(t,\omega)\|^2 \big|\Omega\right] = 0.$$

## III.1.5. Generalized strong large deviations principle of Non-Freidlin-Wentzell type for Colombeau-Ito's SDE.

Let us consider the family $(\mathbf{X}_t^{\varepsilon,\epsilon}(\omega,\omega'))_\epsilon$ of stochastic srocesses (where a pair $(\omega,\omega') \in \Omega \times \Omega', \Omega \cap \Omega' = \varnothing$) which is a solution of the SDE:

$$(d\mathbf{X}_{t,\epsilon}^\varepsilon(\omega,\omega'))_\epsilon = (\mathbf{b}_\epsilon(\mathbf{X}_{t,\epsilon}^\varepsilon(\omega,\omega'),t))_\epsilon + \sqrt{D}\,(d\mathbf{W}_\epsilon(t,\omega))_\epsilon + \sqrt{\varepsilon}\,\mathbf{W}(t,\omega'),$$

$$\epsilon \in (0,1], \qquad (3.1.36)$$

$$(\mathbf{X}_{0,\epsilon}^\varepsilon(\omega,\omega'))_\epsilon = x_0 \in \widetilde{\mathbb{R}}^n, t \in [0,T],$$

where $\mathbf{W}(t,\omega')$ is $n$-dimensional Brownian motion, $(\mathbf{W}_\epsilon(t,\omega))_\epsilon$ is $n$-dimensional Colombeau Brownian motion, $(\mathbf{b}_\epsilon(\circ,t,\omega))_\epsilon : \widetilde{\mathbb{R}}^n \times \mathbb{R}_+ \times \Omega \to \widetilde{\mathbb{R}}^n$ is a polinomial transform, i.e. $(b_{\epsilon,i}(x,t,\omega))_\epsilon = \sum_{|\alpha|}(b_{\epsilon,\alpha}^i(x,t,\omega))_\epsilon x^\alpha, \epsilon \in (0,1], \alpha = (i_1,\ldots,i_k),$

$|\alpha| = \sum_{j=0}^{k} i_j, i = 1,\ldots,n$.

**Definition** 3.1.5. SDE (3.1.36) is dissipative if exist generalized Lyapunov-candidate-function $V_\epsilon(x,t,\omega;f_\epsilon)$ and Colombeau constants $\widetilde{C} = (C_\epsilon)_\epsilon > 0, \widetilde{R} = (R_\epsilon)_\epsilon \geq 0$ such that

$$(\dot{V}_\epsilon(x,t,\omega;\boldsymbol{b}))_\epsilon \leq -\widetilde{C}(V_\epsilon(x,t,\omega))_\epsilon, \|x\| \geq \widetilde{R},$$

$$(\dot{V}_\epsilon(x,t;f_\epsilon))_\epsilon \triangleq \left(\frac{\partial V_\epsilon(x,t,\omega)}{\partial t}\right)_\epsilon + \sum_{i=1}^{n}\left(\frac{\partial V_\epsilon(x,t,\omega)}{\partial x_i}f_\epsilon(x,\omega)\right)_\epsilon \quad (3.1.37)$$

$$\left(\widetilde{V}_\epsilon(x,t,\omega)\right)_\epsilon = \left(\lim_{r\to\infty}\left(\inf_{\|x\|>r} V_\epsilon(x,t,\omega)\right)\right)_\epsilon = \infty.$$

Let us consider now a family $X_t^{\varepsilon,\epsilon}(\omega,\omega')$ of the solutions dissipative CSDE (3.1.36).

**Theorem** 3.1.5.( **Generalized Strong large deviations principle**). For the all solutions $(\boldsymbol{X}_t^{\varepsilon,\epsilon}(\omega,\omega'))_\epsilon = $
$= \left((X_{1,t}^{\varepsilon,\epsilon}(\omega,\omega'))_\epsilon,\ldots,(X_{n,t}^{\varepsilon,\epsilon}(\omega,\omega'))_\epsilon\right)$ dissipative **CSDE** (3.1.36) and $\widetilde{\mathbb{R}}$ valued parameters $\widetilde{\lambda}_1,\ldots,\widetilde{\lambda}_n$, $\widetilde{\lambda} = (\widetilde{\lambda}_1,\ldots,\widetilde{\lambda}_n) \in \widetilde{\mathbb{R}}^n$, there exists Colombeau constant $\widetilde{C} = (C_{\varepsilon(\epsilon)})_{\varepsilon(\epsilon)} \in \widetilde{\mathbb{R}}, (C_{\varepsilon(\epsilon)})_{\varepsilon(\epsilon)} \geq 0,$ such that:

$$\left(\liminf_{\varepsilon\to 0}\boldsymbol{E}_{\Omega'}\left[\|\boldsymbol{X}_t^{\varepsilon,\epsilon}(\omega,\omega') - \lambda\|^2|\Omega\right]\right)_\epsilon \leq (C_{\varepsilon(\epsilon)})_{\varepsilon(\epsilon)}(\|\boldsymbol{U}_\epsilon(t,\lambda,\omega)\|)_\epsilon^2$$
$$(3.1.38)$$

$$\lambda = (\lambda_1,\ldots,\lambda_n) \in \widetilde{\mathbb{R}}^n$$

where $(\boldsymbol{U}_\epsilon(t,\lambda,\omega))_\epsilon = ((U_{\epsilon,1}(t,\lambda,\omega))_\epsilon,\ldots,(U_{\epsilon,n}(t,\lambda,\omega))_\epsilon)$ the solution of the linear differential master equation:

$$\frac{d\mathbf{U}_\epsilon(t,\lambda,\omega)}{dt} = \mathbf{J}[\mathbf{b}_\epsilon(\lambda,t,\omega)]\mathbf{U}_\epsilon + \mathbf{b}_\epsilon(\lambda,t,\omega) + \sqrt{D}\,(d\mathbf{W}_\epsilon(t,\omega))_\epsilon,$$

(3.1.31)

$$\mathbf{U}_\epsilon(0,\lambda,\omega) = x_0 - \lambda,$$

where $\mathbf{J}_\epsilon = \mathbf{J}_\epsilon[\mathbf{b}_\epsilon(\lambda,t,\omega)]$ the Jacobian, i.e. $\mathbf{J}_\epsilon$ is a $n \times n$-matrix:

$$\mathbf{J}_\epsilon[\mathbf{b}_\epsilon(\lambda,t,\omega)] = \mathbf{J}_\epsilon[\mathbf{b}_\epsilon(x,t,\omega)]|_{x=\lambda} =$$

$$= \begin{bmatrix} \frac{\partial b_{\epsilon,1}(x,t,\omega)}{\partial x_1} & \cdots & \frac{\partial b_{\epsilon,1}(x,t,\omega)}{\partial x_n} \\ \cdot & \cdots & \cdot \\ \cdot & \cdots & \cdot \\ \cdot & \cdots & \cdot \\ \frac{\partial b_{\epsilon,n}(x,t,\omega)}{\partial x_1} & \cdots & \frac{\partial b_{\epsilon,n}(x,t,\omega)}{\partial x_n} \end{bmatrix}\Bigg|_{x=\lambda}$$

(3.1.32)

**Corollary** *3.1.4. Assume the conditions of the Theorem 3.1.4 for any*

$$\lambda = (\lambda_1,\ldots,\lambda_n) \in \mathbb{R}^n, t \in [0,T]:$$

$$\|\mathbf{U}_\epsilon(t,\lambda,\omega)\| = 0 \Rightarrow \liminf_{\varepsilon \to 0} \mathbf{E}_{\Omega'}\Big[\|\mathbf{X}_t^{\varepsilon,\epsilon}(\omega,\omega') - \lambda\|^2 \big| \Omega\Big] = 0 \quad (3.1.33)$$

*More precisely, for any $t \in [0,T]$ and*
*$\lambda = \lambda(t) = (\lambda_1(t),\ldots,\lambda_n(t)) \in \widetilde{\mathbb{R}}^n$ sutch*
*that*

$$U_{\epsilon,1}(t,\lambda_{\epsilon,1}(t,\omega),\ldots,\lambda_{\epsilon,n}(t,\omega),\omega) = 0,$$

$$\ldots\ldots\ldots\ldots\ldots\ldots \tag{3.1.34}$$

$$U_{\epsilon,n}(t,\lambda_{\epsilon,1}(t,\omega),\ldots,\lambda_{\epsilon,n}(t,\omega),\omega) = 0,$$

*the equalities (3.1.35) is satisfaed*

$$\liminf_{\varepsilon\to 0} \mathbf{E}_{\Omega'}\left[\|X_{1,t}^{\varepsilon,\epsilon}(\omega,\omega') - \lambda_{\epsilon,1}(t,\omega)\|^2 \big| \Omega\right] = 0,$$

$$\ldots\ldots\ldots\ldots\ldots\ldots \tag{3.1.35}$$

$$\liminf_{\varepsilon\to 0} \mathbf{E}_{\Omega'}\left[\|X_{n,t}^{\varepsilon,\epsilon}(\omega,\omega') - \lambda_n(t,\omega)\|^2 \big| \Omega\right] = 0.$$

**Remark.** Note that $(\delta_\epsilon(t,\omega))_\epsilon \triangleq \left(\liminf_{\varepsilon\to 0} \mathbf{E}_{\Omega'}\left[\|(\mathbf{X}_t^{\varepsilon,\epsilon})_\epsilon - (\mathbf{X}_t^{0,\epsilon})_\epsilon\| \big| \Omega\right]\right)_\epsilon \neq 0.$

## III.1.6. Generalized strong large deviations principle (GLDP) of Non-Freidlin-Wentzell type for Ito's type SDE. Numerical simulation.Comparison of stochastic dynamics with non-perturbative quasiclassical stochastic dynamics.

In this section we have compared by stochastic norm $\delta(t,\omega)$ :

$$\delta(t,\omega) \triangleq \left( \liminf_{\varepsilon \to 0, \epsilon \to 0} \mathbf{E}_{\Omega'}[\|(\mathbf{X}_t^{\varepsilon,\epsilon}(\omega,\omega'))_\epsilon - (\mathbf{X}_t(\omega))\| |\Omega] \right) \quad (3.1.36)$$

the above analytical predictions for the 'ε-limit' $\lambda(t,\omega)$ given by Eq.(3.1.38) for the solution Ito's type SDE (3.1.27) ( 'ε-limit' stochastic dynamics) with numerical results for the solution of the Ito's SDE (3.1.36):

$$(d\mathbf{X}_t(\omega))_\epsilon = \bar{\mathbf{b}}((\mathbf{X}_t(\omega), t; \omega))dt + \sqrt{D}\, W(t,\omega)$$

$$\mathbf{X}_0(\omega,) = x_0 \in \mathbb{R}^n, t \in [0,T], \quad (3.1.37)$$

**Example.1.**

Ito's type SDE:

$$(\dot{X}_t^{\varepsilon,\epsilon}(\omega,\omega'))_\epsilon = -a(X_t^{\varepsilon,\epsilon}(\omega,\omega'))_\epsilon^3 + b(X_t^{\varepsilon,\epsilon}(\omega,\omega'))_\epsilon + A\sin(\Omega \cdot t) +$$

$$+ B\cos(\Theta \cdot t) + \sqrt{D}\,(W_\epsilon(t,\omega))_\epsilon + \sqrt{\varepsilon}\,W(t,\omega')$$

$$a > 0, \varepsilon \ll 1,$$

$$(X_0^{\varepsilon,\epsilon}(\omega,\omega'))_\epsilon = x_0.$$

(3.1.38)

Ito's SDE:

$$\dot{X}_t(\omega) = -aX_t^3(\omega) + bX_t^2(\omega) + cX_t(\omega) + A\sin(\Omega \cdot t) +$$

$$+B\cos(\Theta \cdot t) + \sqrt{D}\,W(t,\omega),$$

(3.1.39)

$$a > 0, \varepsilon \ll 1,$$

$$X_0(\omega, \omega') = x_0.$$

From general stochastic master equation (3.1.31) one obtain the next Ito's linear master equation:

$$\dot{u}(t) = -(3a\lambda^2 - 2b\lambda - c)u(t) - (a\lambda^3 - b\lambda^2 - c\lambda) +$$
$$+A\sin(\Omega \cdot t) + +B\cos(\Theta \cdot t) + \sqrt{D}\,W(t,\omega),$$

(3.1.40)

$$u(t) = x_0 - \lambda.$$

From the Ito's linear master equation (3.1.40) one obtain the next stochastic transcendental master equation:

$$(x_0 - \lambda(t,\omega))\exp[-(3a\lambda^2(t,\omega) - 2b\lambda(t,\omega) - c) \cdot t] -$$

$$-(a\lambda^3(t,\omega) - b\lambda^2(t,\omega) - c\lambda)\int_0^t \exp[-(3a\lambda^2(t,\omega) - 2b\lambda(t,\omega) - c)(t-\tau)]d\tau +$$

(3.1.41)

$$+\int_0^t [A\sin(\Omega \cdot \tau) + B\cos(\Theta \cdot \tau)]\exp[-(3a\lambda^2(t,\omega) - 2b\lambda(t,\omega) - c)(t-\tau)]d\tau +$$

$$+\sqrt{D}\,W(t,\omega).$$

**Numerical simulation. Example.1.** $a = 1, b = 0, c = 0, A = 0.5, \Omega = 5,$

$\Theta = 0, D = 10^{-3}.$

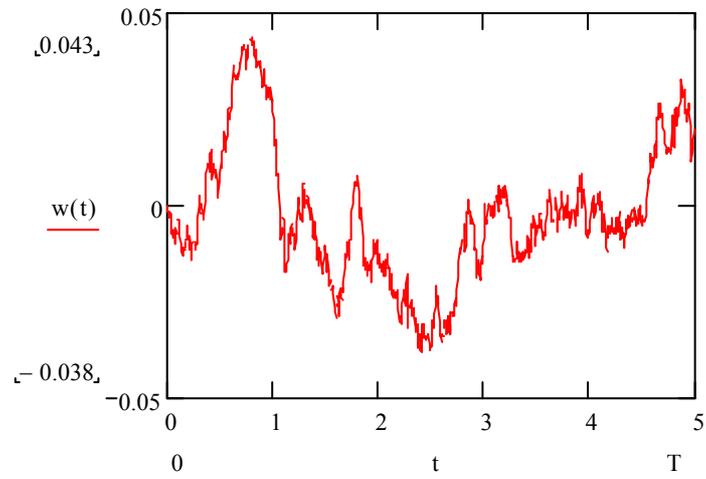

**Pic.1.1.** The realization of a Wiener process $w(t) = \sqrt{D}\,W(t,\omega), D = 10^{-3}$ where $W(t,\omega)$ is a standard Wiener process.

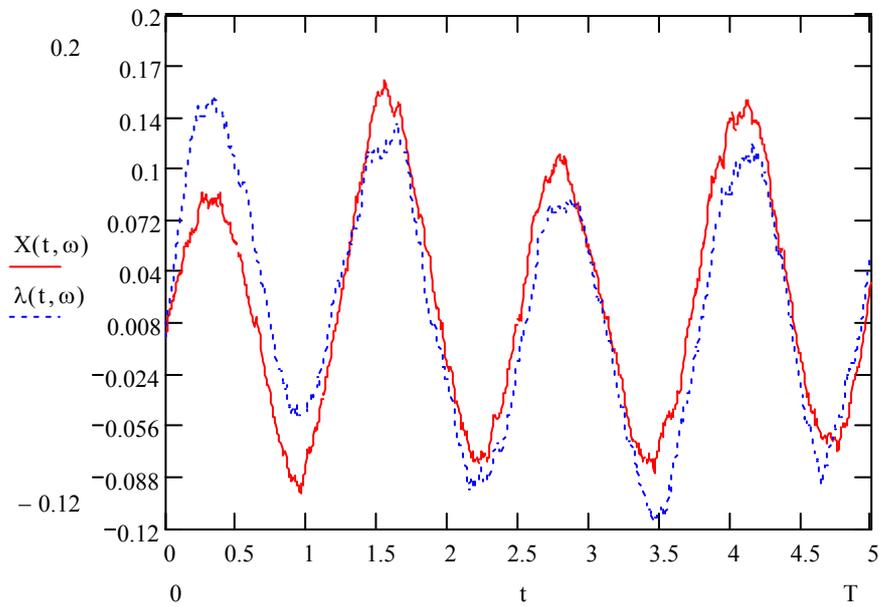

**Pic.1.2.** Numerical solution $X(t,\omega) = X_t(\omega)$ (red curve) in comparison with '$\varepsilon$-limit' stochastic dynamics $\lambda(t,\omega)$ given from **GLDP** (blue curve)

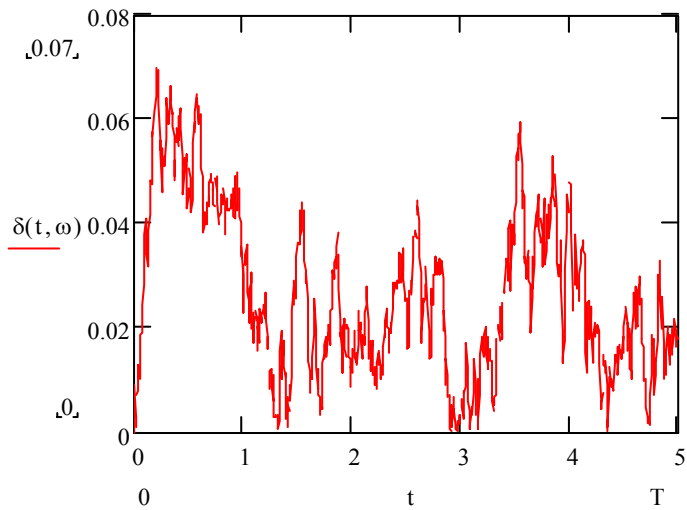

**Pic.1.3.** Stochastic norm: $\delta(t,\omega) = |X_t(\omega) - \lambda(t,\omega)|$.

**Numerical simulation. Example.2.** $a = 1, b = 0, c = 0, A = 0.5, \Omega = 5,$

$\Theta = 0, D = 10^{-3}$.

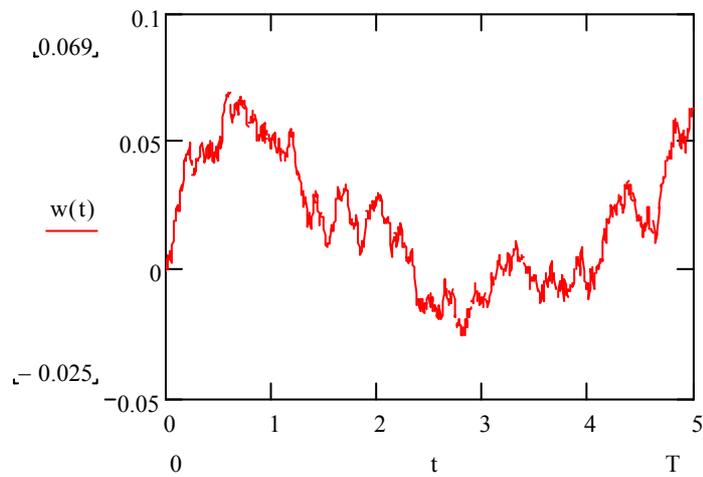

**Pic.2.1.** The realization of a Wiener process
$w(t) = \sqrt{D}\, W(t,\omega), D = 10^{-3}$ where $W(t,\omega)$ is a standard Wiener process.

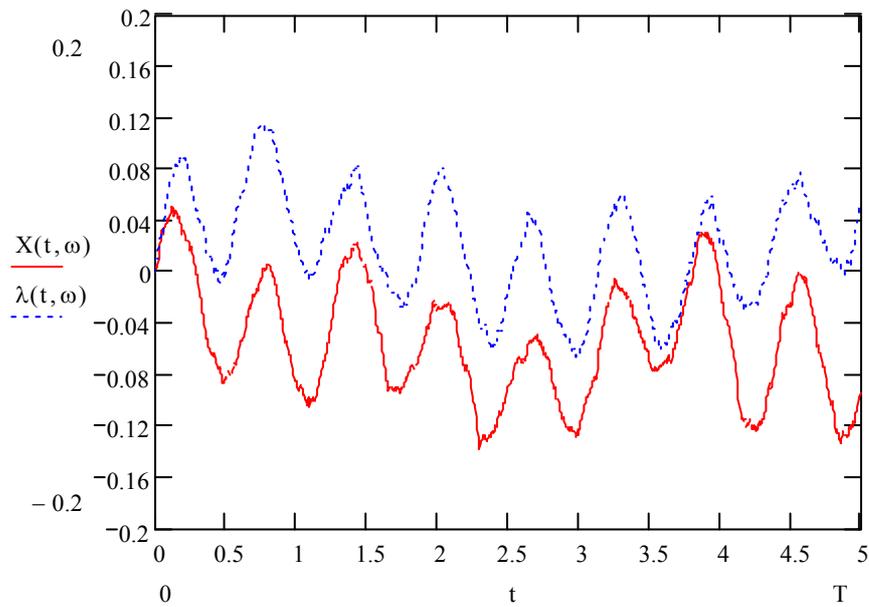

**Pic.2.2.** Numerical solution $X(t,\omega) = X_t(\omega)$ (red curve) in comparison with 'ε-limit' stochastic dynamics $\lambda(t,\omega)$ given from **GLDP** (blue curve)

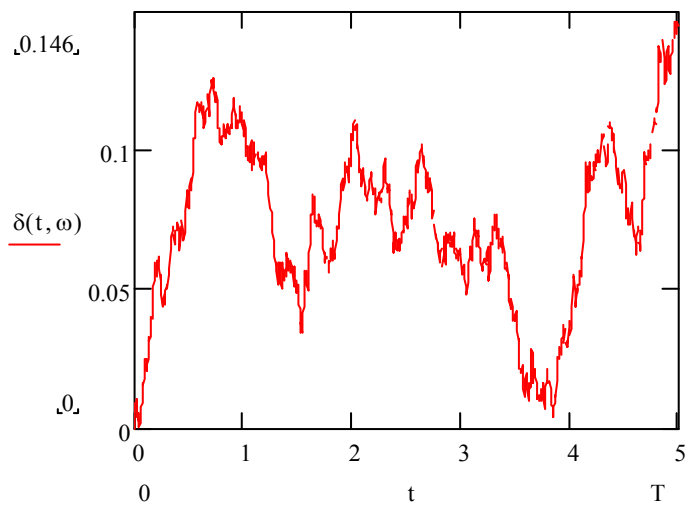

**Pic.2.3.** Stochastic norm: $\delta(t,\omega) = |X_t(\omega) - \lambda(t,\omega)|$.

**Numerical simulation. Example.3.** $a = 1, b = 0, c = -1, A = 0.5, \Omega = 5,$

$\Theta = 0, D = 10^{-3}$.

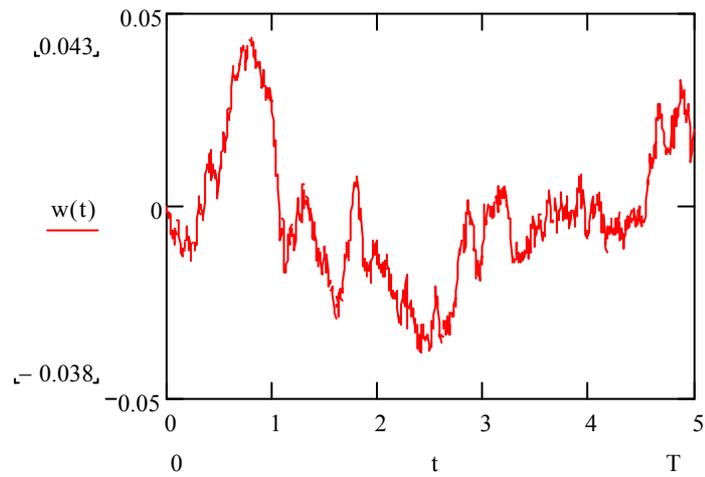

**Pic**.**3**.**1**.The realization of a Wiener process
$w(t) = \sqrt{D}\, W(t,\omega), D = 10^{-3}$ where $W(t,\omega)$ is a standard Wiener process.

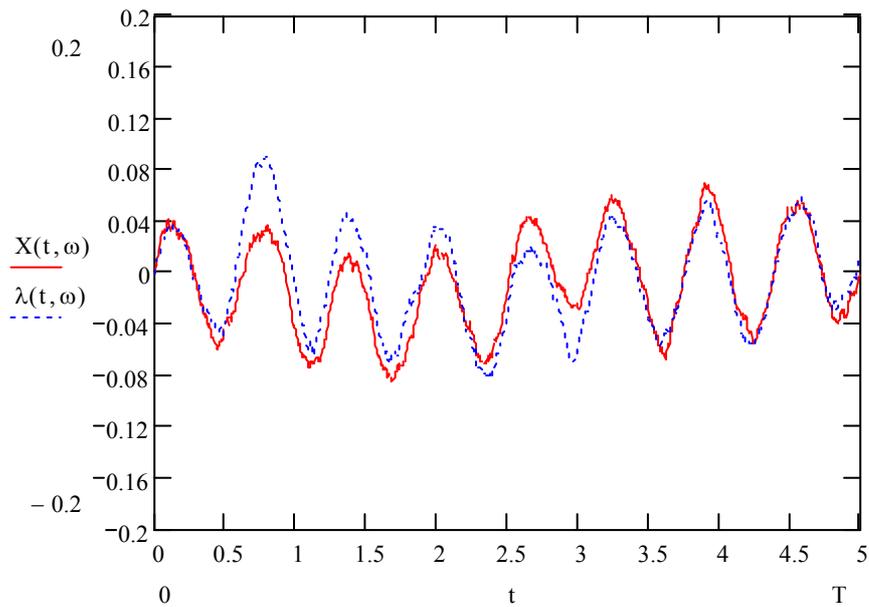

**Pic**.**3**.**2**.Numerical solution $X(t,\omega) = X_t(\omega)$ (red curve) in comparison with 'ε-limit' stochastic dynamics $\lambda(t,\omega)$ given from **GLDP** (blue curve)

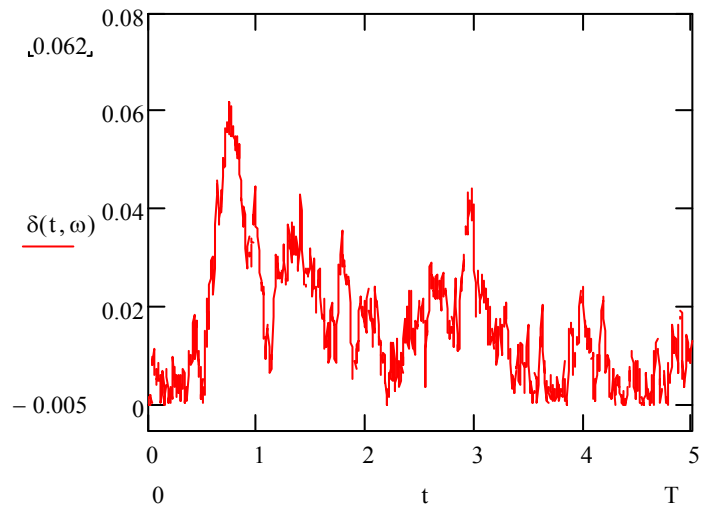

**Pic.3.3.** Stochastic norm: $\delta(t,\omega) = |X_t(\omega) - \lambda(t,\omega)|$.

**Numerical simulation. Example.4.** $a = 1, b = 0, c = -1, A = 0.5, \Omega = 5,$ $\Theta = 0, D = 10^{-2}$.

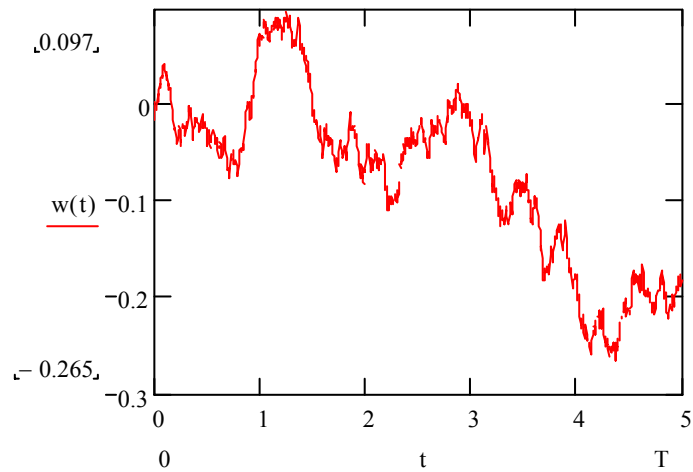

**Pic.4.1.** The realization of a Wiener process $w(t) = \sqrt{D}\,W(t,\omega), D = 10^{-2}$ where $W(t,\omega)$ is a standard Wiener process.

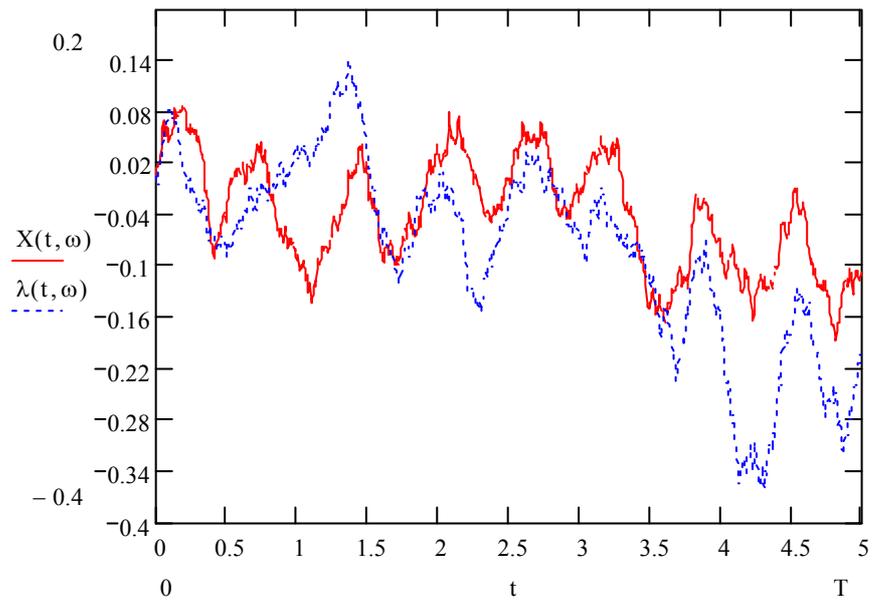

**Pic.4.2.** Numerical solution $X(t,\omega) = X_t(\omega)$ (red curve) in comparison with 'ε-limit' stochastic dynamics $\lambda(t,\omega)$ given from **GLDP** (blue curve).

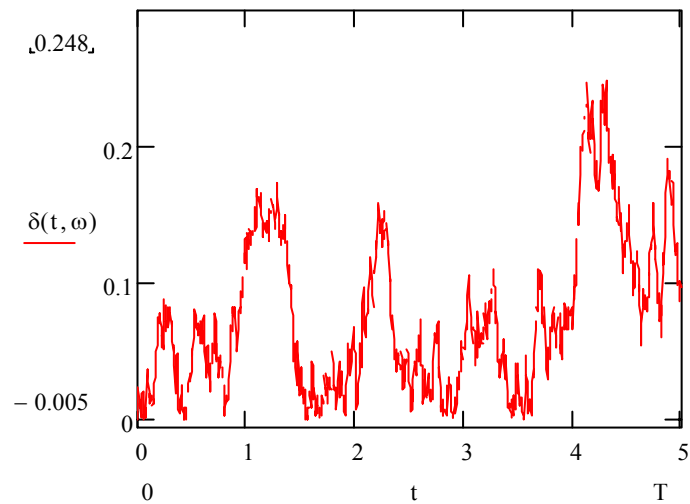

**Pic.4.3.** Stochastic norm: $\delta(t,\omega) = |X_t(\omega) - \lambda(t,\omega)|$.

**Numerical simulation. Example.5.** $a = 1, b = 0, c = -1, A = 0.5, \Omega = 5,$ $\Theta = 0, D = 10^{-1}$.

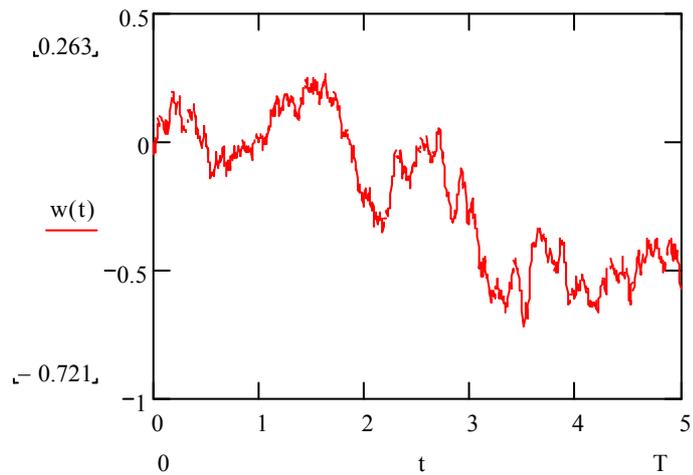

**Pic.5.1.** The realization of a Wiener process $w(t) = \sqrt{D}\, W(t,\omega), D = 10^{-1}$ where $W(t,\omega)$ is a standard Wiener process.

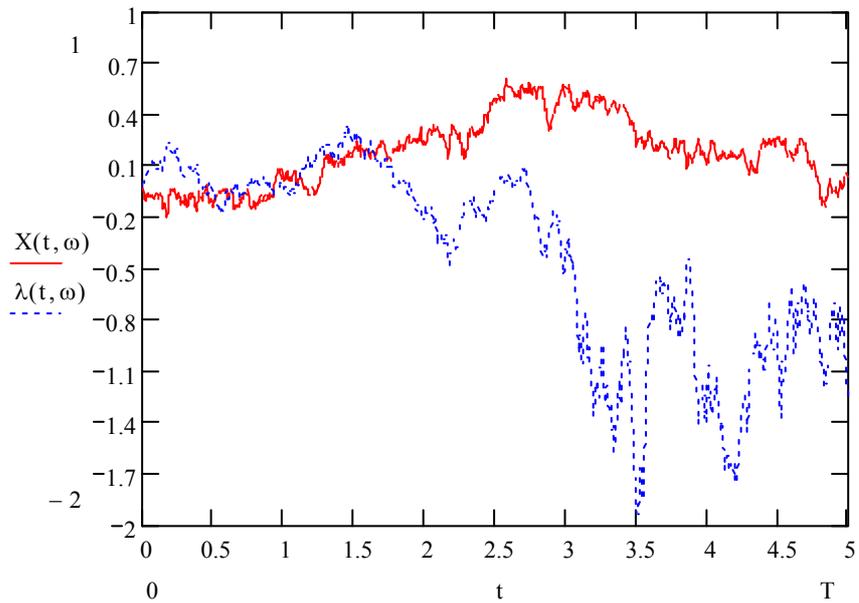

**Pic.5.2.** Numerical solution $X(t,\omega) = X_t(\omega)$ (red curve) in comparison with 'ε-limit' stochastic dynamics $\lambda(t,\omega)$ given from **GLDP** (blue curve).

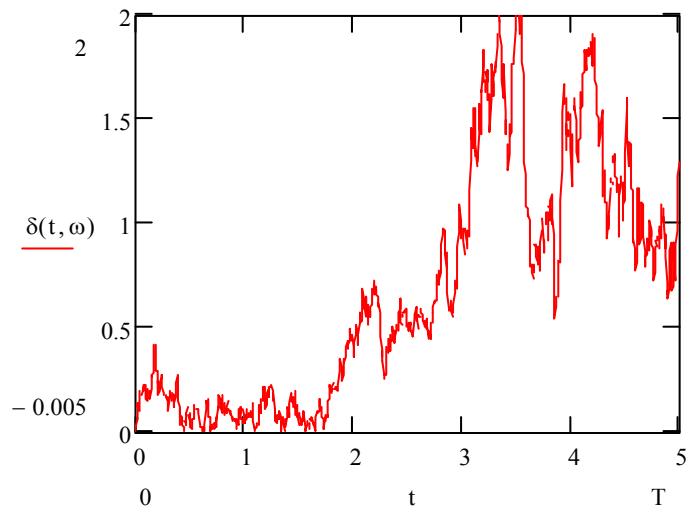

**Pic.5.3.** Stochastic norm: $\delta(t, \omega) = |X_t(\omega) - \lambda(t, \omega)|$.

**Numerical simulation. Example.6.** $a = 1, b = 0, c = -1, A = 0.5, \Omega = 5,$ $\Theta = 0, D = 1.$

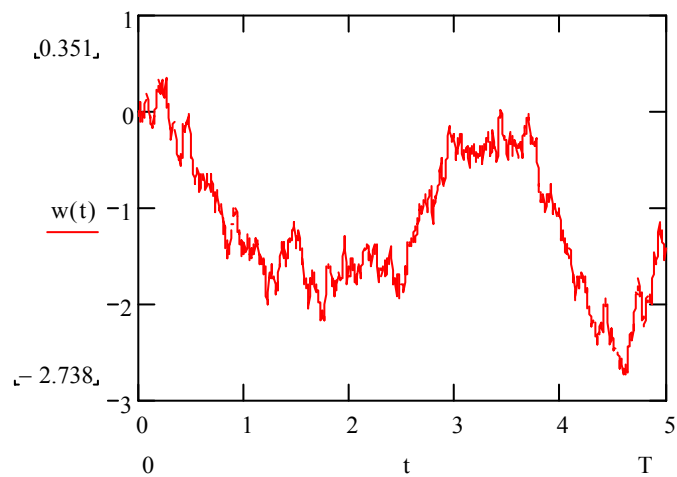

**Pic.6.1.** The realization of a Wiener process $w(t) = \sqrt{D}\,W(t, \omega), D = 10^{-1}$ where $W(t, \omega)$ is a standard Wiener process.

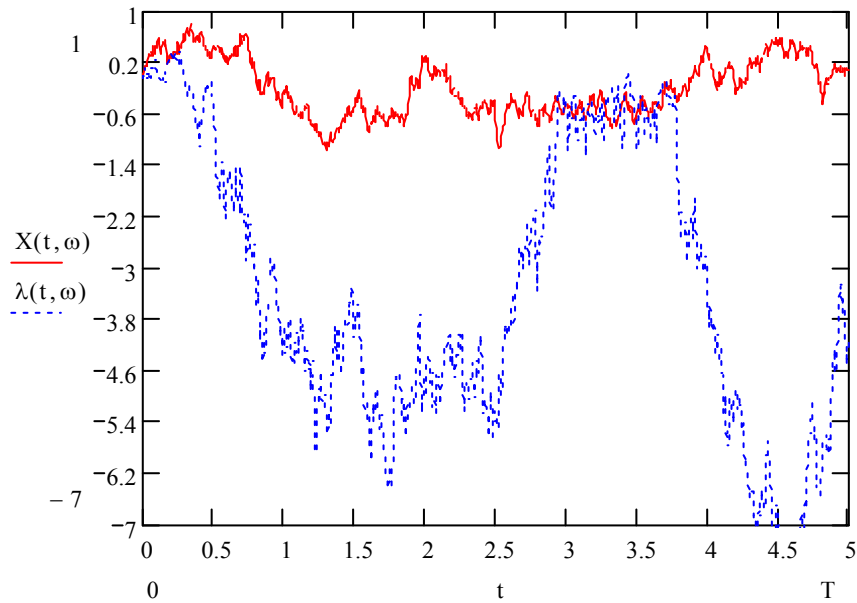

**Pic.6.2.** Numerical solution $X(t,\omega) = X_t(\omega)$ (red curve) in comparison with 'ε-limit' stochastic dynamics $\lambda(t,\omega)$ given from **GLDP** (blue curve).

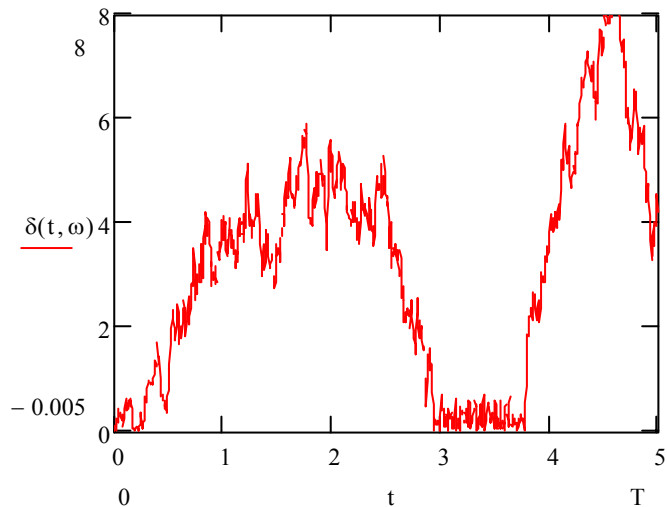

**Pic.6.3.** Stochastic norm: $\delta(t,\omega) = |X_t(\omega) - \lambda(t,\omega)|$.

## III.2. $m$-Persons antagonistic differential game $\mathbf{IDG}_{m;T}^{\#}(\mathbf{f}, g, \mathbf{M}, \mathbf{y})$, with non-linear dynamics and with incomplete or imperfect information about the system.

In this section, we study a differential game of incomplete and imperfect information.Imperfect information relates to the fact that the information we have on the underlying dynamics can not be measured perfectly, i.e. is corrupted by some form of noise.The notion of incomplete information reflects the common fact that we can only observe the dynamic system partially. In the context of linear quadratic differential games was considered the case of incomplete information (see [1], [2],[3]),where was concluded that it might be more interesting to introduce imperfect information.The game we consider is a direct generalization of the **NDG** optimal control problem to a two player differential game.

An approach is proposed for solving problems of positional control of a dynamical system under conditions of incomplete information about its current phase states.

## III.2.1. 2-Persons antagonistic differential game $\text{IDG}_{2;T}^{\#}(\mathbf{f},g,\mathbf{y},\widetilde{\mathbf{w}})$, with linear dynamics and imperfect information about the system.

Let us consider the linear controlled system

$$\dot{x}(t) = \alpha_1(t) - \alpha_2(t) + \widetilde{w}(t),$$

$$\alpha_1(t) \in A_{1,t}, \alpha_2(t) \in A_{2,t}.$$

(3.2.1)

Here, $x$ is the $n$-dimensional phase vector, $\alpha_1(t)$ and $\alpha_2(t)$ are the control vectors of the first and second player, respectively; $\widetilde{w}(t)$ is a summable vector function, i.e. $\int_0^T |\widetilde{w}(t)|\,dt < \infty$, $A_{1,t}$ and $A_{2,t}$ are convex compacta, continuous with respect to $t \in [t_0, T]$ and the instants $t_0$ and $T > t_0$ are given. Notice that any linear controlled system can be reduced to the form (3.2.1) by a nonsingular linear transformation [7].

In the phase space we are given the convex compacta **M** and $\mathbf{R}_0$, where $\mathbf{R}_0$ contains the initial state $x_0$ of the system (at the instant $t_0$). The first player tries to move the system at the instant $T$ into the least possible neighborhood of **M**. The information on the basis of which he constructs his control $\alpha_1(t)$ is as follows:

**(i)** $\mathbf{R}_{t_0} = \mathbf{R}_0$,

**(ii)** At every instant $t > t_0$ he knows some convex compactum $R_t$, containing the phase vector $x(t)$, and he remembers all the $\mathbf{R}_\tau$, $\tau \in [t_0, t)$ (if $t > t_0$),
**(iii)** In addition, he knows the control resources $A_{i,t}, i = 1, 2$ and the law of variation of the set $\mathbf{R}_t$ with respect to $t$, this law amounts to the following:
  **(a)** if, in the interval $[t_1, t_2]$ the controls $\alpha_i(t), i = 1, 2$ are realized, then

$$\mathbf{R}_{t_2} \subset \mathbf{R}_{t_1} + \int_{t_1}^{t_2} [\alpha_1(\xi) - \alpha_2(\xi) + \tilde{w}(t)] d\xi, \tag{3.2.2}$$

**(b)** $\mathbf{R}_t$ is right continuous with respect to $t$. During the motion, the first player may encounter any conceivable realization of this law.

Henceforth, instead of the sets $\mathbf{R}_t$, we shall operate with their support functionals

$$h_t(l) \triangleq \max_{y \in \mathbf{R}_t} (y'l). \tag{3.2.3}$$

Let us refine the statement of the problem [40].

**Definition** *3.2.1. We define a position as a pair $p = \{t, h_\tau(l)\}$, where $t \in [t_0, T]$ and $h_\tau(l)$ is a functional of $\tau \in [t_0, t]$ and $l \in \mathbb{R}^n$, $\|l\| \leq 1$; this functional is linearly convex with respect to $l$, right continuous with respect to $\tau$, uniformly with respect to $l$, and is such that, for any*

$$\tau, \delta, t_0 \leq \tau \leq \tau + \delta \leq t, h_{\tau+\delta}(l) \leq h_\tau(l) \leq h_\tau(l) + \varkappa(\delta),$$

*where*

$$\varkappa(\delta) = \max_{\xi \in [t_0, T]} \int_\xi^{\xi+\delta} \max_{\alpha_1(\xi) \in A_1, \alpha_2(\xi) \in A_2} [\|\alpha_1(\xi) - \alpha_2(\xi) + \tilde{w}(t)\|] d\xi \tag{3.2.4}$$

for all $l$ such that $\|l\| \leq 1$.

**Definition 3.2.2.** *The sequence of positions $p_i = \{t, h^i_\tau(l)\}$ is convergent from the right to the position $p = \{t, h_\tau(l)\}$ ( $p_i \to p$ from the right ), if $t_i \to t + 0$, and, for any $\varepsilon > 0$,*

$$\max_{\tau \in [t_0, t], \|l\| \leq 1} |h^i_\tau(l) - h_\tau(l)| < \varepsilon$$

$$\max_{\tau \in [t_0, t_i], \|l\| \leq 1} |h^i_\tau(l) - h_\tau(l)| < \varepsilon$$

(3.2.5)

*for all sufficiently large $i$.*

**Definition 3.2.3.** *We define a first player's strategy $\mathbf{U} = \mathbf{U}(\circ)$ as a rule whereby, with every position $p = \{t, h_\tau(l)\}$ is associated a set*

$$\mathbf{U}(p) = \mathbf{U}(t, h_\tau(l)) \subset A_{1,t}.$$

*A strategy $\mathbf{U}$ is admissible if the sets $\mathbf{U}(p)$ are convex, closed, and upper semicontinuous, i.e., are such that, if $p_i \to p$ from the right, then, given any $\varepsilon > 0$, we have $\mathbf{U}(p_i)$ a $U^\varepsilon(p)$ for all sufficiently large $i$, ($\mathbf{U}^\varepsilon(p))$ is the $\varepsilon$-neighborhood of $\mathbf{U}(p))$.*

**Definition 3.2.4.** *We define the movement from the position $p^* = \{t^*, h^*\}$ corresponding to the strategy $U(p)$ of the first player, as the functional $h_t = h_t(l) = h_t(l; p^*, U)$ of $t \in [t_0, T]$ and $l \in \mathbb{R}^n, \|l\| < 1$, which is linearly convex with respect to $l$ and satisfies the conditions:*

*(a) for $t \in [t_0, t^*], h_t(l) = h^*_t(l);$*

*(b) $h_t(l)$ is right continuous with respect to $t$ uniformly with respect to $l$;*

*(c) there exist summable functions $\alpha^*_1[\xi], \alpha^*_2[\xi], \xi \in [t^*, T]$ such that, for almost all $\xi \in [t^*, T], \alpha^*_1[\xi] \in U(t, h_\tau(l)), \alpha^*_2[\xi] \in A_{2,\xi}$, and for arbitrary $t_1, t_2 \in [t^*, T], t_1 \leq t_2$, and for all $l$, such that*

$\|l\| < 1$, *we have*

$$h_{t_2}(l) \leq h_{t_1}(l) + \int_{t_1}^{t_2}[\alpha_1(\xi) - \alpha_2(\xi) + \widetilde{w}(t)]d\xi.$$

**Definition** 3.2.5. *We call $\alpha_1^*[\xi]$ and $\alpha_2^*[\xi]$ the controls realized during the motion. We denote the set of the movements thus defined by*

$$\mathbf{D}(p^*, \mathbf{U}).$$

**Theorem** 3.2.1. *[40] Given any position $p^*$ and first player's admissible strategy $\mathbf{U}$, the set $\mathbf{D}(p^*, \mathbf{U})$ is nonempty.*

*One can prove this theorem by introducing a "third player's" strategy, associating, with every segment*
$$[x(\tau)] \triangleq \{x(\tau)|t_0 < \tau < t\},$$
*of the phase vector trajectory, a family of convex compacta $\mathbf{R}_\tau$ such that $x(\tau) \in \mathbf{R}_\tau, t_0 < \tau < t$, and thereby a position $p_t = \{t, h_\tau(l)\}$, where $h_\tau(l)$ is the support functional of $\mathbf{R}_\tau$; the strategy $\mathbf{U}$ can here be treated as a strategy in a game with perfect information and memory, and the usual method for proving existence theorems of a similar type can be employed.*

**Definition** 3.2.6. *In the set of all measurable functions $h(l), l \in \mathbb{R}^n$, $\|l\| < 1$, we define a functional*

$$\sigma(h) = max\left\{0, \mathbf{vrai}\max_{\|l\|<1} [h(l) - \mu(l)]\right\}, \qquad (3.2.6)$$

*where $\mu(l)$ is the support functional of $\mathbf{M}$. For a first player's admissible strategy $\mathbf{U}$ and a position $p$, we put*

$$\gamma(p, \mathbf{U}) = \sup_{h_t \in \mathbf{D}(p, \mathbf{U})} \sigma(h_T) \qquad (3.2.7)$$

**Problem** 3.2.1. Given the initial position $p_0 = \{t_0, h^0(l)\}$, the instant $T > t_0$, and the convex compactum $M \subset \mathbb{R}^n$. It is required to construct an admissible strategy $U^0$ for the first player, such that, given any movement $h_t \in D(p^0, U^0)$, the following condition is satisfied:
whatever the instant $t^* \in [t_0, T]$ and first player's admissible strategy $U$, we have $\gamma(p_{t^*}, U) \leq \gamma(p_{t^*}, U^0)$, where $p_{t^*} = \{t^*, h_\tau(l)\}$.

In this section we define the first player's extremal strategy. It has been proved that it is admissible under certain assumptions [40].
Let $H$ be Hilbert space of functions $h = h(l)$, $l \in \mathbb{R}^n$, $\|l\| < 1$ measurable and square summable in $S = \{l \mid \|l\| < l\}$, and with the norm $\|h\|^* = \left(\int_S h(l)dl\right)^{1/2}$, where the integration is Lebesgue; $<h, g>$ is the scalar product in $H$. For $\alpha > 0$, we put

$$L_\alpha = \{g \mid g \in H, \sigma(g) < \alpha\}. \qquad (3.2.8)$$

**Definition** 3.2.7. The set $W(t, \alpha)$ of programmed $\alpha$-absorption of the target $M$ from the instant $t < T$ will be defined as the collection of all elements $h \in H$ for which the following condition holds: whatever the second player's control $\alpha_2(\xi)$, there exists a first player's control $\alpha_1(\xi)$ such that the element

$$g(l) = h(l) + l' \int_t^T [\alpha_1(\xi) - \alpha_2(\xi) + \widetilde{w}(t)]d\xi \qquad (3.2.9)$$

lies in $L_\alpha$.

Here and below, by the first (second) player's control we understand the measurable functions $\alpha_i(\xi), i = 1, 2$, $t_0 \leq \xi \leq T$, which satisfies, for almost all $l$, the condition $\alpha_1(\xi) \in A_{1,\xi}, \alpha_2(\xi) \in A_{2,\xi}$.

**Definition** 3.2.8. We say that a first player's strategy $U^e$ is extremal if it is specified by the following sets $U^e(p)$: for $p = \{t, g_\tau(l)\}$

$$\mathbf{U}^e \triangleq \left\{ \alpha_1^*(t) \in A_{1,t} | q'(p)\alpha_1^* = \max_{\alpha_1(t) \in A_{1,t}} q'(p)\alpha_1 \right\} \qquad (3.2.10)$$

where

$$q'(p) = \int_S [e(l;p) - g_t(l)] l' dl; \qquad (3.2.11)$$

$e(l;p)$ is the element, closest to $g_t(l)$ in $H$, of the set $\mathbf{W}(t, \beta(p))$, $\beta(p) = \inf\{A(p)\}$, where the set $\mathbf{A}(p)$ is such that: $\alpha \in \mathbf{A}(p)$ if and only if, for some $\tau \in [t_0, t], g_\tau(l) \in \mathbf{W}(t, \alpha)$.

**Definition** 3.2.9. The sets $\mathbf{W}(t, \alpha)$ are called stable in $[t_*, T], t_* \geq t_0$ for $\alpha = \alpha_*$, if, for any instants $t_1 \in [t_*, T], t_2 \in [t_1, T]$, any element $h \in W(t, \alpha)$, and any second player's control $\alpha_2(\xi)$, there exists a first player's control $\alpha_1(\xi)$ such that the element

$$g(l) = h(l) + \int_{t_1}^{t_2} [\alpha_1(\xi) - \alpha_2(\xi) + \widetilde{w}(t)] d\xi$$

lies in $\mathbf{W}(t_2, \alpha)$.

It will be assumed throughout, that the following condition holds.

**Condition** 3.2.1. If, for $\alpha_* \geq 0$ and $t_* \geq t_0$, the set $\mathbf{W}\{t_*, \alpha\}$ is nonempty, then the sets $\mathbf{W}(t, \alpha)$ are stable in $[t_*, T]$ for $\alpha = \alpha_*$.
Let $p_j = \{t_j, g_\tau^j(l)\}, j = 1, \ldots, p^* = \{t^*, g_\tau^*(l)\}$ be positions, $p_j \to p^*$, from the right, $\alpha_j = \inf\{A(p_j)\}, j = 1, \ldots, \alpha^* = \inf\{A(p^*)\}$; let $e^j$ be the element, closest to $g_{t_j}^j$ in $H$, belong to $W(t_j, \alpha_j)$ ,$j = 1, \ldots$; and $e^*$ the element, closest to $g_{t^*}^*$ in $H$, belonging to $W(t^*, \alpha^*)$. Since $W(t, \alpha)$ is convex and closed, and the sphere in Hilbert space is strictly convex, the elements $e_i$ and $e^*$ exist and are unique.

**Theorem** 3.2.2.[40]. Under Condition 3.2.1 the strategy $U^e$ is admissible.
**Theorem** 3.2.3.[40]. Under Condition 3.2.1, the strategy $U^e$ solves Problem 3.2.1.

# III.3.1. 2-Persons antagonistic differential game $DG^{\#}_{2;T}(\mathbf{f}, g, \mathbf{y}, , \widetilde{\mathbf{w}})$, with non-linear dynamics and imperfect information about the system.

Let us consider the non-linear dissipative controlled system:

$$\dot{x}(t) = f(x + \widetilde{w}(t), t) + \alpha_1(t) - \alpha_2(t), \qquad (3.3.1)$$

$$\alpha_1(t) \in A_{1,t}, \alpha_2(t) \in A_{2,t}.$$

Here, $x$ is the $n$-dimensional phase vector, $\alpha_1(t)$ and $\alpha_2(t)$ are the control vectors of the first and second player, respectively; $\widetilde{w}(t)$ is a summable vector function, (i.e. $\int_0^T |\widetilde{w}(t)| dt < \infty$) such that $\int_0^T \|f(\widetilde{w}(t), t)\| dt < \infty$ and $A_{1,t}, A_{2,t}$ are convex compacta, continuous with respect to $t \in [t_0, T]$ and the instants $t_0$ and $T > t_0$ are given.

In the phase space we are given the convex compacta $\mathbf{M}$ and $\mathbf{R}_0$, where $\mathbf{R}_0$ contains the initial state $x_0$ of the system (at the instant $t_0$). The first player tries to move the system at the instant $T$ into the least possible neighborhood of $\mathbf{M}$. The information on the basis of which he constructs his control $\alpha_1(t)$ is as follows:

(i) $\mathbf{R}_{t_0} = \mathbf{R}_0$,
(ii) At every instant $t > t_0$ he knows some convex compactum $R_t$, containing the phase vector $x(t)$, and he remembers all the $\mathbf{R}_\tau$, $\tau \in [t_0, t)$ (if $t > t_0$),
(iii) In addition, he knows the control resources $A_{i,t}, i = 1, 2$ and the law of variation of the set $\mathbf{R}_t$ with respect to $t$, this law amounts to the following:
   **(a)** if, in the interval $[t_1, t_2]$ the controls $\alpha_i(t), i = 1, 2$ are realized, then

$$\mathbf{R}_{t_2} \subset \mathbf{R}_{t_1} + \int_{t_1}^{t_2} [f(x(\xi) + \tilde{w}(\xi), \xi) + \alpha_1(\xi) - \alpha_2(\xi)] d\xi, \tag{3.3.2}$$

**(b)** $\mathbf{R}_t$ is right continuous with respect to $t$. During the motion, the first player may encounter any conceivable realization of this law.

Henceforth, instead of the sets $\mathbf{R}_t$, we shall operate with their support functionals

$$h_t(l) \triangleq \max_{y \in \mathbf{R}_t} (y'l). \tag{3.3.3}$$

Let us refine the statement of the problem.

**Definition 3.3.1.** *We define a position as a pair $p = \{t, h_\tau(l)\}$, where $t \in [t_0, T]$ and $h_\tau(l)$ is a functional of $\tau \in [t_0, t]$ and $l \in \mathbb{R}^n$, $\|l\| \leq 1$; this functional is linearly convex with respect to $l$, right continuous with respect to $\tau$, uniformly with respect to $l$, and is such that, for any*

$$\tau, \delta, t_0 \leq \tau \leq \tau + \delta \leq t, h_{\tau+\delta}(l) \leq h_\tau(l) \leq h_\tau(l) + x(\delta),$$

*where*

$$x(\delta) =$$

$$\max_{\xi \in [t_0, T]} \int_\xi^{\xi+\delta} \max_{\alpha_1(\xi) \in A_1, \alpha_2(\xi) \in A_2} [\|f(x(\xi) + \tilde{w}(\xi), \xi) + \alpha_1(\xi) - \alpha_2(\xi) + \tilde{w}(t)\|] d\xi \tag{3.3.4}$$

for all $l$ such that $\|l\| \leq 1$.

**Definition** 3.3.2. *The sequence of positions $p_i = \{t, h_\tau^i(l)\}$ is convergent from the right to the position $p = \{t, h_\tau(l)\}$ ($p_i \to p$ from the right), if $t_i \to t + 0$, and, for any $\varepsilon > 0$,*

$$\max_{\tau \in [t_0, t], \|l\| \leq 1} |h_\tau^i(l) - h_\tau(l)| < \varepsilon$$

$$\max_{\tau \in [t_0, t_i], \|l\| \leq 1} |h_\tau^i(l) - h_\tau(l)| < \varepsilon$$

(3.3.5)

*for all sufficiently large $i$.*

**Definition** 3.3.3. *We define a first player's strategy $\mathbf{U} = \mathbf{U}(\circ)$ as a rule whereby, with every position $p = \{t, h_\tau(l)\}$ is associated a set*

$$\mathbf{U}(p) = \mathbf{U}(t, h_\tau(l)) \subset A_{1,t}.$$

*A strategy $\mathbf{U}$ is admissible if the sets $\mathbf{U}(p)$ are convex, closed, and upper semicontinuous, i.e., are such that, if $p_i \to p$ from the right, then, given any $\varepsilon > 0$, we have $\mathbf{U}(p_i)$ a $\mathbf{U}^\varepsilon(p)$ for all sufficiently large $i$, ($\mathbf{U}^\varepsilon(p)$) is the $\varepsilon$-neighborhood of $\mathbf{U}(p)$).*

**Definition** 3.3.4. *We define the movement from the position $p^* = \{t^*, h^*\}$ corresponding to the strategy $U(p)$ of the first player, as the functional $h_t = h_t(l) = h_t(l; p^*, U)$ of $t \in [t_0, T]$ and $l \in \mathbb{R}^n, \|l\| < 1$, which is linearly convex with respect to $l$ and satisfies the conditions:*
(a) *for $t \in [t_0, t^*], h_t(l) = h_t^*(l)$;*

(b) *$h_t(l)$ is right continuous with respect to $t$ uniformly with respect to $l$;*
(c) *there exist summable functions $\alpha_1^*[\xi], \alpha_2^*[\xi], \xi \in [t^*, T]$ such that, for almost all $\xi \in [t^*, T]$, $\alpha_1^*[\xi] \in U(t, h_\tau(l))$, $\alpha_2^*[\xi] \in A_{2,\xi}$, and for arbitrary $t_1, t_2 \in [t^*, T], t_1 \leq t_2$, and for all $l$, such that $\|l\| < 1$, we have:*

$$h_{t_2}(l) \leq h_{t_1}(l) + \int_{t_1}^{t_2}[f(x(\xi) + \widetilde{w}(\xi),\xi) + \alpha_1(\xi) - \alpha_2(\xi) + \widetilde{w}(t)]d\xi.$$

**Definition** 3.3.5. We call $\alpha_1^*[\xi]$ and $\alpha_2^*[\xi]$ the controls realized during the motion. We denote the set of the movements thus defined by **D**($p^*$, **U**).

**Definition** 3.3.6. In the set of all measurable functions $h(l), l \in \mathbb{R}^n, \|l\| < 1$, we define a functional

$$\sigma(h) = max\left\{0, \textbf{vrai}\max_{\|l\|<1}[h(l) - \mu(l)]\right\}, \qquad (3.3.6)$$

where $\mu(l)$ is the support functional of **M**. For a first player's admissible strategy **U** and a position $p$, we put

$$\gamma(p, \textbf{U}) = \sup_{h_t \in \textbf{D}(p,\textbf{U})} \sigma(h_T) \qquad (3.3.7)$$

**Theorem** 3.3.1. Given the initial position $p_0 = \{t_0, h^0(l)\}$, the instant $T > t_0$, and the convex compactum $\textbf{M} \subset \mathbb{R}^n$. It is required to construct an admissible strategy $\textbf{U}^0$ for the first player, such that, given any movement $h_t \in \textbf{D}(p^0, \textbf{U}^0)$, the following condition is satisfied: whatever the instant $t^* \in [t_0, T]$ and first player's admissible

**Definition** **Theorem** strategy **U**, we have $\gamma(p_{t^*}, \textbf{U}) \leq \gamma(p_{t^*}, \textbf{U}^0)$, where $p_{t^*} = \{t^*, h_\tau(l)\}$.

In this section we define the first player's extremal strategy. It has been proved that it is admissible under certain assumptions.
Let **H** be Hilbert space of functions $h = h(l), l \in \mathbb{R}^n, \|l\| < 1$ measurable and square summable in $\textbf{S} = \{l \mid \|l\| < 1\}$, and with the norm $\|h\|^* = \left(\int_\textbf{S} h(l)dl\right)^{1/2}$, where the integration is Lebesgue; $<h, g>$ is the scalar product in **H**.
For $\alpha > 0$, we put

$$L_\alpha = \{g | g \in H, \sigma(g) < \alpha\}. \tag{3.3.8}$$

**Definition** 3.3.7. *The set $W(t,\alpha)$ of programmed $\alpha$-absorption of the target $M$ from the instant $t < T$ will be defined as the collection of all elements $h \in H$ for which the following condition holds: whatever the second player's control $\alpha_2(\xi)$, there exists a first player's control $\alpha_1(\xi)$ such that the element*

$$g(l) = h(l) + l' \int_t^T [f(x(\xi) + \widetilde{w}(\xi), \xi) + \alpha_1(\xi) - \alpha_2(\xi) + \widetilde{w}(t)] d\xi \tag{3.3.9}$$

lies in $L_\alpha$. Here and below, by the first (second) player's control we understand the measurable functions $\alpha_i(\xi), i = 1, 2, t_0 \leq \xi \leq T$, which satisfies, for almost all $l$, the condition $\alpha_1(\xi) \in A_{1,\xi}, \alpha_2(\xi) \in A_{2,\xi}$.

**Definition** 3.3.8. *We say that a first player's strategy $U^e$ is extremal if it is specified by the following sets $U^e(p)$: for $p = \{t, g_\tau(l)\}$*

$$U^e \triangleq \left\{ \alpha_1^*(t) \in A_{1,t} | q'(p)\alpha_1^* = \max_{\alpha_1(t) \in A_{1,t}} q'(p)\alpha_1 \right\} \tag{3.3.10}$$

where

$$q'(p) = \int_S [e(l;p) - g_t(l)] l' dl; \tag{3.3.11}$$

$e(l;p)$ *is the element, closest to $g_t(l)$ in $H$, of the set $W(t, \beta(p))$,*

$\beta(p) = \inf\{A(p)\},$ *where the set $A(p)$ is such that: $\alpha \in A(p)$ if and only if, for some $\tau \in [t_0, t], g_\tau(l) \in W(t, \alpha)$.*

**Definition** 3.3.9. *The sets $W(t, \alpha)$ are called stable in $[t_*, T], t_* \geq t_0$ for $\alpha = \alpha_*$, if, for any instants $t_1 \in [t_*, T], t_2 \in [t_1, T]$, any element $h \in W(t, \alpha)$,* and any second player's control $\alpha_2(\xi)$, there exists a first player's control $\alpha_1(\xi)$ such that the element

$$g(l) = h(l) + \int_{t_1}^{t_2} [f(x(\xi) + \widetilde{w}(\xi), \xi) + \alpha_1(\xi) - \alpha_2(\xi) + \widetilde{w}(t)] d\xi$$

lies in $W(t_2, \alpha)$.

It will be assumed throughout, that the following condition holds.

**Condition** 3.3.1. If, for $\alpha_* \geq 0$ and $t_* \geq t_0$, the set $W\{t_*, \alpha\}$ is nonempty, then the sets $W(t, \alpha)$ are stable in $[t_*, T]$ for $\alpha = \alpha_*$. Let

$p_j = \{t_j, g_\tau^j(l)\}, j = 1, \ldots, p^* = \{t^*, g_\tau^*(l)\}$ be positions, $p_j \to p^*$, from the right, $\alpha_j = inf\{A(p_j)\}, j = 1, \ldots, \alpha^* = inf\{A(p^*)\}$; let $e^j$ be the element, closest to $g_{t_j}^j$ in $H$, belong to $W(t_j, \alpha_j), j = 1, \ldots$; and $e^*$ the element, closest to $g_{t^*}^*$ in $H$, belonging to $W(t^*, \alpha^*)$. Since $W(t, \alpha)$ is convex and closed, and the sphere in Hilbert space is strictly convex, the elements $e_i$ and $e^*$ exist and are unique.

**Theorem** 3.3.2. Under Condition 3.3.1 the strategy $U^e$ is admissible.

**Theorem** 3.3.3. Under Condition 3.3.1, the strategy $U^e$ solves Problem 3.3.1.

# III.3.2. 2-Persons antagonistic differential game $IDG_{2;T}^{\#}(f, g, y, \widetilde{w})$, with non-linear dynamics and imperfect information about the system imbeded into a small white noise. "Step-by-step" strategy.

# III.3.2.1. Strong large deviations principle of Non-Freidlin-Wentzell type for infinitesimal stochastic differential game with non-linear dynamics and imperfect information about the system.

Let us consider the non-linear dissipative controlled system:

$$\dot{x}(t) = f(x + \widetilde{w}(t), t) + \alpha_1(t) - \alpha_2(t),$$

(3.3.12)

$$\alpha_1(t) \in A_{1,t}, \alpha_2(t) \in A_{2,t}.$$

Here, $x$ is the $n$-dimensional phase vector, $\alpha_1(t)$ and $\alpha_2(t)$ are the control vectors of the first and second player, respectively; $\widetilde{w}(t)$ is a summable vector function, (i.e. $\int_0^T |\widetilde{w}(t)| dt < \infty$) such that $\int_0^T \|f(\widetilde{w}(t), t)\| dt < \infty$ and $A_{1,t}, A_{2,t}$ are convex compacta, continuous with respect to $t \in [t_0, T]$ and the instants $t_0$ and $T > t_0$ are given.

Here $t \to \alpha_i(t)$ is the determined control chosen by the $i$-th player, within a set of admissible control values $U_i$, and the payoff for the

$i$-th player:

$$\mathbf{J}_i = \int_0^T g_i(x_1, \ldots, x_m; \alpha_1, \ldots, \alpha_m) dt + \sum_{i=1}^m [x_i(T) - y_{1,i}]^2.$$

(3.3.13)

where $t \mapsto x(t)$ is the trajectory of (3.3.12). Optimal control problem for the $i$-th player:

$$\min_{\alpha_i(t)} \left( \max_{\alpha_j(t), j \neq i} \mathbf{J}_i \right).$$

(3.3.14)

Let us consider an $m$-persons stochastic differential game $SIDG_{m;T}(\mathbf{f}, g, \mathbf{y}, \widetilde{\mathbf{w}})$, with nonlinear dynamics and with imperfect information about the sistem:

$$\dot{x}(t) = f(x + \widetilde{w}(t), t) + \alpha_1(t) - \alpha_2(t) + \sqrt{\varepsilon}\dot{\mathbf{W}}(t), \quad (3.3.15)$$

$$\alpha_1(t) \in A_{1,t}, \alpha_2(t) \in A_{2,t}.$$

Here $t \to \alpha_i(t)$ is the determined control chosen by the $i$-th player, within a set of admissible control values $A_{i,t}$, and the payoff for the $i$-th player:

$$\bar{\mathbf{J}}_i = \mathbf{E}\left[\int_0^T g_i(x_1(t,\omega),\ldots,x_m(t,\omega);\alpha_1(t),\ldots,\alpha_m(t))dt\right] +$$

$$+\mathbf{E}\left[\sum_{i=1}^m [x_i(T,\omega) - y_i]^2\right] \quad (3.3.16)$$

where $t \mapsto x(t,\omega)$ is the trajectory of (3.3.15).

**Definition** *3.3.10. Stochastic differential game $SIDG_{m;T}(\mathbf{f}, g, \mathbf{y}, \widetilde{\mathbf{w}})$ with imperfect information is the determined differential game $IDG_{m;T}^{\#}(\mathbf{f}, g, \mathbf{y}, \widetilde{\mathbf{w}})$ with imperfect information about the system imbeded into a 'small' white noise.*

Let us consider an $m$-persons stochastic differential game $SIDG_{m;\bar{t}}^{\#}(\mathbf{f}, g, \mathbf{y}_1, \widetilde{\mathbf{w}})$, with nonlinear dynamics and imperfect information about the system :

$$\frac{dx_i(t)}{dt} = f_i(x_1 + \widetilde{w}_1, \ldots, x_n + \widetilde{w}_n; \alpha_1, \ldots, \alpha_m) + \sqrt{\varepsilon}\dot{W}(t),$$

$$x(t) \in \mathbb{R}^n,$$

$$\alpha_i(t) \in U_i, i = 1,\ldots,m. \quad (3.3.17)$$

$$t \in [0,T],$$

$$\varepsilon \ll 1.$$

**Theorem** *3.3.4. For the all solutions $\{\mathbf{X}_t^\varepsilon, \check{\alpha}(t,\lambda)\} = \{(X_{1,t}^\varepsilon,\ldots,X_{m,t}^\varepsilon),(\check{\alpha}_1(t,\lambda),\ldots,\check{\alpha}_m(t,\lambda))\}$ dissipative SIDG (3.3.17) and $\mathbb{R}$ valued*

*parameters* $\lambda = (\lambda_1, \ldots, \lambda_m) \in \mathbb{R}^n$, *there exists a Colombeau constant* $(C_\varepsilon)_\varepsilon \in \widetilde{\mathbb{R}}, (C_\varepsilon)_\varepsilon \geq 0$, *such that:*

$$\liminf_{\varepsilon \to 0} \mathbf{E}\left[\|\mathbf{X}_t^\varepsilon - \lambda\|^2\right] \leq (C_\varepsilon)_\varepsilon \|\mathbf{U}(t,\lambda)\|^2,$$

$$\lambda = (\lambda_1, \ldots, \lambda_n) \in \mathbb{R}^n,$$

$$t \in [0, T],$$

(3.3.18)

*where* $\mathbf{U}(t,\lambda) = (U_1(t,\lambda), \ldots, U_n(t,\lambda))$ *is the trajectory of the linear differential master game with imperfect information about the system:*

$$\frac{d\mathbf{U}(t,\lambda)}{dt} = \mathbf{J}[\mathbf{f}(\lambda + \widetilde{\mathbf{w}}, \check{\alpha}(t,\lambda))]\mathbf{U} + \mathbf{f}(\lambda + \widetilde{\mathbf{w}}, \check{\alpha}(t,\lambda)),$$

$$\mathbf{U}(0,\lambda) = x_0,$$

$$\mathbf{J}_i = \|\mathbf{U}(T)\|^2 = \sum_{i=1}^{m} [U_i(T)]^2.$$

$$\min_{\alpha_i(t)} \left(\max_{\alpha_j(t), j \neq i} \mathbf{J}_i\right) = 0.$$

(3.3.19)

Where $\mathbf{J} = \mathbf{J}[\mathbf{f}(\lambda + \widetilde{\mathbf{w}}, \check{\alpha}(t,\lambda))]$ the *imperfect* Jacobian, i.e. $\mathbf{J}$ is a $n \times n$-matrix:

$$\mathbf{J}[\mathbf{f}(\lambda + \widetilde{\mathbf{w}}, \check{\alpha}(t,\lambda))] = \mathbf{J}[\mathbf{f}(x, \check{\alpha}(t,z))]|_{x=\lambda+\widetilde{\mathbf{w}}, z=\lambda} =$$

$$= \begin{bmatrix} \dfrac{\partial f_1(x, \check{\alpha}(t,z))}{\partial x_1} & \cdots & \dfrac{\partial f_1(x, \check{\alpha}(t,z))}{\partial x_n} \\ \cdot & \cdots & \cdot \\ \cdot & \cdots & \cdot \\ \cdot & \cdots & \cdot \\ \dfrac{\partial f_n(x, \check{\alpha}(t,z))}{\partial x_1} & \cdots & \dfrac{\partial f_n(x, \check{\alpha}(t,z))}{\partial x_n} \end{bmatrix}_{x=\lambda+\widetilde{\mathbf{w}}, z=\lambda} \quad (3.3.20)$$

**Theorem** 3.3.5. Assume the conditions of the Theorem 3.3.4 for

**Corollary** any $\lambda = (\lambda_1, \ldots, \lambda_n) \in \mathbb{R}^n, t \in [0, T]$ :

$$\|\mathbf{U}(t, \lambda)\| = 0 \Rightarrow$$

$$\liminf_{\varepsilon \to 0} \mathbf{E}\left[\|\mathbf{X}_t^\varepsilon - \lambda\|^2\right] = 0. \quad (3.3.21)$$

$$\liminf_{\varepsilon \to 0}\left[\min_{\alpha_i(t)}\left(\max_{\alpha_j(t), j \neq i} \bar{\mathbf{J}}_{i,\varepsilon}\right)\right] = 0.$$

*More precisely, for any $t \in [0, T]$ and*
$\lambda = \lambda(t) = (\lambda_1(t), \ldots, \lambda_n(t)) \in \mathbb{R}^n$ *sutch that*

$$U_1(t, \lambda_1(t), \ldots, \lambda_n(t)) = 0,$$

$$\cdots \cdots \cdots \cdots \cdots \cdots \quad (3.3.22)$$

$$U_n(t, \lambda_1(t), \ldots, \lambda_n(t)) = 0,$$

the equalities is satisfaed

$$\liminf_{\varepsilon \to 0} \mathbf{E}\left[ \|X_{1,t}^{\varepsilon} - \lambda_1(t)\|^2 \right] = 0,$$

$$\dots\dots\dots\dots\dots\dots\dots$$

$$\liminf_{\varepsilon \to 0} \mathbf{E}\left[ \|X_{n,t}^{\varepsilon} - \lambda_n(t)\|^2 \right] = 0,$$

$$\liminf_{\varepsilon \to 0} \left[ \min_{\alpha_i(t)} \left( \max_{\alpha_j(t), j \neq i} \bar{\mathbf{J}}_{i,\varepsilon} \right) \right] = 0.$$

(3.3.23)

## III.3.3. 2-Persons stochastic differential games with non-linear dinamic and noise corrupted information.   III.3.3.1. Bellman equation for 2-Persons Ito's stochastic differential games with non-linear dinamic and noise corrupted information.

Let us consider stochastic two-player dynamic system in $\mathbb{R}^n$ be given by a diffusion

$$d\mathbf{X}_{t,D_1} = \mathbf{f}(t, \mathbf{X}_{t,D_1}; \boldsymbol{\alpha}_1, \boldsymbol{\alpha}_2)dt + \sqrt{D_1}\, d\mathbf{W}_1(t) \tag{3.3.24}$$

and observation process in $\mathbb{R}^p$ be given by a diffusion

$$d\mathbf{Y}_{t,D_2} = \mathbf{h}(t, \mathbf{X}_{t,D_1}; \boldsymbol{\alpha}_1, \boldsymbol{\alpha}_2)dt + \sqrt{D_2}\, d\mathbf{W}_2(t) \tag{3.3.25}$$

where $\mathbf{W}_1(t)$ and $\mathbf{W}_2(t)$ are two independant vector brownian motions of covariance coefficient matrices $w_1(t)$ and $w_2(t)$ respectively,

and a performance index be given as

$$\bar{\mathbf{J}} = \mathbf{E}\left[ K(t_1) + \int_{t_0}^{t_1} L(t, \mathbf{X}_{t,D_1}; \alpha_1, \alpha_2) dt \right]. \quad (3.3.26)$$

where $t_0$ and $t_1$ are given time instants.

Standard hypotheses are assumed on $\mathbf{f}(\circ,\circ;\circ,\circ), \mathbf{h}(\circ,\circ;\circ,\circ)$, and the admissible processes $\alpha_1(\circ)$ and $\alpha_2(\circ)$ for $\mathbf{X}_{t,D_1}$ and $\mathbf{Y}_{t,D_2}$ to be well defined, and $K$ and $L$ are supposed to be globally $\mathbf{C}^1$. We consider the game where the minimizer only knows $\mathbf{Y}_{t,D_2}$, with perfect memory, while the maximizer knows $\mathbf{X}_{t,D_1}, \mathbf{Y}_{t,D_2}$ and $\alpha_1(t)$. To stay with saddle points, as opposed to Nash points, we assume that in addition, the minimizer knows $x_0$. We call strategies functions $\breve{\alpha}_1(\xi,t) : \mathbb{R}^n \times \mathbb{R} \to \mathbb{R}^m$ and $\breve{\alpha}_2(x,\xi,t) : \mathbb{R}^n \times \mathbb{R}^n \times \mathbb{R} \to \mathbb{R}^{m'}$ such that, if $\alpha_1(t)$ and $\alpha_2(t)$ are replaced by $\breve{\alpha}_1(\hat{\mathbf{X}}_t, t)$ and $\breve{\alpha}_2(\mathbf{X}_{t,D_1}, \hat{\mathbf{X}}_t, t)$ in Eqs. (3.3.25) and (3.3.27) below, the processes $\mathbf{X}_{t,D_1}$ and $\hat{\mathbf{X}}_t$ are well defined in the Ito sense.

**Theorem**  3.3.6. *If there exist a process $\hat{\mathbf{X}}_t, \mathbf{C}^2$ function*
$V(\mathbf{X}, \hat{\mathbf{X}}, t) : \mathbb{R}^n \times \mathbb{R}^n \times \mathbb{R} \to \mathbb{R},$ *strategies* $\breve{\alpha}_1(\xi, t)$
*and* $\breve{\alpha}_2(x, \xi, t),$ *such that:*

**(i)** the process $\hat{\mathbf{X}}_t$ obeys a diffusion law of the form

$$d\hat{\mathbf{X}}_t = \mathbf{G}_1(\hat{\mathbf{X}}_t, \alpha_1, t) d\mathbf{Y}_{t,D_2} + \mathbf{G}_2(\hat{\mathbf{X}}_t, \alpha_1, t) dt. \quad (3.3.27)$$

Thus

$$d\hat{\mathbf{X}}_t = \mathbf{g}(\mathbf{X}_t, \hat{\mathbf{X}}_t, \alpha_1, t) dt + \mathbf{k}(\mathbf{X}_t, \hat{\mathbf{X}}_t, \alpha_1, t) d\mathbf{W}_2(t) \quad (3.3.28)$$

and is such that, whenever $\alpha_2(t) = \breve{\alpha}_2(\mathbf{X}_t, \hat{\mathbf{X}}_t, t)$, the conditional law of $\mathbf{X}_t$ given the information algebra $\mathfrak{I}_t^y$ generated by $\{y(s)|s < t\}$, is of the form $\pi_{\hat{\mathbf{X}}}(\mathbf{X}) = \pi(\mathbf{X} - \hat{\mathbf{X}}, t)$, where $\pi$ it is a fixed zero-mean law, i.e. independant of the control process $\alpha_1(\circ)$.

(ii) $\forall (x, \hat{x}, t) \in \mathbb{R}^n \times \mathbb{R}^n \times [t_0, t_1]$ :

$$\frac{\partial V}{\partial t} + \frac{\partial V}{\partial x}f(t,x,\check{\alpha}_1,\check{\alpha}_2) + \frac{\partial V}{\partial \hat{x}}g(x,\hat{x},\alpha_1,t) + \frac{D_1}{2}\mathbf{tr}\left(\frac{\partial^2 V}{\partial x^2}w_1(t)\right) +$$

$$+\frac{D_2}{2}\mathbf{tr}\left(\frac{\partial^2 V}{\partial \hat{x}^2}w_2(t)\right) + L(t,x;\alpha_1,\alpha_2) = 0 \qquad (3.3.29)$$

$$\forall (x,\hat{x}) \in \mathbb{R}^n \times \mathbb{R}^n : V(x,\hat{x},t_1) = K(x).$$

(iii) $\forall (x,\hat{x},t) \in \mathbb{R}^n \times \mathbb{R}^n \times [t_0,t_1]$ and $\forall \alpha_2 \in \mathbb{R}^{m'}$ :

$$\frac{\partial V}{\partial x}f(t,x,\check{\alpha}_1,\check{\alpha}_2) + L(t,x;\check{\alpha}_1,\check{\alpha}_2) \geq \frac{\partial V}{\partial x}f(t,x,\check{\alpha}_1,\alpha_2) + L(t,x;\check{\alpha}_1,\alpha_2) \qquad (3.3.30)$$

(iv) $\forall (\hat{x},t) \in \mathbb{R}^n \times [t_0,t_1]$ and $\forall \alpha_1 \in \mathbb{R}^m$ :

$$\int_{\mathbb{R}^n}\left[\frac{\partial V}{\partial x}f(t,x,\check{\alpha}_1,\check{\alpha}_2) + \frac{\partial V}{\partial \hat{x}}g(x,\hat{x},\check{\alpha}_1,t) + L(t,x;\check{\alpha}_1,\check{\alpha}_2)\right]d\pi_{\hat{x}}(x) \leq$$

$$\leq \int_{\mathbb{R}^n}\left[\frac{\partial V}{\partial x}f(t,x,\check{\alpha}_1,\alpha_2) + \frac{\partial V}{\partial \hat{x}}g(x,\hat{x},\alpha_1,t) + L(t,x;\check{\alpha}_1,\alpha_2)\right]d\pi_{\hat{x}}(x). \qquad (3.3.31)$$

Then, $(\check{\alpha}_1(\hat{x},t),\check{\alpha}_2(x,\hat{x},t))$ is a saddle point strategy pair, and the value of the game is $V(x_0,x_0,t_0)$.

**Corollary** 3.3.2. Let Eqs.(3.3.24)-(3.3.26) be in particular respectively

$$dX_t = (FX_t + G\alpha_1 + E\alpha_2)dt + \sqrt{D_1}\,dW_1(t),$$

$$dY_t = HX_t dt + \sqrt{D_2}\,dW_2(t). \tag{3.3.32}$$

$$J = E\left[X_{t_1}^T A X_{t_1} + \int_{t_0}^{t_1}(X_t^T Q X_t + \alpha_1^T(t)R\alpha_1(t) - \alpha_2^T B\alpha_2(t))dt\right]$$

over

*Suppose that Eqs. (3.3.34) and (3.3.35) below have a solution $[t_0, t_1]$ Then the above game, with the partial information specified, has a saddle point, given by*

$$d\hat{x} = [(F + EB^{-1}E^T P)\hat{x} + G\alpha_1 - \Sigma H^T w_2(t)H\hat{x}]dt + K dy,$$

$$\check{\alpha}_1(\hat{x}, t) = -R^{-1}G^T P\hat{x}, \tag{3.3.33}$$

$$\check{\alpha}_2(x, \hat{x}, t) = B^{-1}E^T[\Pi x + (P - \Pi\hat{x})] = B^{-1}E^T(P\hat{x} + \Pi\check{x}),$$

$$\check{x} = x - \check{x},$$

*where:*
*(i) $P, \Pi$ and $\Sigma$ the symmetric matrices solutions of the classical game theoretic Riccati equation for $P$:*

$$\dot{P} + PF + F^T P - PGR^{-1}G^T P + PEB^{-1}E^T P + Q = 0,$$

$$P(t_1) = A. \tag{3.3.34}$$

*(ii) a pair of coupled Riccati equations, with two-point boundary values for $\Pi$ and $\Sigma$:*

$$\dot{\Pi} + P(F - \Sigma H^{\mathsf{T}} w_2^{-1} H) + (F^{\mathsf{T}} - H^{\mathsf{T}} w_2^{-1} H \Sigma)\Pi + \Pi E B^{-1} E^{\mathsf{T}} P +$$
$$+ P\Sigma H^T w_2^{-1} H + H^T w_2^{-1} H \Sigma P + Q = 0,$$

$$\Pi(t_1) = A. \tag{3.3.35}$$

$$\dot{\Sigma} - (F - EB^{-1}E^{\mathsf{T}}P)\Sigma - \Sigma(F^{\mathsf{T}} - \Pi EB^{-1}E^{\mathsf{T}}) + \Sigma H^{\mathsf{T}} w_2^{-1} H \Sigma - w_1 = 0,$$

$$\Sigma(t_0) = 0.$$

### III.3.3.2. Bellman equation for 2-Persons Colombeau-Ito's stochastic differential games with non-linear dinamic and noise corrupted information about the measuremets.

Let us consider Colombeau-Ito's type stochastic two-player dynamic system in $\mathbb{R}^n$ be given by a diffusion:

$$(d\mathbf{X}_{t,D_1}^{\varepsilon,\epsilon}(\omega,\omega'))_\epsilon = (\mathbf{f}_\epsilon(t, \mathbf{X}_{t,D_1}^{\varepsilon,\epsilon}(\omega,\omega'); \boldsymbol{\alpha}_{\epsilon,1}, \boldsymbol{\alpha}_{\epsilon,2}))_\epsilon dt + \sqrt{D_1}(d\mathbf{W}_{\epsilon,1}(t,\omega))_\epsilon +$$
$$+ \sqrt{\varepsilon} \, d\mathbf{W}(t,\omega'), \tag{3.3.36}$$

$$\varepsilon \ll 1,$$

(where pair $(\omega, \omega') \in \Omega \times \Omega', \Omega \cap \Omega' = \varnothing$) and observation process in $\mathbb{R}^p$ be given by a diffusion:

$$(d\mathbf{Y}^{\varepsilon,\epsilon}_{t,D_2}(\omega,\omega'))_\epsilon = (\mathbf{h}_\epsilon(t,\mathbf{X}^{\varepsilon,\epsilon}_{t,D_1}(\omega,\omega');\boldsymbol{\alpha}_{\epsilon,1},\boldsymbol{\alpha}_{\epsilon,2}))_\epsilon dt + \sqrt{D_2}\,(d\mathbf{W}_{\epsilon,2}(t,\omega))_\epsilon +$$
$$+ \sqrt{\varepsilon}\,d\mathbf{W}(t,\omega'),$$
(3.3.37)

$$\varepsilon \ll 1$$

(where pair $(\omega,\omega') \in \Omega \times \Omega', \Omega \cap \Omega' = \varnothing$) where $(\mathbf{W}_{\epsilon,1}(t))_\epsilon$ and $(\mathbf{W}_{\epsilon,2}(t))_\epsilon$ are two independant Colombeau vector brownian motions of covariance coefficient matrices $(w_{\epsilon,1}(t))_\epsilon$ and $(w_{\epsilon,2}(t))_\epsilon$ respectively, $\mathbf{W}(t,\omega')$ vector brownian motions of covariance coefficient matrice $w(t)$ and a performance index be given as:

$$(\bar{\mathbf{J}}^\varepsilon_\epsilon)_\epsilon = \left(\mathbf{E}\left[K_\epsilon(t_1) + \int_{t_0}^{t_1} L_\epsilon(t,\mathbf{X}^{\varepsilon,\epsilon}_{t,D_1};\boldsymbol{\alpha}_1,\boldsymbol{\alpha}_2)dt\right]\right)_\epsilon.$$
(3.3.38)

where $t_0$ and $t_1$ are given time instants.

Standard hypotheses are assumed on $(\mathbf{f}_\epsilon(\circ,\circ;\circ,\circ))_\epsilon, (\mathbf{h}_\epsilon(\circ,\circ;\circ,\circ))_\epsilon,$ and the admissible processes $(\boldsymbol{\alpha}_{\epsilon,1}(\circ))_\epsilon$ and $(\boldsymbol{\alpha}_{\epsilon,2}(\circ))_\epsilon$ for $(\mathbf{X}^{\varepsilon,\epsilon}_{t,D_1})_\epsilon$ and $(\mathbf{Y}^{\varepsilon,\epsilon}_{t,D_2})_\epsilon$ to be well defined, and $K_\epsilon$ and $L_\epsilon, \epsilon \in (0,1]$ are supposed to be globally $\mathbf{C}^1$. We consider the game where the minimizer only knows $(\mathbf{Y}^{\varepsilon,\epsilon}_{t,D_2})_\epsilon$, with perfect memory, while the maximizer knows $(\mathbf{X}^{\varepsilon,\epsilon}_{t,D_1})_\epsilon, (\mathbf{Y}^{\varepsilon,\epsilon}_{t,D_2})_\epsilon$ and $(\boldsymbol{\alpha}_{\epsilon,1}(t))_\epsilon$. To stay with saddle points, as opposed to Nash points, we assume that in addition, the minimizer knows $x_0$. We call strategies functions $(\check{\alpha}_{\epsilon,1}(\xi,t))_\epsilon : \widetilde{\mathbb{R}}^n \times \mathbb{R} \to \widetilde{\mathbb{R}}^m$ and $\left(\check{\alpha}_{\epsilon,2}(x,\xi,t)\right)_\epsilon : \widetilde{\mathbb{R}}^n \times \widetilde{\mathbb{R}}^n \times \mathbb{R} \to \widetilde{\mathbb{R}}^{m'}$ such that, if $(\boldsymbol{\alpha}_{\epsilon,1}(t))_\epsilon$ and $(\boldsymbol{\alpha}_{\epsilon,2}(t))_\epsilon$ are replaced by $\check{\alpha}_{\epsilon,1}(\hat{\mathbf{X}}^{\varepsilon,\epsilon}_t,t)$ and $\check{\alpha}_{\epsilon,2}(\mathbf{X}^{\varepsilon,\epsilon}_{t,D_1},\hat{\mathbf{X}}^{\varepsilon,\epsilon}_t,t)$ in Eqs. (3.3.36) and (3.3.39) below, the processes $(\mathbf{X}^{\varepsilon,\epsilon}_{t,D_1})_\epsilon$ and $(\hat{\mathbf{X}}^{\varepsilon,\epsilon}_t)_\epsilon$ are well defined in the Colombeau-Ito's sense.

**Theorem** 3.3.7. *If there exist a process $(\hat{\mathbf{X}}^{\varepsilon,\epsilon}_t)_\epsilon$, Colombeau generalyzed function $V_\epsilon(\mathbf{X},\hat{\mathbf{X}},t) : \widetilde{\mathbb{R}}^n \times \widetilde{\mathbb{R}}^n \times \mathbb{R} \to \mathbb{R}, \forall \epsilon \in (0,1]$ : $V_\epsilon(\mathbf{X},\hat{\mathbf{X}},t) \in \mathbf{C}^2$, strategies $(\check{\alpha}_{\epsilon,1}(\xi,t))_\epsilon$ and $(\check{\alpha}_{\epsilon,2}(x,\xi,t))_\epsilon$, such that:*

**(i)** the process $(\hat{\mathbf{X}}^{\varepsilon,\epsilon}_t)_\epsilon$ obeys Colombeau diffusion law of the form

$$\left(d\hat{\mathbf{X}}_t^{\varepsilon,\epsilon}\right)_\epsilon = \left(\mathbf{G}_{\epsilon,1}\left(\hat{\mathbf{X}}_t^{\varepsilon,\epsilon},\boldsymbol{\alpha}_{\epsilon,1},t\right)d\mathbf{Y}_{t,D_2}^{\varepsilon,\epsilon}\right)_\epsilon + \left(\mathbf{G}_{\epsilon,2}\left(\hat{\mathbf{X}}_t^{\varepsilon,\epsilon},\boldsymbol{\alpha}_{\epsilon,1},t\right)\right)_\epsilon dt +$$

$$+ \sqrt{\varepsilon}\, d\mathbf{W}(t,\omega'). \tag{3.3.39}$$

Thus

$$\left(d\hat{\mathbf{X}}_t\right)_\epsilon = \left(\mathbf{g}_\epsilon\left(\mathbf{X}_t^{\varepsilon,\epsilon},\hat{\mathbf{X}}_t^{\varepsilon,\epsilon},\boldsymbol{\alpha}_{\epsilon,1},t\right)\right)_\epsilon dt + \left(\left(\mathbf{k}_\epsilon\left(\mathbf{X}_t^{\varepsilon,\epsilon},\hat{\mathbf{X}}_t^{\varepsilon,\epsilon},\boldsymbol{\alpha}_{\epsilon,1},t\right)d\mathbf{W}_{\epsilon,2}(t)\right)\right)_\epsilon +$$

$$+ \sqrt{\varepsilon}\, d\mathbf{W}(t,\omega') \tag{3.3.40}$$

and is such that, whenever $(\alpha_{\epsilon,2}(t)) = \left(\check{\boldsymbol{\alpha}}_2\left(\mathbf{X}_t^{\varepsilon,\epsilon},\hat{\mathbf{X}}_t^{\varepsilon,\epsilon},t\right)\right)$, the conditional law of $\left(\mathbf{X}_t^{\varepsilon,\epsilon}\right)_\epsilon$ given the Colombeau information algebra $\left(\mathfrak{I}_{\epsilon,t}^y\right)_\epsilon$ generated by $\{(y_\epsilon(s))_\epsilon | s < t\}$, is of the form $\left(\pi_{\hat{\mathbf{X}}}^\epsilon(\mathbf{X})\right)_\epsilon = \left(\pi_\epsilon(\mathbf{X}-\hat{\mathbf{X}},t)\right)_\epsilon$, where $(\pi_\epsilon)_\epsilon$ it is a fixed zero-mean law, i.e. independant of the control process $(\alpha_{\epsilon,1}(\circ))_\epsilon$.

**(ii)** $\forall (x,\hat{x},t) \in \widetilde{\mathbb{R}}^n \times \widetilde{\mathbb{R}}^n \times [t_0,t_1]$ :

$$\left(\frac{\partial V_\epsilon}{\partial t}\right)_\epsilon + \left[\left(\frac{\partial V_\epsilon}{\partial x}\right)_\epsilon\right] \cdot (f_\epsilon(t,x,\check{\boldsymbol{\alpha}}_{\epsilon,1},\check{\boldsymbol{\alpha}}_{\epsilon,2}))_\epsilon + \left[\left(\frac{\partial V_\epsilon}{\partial \hat{x}}\right)_\epsilon\right] \cdot (g_\epsilon(x,\hat{x},\boldsymbol{\alpha}_{\epsilon,1},t))_\epsilon +$$

$$+ \frac{D_1}{2}\left(\mathrm{tr}\left(\frac{\partial^2 V_\epsilon}{\partial x^2}w_{\epsilon,1}(t)\right)\right)_\epsilon + \frac{D_2}{2}\left(\mathrm{tr}\left(\frac{\partial^2 V}{\partial \hat{x}^2}w_{\epsilon,2}(t)\right)\right)_\epsilon +$$

$$+ \frac{\varepsilon}{2}\left(\mathrm{tr}\left(\frac{\partial^2 V_\epsilon}{\partial x^2}w(t)\right)\right)_\epsilon + \frac{\varepsilon}{2}\left(\mathrm{tr}\left(\frac{\partial^2 V}{\partial \hat{x}^2}w(t)\right)\right)_\epsilon + \tag{3.3.41}$$

$$+(L_\epsilon(t,x;\boldsymbol{\alpha}_{\epsilon,1},\boldsymbol{\alpha}_{\epsilon,2}))_\epsilon = 0$$

$$\forall (x,\hat{x}) \in \widetilde{\mathbb{R}}^n \times \widetilde{\mathbb{R}}^n : (V_\epsilon(x,\hat{x},t_1))_\epsilon = (K_\epsilon(x))_\epsilon.$$

**(iii)** $\forall (x,\hat{x},t) \in \widetilde{\mathbb{R}}^n \times \widetilde{\mathbb{R}}^n \times [t_0,t_1]$ and $\forall (\boldsymbol{\alpha}_{\epsilon,2})_\epsilon \in \widetilde{\mathbb{R}}^{m'}$ :

$$\left[\left(\frac{\partial V_\epsilon}{\partial x}\right)_\epsilon\right] \cdot (f_\epsilon(t,x,(\check{\boldsymbol{\alpha}}_{\epsilon,1})_\epsilon,(\check{\boldsymbol{\alpha}}_{\epsilon,2})_\epsilon))_\epsilon + (L_\epsilon(t,x;\check{\boldsymbol{\alpha}}_{\epsilon,1},\check{\boldsymbol{\alpha}}_{\varepsilon,2}))_\epsilon \geq$$

$$\geq \left[\left(\frac{\partial V_\epsilon}{\partial x}\right)_\epsilon\right] \cdot (f_\epsilon(t,x,\check{\boldsymbol{\alpha}}_{\epsilon,1},\alpha_{\epsilon,2}))_\epsilon + (L_\epsilon(t,x;\check{\boldsymbol{\alpha}}_{\epsilon,1},\boldsymbol{\alpha}_{\epsilon,2}))_\epsilon, \quad (3.3.42)$$

$$\epsilon \in (0,1].$$

**(iv)** $\forall (\hat{x},t) \in \widetilde{\mathbb{R}}^n \times [t_0,t_1]$ and $\forall (\boldsymbol{\alpha}_{\epsilon,1})_\epsilon \in \widetilde{\mathbb{R}}^m$ :

$$\left(\int_{\mathbb{R}^n}\left[\frac{\partial V_\epsilon}{\partial x} \cdot f_\epsilon(t,x,\check{\boldsymbol{\alpha}}_{\epsilon,1},\check{\boldsymbol{\alpha}}_{\epsilon,2}) + \frac{\partial V_\epsilon}{\partial \hat{x}} \cdot g_\epsilon(x,\hat{x},\check{\boldsymbol{\alpha}}_{\epsilon,1},t) + \right.\right.$$

$$\left.\left. + L_\epsilon(t,x;\check{\boldsymbol{\alpha}}_{\epsilon,1},\check{\boldsymbol{\alpha}}_{\epsilon,2})\right]d\pi_{\hat{x}}^\epsilon(x)_\epsilon\right)_\epsilon \leq$$

$$\leq \left(\int_{\mathbb{R}^n}\left[\frac{\partial V_\epsilon}{\partial x} \cdot f_\epsilon(t,x,\check{\boldsymbol{\alpha}}_1,\alpha_2) + \frac{\partial V_\epsilon}{\partial \hat{x}} \cdot g_\epsilon(x,\hat{x},\boldsymbol{\alpha}_{\epsilon,1},t) + \right.\right. \quad (3.3.43)$$

$$\left.\left. + L_\epsilon(t,x;\check{\boldsymbol{\alpha}}_{\epsilon,1},\boldsymbol{\alpha}_{\epsilon,2})\right]d\pi_{\hat{x}}^\epsilon(x)\right)_\epsilon,$$

$$\epsilon \in (0,1].$$

Then, $\{(\check{\boldsymbol{\alpha}}_{\epsilon,1}(\hat{x},t))_\epsilon, (\check{\boldsymbol{\alpha}}_{\epsilon,2}(x,\hat{x},t))_\epsilon\}$ is a saddle point strategy pair, and the value of the game is $(V_\epsilon(x_0,x_0,t_0))_\epsilon$.

**Theorem** 3.3.8. *Let Eqs.(3.3.36)-(3.3.38) be in particular respectively:*

$$(d\boldsymbol{X}_t^{\varepsilon,\epsilon}(\omega,\omega'))_\epsilon = ((F_\epsilon)_\epsilon \cdot (\boldsymbol{X}_t^{\varepsilon,\epsilon}(\omega,\omega'))_\epsilon + (G_\epsilon)_\epsilon \cdot (\boldsymbol{\alpha}_{\epsilon,1})_\epsilon + (E_\epsilon)_\epsilon \cdot (\boldsymbol{\alpha}_{\epsilon,2})_\epsilon)dt +$$
$$\sqrt{D_1}\,(d\boldsymbol{W}_{\epsilon,1}(t,\omega))_\epsilon + \sqrt{\varepsilon}\,d\boldsymbol{W}(t,\omega'),$$

$$(d\boldsymbol{Y}_t^{\varepsilon,\epsilon}(\omega,\omega'))_\epsilon = (H_\epsilon)_\epsilon \cdot (\boldsymbol{X}_t^{\varepsilon,\epsilon}(\omega,\omega'))_\epsilon dt + \sqrt{D_2}\,(d\boldsymbol{W}_{\epsilon,2}(t,\omega))_\epsilon + \sqrt{\varepsilon}\,d\boldsymbol{W}(t,\omega'),$$

(3.3.44)

$$(\boldsymbol{J}_\epsilon)_\epsilon = \left( \boldsymbol{E}\left[ \boldsymbol{X}_{t_1}^{\varepsilon,\epsilon T}A_\epsilon \boldsymbol{X}_{t_1}^{\varepsilon,\epsilon} + \int_{t_0}^{t_1} \left( \boldsymbol{X}_t^{\varepsilon,\epsilon T}Q_\epsilon \boldsymbol{X}_t^{\varepsilon,\epsilon} + \boldsymbol{\alpha}_{\epsilon,1}^T(t)R_\epsilon \boldsymbol{\alpha}_{\epsilon,1}(t) - \boldsymbol{\alpha}_{\epsilon,2}^T B_\epsilon \boldsymbol{\alpha}_{\epsilon,2}(t) \right) dt \right] \right)_\epsilon,$$

$$\varepsilon \ll 1.$$

Suppose that Eqs. (3.3.45) and (3.3.46) below have a solution over $[t_0, t_1]$ Then the above game, with the partial information specified, has a saddle point, given by

$$(d\hat{x}_\epsilon)_\epsilon = \left[ (F_\epsilon + E_\epsilon B_\epsilon^{-1} E_\epsilon^T P_\epsilon)_\epsilon \cdot (\hat{x}_\epsilon)_\epsilon + (G_\epsilon)_\epsilon \cdot (\boldsymbol{\alpha}_{\epsilon,1})_\epsilon - (\Sigma_\epsilon H_\epsilon^T w_{\epsilon,2}(t) H_\epsilon \hat{x}_\epsilon)_\epsilon \right]dt +$$

$$+(K_\epsilon)_\epsilon \cdot (dy_\epsilon)_\epsilon,$$

$$(\check{\boldsymbol{\alpha}}_{\epsilon,1}(\hat{x}_\epsilon,t))_\epsilon = -(R_\epsilon^{-1} G_\epsilon^T P_\epsilon \hat{x}_\epsilon)_\epsilon,$$

(3.3.45)

$$(\check{\boldsymbol{\alpha}}_{\epsilon,2}(x_\epsilon,\hat{x}_\epsilon,t))_\epsilon = (B_\epsilon^{-1} E_\epsilon^T [\Pi_\epsilon x_\epsilon + (P_\epsilon - \Pi_\epsilon \hat{x}_\epsilon)])_\epsilon =$$

$$= (B_\epsilon^{-1} E_\epsilon^T (P_\epsilon \hat{x}_\epsilon + \Pi_\epsilon \check{x}_\epsilon))_\epsilon,$$

$$\check{x}_\epsilon = x_\epsilon - \hat{x}_\epsilon,$$

where:
> **(i)** $(P_\epsilon)_\epsilon, (\Pi_\epsilon)_\epsilon$ and $(\Sigma_\epsilon)_\epsilon$ the symmetric matrices solutions of the game theoretic Colombeau-Riccati equation for $(P_\epsilon)_\epsilon$:

$$(\dot{P}_\epsilon)_\epsilon + (P_\epsilon)_\epsilon \cdot (F_\epsilon)_\epsilon + (F_\epsilon^T)_\epsilon \cdot (P_\epsilon)_\epsilon - (P_\epsilon)_\epsilon \cdot (G_\epsilon)_\epsilon \cdot (R_\epsilon^{-1})_\epsilon \cdot (G_\epsilon^T)_\epsilon \cdot (P_\epsilon)_\epsilon +$$

$$+(P_\epsilon)_\epsilon \cdot (E_\epsilon)_\epsilon \cdot (B_\epsilon^{-1})_\epsilon \cdot (E_\epsilon^T)_\epsilon \cdot (P_\epsilon)_\epsilon + (Q_\epsilon)_\epsilon = 0, \qquad (3.3.46)$$

$$(P_\epsilon(t_1))_\epsilon = (A_\epsilon)_\epsilon.$$

***(ii)** a pair of coupled Riccati equations, with two-point boundary values for $(\Pi_\epsilon)_\epsilon$ and $(\Sigma_\epsilon)_\epsilon$:*

$$(\dot{\Pi}_\epsilon)_\epsilon + (P_\epsilon)_\epsilon \cdot \left((F_\epsilon)_\epsilon - (\Sigma_\epsilon)_\epsilon \cdot (H_\epsilon^T)_\epsilon \cdot (w_{\epsilon,2}^{-1}) \cdot (H_\epsilon)\right) +$$

$$+\left((F_\epsilon^T)_\epsilon - (H_\epsilon^T)_\epsilon (w_{\epsilon,2}^{-1})_\epsilon (H_\epsilon) \cdot (\Sigma_\epsilon)_\epsilon\right) \cdot (\Pi_\epsilon)_\epsilon +$$

$$+(\Pi_\epsilon)_\epsilon \cdot (E_\epsilon)_\epsilon \cdot (B_\epsilon^{-1}) \cdot (E_\epsilon^T)_\epsilon \cdot (P_\epsilon)_\epsilon +$$

$$+(P_\epsilon)_\epsilon (\Sigma_\epsilon)_\epsilon (H_\epsilon^T)_\epsilon (w_{\epsilon,2}^{-1})_\epsilon \cdot (H_\epsilon)_\epsilon +$$

$$+(H_\epsilon^T)_\epsilon \cdot (w_{\epsilon,2}^{-1})_\epsilon \cdot (H_\epsilon)_\epsilon \cdot (\Sigma_\epsilon)_\epsilon \cdot (P_\epsilon)_\epsilon + (Q_\epsilon)_\epsilon = 0, \qquad (3.3.47)$$

$$(\Pi_\epsilon(t_1))_\epsilon = (A_\epsilon)_\epsilon.$$

$$(\dot{\Sigma}_\epsilon)_\epsilon - \left((F_\epsilon)_\epsilon - (E_\epsilon)_\epsilon \cdot (B_\epsilon^{-1})_\epsilon \cdot (E_\epsilon^T)_\epsilon P_\epsilon\right)\Sigma_\epsilon - \Sigma_\epsilon(F_\epsilon^T - \Pi_\epsilon E_\epsilon B_\epsilon^{-1} E_\epsilon^T)$$

$$+\Sigma_\epsilon H_\epsilon^T w_{\epsilon,2}^{-1} H_\epsilon \Sigma_\epsilon - (w_{\epsilon 1})_\epsilon = 0,$$

$$\Sigma(t_0) = 0.$$

### III.3.3.3. Strong large deviations principle of Non-Freidlin-Wentzell type for Colombeau-Ito's

# type stochastic differential game with non-linear dynamics and noise corrupted information about the measuremets.

Let us consider Colombeau-Ito's type stochastic two-player dynamic system in $\mathbb{R}^n$ be given by a diffusion:

$$(d\mathbf{X}_{t,D_1}^{\varepsilon,\epsilon}(\omega,\omega'))_\epsilon = (\mathbf{f}_\epsilon(t,\mathbf{X}_{t,D_1}^{\varepsilon,\epsilon}(\omega,\omega');\boldsymbol{\alpha}_1,\boldsymbol{\alpha}_2))_\epsilon dt + \sqrt{D_1}\,(d\mathbf{W}_{\epsilon,1}(t,\omega))_\epsilon +$$

$$+ \sqrt{\varepsilon}\,d\mathbf{W}(t,\omega'), \quad (3.3.48)$$

$$\varepsilon \ll 1,$$

(where pair $(\omega,\omega') \in \Omega \times \Omega', \Omega \cap \Omega' = \varnothing$) and observation process in $\mathbb{R}^p$ be given by a diffusion:

$$(d\mathbf{Y}_{t,D_2}^{\varepsilon,\epsilon}(\omega,\omega'))_\epsilon = (\mathbf{h}_\epsilon(t,\mathbf{X}_{t,D_1}^{\varepsilon,\epsilon}(\omega,\omega');\boldsymbol{\alpha}_1,\boldsymbol{\alpha}_2))_\epsilon dt + \sqrt{D_2}\,(d\mathbf{W}_{\epsilon,2}(t,\omega))_\epsilon +$$

$$+ \sqrt{\varepsilon}\,d\mathbf{W}(t,\omega'), \quad (3.3.37)$$

$$\varepsilon \ll 1$$

(where pair $(\omega,\omega') \in \Omega \times \Omega', \Omega \cap \Omega' = \varnothing$) where $(\mathbf{W}_{\epsilon,1}(t))_\epsilon$ and $(\mathbf{W}_{\epsilon,2}(t))_\epsilon$ are two independant Colombeau vector brownian motions of covariance coefficient matrices $(w_{\epsilon,1}(t))_\epsilon$ and $(w_{\epsilon,2}(t))_\epsilon$ respectively, and a performance index be given as:

$$(\bar{\mathbf{J}}_\epsilon^\varepsilon)_\epsilon = \left(\mathbf{E}\left[K_\epsilon(t_1) + \int_{t_0}^{t_1} L_\epsilon(t,\mathbf{X}_{t,D_1}^{\varepsilon,\epsilon}(\omega,\omega');\boldsymbol{\alpha}_{\epsilon,1},\boldsymbol{\alpha}_{\epsilon 2})dt\right]\right)_\epsilon. \quad (3.3.38)$$

where $t_0$ and $t_1$ are given time instants.

**Definition** *3.3.11. Colombeau-Ito's stochastic differential game $CSDG_{m;T}(f,h,L,K,(w_{\epsilon,1})_\epsilon,(w_{\epsilon,2})_\epsilon)$ is the Ito's stochastic differential game $SDG_{m;T}(f,h,L,K,w_1,w_2)$ imbeded into a*

*'small' white noise.*

Let us consider *dissipative* **CSDG** (3.3.36)-(3.3.38) with performance index be given as:

$$(\mathbf{J}_\epsilon^\varepsilon)_\epsilon = \Big(\mathbf{E}\Big[\mathbf{X}_{t_1}^{\varepsilon,\epsilon\mathsf{T}}(\omega,\omega')A\mathbf{X}_{t_1}^{\varepsilon,\epsilon}(\omega,\omega') +$$

$$+ \int_{t_0}^{t_1}\Big(\mathbf{X}_t^{\varepsilon,\epsilon\mathsf{T}}(\omega,\omega')Q_\epsilon\mathbf{X}_t^{\varepsilon,\epsilon}(\omega,\omega') + \boldsymbol{\alpha}_{\epsilon,1}^{\mathsf{T}}(t)R_\epsilon\boldsymbol{\alpha}_{\epsilon,1}(t) - \boldsymbol{\alpha}_{\epsilon,2}^{\mathsf{T}}B_\epsilon\boldsymbol{\alpha}_{\epsilon,2}(t)\Big)dt\Big]\Big)_\epsilon \quad ((3.3.40))$$

**Theorem** 3.3.9. *For the all solution* $\{\mathbf{X}_t^{\varepsilon,\epsilon}(\omega,\omega'),\check{\alpha}_{\epsilon,1}(t),\check{\alpha}_{\epsilon,2}(t)\} =$
$= \{(X_{1,t}^{\varepsilon,\epsilon}(\omega,\omega'),\ldots,X_{n,t}^{\varepsilon,\epsilon}(\omega,\omega')),(\check{\alpha}_{\epsilon,1}(t),\check{\alpha}_{\epsilon,2}(t))\}$
*dissipative* **CSDG** *(3.3.36)-(3.3.38) and* $\widetilde{\mathbb{R}}^n$ *valued parameters* $\lambda = (\lambda_1,\ldots,\lambda_n) \in \widetilde{\mathbb{R}}^n$, *there exists a Colombeau constant* $(C_{\varepsilon(\epsilon)})_{\varepsilon(\epsilon)} \in \widetilde{\mathbb{R}}, (C_\varepsilon)_\varepsilon \geq 0$, *such that:*

$$\Big(\liminf_{\varepsilon\to 0}\mathbf{E}_{\Omega'}\Big[\|\mathbf{X}_t^{\varepsilon,\epsilon}(\omega,\omega') - \lambda\|^2|\Omega\Big]\Big)_\epsilon \leq (C_{\varepsilon(\epsilon)})_{\varepsilon(\epsilon)}\|(\mathbf{U}_\epsilon(t,\lambda,\omega))_\epsilon\|^2,$$

$$\lambda = (\lambda_1,\ldots,\lambda_n) \in \widetilde{\mathbb{R}}^n,$$

$$t \in [0,T],$$

(3.3.41)

where $(\mathbf{U}_\epsilon(t,\lambda,\omega))_\epsilon = ((U_{\epsilon,1}(t,\lambda,\omega))_\epsilon,\ldots,(U_{\epsilon,n}(t,\lambda,\omega))_\epsilon)$ *is the trajectory of the linear differential master game with noise corrupted information about the measuremets:*

$$\left(\frac{d\mathbf{U}_\epsilon(t,\lambda)}{dt}\right)_\epsilon = (\mathbf{J}_{1,\epsilon}[\mathbf{f}_\epsilon(\lambda,\breve{\alpha}_{1,\epsilon}(t,\lambda),\breve{\alpha}_{\epsilon,2}(t))]\mathbf{U}_\epsilon)_\epsilon + (\bar{\mathbf{f}}_\epsilon(\lambda,\breve{\alpha}_{\epsilon,1}(t,\lambda),\breve{\alpha}_{\epsilon,2}(t)))_\epsilon,$$

$$\bar{\mathbf{f}}_\epsilon(x,\breve{\alpha}(t,\lambda)) = \mathbf{f}_\epsilon(\lambda,\breve{\alpha}(t,\lambda)) + \sqrt{D_1}\,(d\mathbf{W}_{\epsilon,1}(t,\omega))_\epsilon, \breve{\alpha}(t,\lambda) = \{\breve{\alpha}_{\epsilon,1}(t,\lambda),\breve{\alpha}_{\epsilon,2}(t)\},$$

$$(\mathbf{U}_\epsilon(0,\lambda))_\epsilon = x_0 - \lambda,$$

$$\left(\frac{d\mathbf{H}_\epsilon(t,\lambda)}{dt}\right)_\epsilon = (\mathbf{J}_{2,\epsilon}[\mathbf{h}_\epsilon(\lambda,\breve{\alpha}_{1,\epsilon}(t,\lambda),\breve{\alpha}_{\epsilon,2}(t))]\mathbf{H}_\epsilon)_\epsilon + (\bar{\mathbf{h}}_\epsilon(\lambda,\breve{\alpha}_{\epsilon,1}(t,\lambda),\breve{\alpha}_{\epsilon,2}(t)))_\epsilon,$$

(3.3.42)

$$\bar{\mathbf{h}}_\epsilon(x,\breve{\alpha}(t,\lambda)) = \mathbf{h}_\epsilon(\lambda,\breve{\alpha}(t,\lambda)) + \sqrt{D_2}\,(d\mathbf{W}_{\epsilon,2}(t,\omega))_\epsilon,$$

$$(\mathbf{H}_\epsilon(0,\lambda))_\epsilon = y_0 - \lambda,$$

$$(\mathbf{J}_{\epsilon,i})_\epsilon = \mathbf{E}_\Omega\left[\|(\mathbf{U}_\epsilon(T))_\epsilon\|^2\right] = \mathbf{E}_\Omega\left[\sum_{j=1}^n[(U_{\epsilon,i}(T))_\epsilon]^2\right].$$

$$\left(\min_{\alpha_i(t)}\left(\max_{\alpha_j(t),j\neq i}\mathbf{J}_{\epsilon,i}\right)\right)_\epsilon = 0.$$

Where $(\mathbf{J}_{1,\epsilon})_\epsilon = (\mathbf{J}_{1,\epsilon}[\mathbf{f}_\epsilon(\lambda,\breve{\alpha}(t,\lambda))])_\epsilon$ the Jacobian, i.e. $\mathbf{J}_{1,\epsilon}$ is a $n\times n$-matrix:

$$\mathbf{J}_{1,\epsilon}[\mathbf{f}_\epsilon(\lambda,\breve{\alpha}_\epsilon(t,\lambda))] = \mathbf{J}_{2,\epsilon}[\mathbf{f}(x,\breve{\alpha}_\epsilon(t,z))]|_{x=\lambda,z=\lambda} =$$

$$= \begin{bmatrix} \dfrac{\partial f_{\epsilon,1}(x,\breve{\alpha}(t,z))}{\partial x_1} & \cdots & \dfrac{\partial f_{\epsilon,1}(x,\breve{\alpha}(t,z))}{\partial x_n} \\ \cdot & \cdots & \cdot \\ \cdot & \cdots & \cdot \\ \cdot & \cdots & \cdot \\ \dfrac{\partial f_{\epsilon,n}(x,\breve{\alpha}(t,z))}{\partial x_1} & \cdots & \dfrac{\partial f_{\epsilon,n}(x,\breve{\alpha}(t,z))}{\partial x_n} \end{bmatrix}\Bigg|_{x=\lambda,z=\lambda}$$

(3.3.43)

$(\mathbf{J}_{2,\epsilon})_\epsilon = (\mathbf{J}_{2,\epsilon}[\mathbf{h}_\epsilon(\lambda,\breve{\alpha}(t,\lambda))])_\epsilon$ the Jacobian, i.e. $\mathbf{J}_{2,\epsilon}$ is a $m\times n$-matrix:

$$\mathbf{J}_{1,\epsilon}[\mathbf{h}_\epsilon(\lambda,\check{\alpha}_\epsilon(t,\lambda))] = \mathbf{J}_{2,\epsilon}[\mathbf{h}(x,\check{\alpha}_\epsilon(t,z))]|_{x=\lambda,z=\lambda} =$$

$$= \begin{bmatrix} \dfrac{\partial h_{\epsilon,1}(x,\check{\alpha}(t,z))}{\partial x_1} & \cdots & \dfrac{\partial h_{\epsilon,1}(x,\check{\alpha}(t,z))}{\partial x_n} \\ \cdot & \cdots & \cdot \\ \cdot & \cdots & \cdot \\ \cdot & \cdots & \cdot \\ \dfrac{\partial h_{\epsilon,n}(x,\check{\alpha}(t,z))}{\partial x_1} & \cdots & \dfrac{\partial h_{\epsilon,n}(x,\check{\alpha}(t,z))}{\partial x_n} \end{bmatrix}_{x=\lambda,z=\lambda} \quad (3.3.43)$$

**Theorem** *3.3.10. Assume the conditions of the Theorem 3.3.9 for any $\omega \in \Omega$, any $\lambda = (\lambda_1,\ldots,\lambda_n) \in \widetilde{\mathbb{R}}^n, t \in [0,T]$ :*

$$(\|\mathbf{U}_\epsilon(t,\lambda,\omega)\|)_\epsilon = 0 \Rightarrow$$

$$\liminf_{\varepsilon \to 0} \mathbf{E}_{\Omega'}\left[\|(\mathbf{X}_t^{\varepsilon,\epsilon}(\omega))_\epsilon - \lambda\|^2|\Omega'\right] = 0. \quad (3.3.44)$$

$$\left(\liminf_{\varepsilon \to 0}\left[\min_{\alpha_i(t)}\left(\max_{\alpha_j(t),j\neq i} \bar{\mathbf{J}}^{\varepsilon}_{\epsilon,i,}\right)\right]\right)_\epsilon = 0.$$

*More precisely, for any $t \in [0,T]$ and $\lambda = \lambda(t,\omega) = (\lambda_1(t,\omega),\ldots,\lambda_n(t,\omega)) \in \widetilde{\mathbb{R}}^n$ sutch that*

$$(U_{\epsilon,1}(t,\lambda_1(t,\omega),\ldots,\lambda_{\epsilon,n}(t,\omega)))_\epsilon = 0,$$

$$\cdots\cdots\cdots\cdots\cdots\cdots \quad (3.3.45)$$

$$(U_{\epsilon,n}(t,\lambda_{\epsilon,1}(t,\omega),\ldots,\lambda_{\epsilon,n}(t,\omega)))_\epsilon = 0,$$

*the equalities is satisfaed*

$$\liminf_{\varepsilon \to 0} \mathbf{E}_{\Omega'}\left[ \|X^\varepsilon_{1,t}(\omega,\omega') - \lambda_1(t,\omega)\|^2 | \Omega' \right] = 0,$$

$$\cdots\cdots\cdots\cdots\cdots\cdots$$

$$\liminf_{\varepsilon \to 0} \mathbf{E}_{\Omega'}\left[ \|X^\varepsilon_{n,t}(\omega,\omega') - \lambda_n(t,\omega)\|^2 | \Omega' \right] = 0, \tag{3.3.46}$$

$$\liminf_{\varepsilon \to 0} \left[ \min_{\alpha_i(t)} \left( \max_{\alpha_j(t), j \neq i} \bar{J}_{i,\varepsilon} \right) \right] = 0.$$

## III.4. Full controllability of nonlinear systems.

The controllability of linear systems has been studied by many authors. La'Salle used a geometric growth condition to establish criteria for null controllability of linear systems defined on $[t_0, \infty)$, Kalman, Ho, and Narendra, Weiss, and Silverman and Meadows developed three basic necessary and sufficient conditions which are algebraic in nature. Conti used a functional analysis point of view to consider controllability with IP spaces of controls.

On the other hand, the problem of complete controllability for most nonlinear systems remains unsolved.

Let us consider now the continuous-time linear controllable dynamical system

$$\dot{x}_t = \mathbf{b}(x_t, u_t),$$

$$x \in \mathbb{R}^d, a \in \mathbb{R}^m. \tag{3.4.1}$$

**Definition**  3.4.1. *Given a starting point $x_0$, the controlled process for control $(u_t)_{t>0}$ is given by the solution of $\dot{x}_t = b(x_t, u_t)$ for $t > 0$. We say that $\mathbf{b}(\cdot, \cdot)$ is fully controllable in time $t$ if, for all $x_0, x \in \mathbb{R}^d$, there exists a control $(u_s)_{0 \leq s \leq t}$ such that $x_t = x$.*

In this section,we establish sufficient conditions for complete controllability of systems of the general form (3.4.1).

## III.4.1. Local controllability of nonlinear systems.

We consider systems described by

$$\dot{x}_t = \mathbf{b}(x_t, u_t),$$

$$\mathbf{b} : \mathbb{R}^n \times \mathbb{R}^m \mapsto \mathbb{R}^n,$$

(3.4.2)

where **b** is a given continuously differentiable function, and $u_t = u(t)$ is an $m$-dimensional time-varying input to be chosen to steer the solution $x_t = x(t)$ in a desired direction.

**Definition** 3.4.2. Let $U$ be an open subset of $\mathbb{R}^n$, $\bar{x}_0 \in \mathbb{R}^n$. The reachable set for a given $T > 0$ the ($U$-locally) reachable set $\mathbf{R}_U(\bar{x}_0, T)$ is defined as the set of all $x(T) \in \mathbb{R}^n$ such that $x(t)|_{t=T} = x(T)$ where $x(\cdot) : [0, T] \to \mathbb{R}^n$, $u(\cdot) : [0, T] \to \mathbb{R}^m$ is a bounded solution of (3.4.2) such that $x(0) = \bar{x}_0$ and $x(t) \in U$ for all $t \in [0, T]$.

**Problem** 3.4.1. Our task is to find conditions under which $\mathbf{R}_U(\bar{x}_0, T)$ is guaranteed to contain a neigborhod of some point in $\mathbb{R}^n$, or, alternatively, conditions which guarante that $\mathbf{R}_U(\bar{x}_0, T)$ has an empty interior. In particular, when $\bar{x}_0$ is a controlled equilibrium of (3.4.2) i.e. $\mathbf{b}(\bar{x}_0, \bar{u}_0) = 0$ for some $\bar{u}_0 \in \mathbb{R}^m$, complete local controllability of (3.4.2) at $\bar{x}_0$ means that for every $\epsilon > 0$ and $T > 0$ there exists $\delta > 0$ such that $\mathbf{R}_U(\bar{x}_0, T) \supsetneq \mathbf{B}_\delta(\bar{x}_0)$ for every $\bar{x} \in \mathbf{B}_\delta(\bar{x}_0)$, where $U = \mathbf{B}_\epsilon(\bar{x}_0)$ and $\mathbf{B}_r(\bar{x}_0) = \{\bar{x}|\|\bar{x} - \bar{x}_0\| \leq r\}$ denotes the ball of radius $r$ centered at $\bar{x}_0$.

## III.4.2. Controllability of linearized system.

This is well-known that a relatively straightforward case of standard local controllability analysis is defined by systems with controllable linearizations.

Let $x_0(\cdot) : [0, T] \to \mathbb{R}^n, u_0(\cdot) : [0, T] \to \mathbb{R}^m$ is a bounded solution of (3.4.2). The standard linearization of (3.4.2) around the solution

$$\{x_0(t), u_0(t)\}_{t \in [0,T]} \qquad (3.4.3)$$

describes the dependency of small state increments

$$\delta_x(t) = x(t) - x_0(t) + o(\delta_x(t))$$

on small input increments $\delta_u(t) = u(t) - u_0(t)$ :

$$\dot{\delta}_x(t) = \mathbf{A}(t)\, \delta_x(t) + \mathbf{B}(t) \delta_u(t) \qquad (3.4.4)$$

where

$$\mathbf{A}(t) = \left(\frac{\partial \mathbf{b}}{\partial x}\right)\bigg|_{x=x_0(t), u=u_0(t)},$$

$$\mathbf{B}(t) = \left(\frac{\partial \mathbf{b}}{\partial u}\right)\bigg|_{x=x_0(t), u=u_0(t)} \qquad (3.4.5)$$

are bounded measureable matrix-valued functions of time.

**Definition** 3.4.3. System (3.4.4) is controllable on time interval $[0, T]$ if for every $\bar{\delta}_x^0, \bar{\delta}_x^T \in \mathbb{R}^n$ there exists a bounded measureable function $\delta_u(t) : [0, T] \to \mathbb{R}^m$ such that the solution of (3.4.4) with $\delta_x(0) = \bar{\delta}_x^0$ satisfies $\delta_x(T) = \bar{\delta}_x^T$.
The following simple criterion of controllability is well known from the linear system theory.

**Theorem** 3.4.1. System (3.4.4) is controllable on interval $[0, T]$ if and only
**Definition** if the matrix:

$$\mathbf{W}(T) = \int_0^T \mathbf{M}^{-1}(t)\mathbf{B}(t)\mathbf{B}^T(t)\left(\mathbf{M}^T(t)\right)^{-1} dt \qquad (3.4.6)$$

is positive definite, where $\mathbf{M}(t)$ is the evolution matrix of (3.4.4), defined by

$$\dot{\mathbf{M}}(t) = \mathbf{A}(t)\mathbf{M}(t),$$
$$\mathbf{M}(0) = \mathbf{I}. \qquad (3.4.7)$$

The variable change $\delta_x(t) = \mathbf{M}(t)z(t)$ reduces (3.4.4) to

$$\frac{dz(t)}{dt} = \mathbf{M}^{-1}(t)\mathbf{B}(t)\delta_u(t). \qquad (3.4.8)$$

Moreover, since

$$z(T) = \int_0^T \mathbf{M}^{-1}(t)\mathbf{B}(t)\delta_u(t)dt \qquad (3.4.9)$$

is a linear integral dependence, function u can be chosen to belong to any subclass which is dense in $L^1(0,T)$.
Is well known that controllability of linearization implies local controllability. The converse is not true: a nonlinear system with an uncontrollable linearization can easily be controllable.

**Theorem** 3.4.2. *Let $b(\cdot,\cdot): \mathbb{R}^n \times \mathbb{R}^m$ be continuously differentiable. Let*

**Theorem** *$x_0(\cdot) : [0,T] \to \mathbb{R}^n, u_0(\cdot) : [0,T] \to \mathbb{R}^m$ be a bounded solution of (3.4.2). Assume that system (3.4.4), defined by (3.4.5), is controllable over $[0,T]$. Then for every $\varepsilon > 0$ there exists $\delta > 0$ such that for all $\bar{x}_0, \bar{x}_T$ satisfying*

$$\|x_0(0) - \bar{x}_0\| < \delta, \|x_0(T) - \bar{x}_T\| < \delta, \quad (3.4.10)$$

*there exist functions $x(\cdot) : [0,T] \to \mathbb{R}^n, u(\cdot) : [0,T] \to \mathbb{R}^m$ satisfying the ODE and conditions:*

$$x(0) = \bar{x}_0, x(T) = \bar{x}_T,$$
$$(3.4.11)$$
$$\|x(t) - x_0(t)\| < \varepsilon, \|u(t) - u_0(t)\| < \varepsilon, \forall t \in [0,T].$$

**Remark** *3.4.1.In other words, if linearization around a trajectory $(x_0(t), u_0(t))$ is controllable then from every point in a sufficiently small neigborhood of $x_0(0)$ the solution of (3.4.3) can be steered to every point in a sufficiently small neigborhood of $x_0(T)$ by applying a small perturbation $u_\delta(t) = u(t)$ of the nominal control $u_0(t)$.*

**Proof.** The proof of Theorem (3.4.2) is based on the implicit mapping theorem. Let $\mathbf{e}_1, \ldots, \mathbf{e}_n$ be the standard basis in $\mathbb{R}^n$. Let $\delta_u^k$ be the controls which cause the solution of (3.4.4) with $\delta_x(0) = 0$ to reach $\delta_x(T) = \mathbf{e}_k$. For $\varepsilon > 0$ let

$$\mathbf{B}_\varepsilon = \{\bar{x} \in \mathbb{R}^n | \|x\| < \varepsilon\}$$

The function $\Psi : \mathbf{B}_\varepsilon \times \mathbf{B}_\varepsilon \to \mathbb{R}^n$, which maps $\{w, v\} = \{(w_1, w_2, \ldots, w_n), (v_1, v_2, \ldots, v_n)\} \in \mathbf{B}_\varepsilon \times \mathbf{B}_\varepsilon$ to $\Psi(w, v) = x(T)$ where $x(t)$ is the solution of (3.4.2) with $x(0) = x_0(0) + v$ and

$$u(t) = u_0(t) + \sum_{k=1}^{n} w_k \delta_u^k(t) \quad (3.4.12)$$

is well defined and continuously differentiable when $\varepsilon > 0$ is sufficiently small. The derivative of $\Psi$ with respect to $w$ at $w = v = 0$ is identity. Hence, by the implicit mapping theorem, equation $S(w, v) = x$ has a solution $w \approx 0$ whenever $\|v\|$ and $\|\bar{x} - x_0(T)\|$ are small enough.

## III.4.3. Full controllability of nonlinear systems imbeded into a 'small' white noise. Infinitesimal Reduction.

Let us consider ODE described by:

$$\dot{x}_t = \mathbf{b}_1(x_t, t) + \mathbf{b}_2(x_t, t)\mathbf{u}_t$$

$$\mathbf{b}_1 : \mathbb{R}^n \times [0, T] \mapsto \mathbb{R}^n, \qquad (3.4.13)$$

$$\mathbf{b}_2 : \mathbb{R}^n \times [0, T] \mapsto \mathbb{R}^n \times \mathbb{R}^m,$$

where $\mathbf{b}_1(\circ, t)$ and $\mathbf{b}_2(\circ, t)$ is a polinomial transforms, i.e.

$$b_{1,i}(x, t) = \sum_{|\alpha|} b_{1,\alpha}^i(x, t) x^\alpha,$$

$$\alpha = (i_1, \ldots, i_k), |\alpha| = \sum_{l=0}^{k} i_l,$$

$$i = 1, \ldots, n \qquad (3.4.14)$$

$$b_{2,ij}(x, t) = \sum_{|\alpha|} b_{2,\alpha}^{ij}(x, t) x^\alpha,$$

$$i = 1, \ldots, n \; ; \; j = 1, \ldots, m,$$

$\mathbf{u}_t = \mathbf{u}(t)$ is an $m$-dimensional time-varying input.

Let us consider corresponding SDE described by:

$$\dot{x}_t = \mathbf{b}_1(x_t,t) + \mathbf{b}_2(x_t,t)\mathbf{u}_t + \sqrt{\varepsilon}\dot{\mathbf{W}}(t),$$

$$\varepsilon \ll 1,$$

(3.4.15)

$$\mathbf{b}_1 : \mathbb{R}^n \times [0,T] \mapsto \mathbb{R}^n,$$

$$\mathbf{b}_2 : \mathbb{R}^n \times [0,T] \mapsto \mathbb{R}^n \times \mathbb{R}^m,$$

where $\mathbf{W}(t)$ is $n$-dimensional Brownian motion, $\mathbf{b}_1(\circ,t), \mathbf{b}_2(\circ,t)$ is a polinomial transforms, $\mathbf{u}_t = \mathbf{u}(t)$ is an $m$-dimensional time-varying input.

**Definition** *3.4.4. Stochastic differential system (3.4.15) is the determined differential system (3.4.13) imbeded into a 'small' white noise.*

**Definition** *3.4.5. Let $U$ be an open subset of $\mathbb{R}^n, \bar{x}_0 \in \mathbb{R}^n$. The reachable set for a given $T > 0$ the (U-locally) reachable set $\mathbf{R}_U(\bar{x}_0, T)$ is defined as the set of all $x(T,\omega) \in \mathbb{R}^n$ such that*

$$\mathbf{P} - a.s.: x(t,\omega)|_{t=T} = x(T)$$

*where a pair $\{x(\cdot,\omega), u(\cdot)\} : x(\cdot,\omega) : [0,T] \times \Omega \to \mathbb{R}^n,$ $\mathbf{u}(\cdot) : [0,T] \to \mathbb{R}^m$ is the admissible) solution of (3.4.15), i.e. such that:*
*(1) $x(0) = \bar{x}_0$,*

*(2) $\mathbf{P} - \mathbf{a.s.}$, for all $t \in [0,T] : x(t,\omega) \in U$,*

*(3) There exists a constant $C = C(\bar{x}_0, T, U)$ such that,*

$$\sup_{t\in[0,T]} \|\mathbf{u}(t)\| \leq C < \infty.$$

**Problem** *3.4.2. Our task is to find conditions under which $\mathbf{R}_U(\bar{x}_0, T)$ is guaranteed to contain a neigborhod of some point in $\mathbb{R}^n$, or, alternatively, conditions which guarante that $\mathbf{R}_U(\bar{x}_0, T)$ has an empty interior. In particular, when $\bar{x}_0$ is a controlled equilibrium of (3.4.15) i.e. $\mathbf{b}(\bar{x}_0, \bar{u}_0) = 0$ for some $\bar{u}_0 \in \mathbb{R}^m$,*

*complete local controllability of (3.4.15) at $\bar{x}_0$ means that for*

every $\epsilon > 0$ and $T > 0$ there exists $\delta > 0$ such that $\mathbf{R}_U(\bar{x}_0, T) \not\supseteq \mathbf{B}_\delta(\bar{x}_0)$ for every $\bar{x} \in \mathbf{B}_\delta(\bar{x}_0)$, where $U=\mathbf{B}_\epsilon(\bar{x}_0)$ and $\mathbf{B}_\epsilon(\bar{x}_0) = \{\bar{x} | \|\bar{x} - \bar{x}_0\| \leq \epsilon\}$ denotes the ball of radius $r$ centered at $\bar{x}_0$.

**Theorem** 3.4.3. ( **Strong large deviations principle**).For the all solutions $\mathbf{X}_t^\varepsilon = (X_{1,t}^\varepsilon, \ldots, X_{n,t}^\varepsilon)$ dissipative SDE (3.4.15) and $\mathbb{R}$ valued parameters $\lambda_1, \ldots, \lambda_n, \lambda = (\lambda_1, \ldots, \lambda_n) \in \mathbb{R}^n$, there exists Colombeau generalized constant $(C_\varepsilon)_\varepsilon \in \widetilde{\mathbb{R}}$, $(C_\varepsilon)_\varepsilon \geq 0$, such that:

$$\liminf_{\varepsilon \to 0} \mathbf{E}\left[\|\mathbf{X}_t^\varepsilon - \lambda\|^2\right] \leq (C_\varepsilon)_\varepsilon \|\mathbf{U}(t,\lambda)\|^2$$

(3.4.16)

$$\lambda = (\lambda_1, \ldots, \lambda_n) \in \mathbb{R}^n,$$

where $\mathbf{U}(t,\lambda) = (U_1(t,\lambda), \ldots, U_n(t,\lambda))$ the solution of the linear differential master equation:

$$\frac{d\mathbf{U}(t,\lambda)}{dt} = \mathbf{J}[\mathbf{b}(\lambda,t;\mathbf{u}(t))]\mathbf{U} + \mathbf{b}(\lambda,t;\mathbf{u}(t)),$$

$$\mathbf{U}(0,\lambda;\mathbf{u}(0)) = x_0 - \lambda,$$

(3.4.17)

$$\mathbf{b}(\lambda,t;\mathbf{u}(t)) = \mathbf{b}_1(\lambda,t) + \mathbf{b}_2(\lambda,t)\mathbf{u}(t),$$

where $\mathbf{J}[\mathbf{b}(\lambda,t;\mathbf{u}(t))]$ the Jacobian, i.e. $\mathbf{J}$ is a $n \times n$-matrix:

$$\mathbf{J}[\mathbf{b}(\lambda,t;\mathbf{u}(t))] = \mathbf{J}[\mathbf{b}(x,t;\mathbf{u}(t))]|_{x=\lambda} =$$

$$= \begin{bmatrix} \dfrac{\partial b^1(x,t;\mathbf{u}(t))}{\partial x_1} & \cdots & \dfrac{\partial b^1(x,t;\mathbf{u}(t))}{\partial x_n} \\ \cdot & \cdots & \cdot \\ \cdot & \cdots & \cdot \\ \cdot & \cdots & \cdot \\ \dfrac{\partial b^n(x,t;\mathbf{u}(t))}{\partial x_1} & \cdots & \dfrac{\partial b^n(x,t;\mathbf{u}(t))}{\partial x_n} \end{bmatrix}\Bigg|_{x=\lambda} \quad (3.4.18)$$

**Corollary** *3.4.1. Assume the conditions of the Theorem 3.4.3 then for any*

$$\lambda = (\lambda_1,\ldots,\lambda_n) \in \mathbb{R}^n, t \in [0,T]:$$

$$\|\mathbf{U}(t,\lambda)\| = 0 \Rightarrow \liminf_{\varepsilon \to 0} \mathbf{E}\left[\|\mathbf{X}_t^\varepsilon - \lambda\|^2\right] = 0 \quad (3.4.19)$$

*More precisely, for any $t \in [0,T]$ and $\lambda = \lambda(t) = (\lambda_1(t),\ldots,\lambda_n(t)) \in \mathbb{R}^n$ sutch that:*

$$U_1(t,\lambda_1(t),\ldots,\lambda_n(t)) = 0,$$

$$\cdots\cdots\cdots\cdots\cdots\cdots \quad (3.4.20)$$

$$U_n(t,\lambda_1(t),\ldots,\lambda_n(t)) = 0,$$

*the equalities is satisfaed*

$$\liminf_{\varepsilon \to 0} \mathbf{E}\Big[\,\|X_{1,t}^{\varepsilon} - \lambda_1(t)\|^{\,2}\,\Big] = 0,$$

(3.4.21)

$$\cdots\cdots\cdots\cdots\cdots\cdots\cdots$$

$$\liminf_{\varepsilon \to 0} \mathbf{E}\Big[\,\|X_{n,t}^{\varepsilon} - \lambda_n(t)\|^{\,2}\,\Big] = 0.$$

**Definition** 3.4.5. Let us consider ODE described by:

$$\dot{x}_t = \mathbf{A}_1(x_t, t) + \mathbf{A}_2(x_t, t)\mathbf{u}_t$$

$$\mathbf{A}_1 : \mathbb{R}^n \times [0, T] \mapsto \mathbb{R}^n,$$

(3.4.22)

$$\mathbf{A}_2 : \mathbb{R}^n \times [0, T] \mapsto \mathbb{R}^n \times \mathbb{R}^m,$$

where $A_1(\circ, t)$ and $A_2(\circ, t)$ is a linear transforms.

System (3.4.22) is globally null-controllable (**GNC**) in a finite time $T > 0$ if for every initial state $x_0$, there is an admissible control $u(t)$ such that the solution $x_t$ of the system satisfies: $x(0) = x_0, x(T) = 0$.

**Theorem** 3.4.4. Assume the conditions of the Theorem 3.4.3. Supose that a linear system (3.4.16) is globally null-controllable. Then SDE (3.4.15) is globally controllable.

Let us consider ODE described by:

$$\dot{x}_t = \mathbf{A}(x_t, t) + \mathbf{B}(t)\mathbf{u}_t$$

(3.4.23)

$$\mathbf{A} : \mathbb{R}^n \times [0, T] \mapsto \mathbb{R}^n,$$

where $\mathbf{A}(\circ, t)$ is a polinomial transform.

Let us consider corresponding SDE described by:

$$\dot{x}_t = \mathbf{A}(x_t, t) + \mathbf{B}(t)\mathbf{u}_t + \sqrt{\varepsilon}\,\dot{\mathbf{W}}(t),$$

$$\varepsilon \ll 1, \qquad (3.4.24)$$

$$\mathbf{A} : \mathbb{R}^n \times [0, T] \mapsto \mathbb{R}^n,$$

By Theorem (3.4.3) we obtain corresponding master equation:

$$\frac{d\mathbf{U}(t,\lambda)}{dt} = \mathbf{J}[\mathbf{A}(\lambda,t)]\mathbf{U}(\mathbf{t},\lambda) + \mathbf{A}(\lambda,\mathbf{t}) + \mathbf{B}(t)\mathbf{u}(t),$$

$$\mathbf{U}(0,\lambda) = x_0 - \lambda, \qquad (3.4.25)$$

*where $\mathbf{J}[\mathbf{A}(\lambda,t)]$ the Jacobian, i.e. $\mathbf{J}$ is a $n \times n$-matrix:*

$$\mathbf{J}[\mathbf{A}(\lambda,t)] = \mathbf{J}[\mathbf{A}(x,t)]|_{x=\lambda} =$$

$$= \begin{bmatrix} \frac{\partial A_1(x,t)}{\partial x_1} & \cdots & \frac{\partial A_1(x,t)}{\partial x_n} \\ \cdot & \cdots & \cdot \\ \cdot & \cdots & \cdot \\ \cdot & \cdots & \cdot \\ \frac{\partial A_n(x,t)}{\partial x_1} & \cdots & \frac{\partial A_n(x,t)}{\partial x_n} \end{bmatrix}\Bigg|_{x=\lambda} \qquad (3.4.26)$$

**Theorem**   *3.4.5. Supose that for any $\lambda \in \mathbb{R}^n$ the matrix $\mathbf{W}(\lambda, T)$ is positive definite, where*

$$\mathbf{W}(\lambda, T) = \int_0^T \mathbf{M}^{-1}(\lambda,t)\mathbf{B}(t)\mathbf{B}^T(t)\left(\mathbf{M}^T(\lambda,t)\right)^{-1}dt, \qquad (3.4.23)$$

and where **M**($\lambda, t$) is the matrix, defined by

$$\dot{\mathbf{M}}(\lambda, t) = \mathbf{J}[\mathbf{A}(\lambda, t)]\mathbf{M}(\lambda, t),$$

(3.4.24)

$$\mathbf{M}(\lambda, 0) = \mathbf{I}.$$

Then SDE (3.4.24) is globally controllable.

# Chapter IV. Optimal control numerical simulation.
## IV.1. Optimal control numerical simulation. Dissipative systems. Global strategy.

**Example** 4.1.1 Let us consider an 2-persons dissipative differential game $DG_{2;T}(f,0,0)$, with nonlinear dynamics:

$$\dot{x}_1 = x_2,$$

$$\dot{x}_2 = -\kappa x_2^3 + \alpha_1(t) + \alpha_2(t),$$

$$\kappa > 0.$$

(4.1.1)

$$\alpha_1(t) \in [-\rho_1, \rho_1], \alpha_2(t) \in [-\rho_2, \rho_2],$$

$$\mathbf{J}_i = x_1^2(T), i = 1, 2$$

Thus optimal control problem for the first player:

$$\min_{\alpha_1(t)\in[-\rho_u,\rho_u]} \left( \max_{\alpha_2(t)\in[-\rho_v,\rho_v]} [x_1^2(T)] \right) \tag{4.1.2}$$

and optimal control problem for the second player:

$$\max_{\alpha_1(t)\in[-\rho_u,\rho_u]} \left( \min_{\alpha_2(t)\in[-\rho_v,\rho_v]} [x_1^2(T)] \right). \tag{4.1.3}$$

From Eqs.(3.15)-(3.16) we obtain linear master game for the optimal control problem (4.1.1)-(4.1.3):

$$\dot{u}_1 = u_2 + \lambda_2,$$

$$\dot{u}_2 = -3\kappa\lambda_2^2 u_2 - \kappa\lambda_2^3 + \check{\alpha}_1(t) + \check{\alpha}_2(t),$$

$$\kappa > 0, \tag{4.1.4}$$

$$\check{\alpha}_1(t) \in [-\rho_1,\rho_1], \check{\alpha}_2(t) \in [-\rho_2,\rho_2],$$

$$\mathbf{J}_i = u_1^2(T), i = 1,2.$$

Thus optimal control problem for the first player:

$$\min_{\check{\alpha}_1(t)\in[-\rho_u,\rho_u]} \left( \max_{\check{\alpha}_2(t)\in[-\rho_v,\rho_v]} u_1^2(T) \right), \tag{4.1.5}$$

and optimal control problem for the second player:

$$\max_{\check{\alpha}_1(t)\in[-\rho_u,\rho_u]} \left( \min_{\check{\alpha}_2(t)\in[-\rho_v,\rho_v]} u_1^2(T) \right). \tag{4.1.6}$$

From Eq.(A.14) we obtain optimal control $\alpha_1^*(t)$ for the first player and optimal control $\alpha_2^*(t)$ for the second player

$$\alpha_1^*(t) = -\rho_1\text{sign}[x_1(t) + (T-t)x_2(t)],$$

$$\alpha_2^*(t) = \rho_2\text{sign}[x_1(t) + (T-t)x_2(t)].$$

(4.1.7)

Thus for numerical simulation we obtain ODE:

$$\dot{x}_1(t) = x_2(t),$$

$$\dot{x}_2(t) = -\kappa x_2^3(t) - \rho_1 \cdot \text{sign}[x_1(t) + (T-t)x_2(t)] + \alpha_2(t).$$

(4.1.8)

**Definition** 4.1.1.

$$T^* = T^*(\kappa, \rho_1, x_1(0)) = \inf_{T \in (0,\infty)} \{T | x_1(T, \rho_1, \rho_2 = 0) = 0\}. \quad (4.1.9)$$

***Numerical simulation. Example*** *4.1.1.1.*
Control of the second player: $\alpha_2(t) = 0;\ \kappa = 0, 0 \leq t \leq T = T^* = 2.6\,\text{sec}$

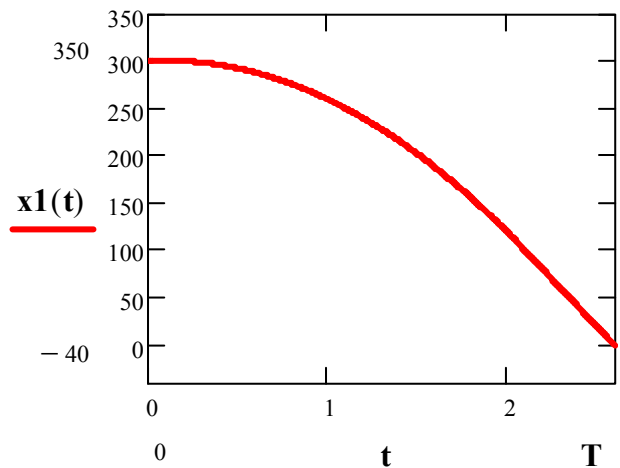

Optimal trajectory

Pic.4.1.1.1.1.Optimal trajectory:$x_1(0) = 300 \cdot 10^2\,\textbf{sm}$.

$$T^*(\kappa, \rho_1, x_1(0)) = 2.6\,\text{sec}, \rho_1 = 100, \rho_2 = 0, \kappa = 0,$$
$$x_1(T) = 5.8 \cdot 10^{-6}\,\text{sm},$$

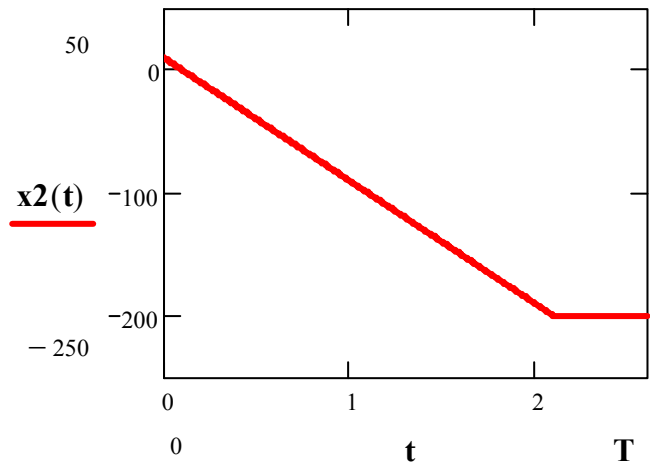

Optimal velocity

Pic.4.1.1.1.2. Optimal velocity $x_2(t) = \dot{x}_1(t)$.

$T^*(\kappa, x_1(0)) = 2.6\,\text{sec}, \rho_1 = 100, \rho_2 = 0,$
$\kappa = 0.$

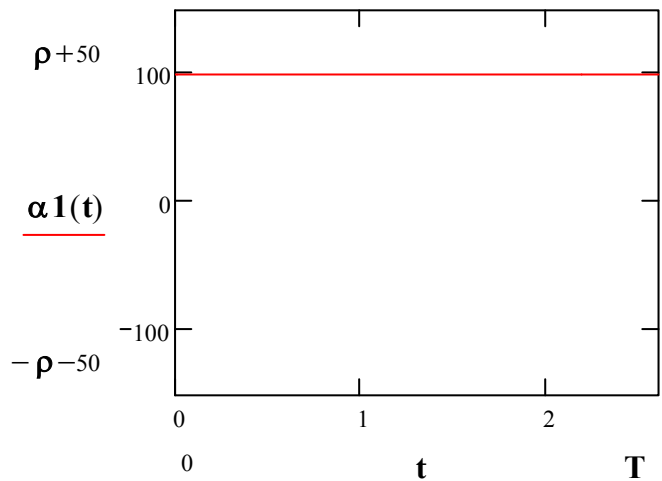

Optimal control for the first player.

Pic.4.1.1.1.3. $\alpha_1(t)$-optimal control for the first player.

$T^*(\kappa, x_1(0)) = 2.6\,\text{sec}, \rho_1 = 100, \rho_2 = 0, \kappa = 0.$

**Numerical simulation. Example** 4.1.1.2.
Control of the second player: $\alpha_2(t) = 0; \kappa = 1, 0 \leq t \leq T = 70\,\text{sec}$

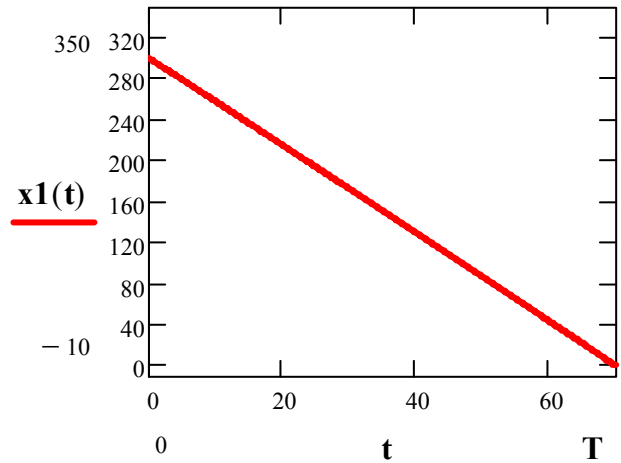

Optimal trajectory

Pic.4.1.1.2.1.Optimal trajectory:$x_1(0) = 300 \cdot 10^2\,\text{sm}$.

$T^*(\kappa, \rho_1, x_1(0)) = 67\,\text{sec}, \rho_1 = 100, \rho_2 = 0, \kappa = 1,$
$x_1(T) = 5.8 \cdot 10^{-2}\,\text{sm},$

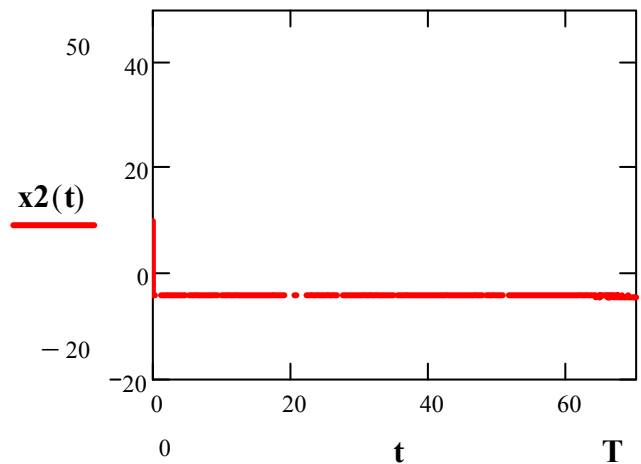

Optimal velocity

Pic.4.1.1.2.2.Optimal velocity $x_2(t) = \dot{x}_1(t)$.

$T^*(\kappa, x_1(0)) = 67\,\text{sec}, \rho_1 = 100, \rho_2 = 0,$
$\kappa = 1.$

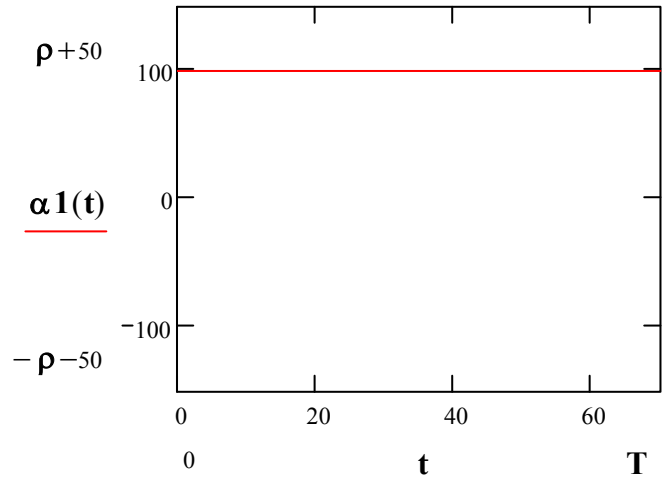

Optimal control for the first player.

Pic.4.1.1.2.3. $\alpha_1(t)$-optimal control for the first player.

$$T^*(\kappa, x_1(0)) = 67\,\text{sec},\ \rho_1 = 100, \rho_2 = 0, \kappa = 1.$$

**Numerical simulation. Example** 4.1.1.3.
Control of the second player: $\alpha_2(t) = 0;\ \kappa = 1, 0 \leq t \leq T = T^* = 77\,\text{sec}$

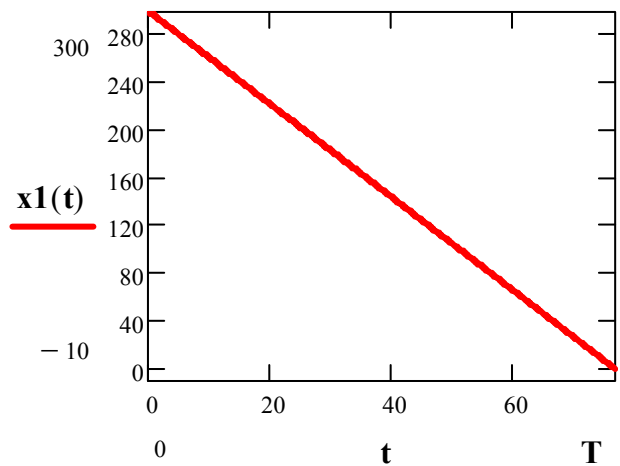

Optimal trajectory

Pic.4.1.1.3.1. Optimal trajectory: $x_1(t)$. $x_1(0) = 300 \cdot 10^2\,\text{sm}$.
$T^*(\kappa, x_1(0)) = 77\,\text{sec}, \rho_1 = 60, \rho_2 = 0, \kappa = 1,$
$x_1(T) = 4.58 \cdot 10^{-2}\,\text{sm},$

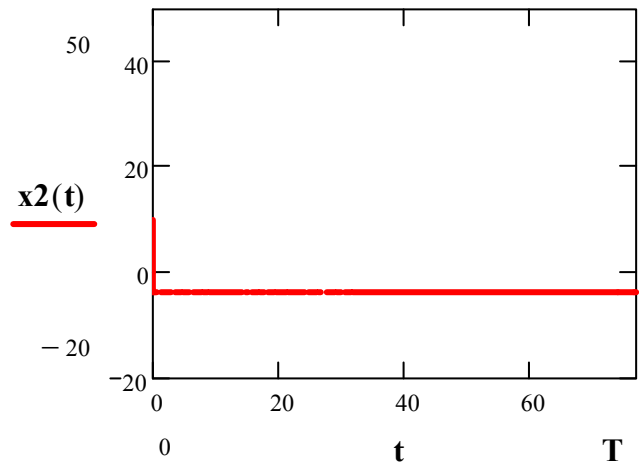

Optimal velocity

Pic.4.1.1.3.2. Optimal velocity $x_2(t) = \dot{x}_1(t)$.

$T^*(\kappa, x_1(0)) = 77\,\text{sec},\ \rho_1 = 60, \rho_2 = 0,$
$\kappa = 1.$

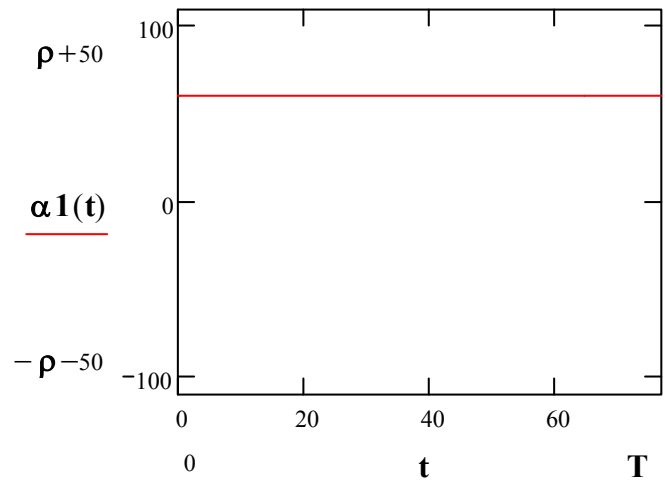

Optimal control for the first player.

Pic.4.1.1.3.3. $T^*(\kappa, \rho_1, x_1(0)) = 77\,\text{sec},\ \rho_1 = 60, \rho_2 = 0,$
$\kappa = 1.$

**Numerical simulation. Example** 4.1.1.4.

Control of the second player: $\alpha_2(t) = A\sin^2(\omega \cdot t), A = 100, \omega = 5;$
$\kappa = 1, \rho_1 = 300, 0 \leq t \leq T = 60\,\text{sec}$

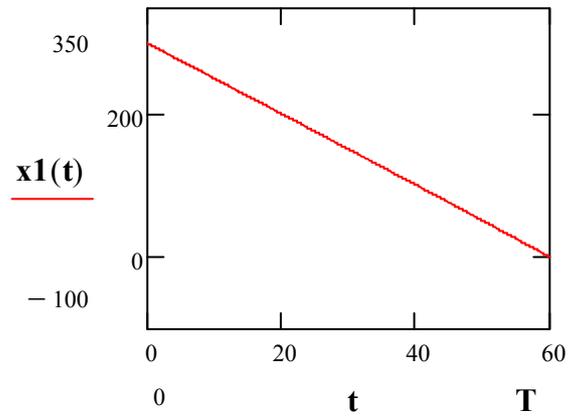

Optimal trajectory

Pic.4.1.1.4.1.Optimal trajectory : $x_1(t).x_1(T) = 0$

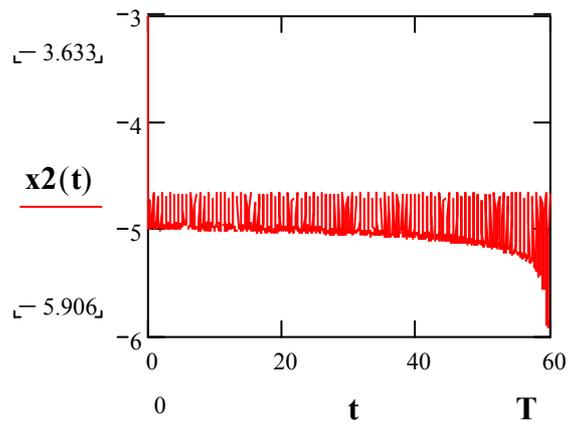

Optimalvelocity

Pic.4.1.1.4.2.Optimal velocity : $x_2(t) = \dot{x}_1(t)$.

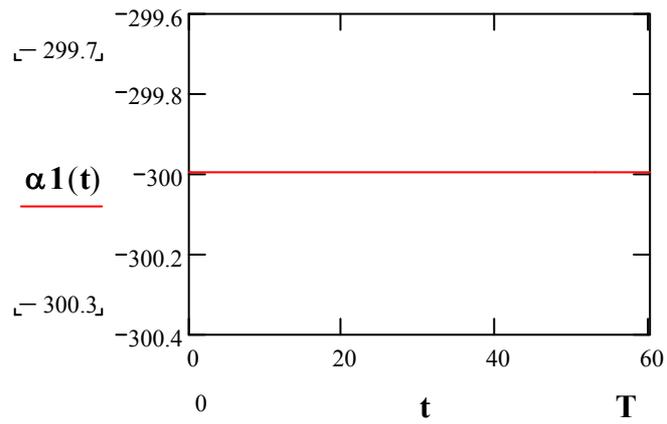

optimal control of the first player.

Pic.4.1.1.4.3. $\alpha_1(t)$ – optimal control of the first player.

$$\rho_1 = 300.$$

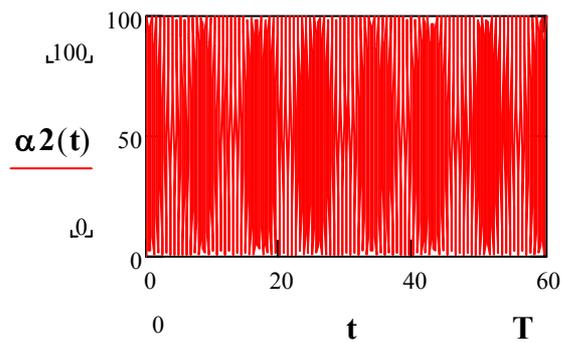

control of the second player.

Pic.4.1.1.4.4. $\alpha_2(t)$ control of the second player.

$$A = 100, \omega = 5.$$

**Numerical simulation. Example** 4.1.1.5.
Control of the second player: $\alpha_2(t) = A\sin(\omega \cdot t), A = 100, \omega = 5;$
$\kappa = 1, \rho_1 = 300, 0 \leq t \leq T = 60\,\text{sec}$

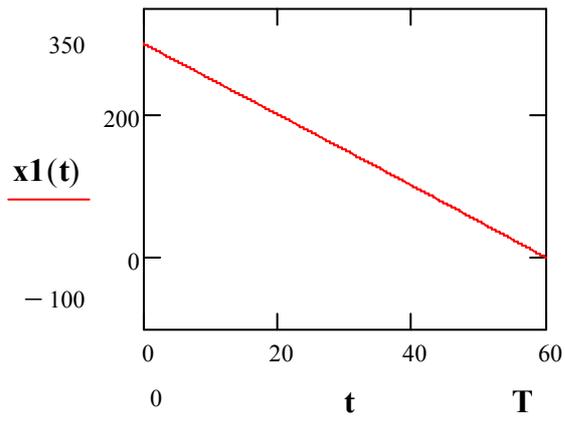

Optimal trajectory

Pic.4.1.1.5.1.Optimal trajectory : $x_1(t). x_1(T) = 0$

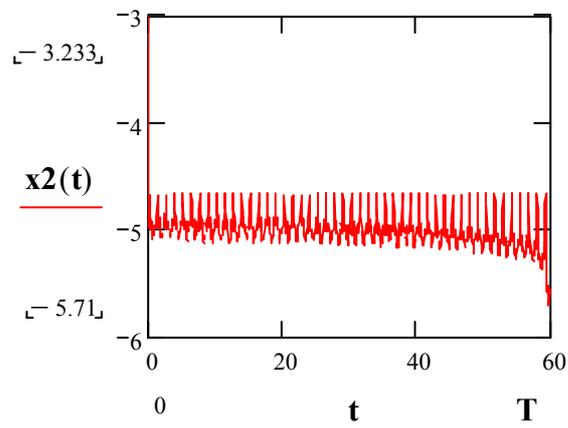

Optimalvelocity

Pic.4.1.1.5.2.Optimal velocity : $x_2(t) = \dot{x}_1(t)$.

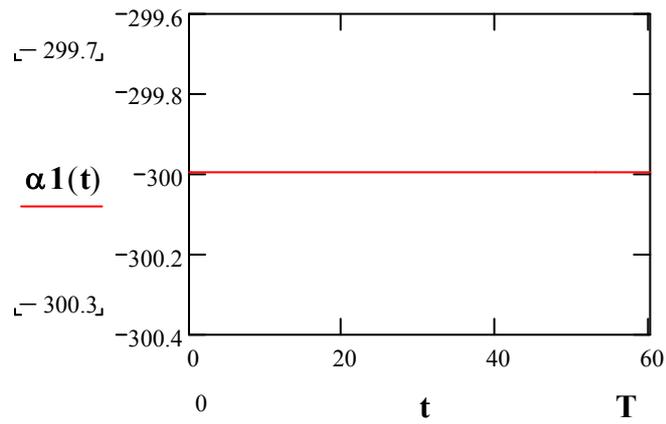

optimal control of the first player.

Pic.4.1.1.5.3. $\alpha_1(t)$ – optimal control of the first player.
$$\rho_1 = 300.$$

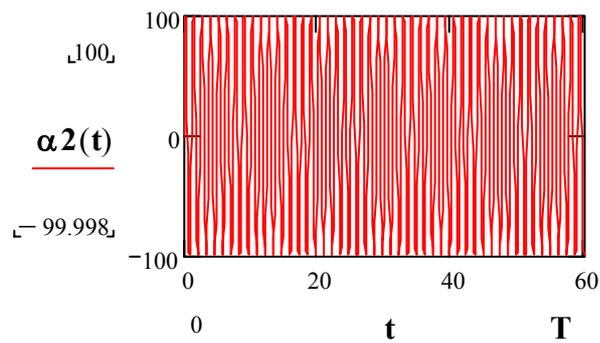

control of the second player.

Pic.4.1.1.5.4. $\alpha_2(t)$ control of the second player.
$$A = 100, \omega = 5.$$

**Numerical simulation. Example** 4.1.1.6.

Control of the second player: $\alpha_2(t) = A\sin(\omega \cdot t), A = 100, \omega = 5;$
$\kappa = 3, \rho_1 = 300, 0 \leq t \leq T = 80\,\text{sec}$

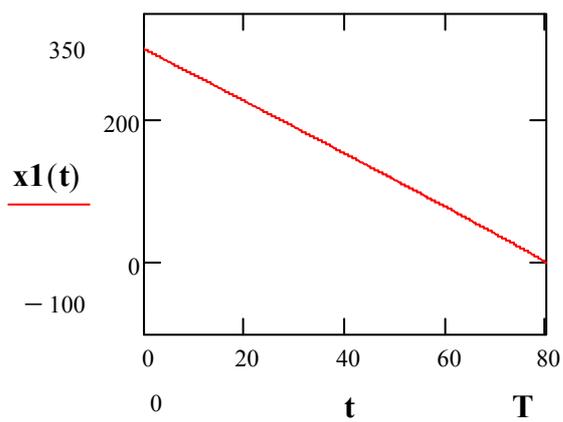

Optimal trajectory

Pic.4.1.1.6.1.Optimal trajectory : $x_1(t). x_1(T) = 0$.

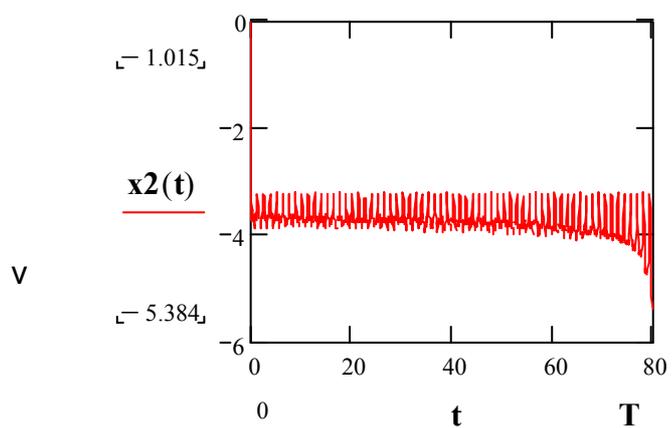

Optimalvelocity

Pic.4.1.1.6.2.Optimal velocity : $x_2(t) = \dot{x}_1(t)$.

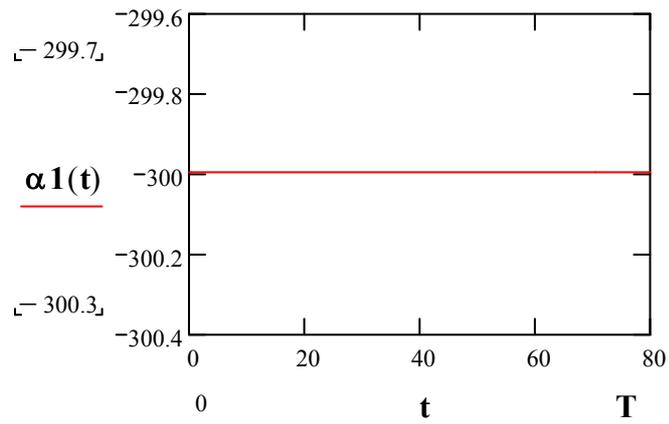

optimal control of the first player.

Pic.4.1.1.6.3.$\alpha_1(t)$ – optimal control of the first player.
$\rho_1 = 300.$

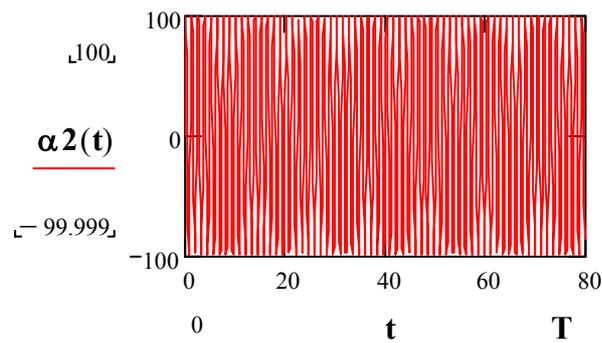

control of the second player.

Pic.4.1.1.6.4.$\alpha_2(t)$ control of the second player.
$A = 100, \omega = 5.$

**Numerical simulation. Example** 4.1.1.7.

Control of the second player: $\alpha_2(t) = A\sin^2(\omega \cdot t), A = 100, \omega = 5;$
$\kappa = 3, \rho_1 = 300, 0 \leq t \leq T = 80\,\text{sec}$

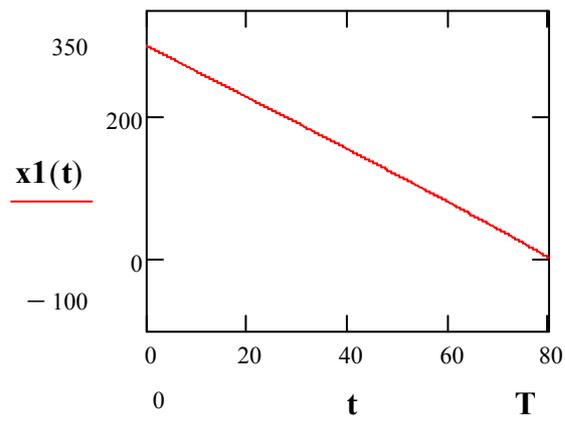

Optimal trajectory

Pic.4.1.1.7.1.Optimal trajectory : $x_1(t). x_1(T) = 0$

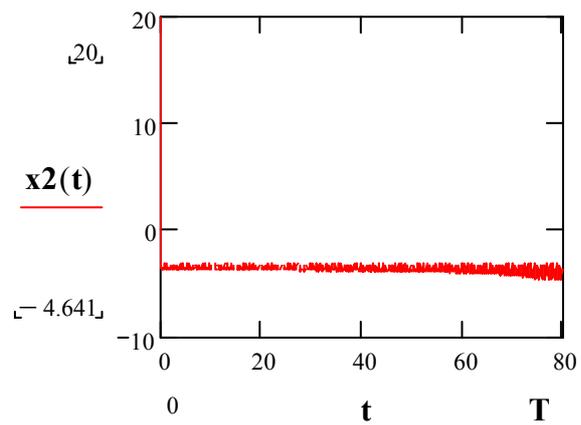

Optimal velocity.

Pic.4.1.1.7.2.Optimal velocity : $x_2(t) = \dot{x}_1(t)$.

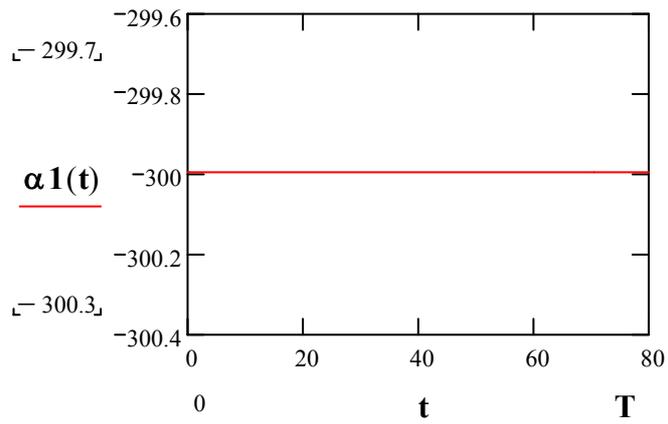

optimal control of the first player.

Pic.4.1.1.7.3. $\alpha_1(t)$ – optimal control of the first player.
$$\rho_1 = 300.$$

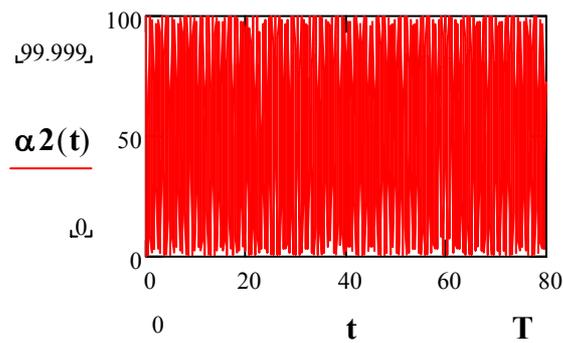

control of the second player.

Pic.4.1.1.7.4. $\alpha_2(t)$ control of the second player.
$$A = 100, \omega = 5.$$

**Example** 4.1.2. Let us consider an 2-persons dissipative differential game $DG_{2;T}(f,0,0)$, with nonlinear dynamics

$$\dot{x}_1 = x_2,$$

$$\dot{x}_2 = -\kappa x_2^3 + \alpha_1(t) + \alpha_2(t),$$

$$\kappa > 0. \qquad (4.1.10)$$

$$\alpha_1(t) \in [-\rho_1, \rho_1], \alpha_2(t) \in [-\rho_2, \rho_2],$$

$$\mathbf{J}_i = x_1^2(T) + x_2^2(T), i = 1, 2$$

Thus optimal control problem for the first player:

$$\min_{\alpha_1(t)\in[-\rho_1,\rho_2]} \left( \max_{\alpha_2(t)\in[-\rho_2,\rho_2]} [x_1^2(T) + x_2^2(T)] \right) \qquad (4.1.11)$$

and optimal control problem for the second player:

$$\max_{\alpha_2(t)\in[-\rho_u,\rho_u]} \left( \min_{\alpha_1(t)\in[-\rho_1,\rho_1]} [x_1^2(T) + x_2^2(T)] \right). \qquad (4.1.12)$$

From Eqs.(3.15)-(3.16) we obtain linear master game for the optimal control problem (4.1.10)-(4.1.12):

$$\dot{u}_1 = u_2 + \lambda_2,$$

$$\dot{u}_2 = -3\kappa\lambda_2^2 u_2 - \kappa\lambda_2^3 + \check{\alpha}_1(t) + \check{\alpha}_2(t),$$

$$\kappa > 0, \qquad (4.1.13)$$

$$\check{\alpha}_1(t) \in [-\rho_1, \rho_1], \check{\alpha}_2(t) \in [-\rho_2, \rho_2],$$

$$\mathbf{\bar{J}}_i = u_1^2(T) + u_2^2(T), i = 1, 2.$$

Thus optimal control problem for the first player:

$$\min_{\breve{\alpha}_1(t)\in[-\rho_1,\rho_1]} \left( \max_{\breve{\alpha}_2(t)\in[-\rho_2,\rho_2]} [u_1^2(T) + u_2^2(T)] \right), \quad (4.1.14)$$

and optimal control problem for the second player:

$$\max_{\breve{\alpha}_2(t)\in[-\rho_2,\rho_2]} \left( \min_{\breve{\alpha}_1(t)\in[-\rho_1,\rho_1]} [u_1^2(T) + u_2^2(T)] \right). \quad (4.1.15)$$

From Eq.(A.14) we obtain optimal control $\alpha_1^*(t)$ for the first player and optimal control $\alpha_2^*(t)$ for the second player:

$$\alpha_1^*(t) = -\rho_1 \text{sign}[x_1(t) + [(T-t) + \exp[-3\kappa x_2^2(t)(T-t)]]x_2(t)],$$

$$(4.1.16)$$

$$\alpha_2^*(t) = \rho_2 \text{sign}[x_1(t) + [(T-t) + \exp[-3\kappa x_2^2(t)(T-t)]]x_2(t)].$$

Thus for numerical simulation we obtain ODE:

$$\dot{x}_1(t) = x_2(t),$$

$$(4.1.17)$$

$$\dot{x}_2(t) = -\kappa x_2^3(t) - \rho_1 \cdot \text{sign}[x_1(t) + [(T-t) + \exp[-3\kappa x_2^2(t)(T-t)]]x_2(t)] + \alpha_2(t).$$

**Numerical simulation. Example** 4.1.2.1.
Control of the second player: $\alpha_2(t) = A\sin^2(\omega \cdot t), A = 100, \omega = 5$;
$\kappa = 1, \rho_1 = 400, 0 \leq t \leq T = 80\sec$

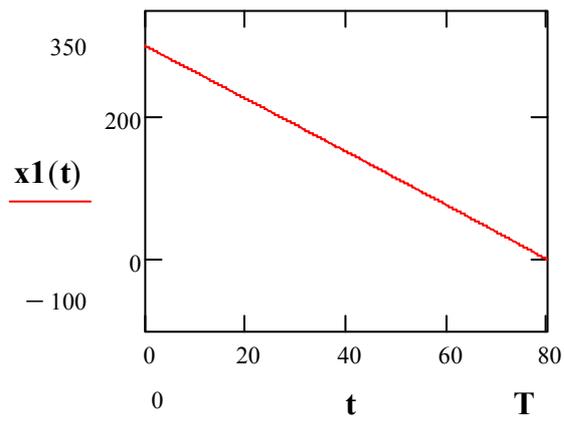

Optimal trajectory

Pic.4.1.2.1.1.Optimal trajectory : $x_1(t)$.
$x_1(0) = 300, x_1(T) = 0.39$.

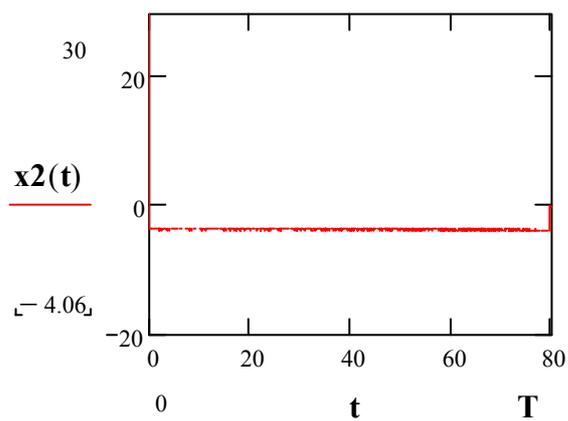

Optimal velocity.

Pic.4.1.2.1.2.Optimal velocity : $x_2(t) = \dot{x}_1(t)$.
$\dot{x}_1(0) = 30, |\dot{x}_1(T)| = 0.25$.

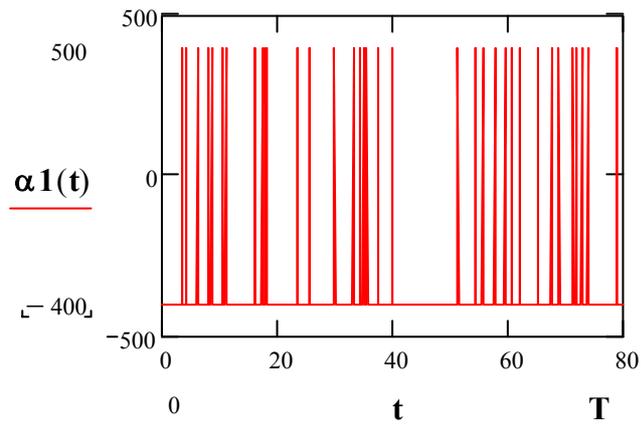

optimal control of the first player.

Pic.4.1.2.1.3. $\alpha_1(t)$ – optimal control of the first player.

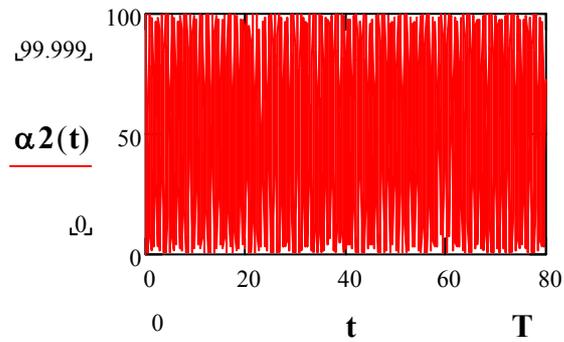

control of the second player.

Pic.4.1.2.1.4. $\alpha_2(t)$ – control of the second player.

**Numerical simulation. Example** 4.1.2.2.

Control of the second player: $\alpha_2(t) = A\sin(\omega \cdot t), A = 100, \omega = 5$;
$\kappa = 1, \rho_1 = 400, 0 \leq t \leq T = 80\sec$

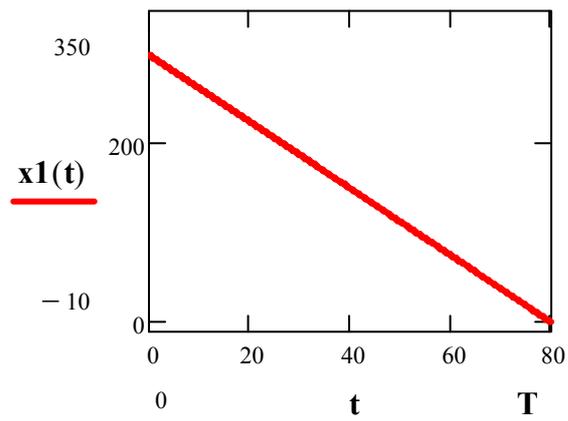

Optimal trajectory

Pic.4.1.2.2.1.Optimal trajectory : $x_1(t)$.
$x_1(0) = 300, x_1(T) = 0.386$.

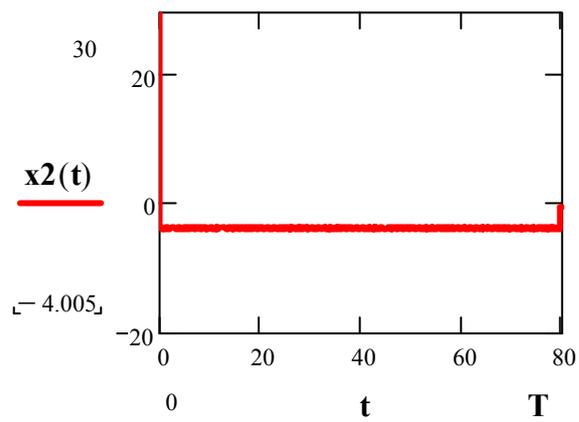

Optimal velocity.

Pic.4.1.2.2.2.Optimal velocity : $x_2(t) = \dot{x}_1(t)$.
$\dot{x}_1(0) = 30, |\dot{x}_1(T)| = 0.408$.

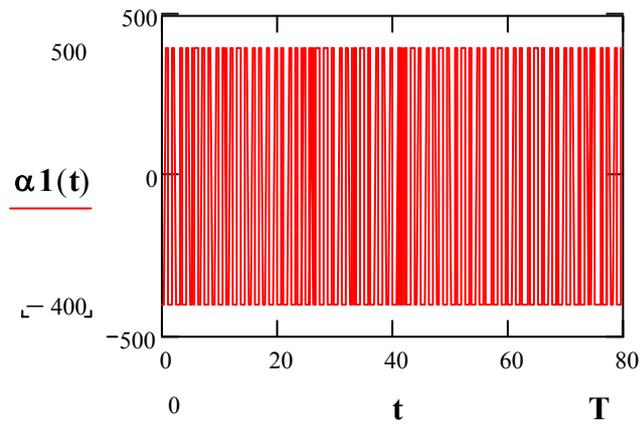

optimal control of the first player.

Pic.4.1.2.2.3.$\alpha_1(t)$ – optimal control of the first player.

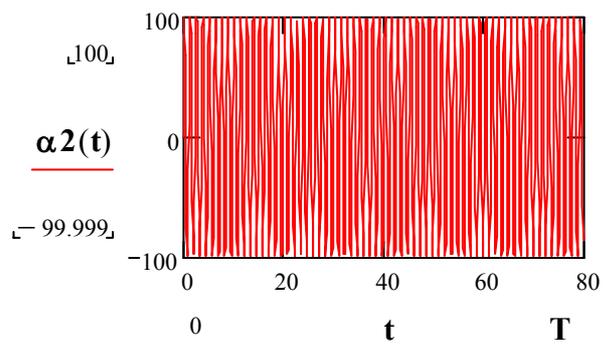

control of the second player.

Pic.4.1.2.2.4.$\alpha_2(t)$ – control of the second player.

**Numerical simulation. Example** 4.1.2.3.

Control of the second player: $\alpha_2(t) = A\sin(\omega \cdot t), A = 100, \omega = 5$; $\kappa = 3, \rho_1 = 400, 0 \leq t \leq T = 80\,\text{sec}$

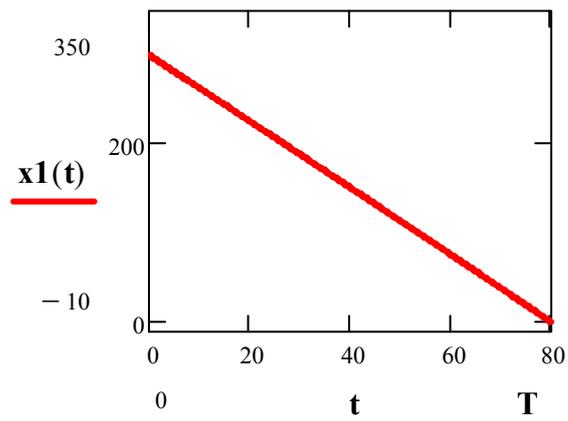

Optimal trajectory

Pic.4.1.2.3.1.Optimal trajectory : $x_1(t)$.
$x_1(0) = 300, x_1(T) = 0.144$.

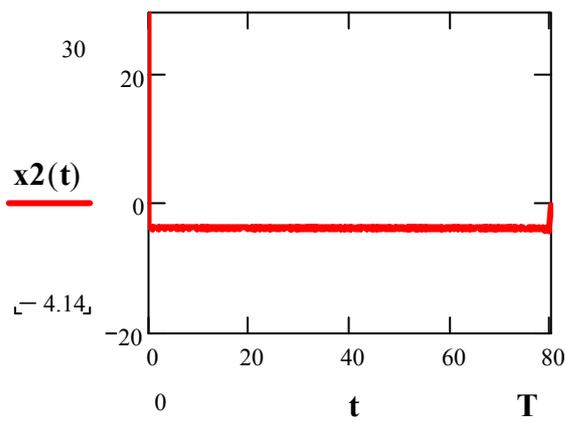

Optimal velocity.

Pic.4.1.2.3.2.Optimal velocity : $x_2(t) = \dot{x}_1(t)$.
$\dot{x}_1(0) = 30, |\dot{x}_1(T)| = 0.221$.

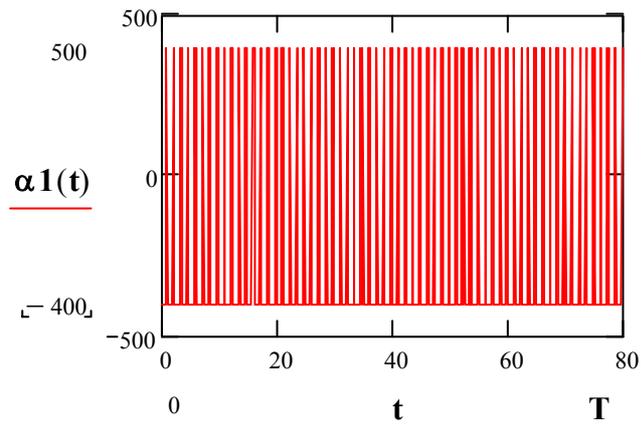

optimal control of the first player.

Pic.4.1.2.3.3. $\alpha_1(t)$ – optimal control of the first player.

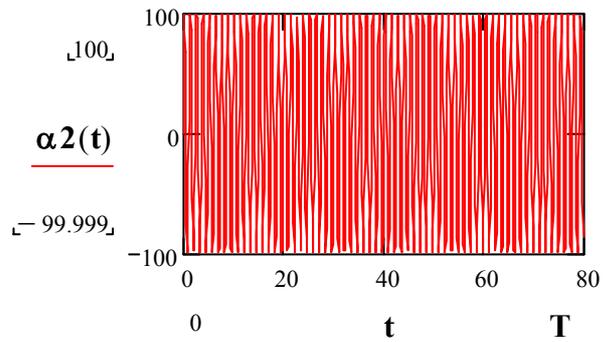

control of the second player.

Pic.4.1.2.3.4. $\alpha_2(t)$ – control of the second player.

**Numerical simulation. Example** 4.1.2.4.

Control of the second player: $\alpha_2(t) = A\sin(\omega \cdot t), A = 100, \omega = 5$; $\kappa = 3, \rho_1 = 200, 0 \leq t \leq T = 135 \sec$

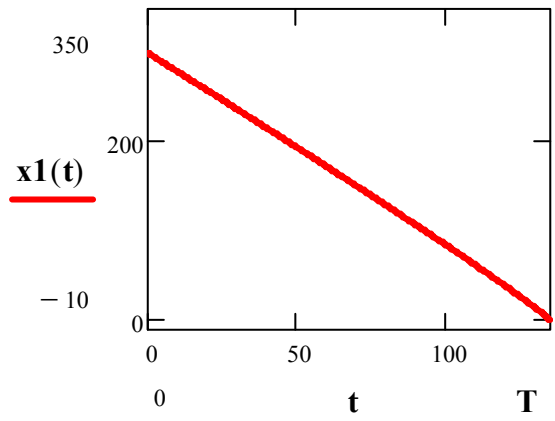

Optimal trajectory

Pic.4.1.2.4.1.Optimal trajectory : $x_1(t)$.
$x_1(0) = 300, x_1(T) = 0.62$.

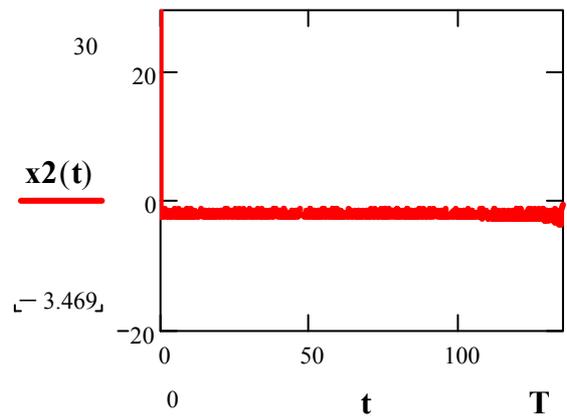

Optimal velocity.

Pic.4.1.2.4.2.Optimal velocity : $x_2(t) = \dot{x}_1(t)$.
$\dot{x}_1(0) = 30, |\dot{x}_1(T)| = 0.605$.

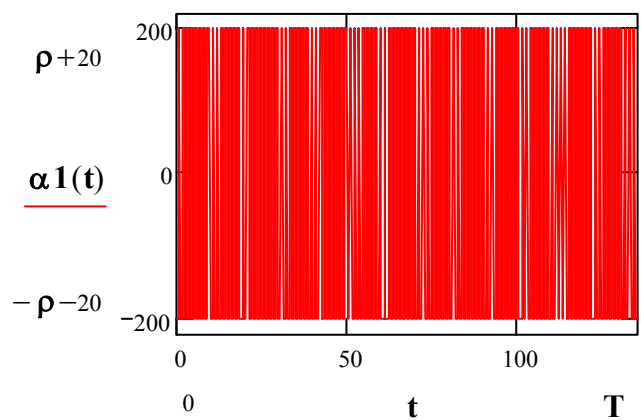

optimal control of the first player.

Pic.4.1.2.4.3. $\alpha_1(t)$ – optimal control of the first player.

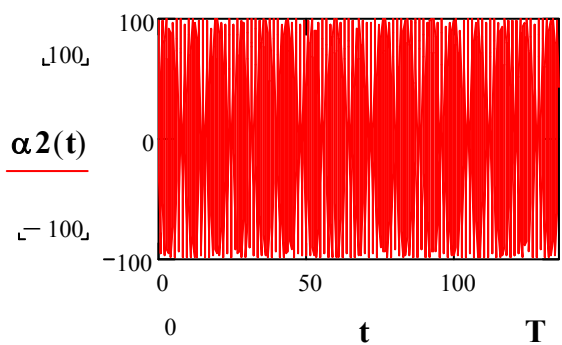

control of the second player.

Pic.4.1.2.4.4. $\alpha_2(t)$ – control of the second player.

## IV.2.1. Optimal control numerical simulation. Dissipative systems. "Step-by-step" strategy.

**Example 4.2.1.** Let us consider an 2-persons dissipative differential game $DG_{2;T}(f,0,0),$ with nonlinear dynamics

$$\dot{x}_1 = x_2,$$

$$\dot{x}_2 = -\kappa x_2^3 + \alpha_1(t) + \alpha_2(t),$$

$$\kappa > 0. \tag{4.2.1}$$

$$\alpha_1(t) \in [-\rho_1, \rho_1], \alpha_2(t) \in [-\rho_2, \rho_2]$$

$$\mathbf{J}_i = x_1^2(T) + \dot{x}_1^2(T), i = 1,2$$

Thus optimal control problem for the first player:

$$\min_{\alpha_1(t) \in [-\rho_u, \rho_u]} \left( \max_{\alpha_2(t) \in [-\rho_v, \rho_v]} [x_1^2(T) + \dot{x}_1^2(T)] \right) \tag{4.2.2}$$

and optimal control problem for the second player:

$$\max_{\alpha_1(t) \in [-\rho_u, \rho_u]} \left( \min_{\alpha_2(t) \in [-\rho_v, \rho_v]} [x_1^2(T) + \dot{x}_1^2(T)] \right). \tag{4.2.3}$$

From Eqs.(3.15)-(3.16) we obtain linear master game for the optimal control problem (4.2.1)-(4.2.3):

$$\dot{u}_1 = u_2 + \lambda_2,$$

$$\dot{u}_2 = -3\kappa \lambda_2^2 u_2 - \kappa \lambda_2^3 + \check{\alpha}_1(t) + \check{\alpha}_2(t),$$

$$\kappa > 0, \tag{4.2.4}$$

$$\check{\alpha}_1(t) \in [-\rho_1, \rho_1], \check{\alpha}_2(t) \in [-\rho_2, \rho_2],$$

$$\mathbf{J}_i = u_1^2(T) + \dot{u}_1^2(T), i = 1,2.$$

Thus optimal control problem for the first player:

$$\min_{\check{\alpha}_1(t) \in [-\rho_u, \rho_u]} \left( \max_{\check{\alpha}_2(t) \in [-\rho_v, \rho_v]} [u_1^2(T) + \dot{u}_1^2(T)] \right), \tag{4.2.5}$$

and optimal control problem for the second player:

$$\max_{\check{\alpha}_1(t) \in [-\rho_u, \rho_u]} \left( \min_{\check{\alpha}_2(t) \in [-\rho_v, \rho_v]} [u_1^2(T) + \dot{u}_1^2(T)] \right). \tag{4.2.6}$$

From Eq.(A.24),Eq.(A.26) we obtain standard solution for the linear optimal control problem (4.2.4)-(4.2.6):

$$\check{\alpha}_1(t) = -\rho_1\mathbf{sign}[x_1(t) + [(T-t) + \xi(T-t,\kappa^*)]x_2(t)],$$

$$\check{\alpha}_2(t) = \rho_2\mathbf{sign}[x_1(t) + [(T-t) + \xi(T-t,\kappa^*)]x_2(t)],$$

$$\kappa^* = -3\kappa\lambda_2^2, \qquad (4.2.7)$$

$$\xi(T-t,\kappa^*) = (\kappa^*)^{-2}[\exp(-\kappa^*(T-t)) + \kappa^*(T-t) - 1] =$$

$$= 3^{-2}\kappa^{-2}\lambda_2^{-4}[\exp(3\kappa\lambda_2^2(T-t)) - 3\kappa\lambda_2^2(T-t) - 1].$$

From Eq.(4.2.7) and Theorem 3.2, we obtain "step by step" feedback optimal control for the nonlinear optimal control problem (4.1)-(4.3). Thus "step by step" optimal control for the first player $\alpha_1^*(t)$ in the next form:

$$\alpha_1^*(t) = -\rho_1\mathbf{sign}[x_1(t_n) + [(t_{n+1}-t) + \xi(t_{n+1}-t,\kappa;x_2(t_n))]x_2(t)],$$

$$\xi(t_{n+1}-t,\kappa;x_2(t_n)) =$$

$$= 3^{-2}\kappa^{-2}x_2^{-4}(t_n)[\exp(3\kappa x_2^2(t_n)(t_{n+1}-t)) - 3\kappa x_2^2(t_n)(t_{n+1}-t) - 1],$$

$$(4.2.8)$$

$$t \in [t_n, t_{n+1}], t_{n+1} - t_n = \frac{T}{N}, n = 1,\ldots,N.$$

and "step by step" optimal control for the second player $\alpha_2^*(t)$ in the next form:

$$\alpha_2^*(t) = \rho_2\mathbf{sign}[x_1(t_n) + [(t_{n+1}-t) + \xi(t_{n+1}-t_n,\kappa;x_2(t_n))]x_2(t)],$$

$$\xi(t_{n+1}-t,\kappa;x_2(t_n)) =$$

$$(4.2.9)$$

$$= 3^{-2}\kappa^{-2}x_2^{-4}(t_n)[\exp(3\kappa x_2^2(t_n)(t_{n+1}-t)) - 3\kappa x_2^2(t_n)(t_{n+1}-t) - 1],$$

$$t \in [t_n, t_{n+1}], t_{n+1} - t_n = \frac{T}{N}, n = 1,\ldots,N.$$

Suppose that $3\kappa x_2^2(t)(t_{n+1}-t_n) \ll 1, t \in [t_n,t_{n+1}], n = 1,\ldots,N,$ from Eq.(A.28) we obtain (quasy) optimal control $\alpha_1^*(t)$ for the first player and optimal control $\alpha_2^*(t)$ for the second player in the next form:

$$\alpha_1^*(t) \simeq -\rho_1\mathbf{sign}\left[x_1(t) + \left[(t_{n+1}-t) + 0.5(t_{n+1}-t)^2\right]x_2(t)\right],$$

$$(4.2.10)$$

$$\alpha_2^*(t) \simeq \rho_2\mathbf{sign}\left[x_1(t) + \left[(t_{n+1}-t) + 0.5(t_{n+1}-t)^2\right]x_2(t)\right].$$

Suppose that $0.5(t_{n+1} - t_n)^2 x_2(t) \ll 1, t \in [t_n, t_{n+1}], n = 1, \ldots, N,$ from Eq.(4.10) we obtain (quasy) optimal control $\alpha_1^*(t)$ for the first player and (quasy) optimal control $\alpha_2^*(t)$ for the second player:

$$\alpha_1^*(t) \simeq -\rho_1 \text{sign}[x_1(t) + (t_{n+1} - t)x_2(t)],$$

$$\alpha_2^*(t) \simeq \rho_2 \text{sign}[x_1(t) + (t_{n+1} - t)x_2(t)].$$

(4.2.11)

**Definition** 4.2.1. Cutting function $\Theta_\tau(t)$ :

$$\theta_\tau(t) \triangleq \tau - t,$$

$$\eta_\tau(t) \triangleq t - \left(\text{ceil}\left(\frac{t}{\tau}\right) - 1\right)$$

(4.2.12)

**ceil**$(x)$ is a part-whole number $x \in \mathbb{R}$,

$$\Theta_\tau(t) = \theta_\tau(\eta_\tau(t)).$$

*From Eqs.(4.2.11)-(4.2.12) we obtain (quasy) optimal control $\alpha_1^*(t)$ for the first player and (quasy) optimal control $\alpha_2^*(t)$ for the second player in the next form:*

$$\alpha_1^*(t) \simeq -\rho_1 \text{sign}[x_1(t) + \Theta_\tau(t)x_2(t)],$$

$$\alpha_2^*(t) \simeq \rho_2 \text{sign}[x_1(t) + \Theta_\tau(t)x_2(t)].$$

(4.2.13)

*Thus for the numerical simulation we obtain ODE:*

$$\dot{x}_1 = x_2,$$

$$\dot{x}_2 = -\kappa x_2^3 - \rho \cdot \text{sign}[x_1 + \Theta_\tau(t)x_2] + \alpha_2(t),$$

(4.2.14)

$$\kappa > 0.$$

***Numerical simulation. Example*** *4.2.1.*

Control of the second player: $\alpha_2(t) = 500\sin(50 \cdot t). \kappa = 1, \tau = 0.1.$

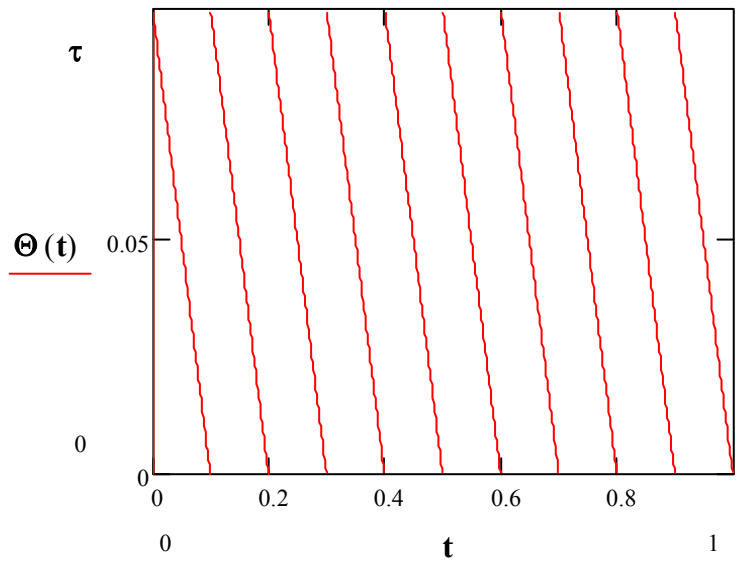

Cutting functuion

Pic.4.2.1.1.Cutting function : $\Theta(t), \tau = 0.1$.

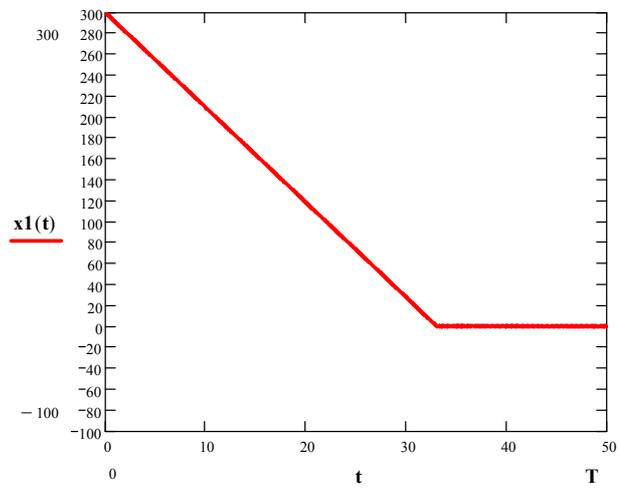

Optimal trajectory

Pic.4.2.1.2.Optimal trajectory : $x_1(t). x_1(T) = -6 \cdot 10^{-4}$, $\rho = 800, \kappa = 1, \tau = 0.1$.

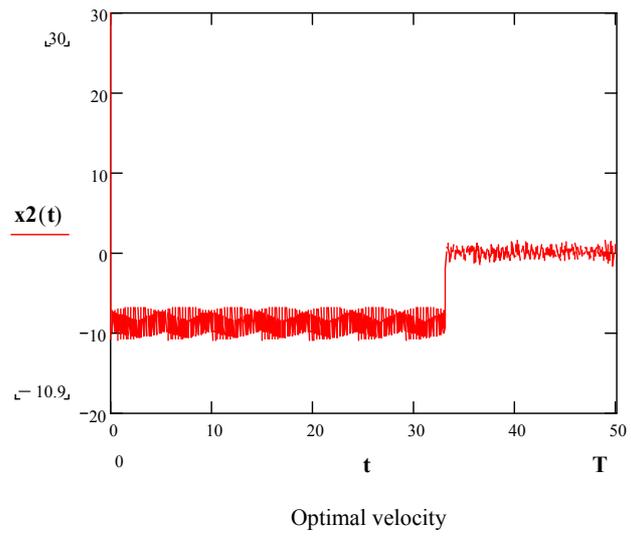

Optimal velocity

Pic.4.2.1.3. Optimal velocity : $x_2(t) = \dot{x}_1(t)$.

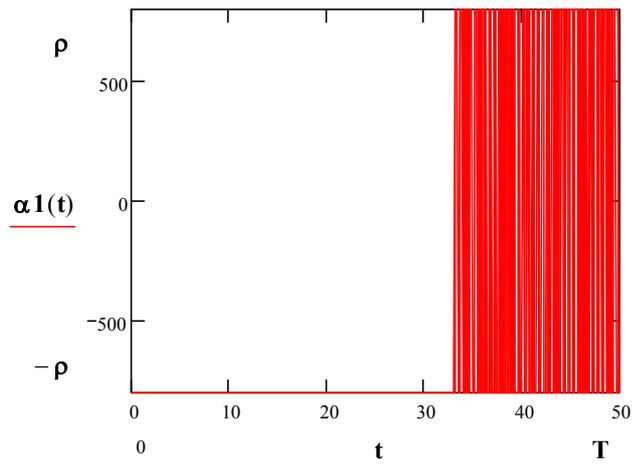

Optimal control for the first player.

Pic4.2.1.4. $\alpha_1(t)$ – optimal control for the firstplayer.

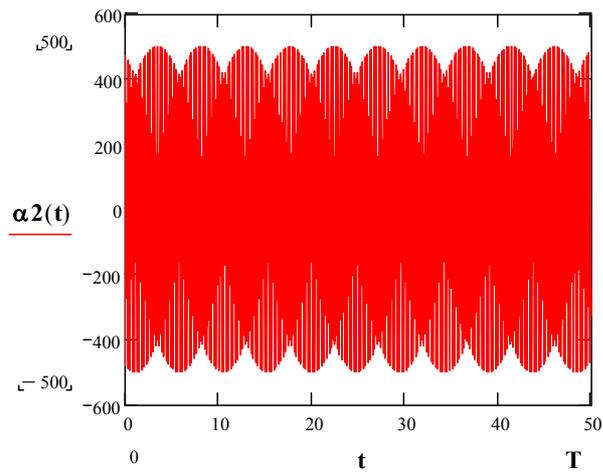

control of the second player

Pic.4.2.1.5. $\alpha_2(t)$ -control of the second player.

**Numerical simulation. Example** 4.2.2.

Control of the second player: $\alpha_2(t) = 500\sin(50 \cdot t). \kappa = 3, \tau = 0.1$.

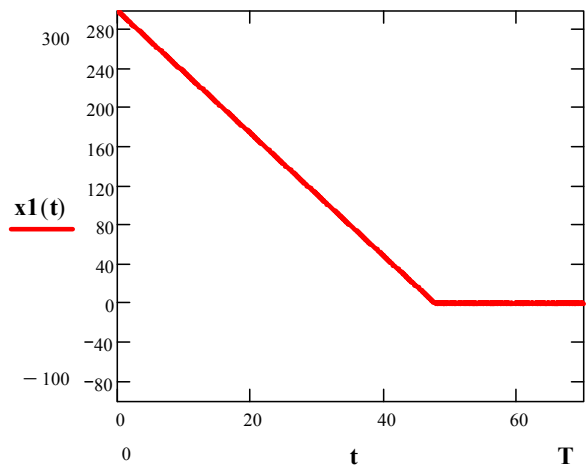

Optimal trajectory

Pic.4.2.2.1. Optimal trajectory : $x_1(t)$.
$\rho = 800, \kappa = 3, x_1(T) = -6 \cdot 10^{-3} sm, \tau = 0.1$

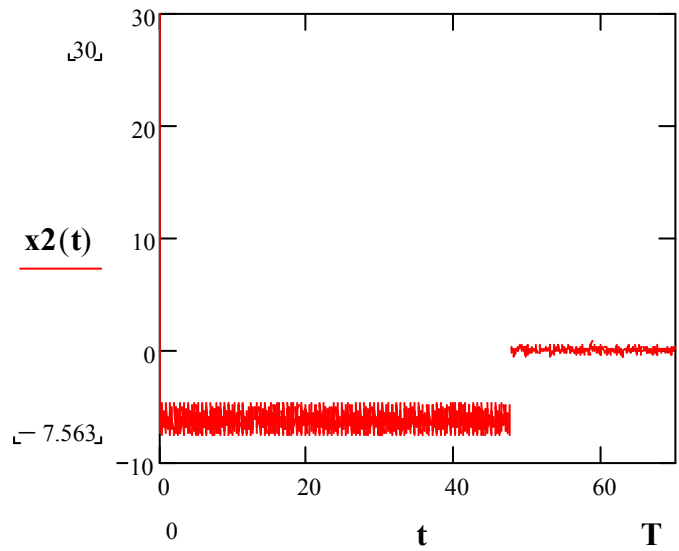

Optimal velocity

Pic.4.2.2.2. Optimal velocity : $x_2(t) = \dot{x}_1(t)$. $\rho = 800, \kappa = 3$.

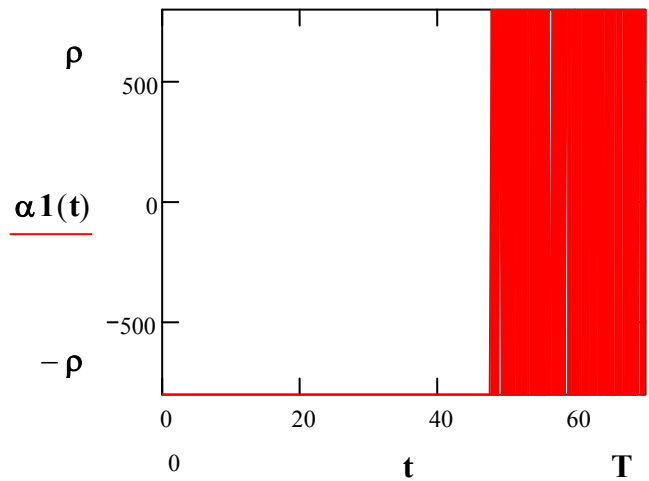

Optimal control for the first player.

Pic.4.2.2.3. $\alpha_1(t)$-optimal control for the first player :

$$\rho = 800, \kappa = 3.$$

**Numerical simulation. Example** 4.2.3.
Control of the second player: $\alpha_2(t) = 500\sin(50 \cdot t)$. $\kappa = 3$.

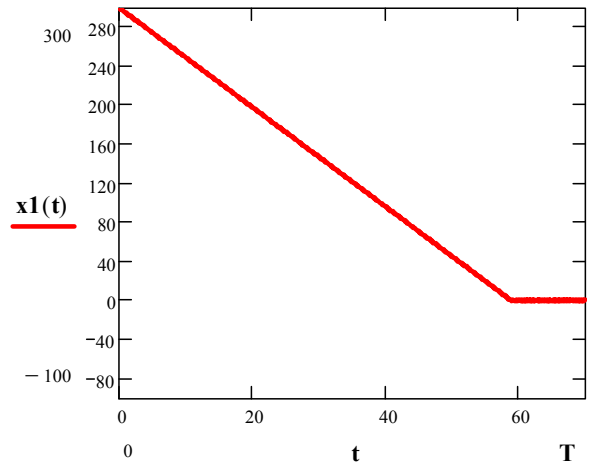

Optimal trajectory

Pic.4.2.3.1.Optimal trajectory: $x_1(t).x_1(T) = 6 \cdot 10^{-3} sm,$ $\rho = 500, \kappa = 3.$

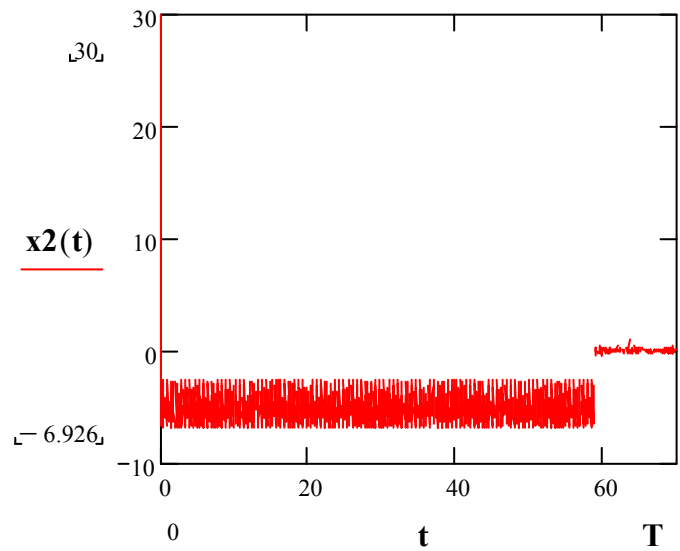

Optimal velocity

Pic.4.2.3.2.Optimal velocity $x_2(t) = \dot{x}_1(t). \rho = 500, \kappa = 3.$

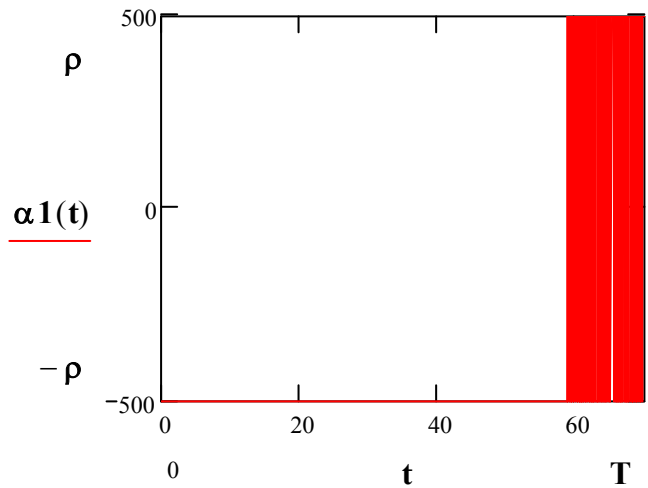

Optimal control for the first player.

Pic.4.2.3.3. $\alpha_1(t)$-optimal control for the first player:

$\rho = 500, \kappa = 3.$

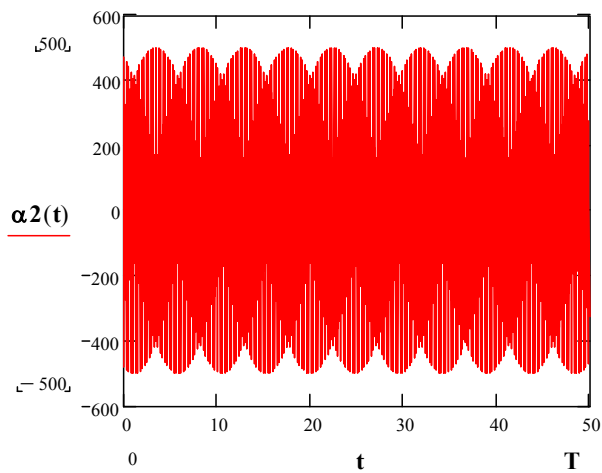

control of the second player

Pic.4.2.3.4. $\alpha_2(t)$ control of the second player.

**Numerical simulation. Example** 4.2.4.

Control of the second player: $\alpha_2(t) = 500\sin(50 \cdot t^2)$.
**Example**    4.2.4.1.

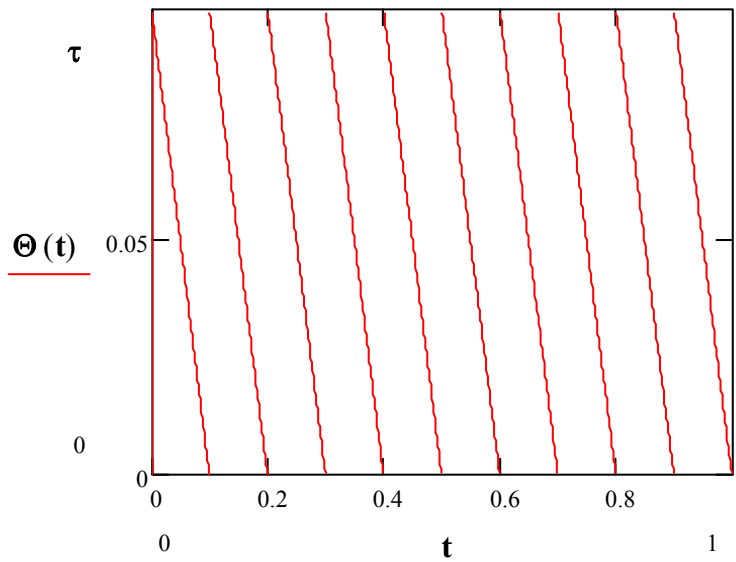

Cutting functuion

*Pic.4.2.4.1.Cutting function: $\Theta(t), \tau = 0.1$.*

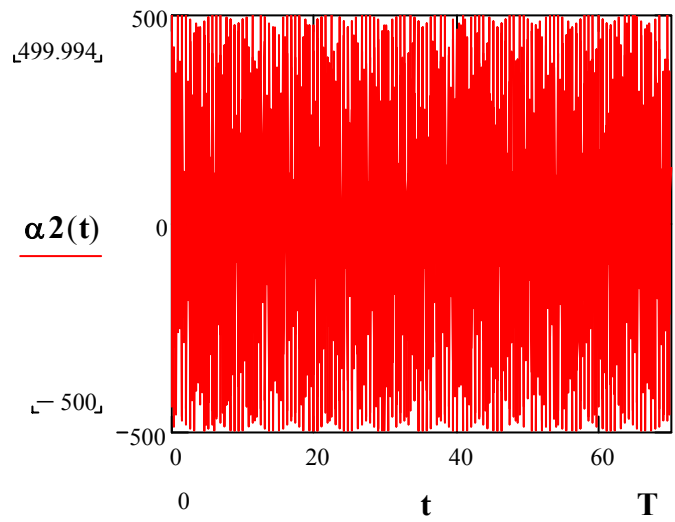

control of the second player

*Pic.4.2.4.2. $\alpha_2(t)$ control of the second player.*

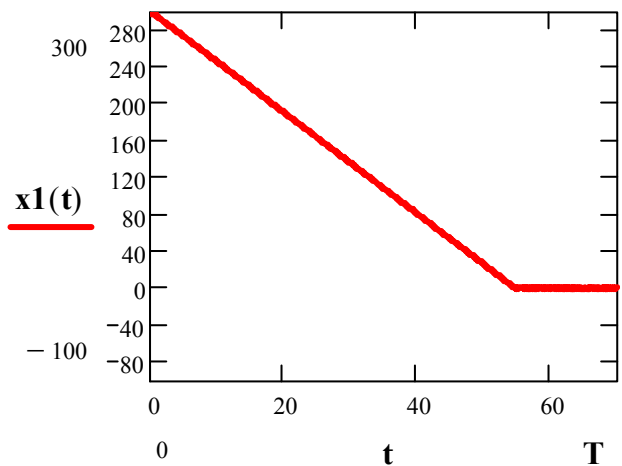

Optimal trajectory

Pic.4.2.4.3.Optimal trajectory: $x_1(t)$. $|x_1(T)| = 2.6 \cdot 10^{-3} sm$, $\rho = 500, \kappa = 3$.

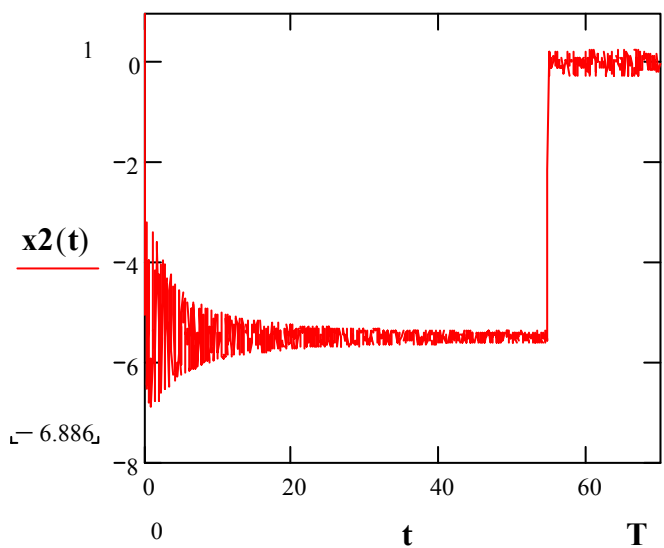

Optimal velocity

Pic.4.2.4.5.Optimal velocity $x_2(t) = \dot{x}_1(t)$. $\rho = 500, \kappa = 3$.

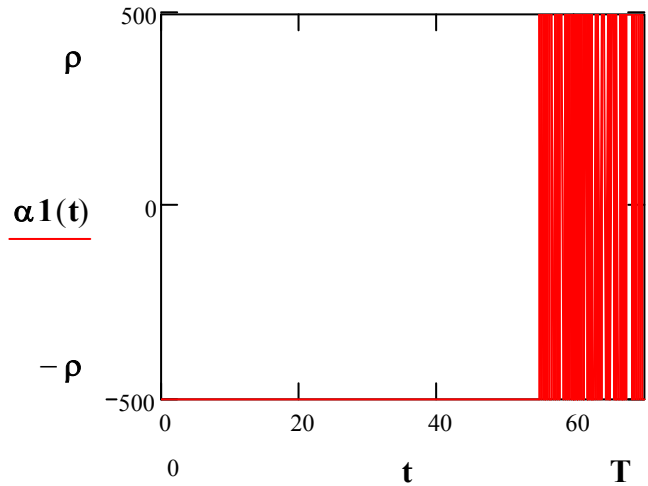

Optimal control for the first player.

Pic.4.2.4.6.$\alpha_1(t)$-optimal control for the first player: $\rho = 500, \kappa = 3$.

## Example 4.2.5.

Let us consider an 2-persons dissipative differential game $DG_{2;T}(\mathbf{f}, \mathbf{0}, \mathbf{0})$, with nonlinear dynamics

$$\dot{x}_1 = x_2,$$

$$\dot{x}_2 = -\kappa_1 x_2^3 + \kappa_2 x_2 + \alpha_1(t) + \alpha_2(t),$$

$$\kappa_1 > 0. \tag{4.2.15}$$

$$\alpha_1(t) \in [-\rho_1, \rho_1], \alpha_2(t) \in [-\rho_2, \rho_2],$$

$$\mathbf{J}_i = x_1^2(T) + \dot{x}_1^2(T), i = 1, 2$$

Thus optimal control problem for the first player:

$$\min_{\alpha_1(t) \in [-\rho_u, \rho_u]} \left( \max_{\alpha_2(t) \in [-\rho_v, \rho_v]} [x_1^2(T) + \dot{x}_1^2(T)] \right) \tag{4.2.16}$$

and optimal control problem for the second player:

$$\max_{\alpha_1(t) \in [-\rho_u, \rho_u]} \left( \min_{\alpha_2(t) \in [-\rho_v, \rho_v]} [x_1^2(T) + \dot{x}_1^2(T)] \right). \tag{4.2.17}$$

For the numerical simulation we obtain ODE:

$$\dot{x}_1 = x_2,$$
$$\dot{x}_2 = -\kappa_1 x_2^3 + \kappa_2 x_2 - \rho \cdot \text{sign}[x_1 + \Theta_\tau(t)x_2] + \alpha_2(t), \qquad (4.2.18)$$
$$\kappa_1 > 0.$$

**Numerical simulation. Example** 4.2.5.

Control of the second player: $\alpha_2(t) = 500\sin(50 \cdot t^2); \kappa_1 = 3, \kappa_2 = 30$.

**Example**

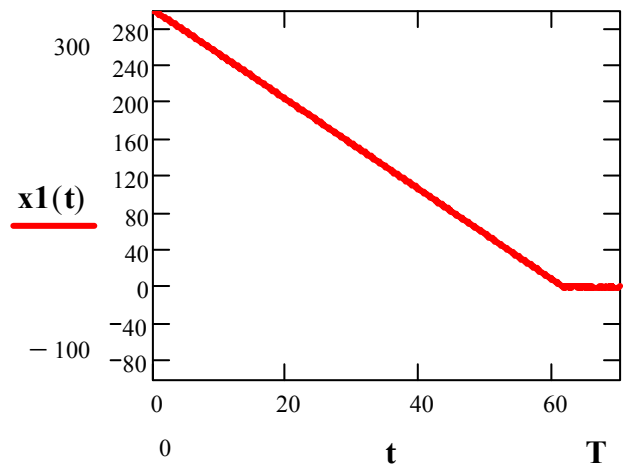

Optimal trajectory

*Pic.4.2.5.1 Optimal trajectory:* $x_1(t). |x_1(T)| = 2 \cdot 10^{-3} sm$, $\rho = 500, \kappa_1 = 3, \kappa_2 = 30$.

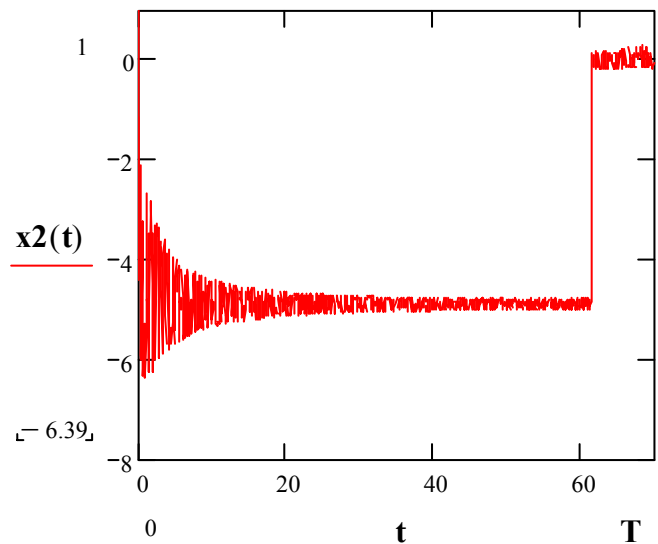

Optimal velocity

*Pic.4.2.5.2. Optimal velocity* $x_2(t) = \dot{x}_1(t). \rho = 500$, $\kappa_1 = 3, \kappa_2 = 30$.

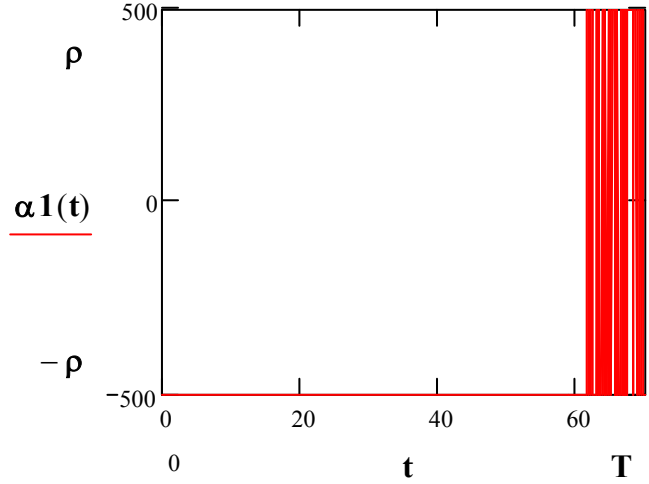

Optimal control for the first player.

*Pic.4.2.5.3.$\alpha_1(t)$-optimal control for the first player:*
$\rho = 500, \kappa_1 = 3, \kappa_2 = 30$.

**Example** 4.2.6. Let us consider an 2-persons dissipative differential game $DG_{2;T}(\mathbf{f},\mathbf{0},\mathbf{0})$, with nonlinear dynamics

$$\dot{x}_1 = x_2,$$

$$\dot{x}_2 = -\kappa_1 x_2^3 + \kappa_2 x_2^2 + \alpha_1(t) + \alpha_2(t),$$

$$\kappa_1 > 0. \tag{4.2.19}$$

$$\alpha_1(t) \in \mathbb{R},$$

$$\mathbf{J}_i = x_1^2(T) + \dot{x}_1^2(T).$$

Thus optimal control problem for the first player:

$$\min_{\alpha_1(t)} \left( \max_{\alpha_2(t)} x_1^2(T) + \dot{x}_2^2(T) \right). \tag{4.2.20}$$

For the numerical simulation we obtain ODE:

$$\dot{x}_1 = x_2,$$

$$\dot{x}_2 = -\kappa_1 x_2^3 + \kappa_2 x_2^2 - \rho \cdot sign[x_1 + \Theta(t)x_2] + \alpha_2(t), \tag{4.2.21}$$

$$\kappa > 0.$$

**Numerical simulation.Example** 4.2.6.

Control of the second player: $\alpha_2(t) = 500\sin(50 \cdot t^2); \kappa_1 = 3, \kappa_2 = 3$.

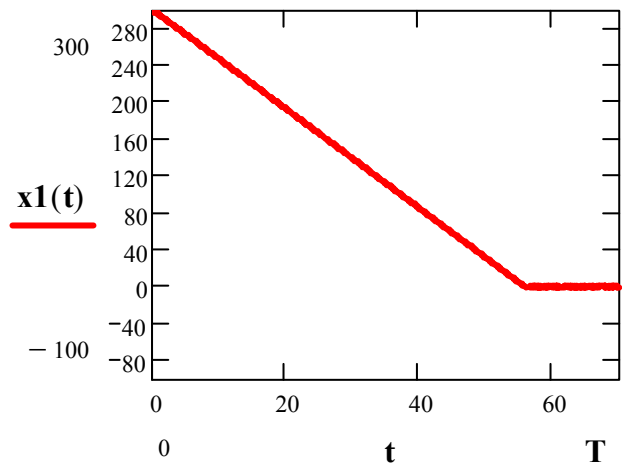

Optimal trajectory

Pic.4.2.6.1. Optimal trajectory: $x_1(t)$. $|x_1(T)| = 10^{-3} sm$, $\rho = 500, \kappa_1 = 3, \kappa_2 = 3$.

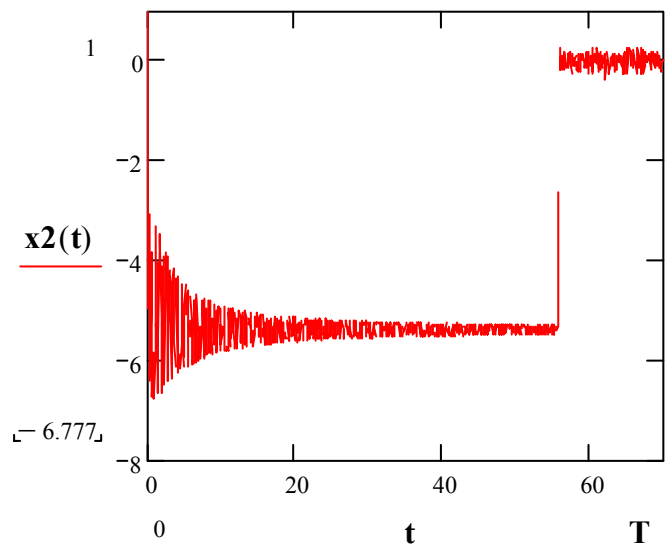

Optimal velocity

Pic.4.2.6.2. Optimal velocity $x_2(t) = \dot{x}_1(t)$.

$\kappa_1 = 3, \kappa_2 = 3$.

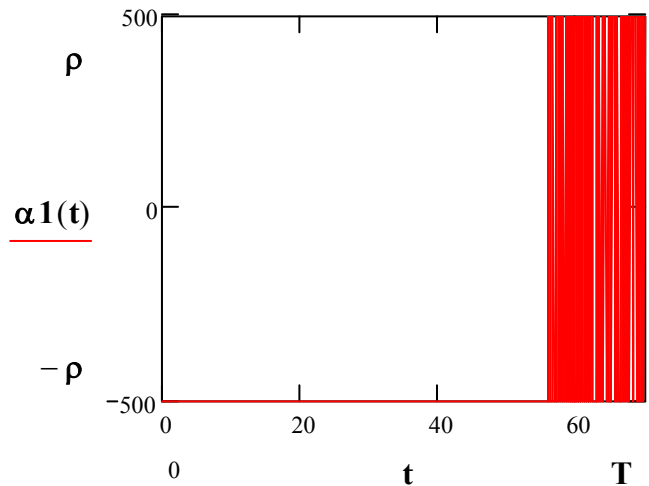

Optimal control for the first player.

Pic.4.2.6.3. $\alpha_1(t)$-optimal control for the first player:

$\rho = 500,\ \kappa_1 = 3, \kappa_2 = 3.\ \kappa = 1.$

# IV.2.2. Optimal control numerical simulation. Non-dissipative systems. Homing missile guidance Laws using "Step-by-step" strategy.

Taking the origin of the reference frame to be the instantaneous position of the missile, the equation of motion in polar form are:

$$\ddot{R} = R\dot{\sigma}^2 + a_M^r(t) + a_T^r(t),$$

$$a_M^r(t) \in [-\bar{a}_M^r, \bar{a}_M^r], a_T^r(t) \in [-\bar{a}_T^r, \bar{a}_T^r].$$

(4.2.22)

$$R\ddot{\sigma} + 2\dot{R}\dot{\sigma} = a_M^\tau(t) + a_T^\tau(t)$$

$$a_M^\tau(t) \in [-\bar{a}_M^\tau, \bar{a}_M^\tau], a_T^\tau(t) \in [-\bar{a}_T^\tau, \bar{a}_T^\tau]$$

1. The variable $R = R(t)$ denotes the target-to-missle range $R_{TM}(t)$.
2. The variable $\sigma = \sigma(t)$ denotes the line-of-sight angle (LOS) i.e.,the angle between the constant reference direction and target-to-missile direction,
4. The variable $a_M^\tau(t)$ denotes the missiles tangent acceleration,i.e.missile acceleration along direction which perpendicularly to line-of-sight direction.
5. The variable $a_M^r(t)$ denotes the missile acceleration along target-to-missile direction.
6. The variable $a_T^\tau(t)$ denotes the target tangent acceleration
7. The variable $a_T^r(t)$ denotes the target acceleration along target-to-missile direction.

Using the replacement $z = R\dot\sigma$ into Eq.(4.2.22) one obtain:

$$\dot\sigma = \frac{z}{R},$$

$$\dot z = \dot R\dot\sigma + R\ddot\sigma, R\ddot\sigma = \dot z - \dot R\dot\sigma, \qquad (4.2.23)$$

$$R\ddot\sigma + 2\dot R\dot\sigma = \dot z + \dot R\dot\sigma = \dot z + \frac{\dot R z}{R}.$$

Substitution Eq.(4.2.23) into Eq.(4.2.22) gives:

$$\ddot R = \frac{z^2}{R} + a_M^r(t) + a_T^r(t),$$

$$a_M^r(t) \in [-\bar a_M^r, \bar a_M^r], a_T^r(t) \in [-\bar a_T^r, \bar a_T^r].$$

$$\dot z = -\frac{\dot R z}{R} + a_M^\tau(t) + a_T^\tau(t), \qquad (4.2.24)$$

$$a_M^\tau(t) \in [-\bar a_M^\tau, \bar a_M^\tau], a_T^\tau(t) \in [-\bar a_T^\tau, \bar a_T^\tau].$$

Using the replacement $\dot R = V_r$ into Eq.(4.2.24) one obtain:

$$\dot{R} = V_r,$$

$$\dot{V}_r = \frac{z^2}{R} + a^r_M(t) + a^r_T(t),$$

$$a^r_M(t) \in [-\bar{a}^r_M, \bar{a}^r_M], a^r_T(t) \in [-\bar{a}^r_T, \bar{a}^r_T]. \qquad (4.2.25)$$

$$\dot{z} = -\frac{V_r z}{R} + a^\tau_M(t) + a^\tau_T(t),$$

$$a^\tau_M(t) \in [-\bar{a}^\tau_M, \bar{a}^\tau_M], a^\tau_T(t) \in [-\bar{a}^\tau_T, \bar{a}^\tau_T].$$

Let us consider the optimal control problem:

$$\dot{R} = V_r,$$

$$\dot{V}_r = \frac{z^2}{R} + \check{a}^r_M(t) + a^r_T(t) - \kappa_1 V_r^3,$$

$$\check{a}^r_M(t) - \kappa_1 V_r^3 \in [-\bar{a}^r_M, \bar{a}^r_M], a^r_T(t) \in [-\bar{a}^r_T, \bar{a}^r_T], 0 < \kappa_1.$$

$$(4.2.26)$$

$$\dot{z} = -\frac{V_r z}{R} + \check{a}^\tau_M(t) + a^\tau_T(t) - \kappa_2 z^3,$$

$$\check{a}^\tau_M(t) - \kappa_2 z^3 \in [-\bar{a}^\tau_M, \bar{a}^\tau_M], a^\tau_T(t) \in [-\bar{a}^\tau_T, \bar{a}^\tau_T], 0 < \kappa_2.$$

$$\mathbf{J}_i = R^2(t_1) + z^2(t_1), i = 1, 2.$$

Optimal control problem for the first player are:

$$\mathbf{J}_1 = \min_{a^r_M(t) \in [-\bar{a}^r_M, \bar{a}^r_M], a^r_T(t) \in [-\bar{a}^r_T, \bar{a}^r_T], 0 < \kappa_1} \left\{ \max_{a^r_T(t) \in [-\bar{a}^r_T, \bar{a}^r_T], a^\tau_T(t) \in [-\bar{a}^\tau_T, \bar{a}^\tau_T]} [R^2(t_1) + z^2(t_1)] \right\}. \qquad (4.2.27)$$

Optimal control problem for the first player are:

$$\mathbf{J}_2 = \max_{a_T^r(t)\in[-\bar{a}_T^r,\bar{a}_T^r], a_T^\tau(t)\in[-\bar{a}_T^\tau,\bar{a}_T^\tau]} \left\{ \min_{a_M^r(t)\in[-\bar{a}_M^r,\bar{a}_M^r], a_T^r(t)\in[-\bar{a}_T^r,\bar{a}_T^r], 0<\kappa_1} [R^2(t_1)+z^2(t_1)] \right\}. \quad (4.2.28)$$

Let us consider the optimal control problem:

$$\dot{r} = v_r,$$

$$\dot{v}_r = a_M^r(t) + a_T^r(t) - \kappa_1 V_r^3,$$

$$a_M^r(t) \in [-\bar{a}_M^r, \bar{a}_M^r], a_T^r(t) \in [-\bar{a}_T^r, \bar{a}_T^r], 0 < \kappa_1.$$

$$(4.2.29)$$

$$\dot{z} = a_M^\tau(t) + a_T^\tau(t) - \kappa_2 z^3,$$

$$a_M^\tau(t) \in [-\bar{a}_M^\tau, \bar{a}_M^\tau], a_T^\tau(t) \in [-\bar{a}_T^\tau, \bar{a}_T^\tau], 0 < \kappa_2.$$

$$\mathbf{J}_i = R^2(t_1) + z^2(t_1), i = 1, 2.$$

**Example 1.** $\tau = 0.01, \kappa_1 = 0.1, \kappa_2 = 0.1, \bar{a}_T^r = 2000 m/\sec^2,$
$\bar{a}_T^\tau = 2000 m/\sec^2, R(0) = 200 m, V_r(0) = 10 m/\sec,$
$z(0) = 30, \dot{z}(0) = 400, a_T^r(t) = \bar{a}_T^r(\sin(\omega \cdot t))^p,$
$a_T^\tau(t) = \bar{a}_T^\tau(\sin(\omega \cdot t))^q, \omega = 5, p = 2, q = 1.$

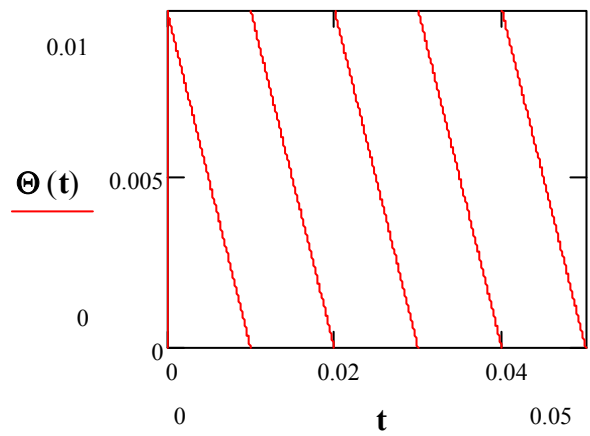

**Pic.1.1.** Cutting function: $\Theta_\tau(t)$.

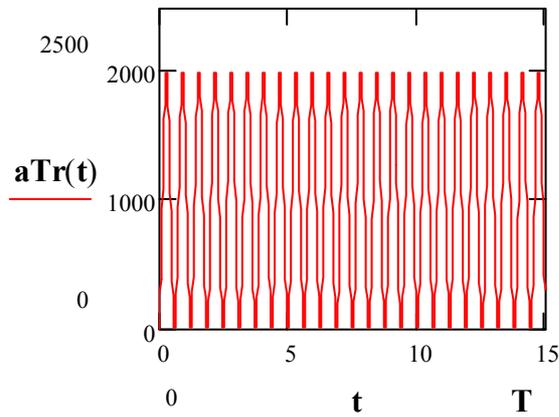

**Pic.1.2.** Target acceleration along target-to-missile direction: $a_T^r(t)$.

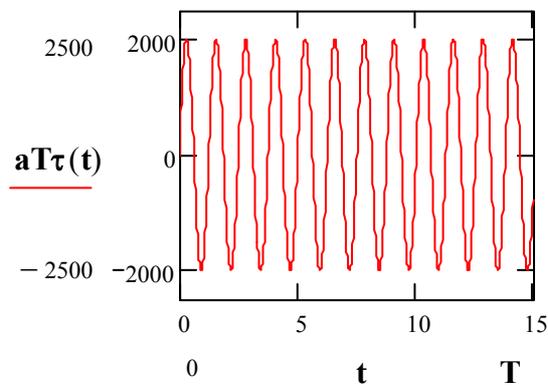

**Pic.1.3.** Target tangent acceleration: $a_T^\tau(t)$.

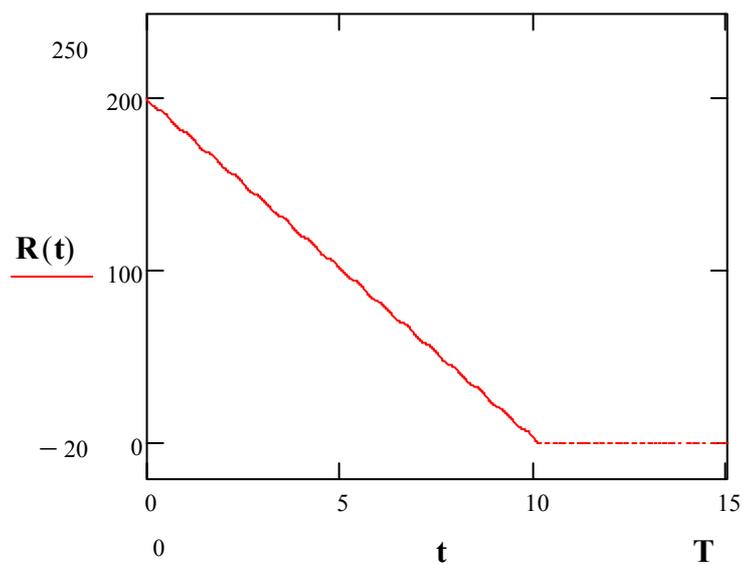

Target-to-missile range R(t)

**Pic.1.4.** Target-to-missile range: $R(t)$.

$$R(T) = 2.94 \times 10^{-7} m$$

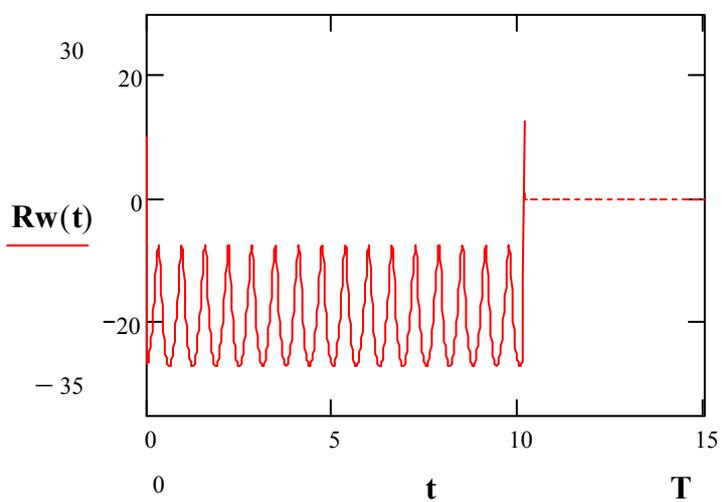

Speed of rapprochement missile-to-target

**Pic.1.5.** Speed of rapprochement missile-to-target: $\dot{R}(t)$.

$$R(T) = -0.064 m/\sec.$$

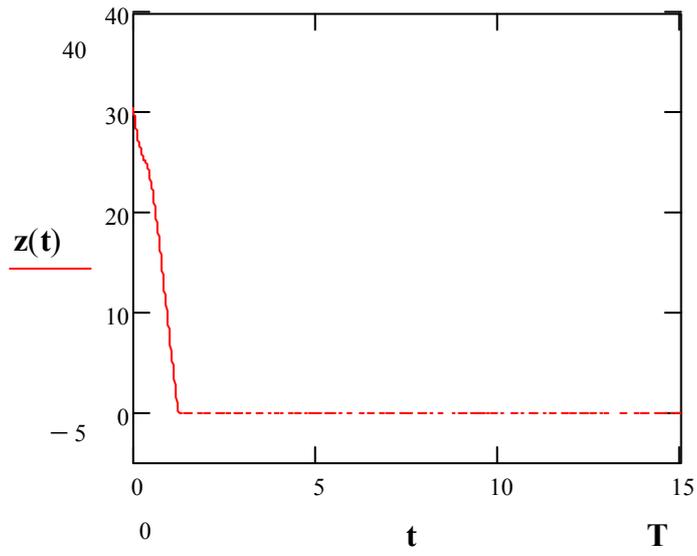

**Pic.1.6.** Variable: $z(t) = R(t)\dot{\sigma}(t)$.
$z(0) = 30, z(T) = -1.041 \times 10^{-6}$.

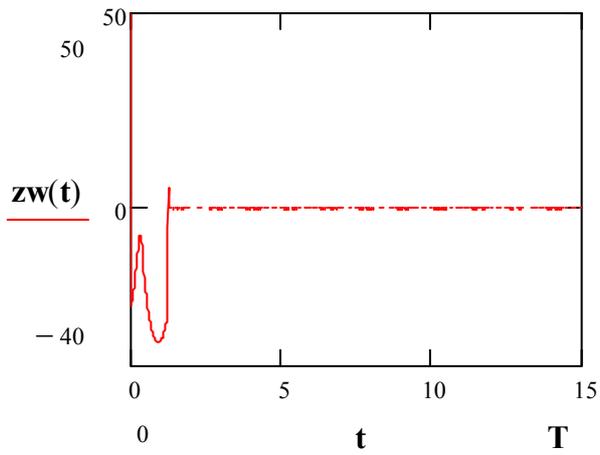

**Pic.1.7.** Variable: $\dot{z}(t) = \dot{R}\dot{\sigma} + R\ddot{\sigma}$.
$\dot{z}(0) = 400, \dot{z}(T) = 0.111$.

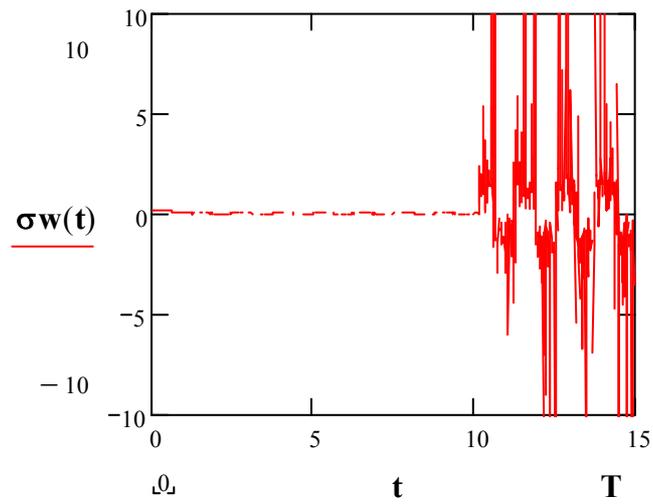

**Pic**.**1**.**8**.Variable:$\dot{\sigma}(t)$.

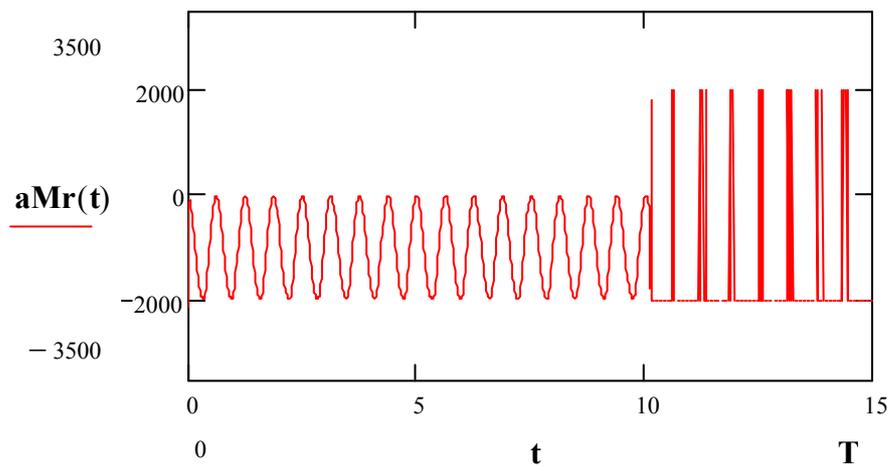

Missile acceleration along target-to-missile direction

**Pic**.**1**.**9**.Missile acceleration along target-to-missile direction:$a_M^r(t)$

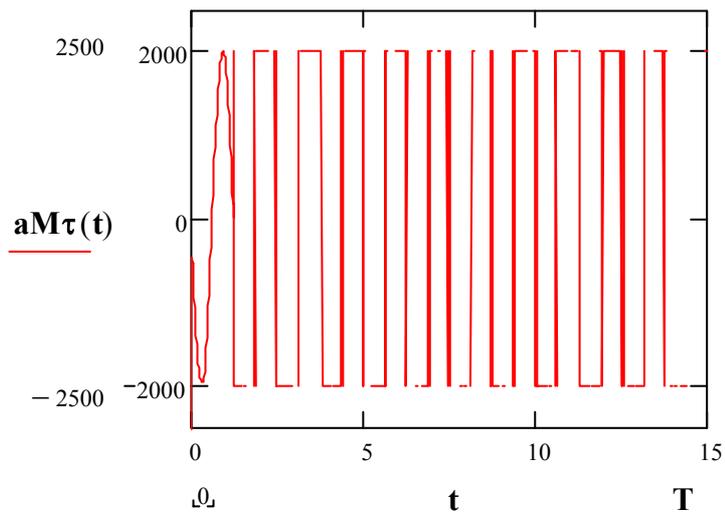

Missile tangent acceleration

**Pic.1.10**.Missile tangent acceleration:$a_M^\tau(t)$.

**Example 2.** $\tau = 0.01, \kappa_1 = 0.1, \kappa_2 = 0.1, \bar{a}_T^r = 2000 m/\sec^2,$
$\bar{a}_T^\tau = 2000 m/\sec^2, R(0) = 200 m, V_r(0) = 10 m/\sec,$
$z(0) = 1000, \dot{z}(0) = 400, a_T^r(t) = \bar{a}_T^r(\sin(\omega \cdot t))^p,$
$a_T^\tau(t) = \bar{a}_T^\tau(\sin(\omega \cdot t))^q, \omega = 5, p = 2, q = 1.$

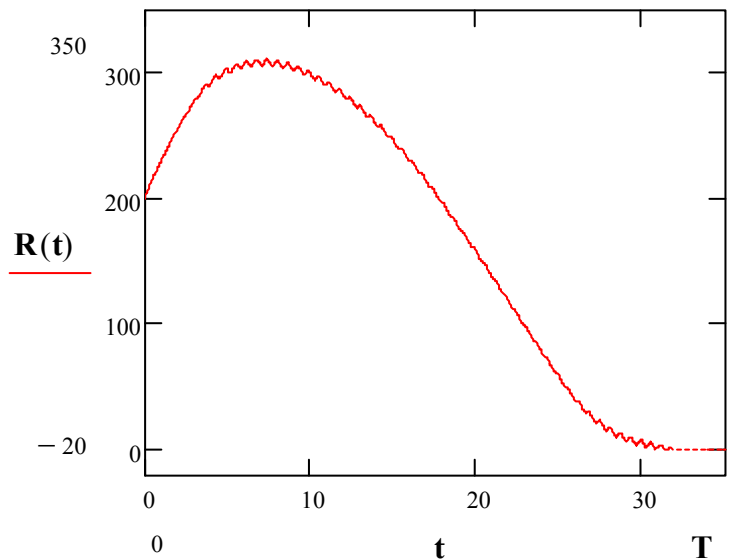

Target-to-missile range R(t)

**Pic.2.1**..Target-to-missile range: $R(t)$.

$R(T) = 2.186 \times 10^{-6} m.$

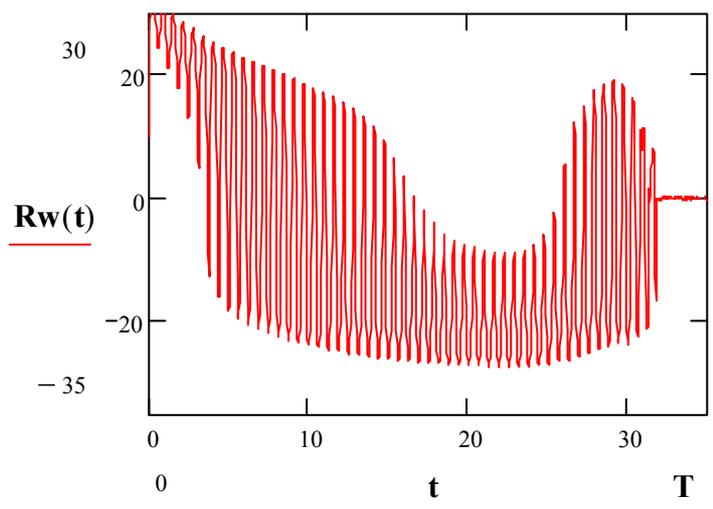

Speed of rapprochement missile-to-target

**Pic.2.2.** Speed of rapprochement missile-to-target: $\dot{R}(t)$.
$$\dot{R}(T) = -0.386 m/\sec.$$

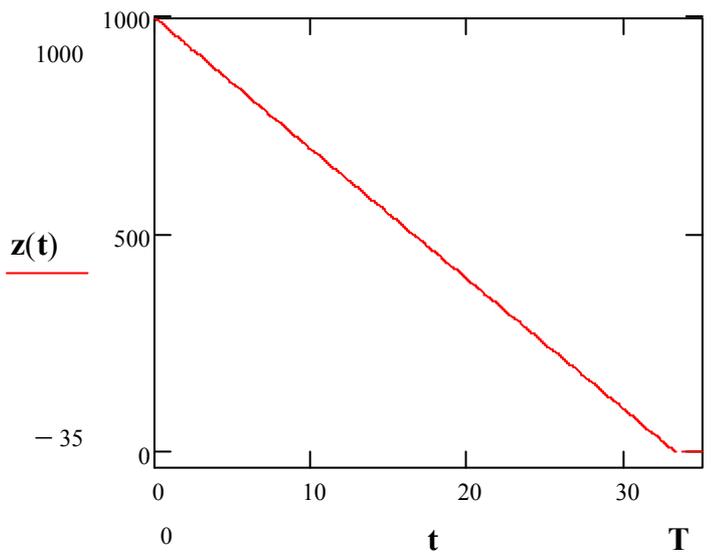

**Pic.2.3.** Variable: $z(t) = R(t)\dot{\sigma}(t)$.
$$z(0) = 1000, z(T) = -9.328 \times 10^{-7}.$$

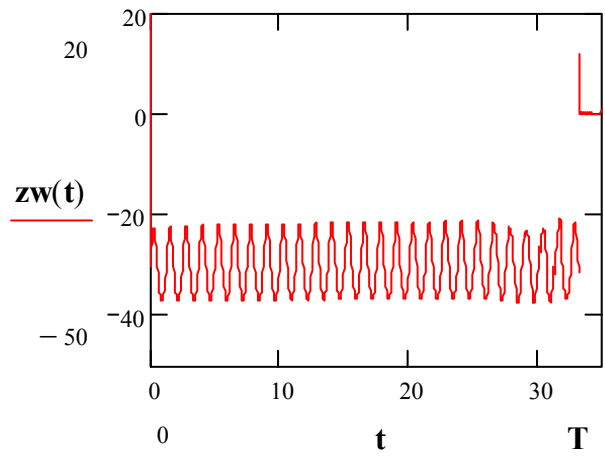

**Pic.2.4**.Variable: $\dot{z}(t) = \dot{R}\dot{\sigma} + R\ddot{\sigma}$.
$\dot{z}(0) = 400, \dot{z}(T) = 0.569$.

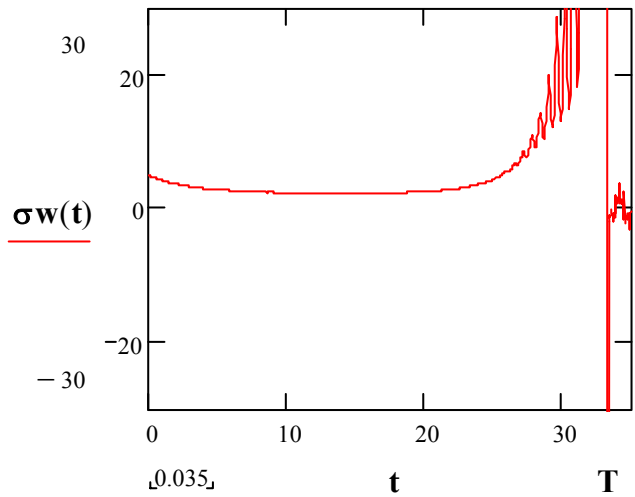

**Pic.2.5**.Variable:$\ddot{\sigma}(t)$.
$\ddot{\sigma}(0) = 5, \ddot{\sigma}(T) = -0.427$.

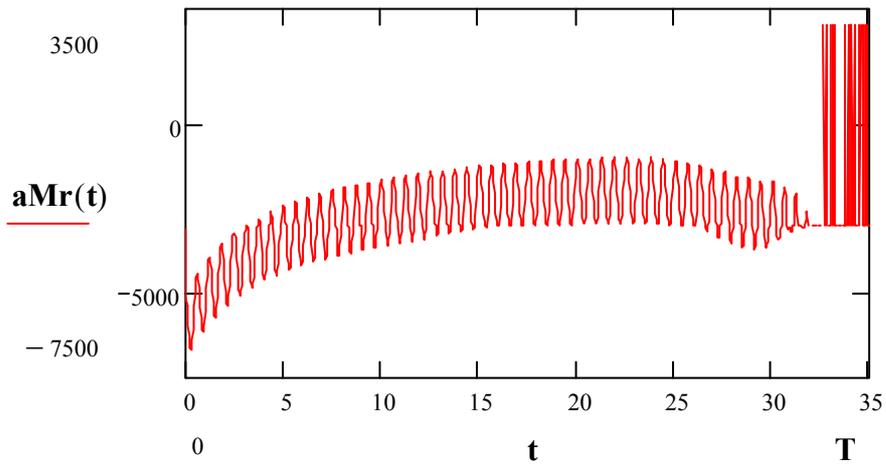

Missile acceleration along target-to-missile direction

**Pic**.2.6.Missile acceleration along target-to-missile direction:$a_M^r(t)$.

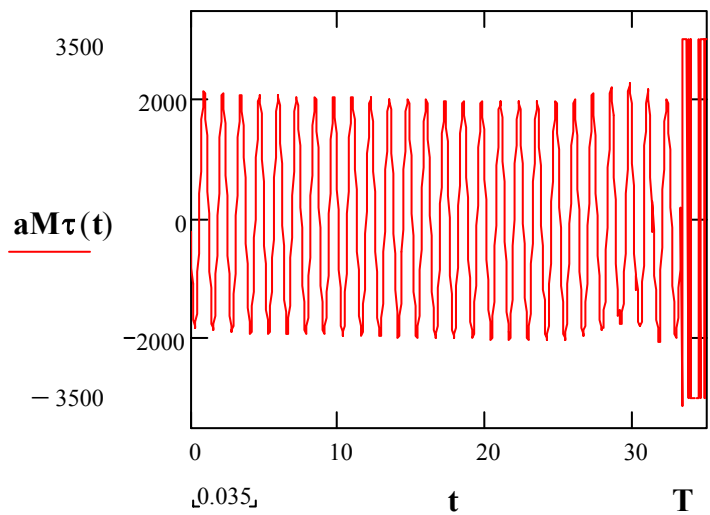

Missile tangent acceleration

**Pic**.2.7.Missile tangent acceleration:$a_M^\tau(t)$.

**Example 3.**$\tau = 0.01, \kappa_1 = 0.1, \kappa_2 = 0.001, \bar{a}_T^r = 20 m/\sec^2$,
$\bar{a}_T^\tau = 20 m/\sec^2$, $R(0) = 200m, V_r(0) = 10m/\sec$,
$z(0) = 100, \dot{z}(0) = 40, a_T^r(t) = \bar{a}_T^r(\sin(\omega \cdot t))^p$,
$a_T^\tau(t) = \bar{a}_T^\tau(\sin(\omega \cdot t))^q, \omega = 5, p = 2, q = 1$.

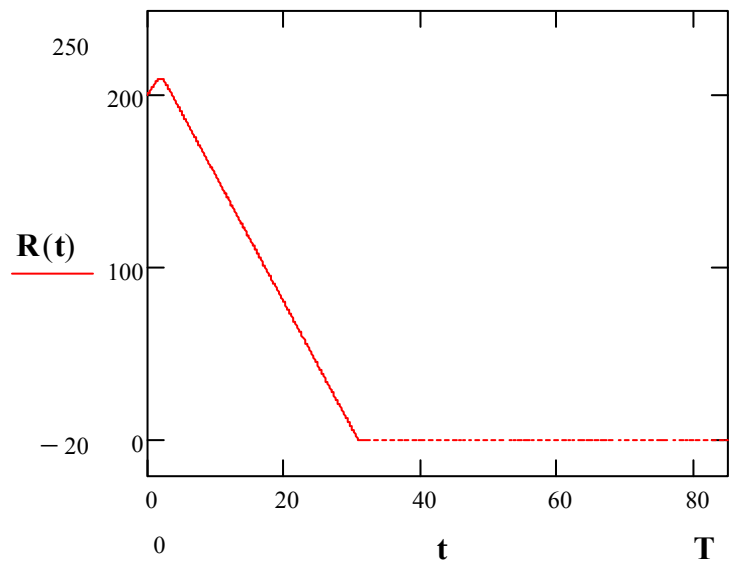

Target-to-missile range R(t)

**Pic.3.1.** Target-to-missile range: $R(t)$.
$R(T) = 9.179 \times 10^{-8} m.$

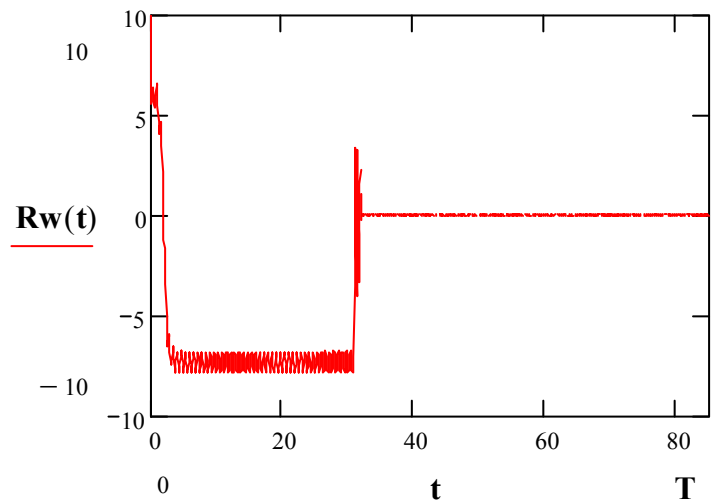

Speed of rapprochement missile-to-target

**Pic.3.2.** Speed of rapprochement missile-to-target: $\dot{R}(t)$.
$\dot{R}(0) = 10 m/\sec, \dot{R}(T) = -5.995 \times 10^{-3} m/\sec.$

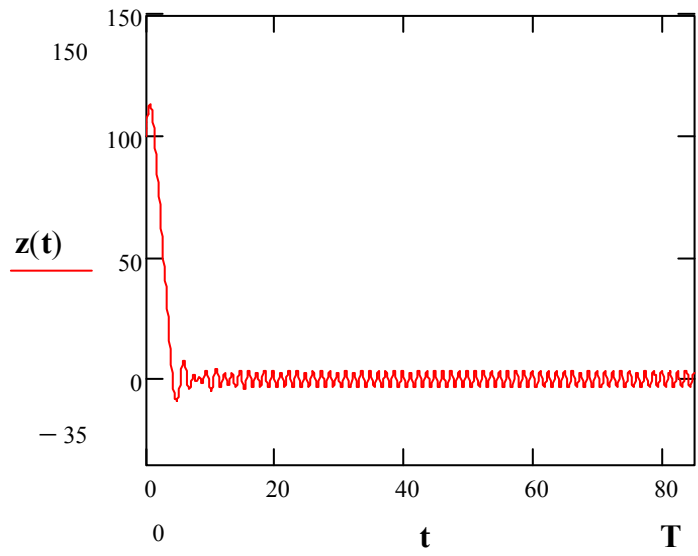

**Pic.3.3**.Variable: $z(t) = R(t)\dot{\sigma}(t)$.
$z(0) = 100, z(T) = 2.922$.

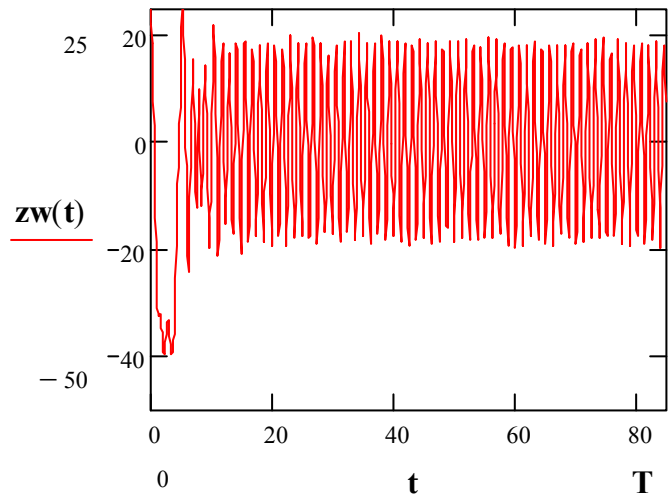

**Pic.3.4**.Variable: $\dot{z}(t) = \dot{R}\dot{\sigma} + R\ddot{\sigma}$.
$\dot{z}(0) = 40$.

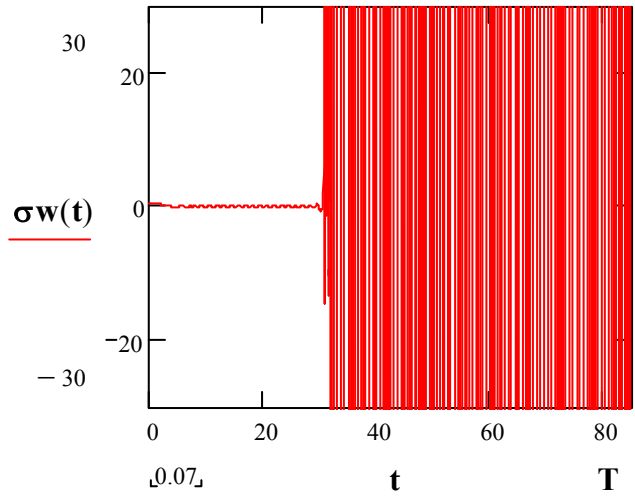

**Pic**.**3**.**5**.Variable:$\dot{\sigma}(t)$.

$\dot{\sigma}(0) = 0.5$.

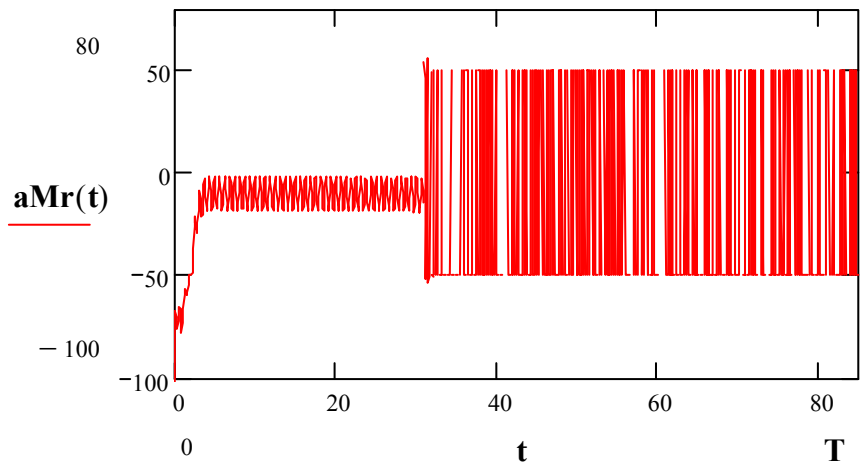

Missile acceleration along target-to-missile direction

**Pic**.**3**.**6**.Missile acceleration along target-to-missile direction:$a_M^r(t)$.

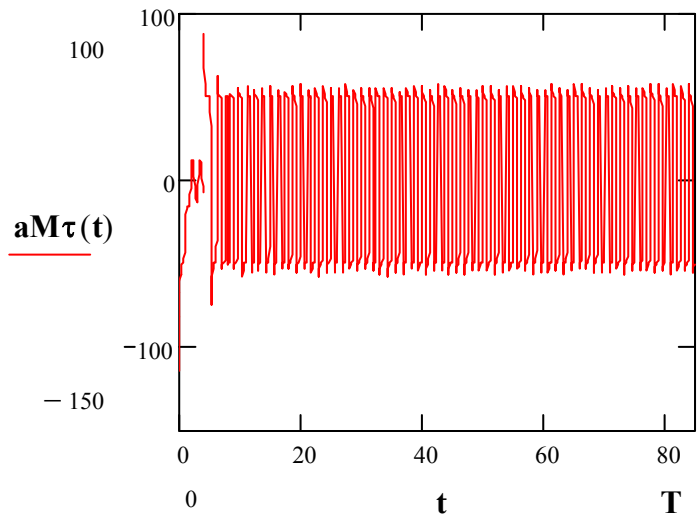

Missile tangent acceleration

**Pic**.**3**.**7**.Missile tangent acceleration:$a_M^\tau(t)$.

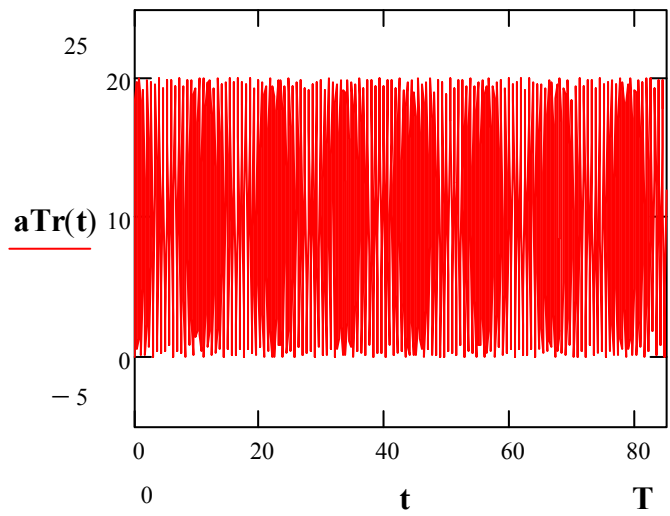

**Pic**.**3**.**8**.Target acceleration along target-to-missile direction:$a_T^r(t)$.

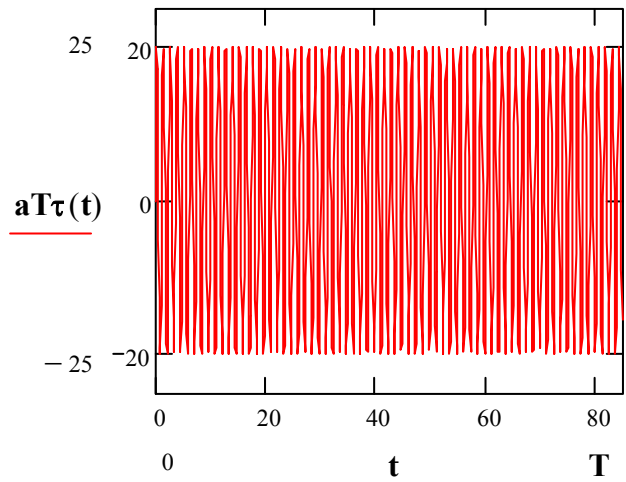

**Pic.3.9**.Target tangent acceleration:$a_T^\tau(t)$.

**Example 4.**$\tau = 0.01, \kappa_1 = 0.1, \kappa_2 = 0.001, \bar{a}_T^r = 20 m/\sec^2$,
$\bar{a}_T^\tau = 20 m/\sec^2$, $R(0) = 1000m, V_r(0) = 10m/\sec$,
$z(0) = 600, \dot{z}(0) = 40, a_T^r(t) = \bar{a}_T^r(\sin(\omega \cdot t))^p$,
$a_T^\tau(t) = \bar{a}_T^\tau(\sin(\omega \cdot t))^q, \omega = 5, p = 2, q = 1$.

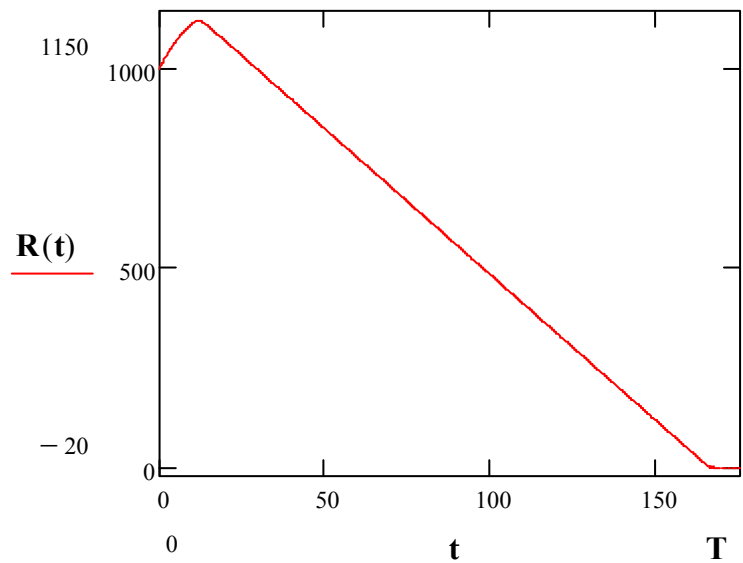

Target-to-missile range R(t)

**Pic.4.1**..Target-to-missile range: $R(t)$.
$$R(T) = 5.87 \times 10^{-7} m.$$

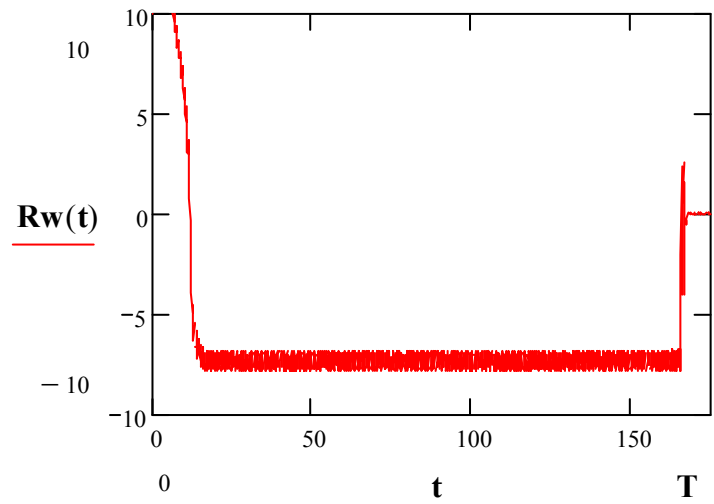

Speed of rapprochement missile-to-target

**Pic.4.2.** Speed of rapprochement missile-to-target: $\dot{R}(t)$.
$$\dot{R}(T) = -5.989 \times 10^{-3} m/\sec.$$

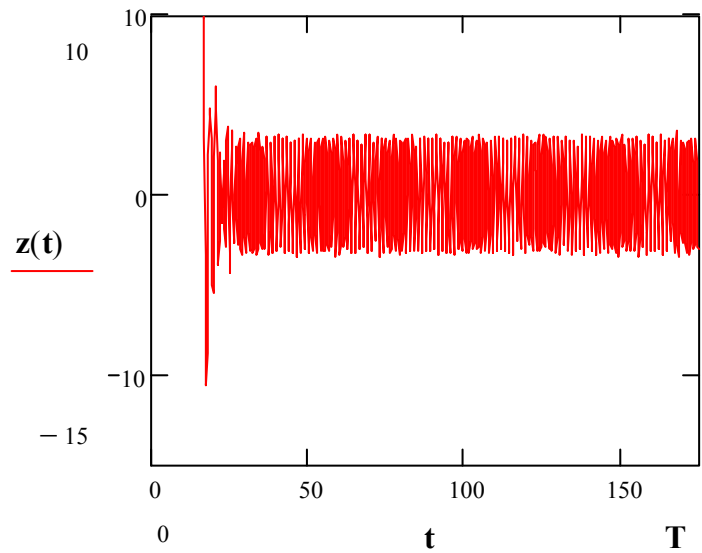

**Pic.4.3.** Variable: $z(t) = R(t)\dot{\sigma}(t)$.
$$z(0) = 600, z(T) = -2.976.$$

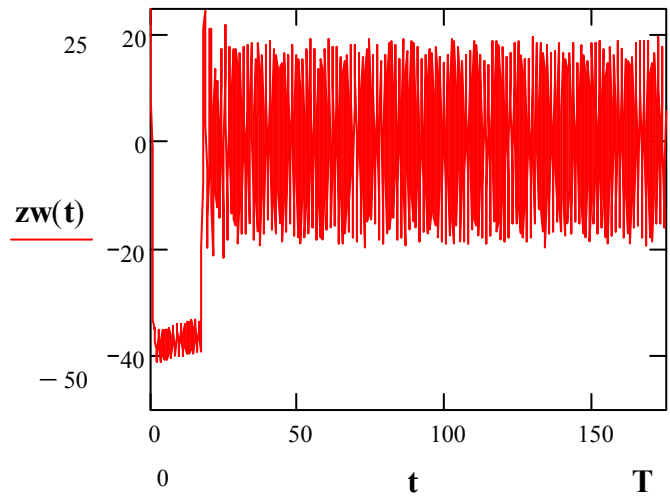

**Pic.4.4**.Variable: $\dot{z}(t) = \dot{R}\dot{\sigma} + R\ddot{\sigma}$.
$\dot{z}(0) = 40, \dot{z}(T) = 5.826$

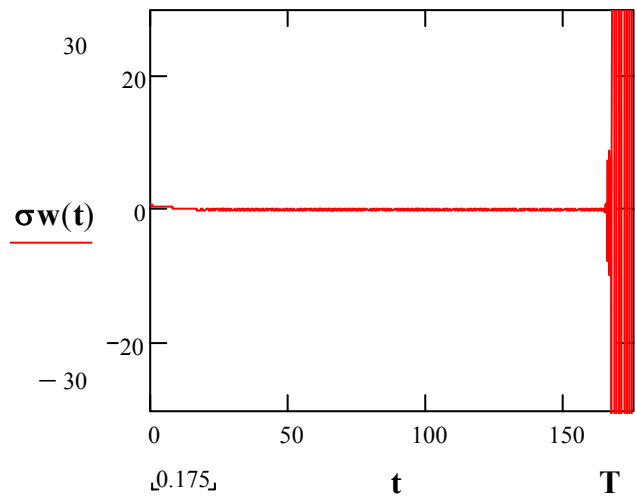

**Pic.4.5**.Variable:$\dot{\sigma}(t)$.
$\dot{\sigma}(0) = 0.6$.

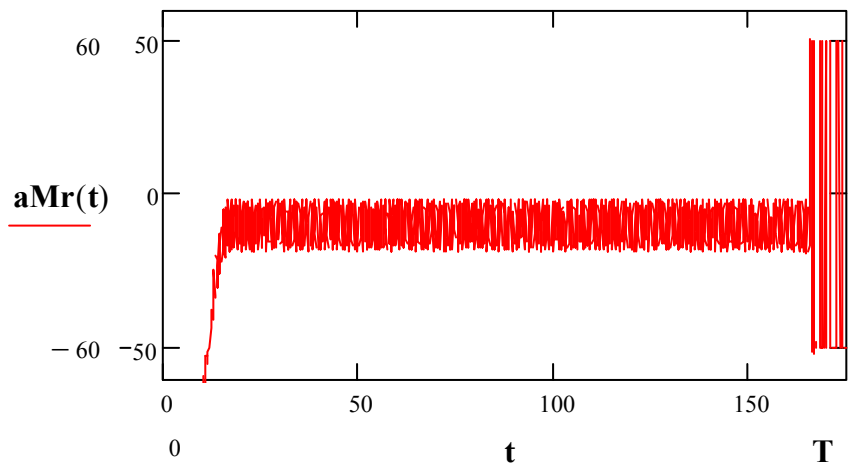

Missile acceleration along target-to-missile direction

**Pic.4.6.** Missile acceleration along target-to-missile direction: $a_M^r(t)$.

$$a_M^r(0) = -150 m/\sec^2.$$

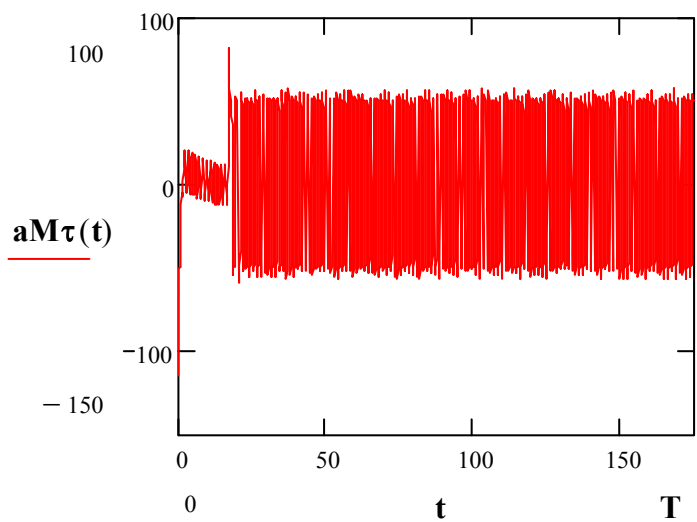

Missile tangent acceleration

**Pic.4.7.** Missile tangent acceleration: $a_M^\tau(t)$.

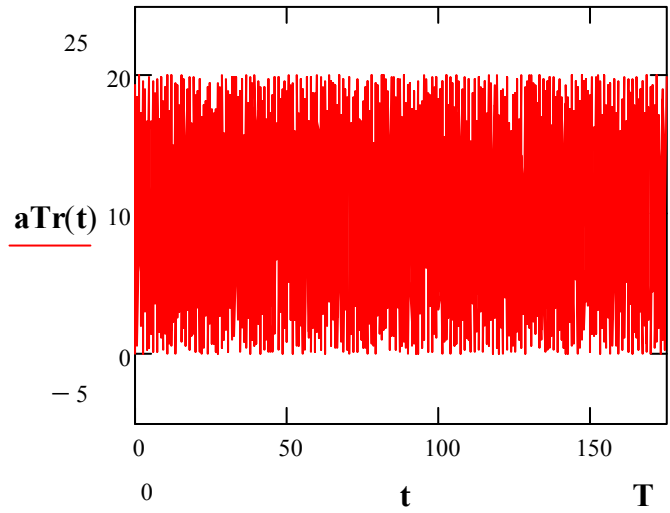

**Pic.4.8.** Target acceleration along target-to-missile direction: $a_T^r(t)$.

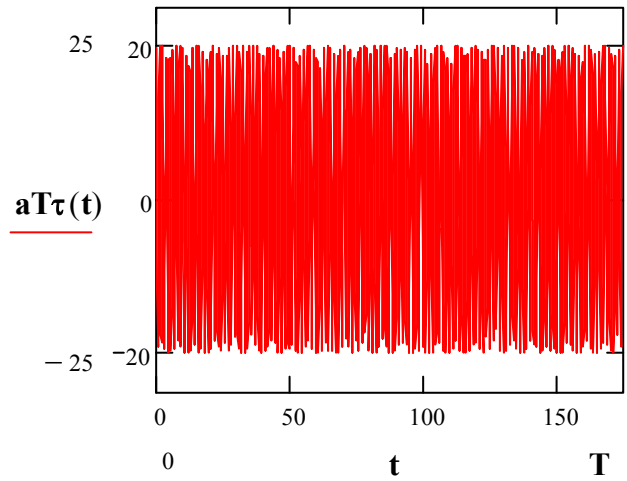

**Pic.4.9.** Target tangent acceleration: $a_T^\tau(t)$.

**Example 5.** $\tau = 0.01, \kappa_1 = 10^{-5}, \kappa_2 = 0.001, \bar{a}_T^r = 20 m/\sec^2$,
$\bar{a}_T^\tau = 20 m/\sec^2, R(0) = 1000 m, V_r(0) = 100 m/\sec$,
$z(0) = 600, \dot{z}(0) = 40, a_T^r(t) = \bar{a}_T^r (\sin(\omega \cdot t))^p$,
$a_T^\tau(t) = \bar{a}_T^\tau (\sin(\omega \cdot t))^q, \omega = 5, p = 2, q = 1$.

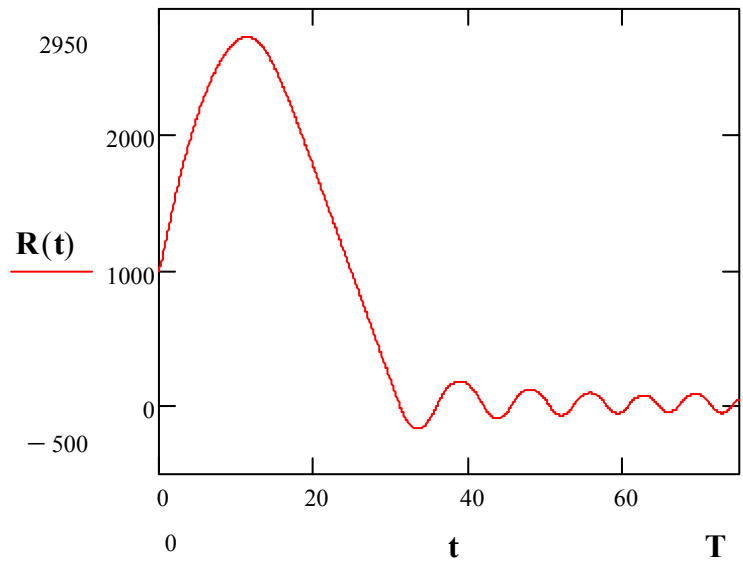

Target-to-missile range R(t)

**Pic.5.1.** Target-to-missile range: $R(t)$.

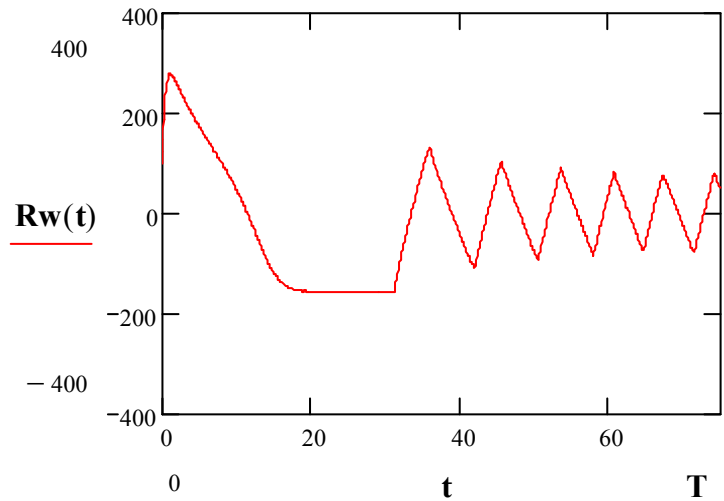

Speed of rapprochement missile-to-target

**Pic.5.2.** Speed of rapprochement missile-to-target: $\dot{R}(t)$.
$$\dot{R}(t) = 100 m/\sec^2.$$

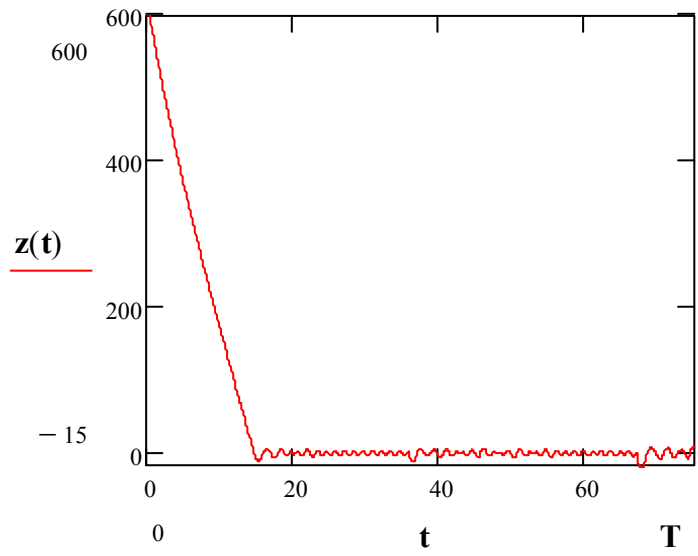

**Pic.5.3.1**.Variable: $z(t) = R(t)\dot{\sigma}(t)$.
$z(0) = 600, z(T) =$

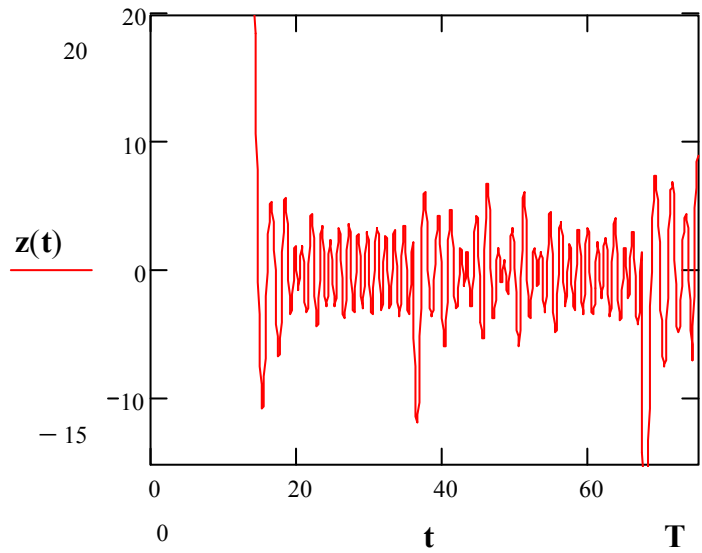

**Pic.5.3.2**.Variable: $z(t) = R(t)\dot{\sigma}(t)$.
$z(0) = 600, z(T) =$

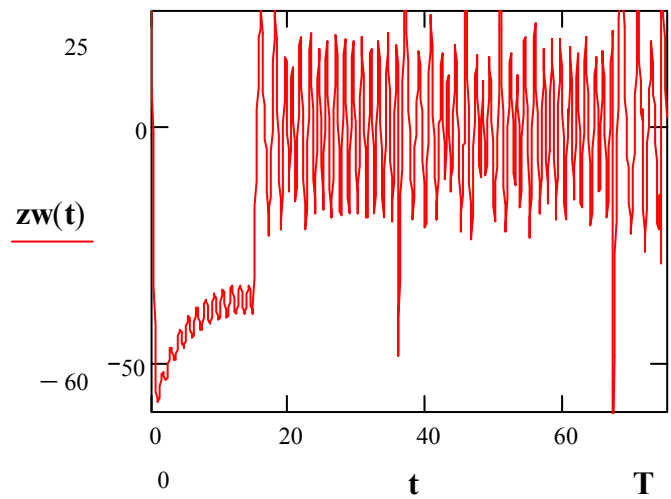

**Pic.5.4.** Variable: $\dot{z}(t) = \dot{R}\dot{\sigma} + R\ddot{\sigma}$.
$\dot{z}(0) = 40.$

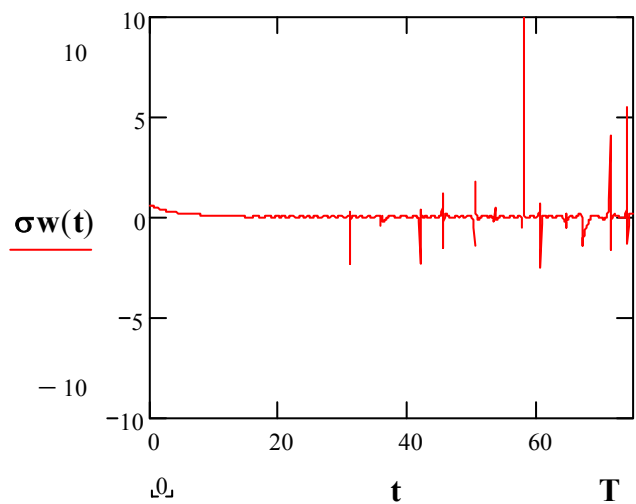

**Pic.5.5.** Variable: $\dot{\sigma}(t)$.
$\dot{\sigma}(0) = 0.6.$

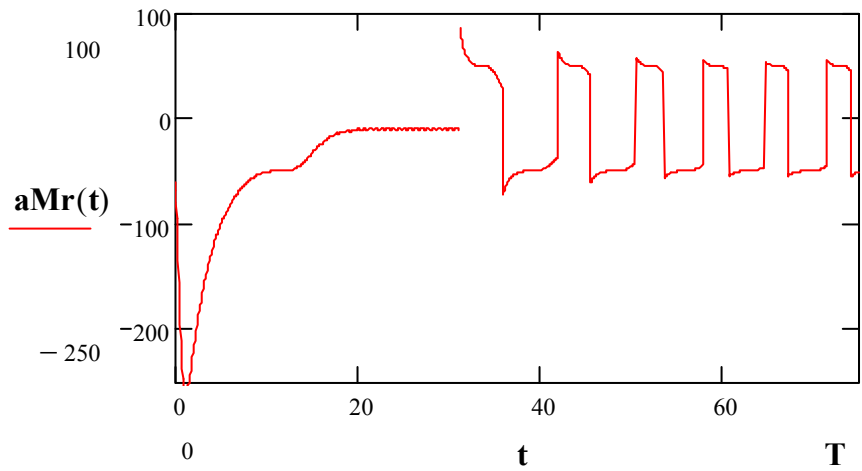

Missile acceleration along target-to-missile direction

**Pic**.5.6.Missile acceleration along target-to-missile direction:$a_M^r(t)$

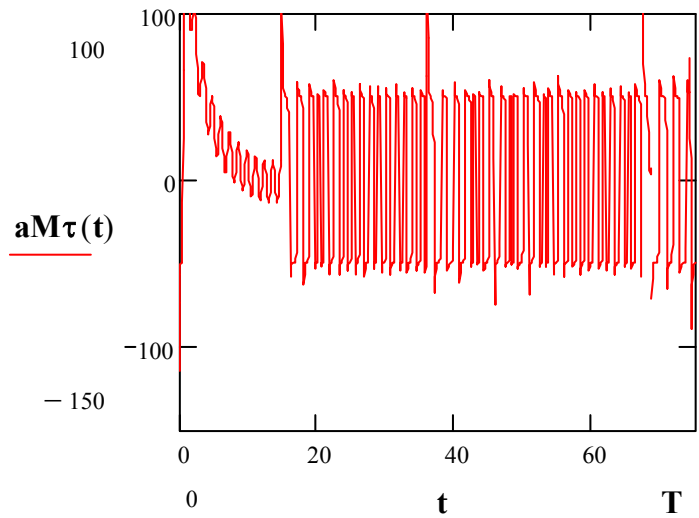

Missile tangent acceleration

**Pic**.5.7.Missile tangent acceleration:$a_M^\tau(t)$.

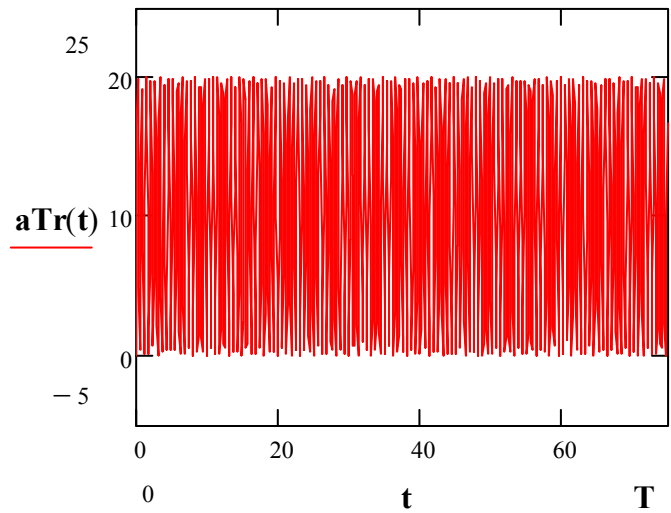

**Pic.5.8.** Target acceleration along target-to-missile direction: $a_T^r(t)$

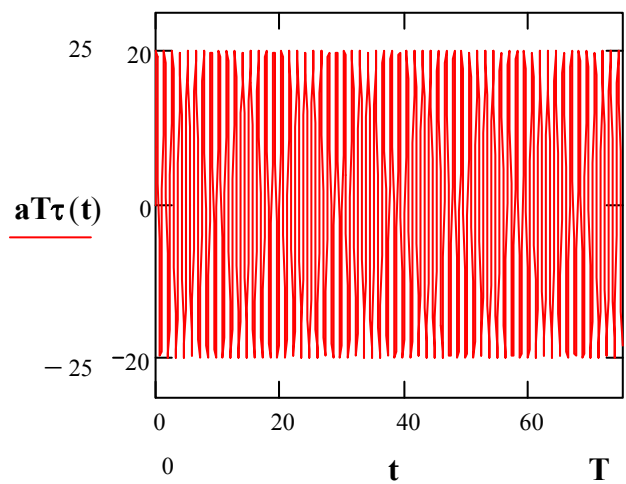

**Pic.5.9.** Target tangent acceleration: $a_T^\tau(t)$.

**Example 6.** $\tau = 0.01, \kappa_1 = 10^{-3}, \kappa_2 = 0.001, \bar{a}_T^r = 20 m/\sec^2,$
$\bar{a}_T^\tau = 20 m/\sec^2,\ R(0) = 1000 m, V_r(0) = 100 m/\sec,$
$z(0) = 600, \dot{z}(0) = 40, a_T^r(t) = \bar{a}_T^r(\sin(\omega \cdot t))^p,$
$a_T^\tau(t) = \bar{a}_T^\tau(\sin(\omega \cdot t))^q, \omega = 5, p = 2, q = 1.$

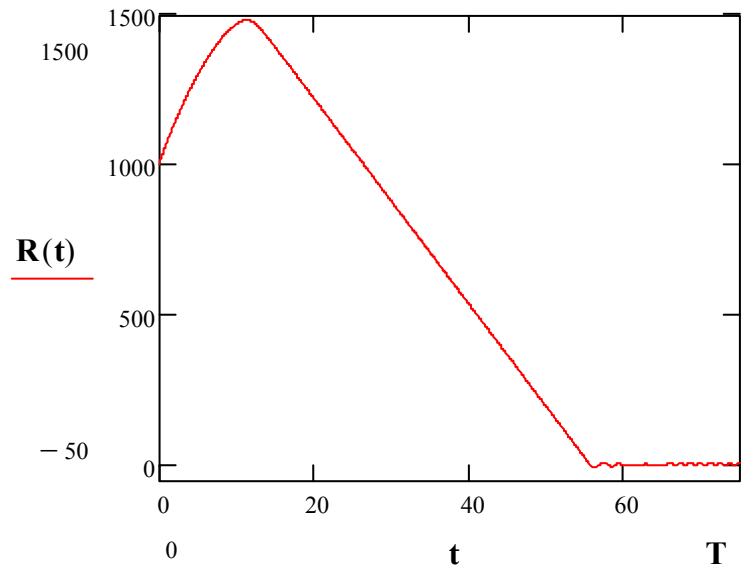

Target-to-missile range R(t)

**Pic.6.1.** Target-to-missile range: $R(t)$.

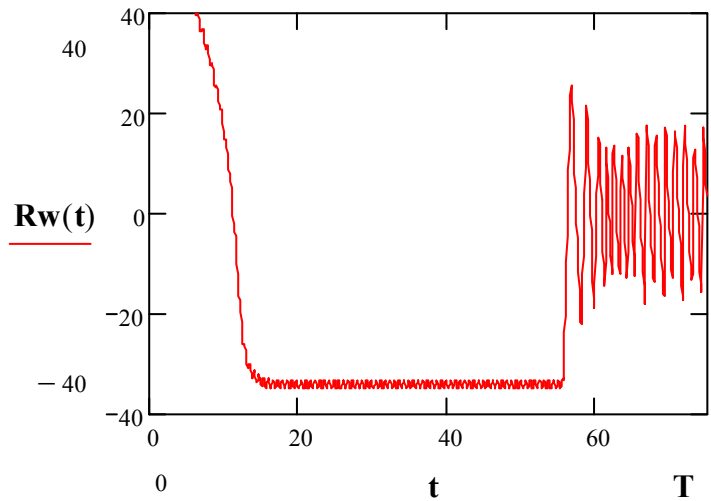

Speed of rapprochement missile-to-target

**Pic.6.2.** Speed of rapprochement missile-to-target: $\dot{R}(t)$.

$$\dot{R}(0) = 100 m/\sec.$$

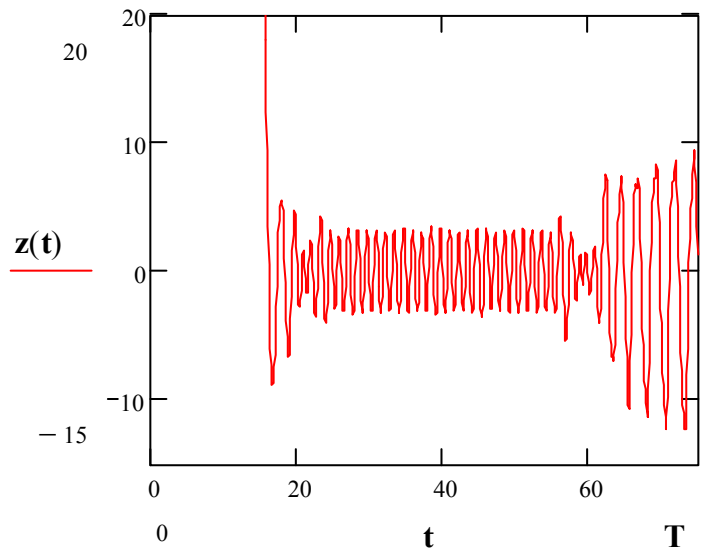

**Pic.6.3.**Variable: $z(t) = R(t)\dot{\sigma}(t)$.
$z(0) = 600$.

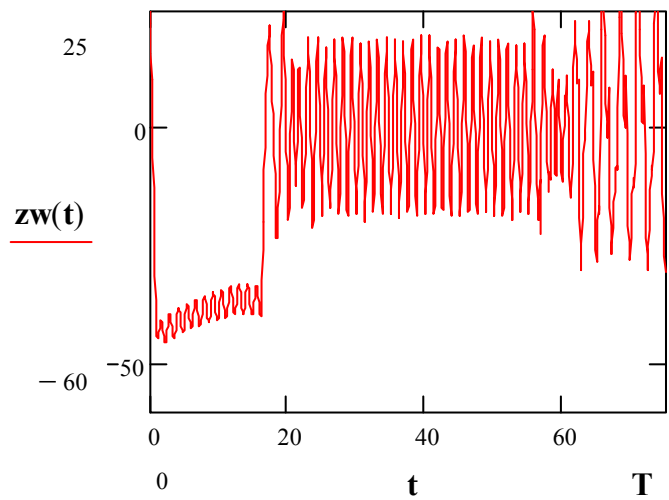

**Pic.6.4.**Variable: $\dot{z}(t) = \dot{R}\dot{\sigma} + R\ddot{\sigma}$.
$\dot{z}(0) = 400, \dot{z}(T) =$

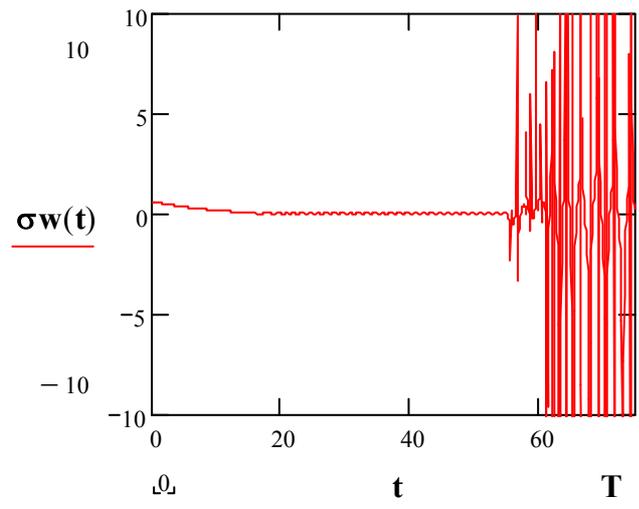

**Pic.6.5.** Variable: $\dot{\sigma}(t)$.

$\dot{\sigma}(0) = 0.6$.

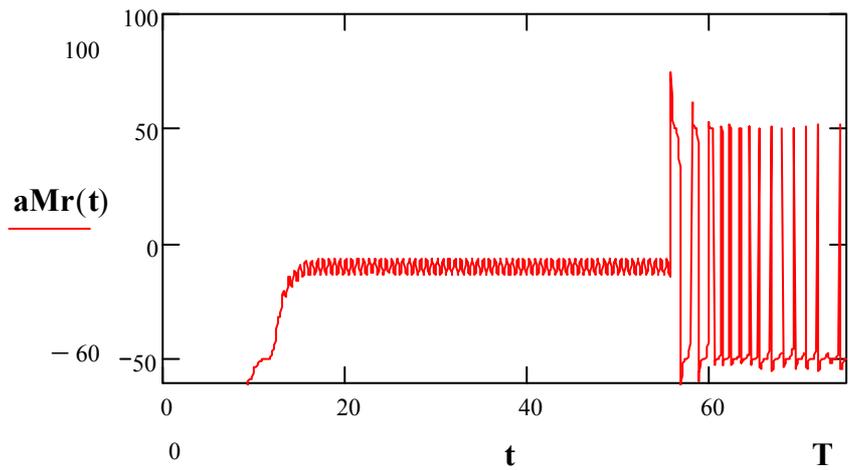

Missile acceleration along target-to-missile direction

**Pic.6.6.** Missile acceleration along target-to-missile direction: $a_M^r(t)$.

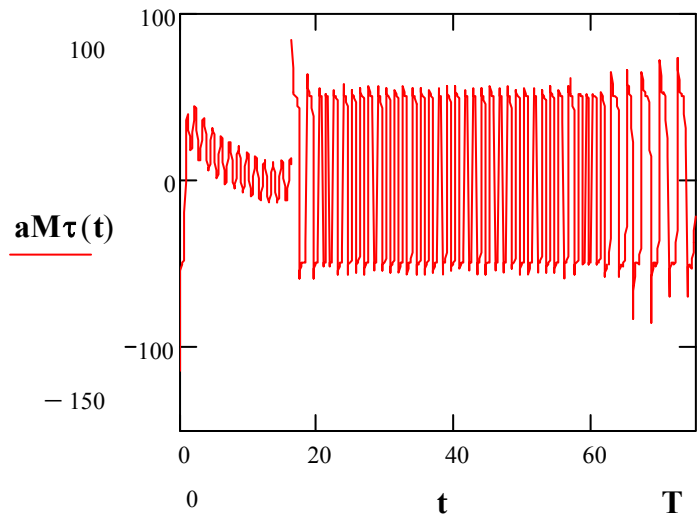

Missile tangent acceleration

**Pic**.**6**.**7**.Missile tangent acceleration:$a_M^\tau(t)$.

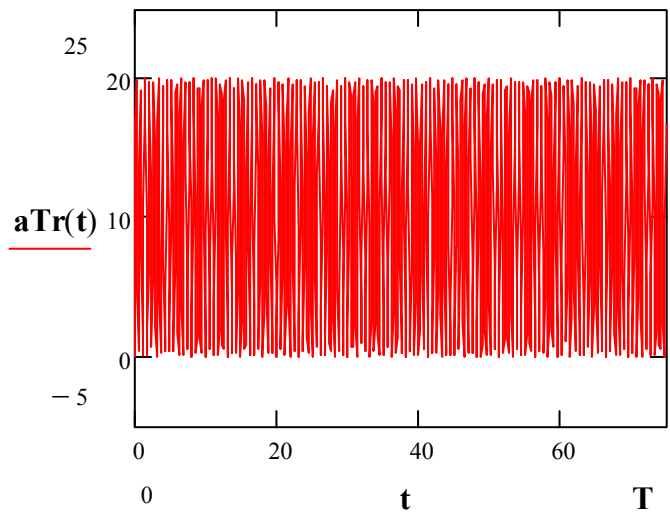

**Pic**.**6**.**8**.Target acceleration along target-to-missile direction:$a_T^r(t)$

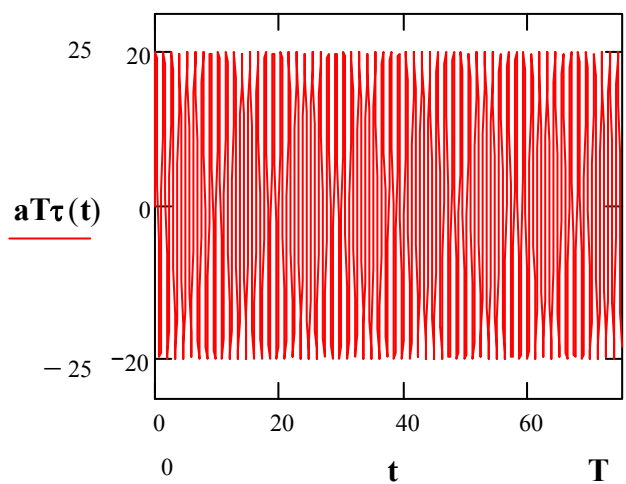

**Pic.6.9**.Target tangent acceleration:$a_T^\tau(t)$.

**Example 7.** $\tau = 0.001, \kappa_1 = 10^{-3}, \kappa_2 = 0.001, \bar{a}_T^r = 20m/\sec^2$,
$\bar{a}_T^\tau = 20m/\sec^2, R(0) = 1000m, V_r(0) = 100m/\sec$,
$z(0) = 600, \dot{z}(0) = 40, a_T^r(t) = \bar{a}_T^r(\sin(\omega \cdot t))^p$,
$a_T^\tau(t) = \bar{a}_T^\tau(\sin(\omega \cdot t))^q, \omega = 50, p = 2, q = 1$.

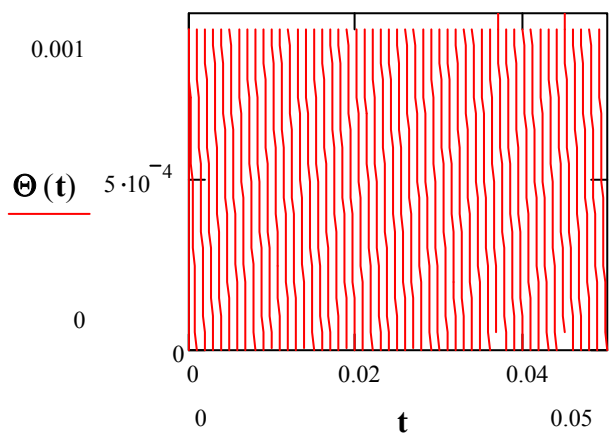

**Pic.7.0**.Cutting function:$\Theta_\tau(t)$.

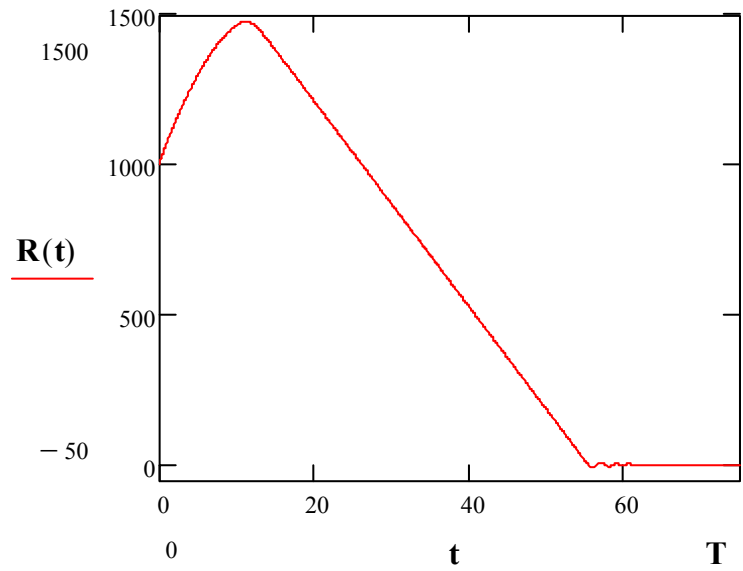

Target-to-missile range R(t)

**Pic.7.1.** Target-to-missile range: $R(t)$.

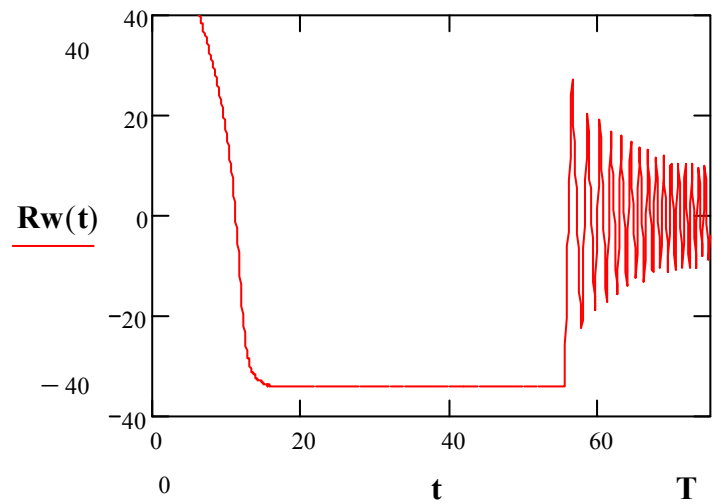

Speed of rapprochement missile-to-target

**Pic.7.2.** Speed of rapprochement missile-to-target: $\dot{R}(t)$.

$$\dot{R}(0) = 100 m/\sec.$$

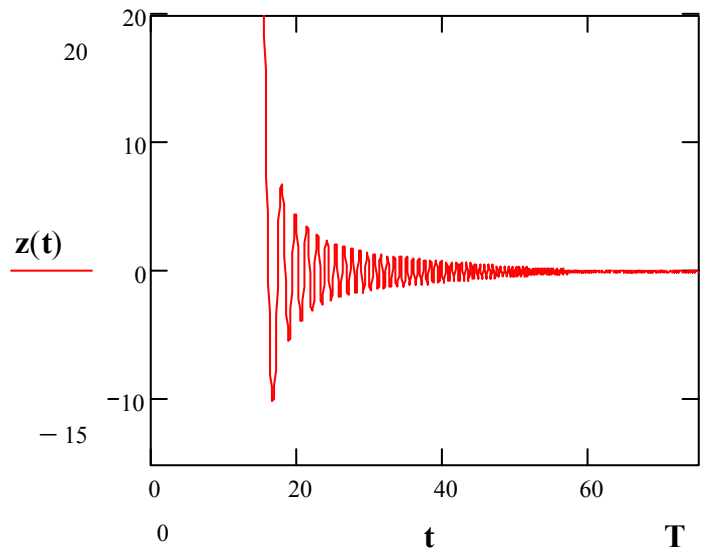

**Pic.7.3**.Variable: $z(t) = R(t)\dot{\sigma}(t)$.
$z(0) = 600, z(T) = -7.76 \times 10^{-8}$.

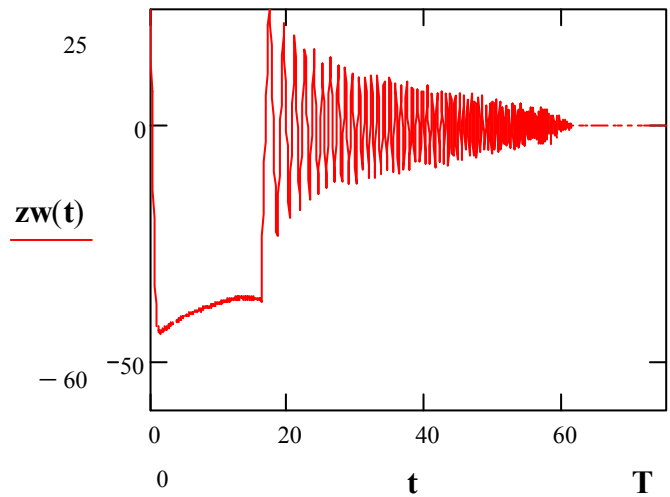

**Pic.7.4**.Variable: $\dot{z}(t) = \dot{R}\dot{\sigma} + R\ddot{\sigma}$.
$\dot{z}(0) = 40, \dot{z}(T) = 6.17 \times 10^{-4}$.

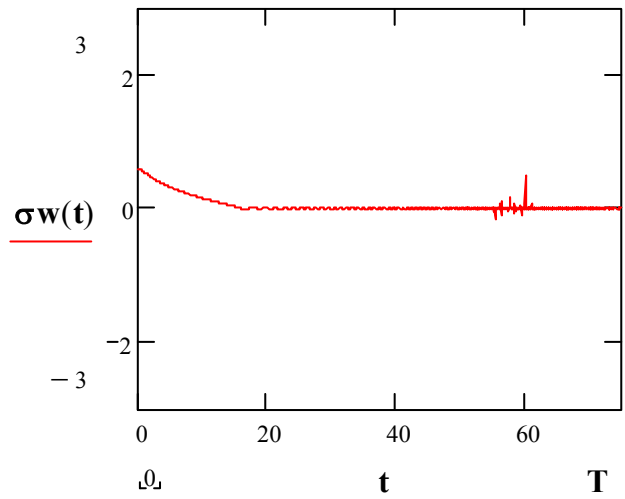

**Pic.7.5.**Variable:$\dot{\sigma}(t)$.

$\dot{\sigma}(0) = 0.6, \dot{\sigma}(T) = 1.169 \times 10^{-7}$.

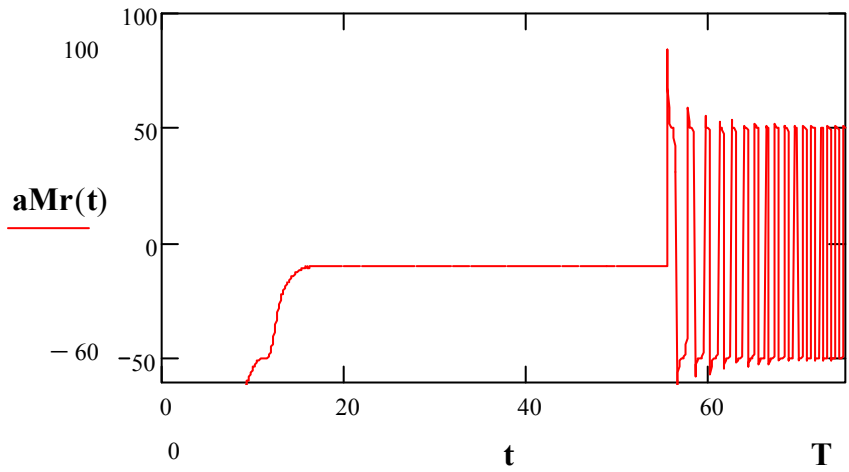

Missile acceleration along target-to-missile direction

**Pic**.**7**.**6**.Missile acceleration along target-to-missile direction:$a_M^r(t)$.

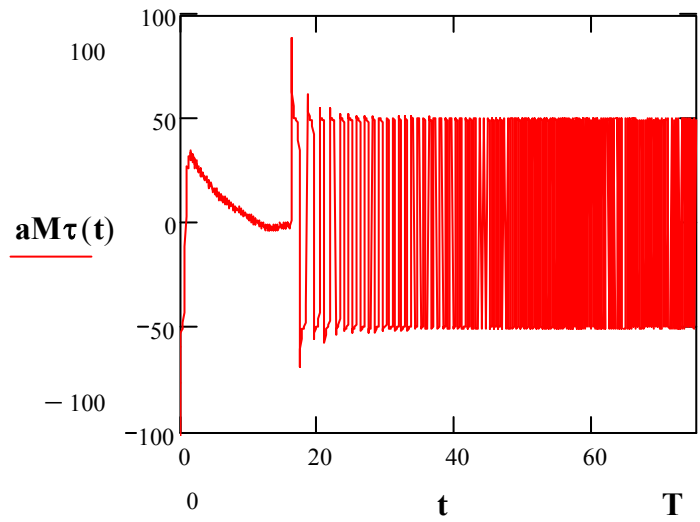

Missile tangent acceleration

**Pic**.7.7.Missile tangent acceleration: $a_M^\tau(t)$.

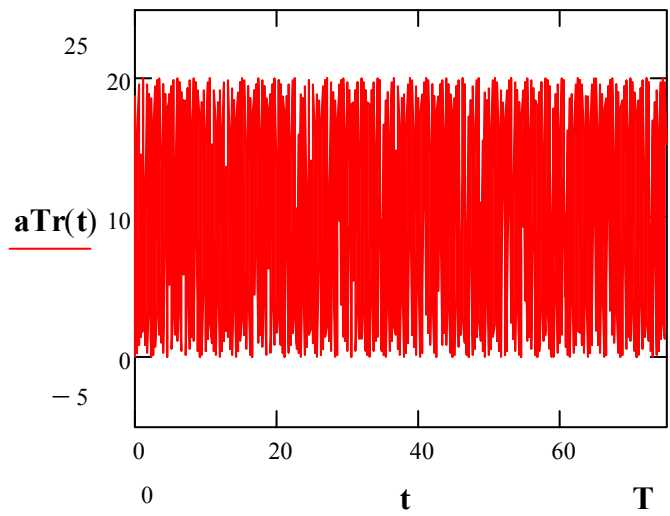

**Pic**.7.8.Target acceleration along target-to-missile direction: $a_T^r(t)$.

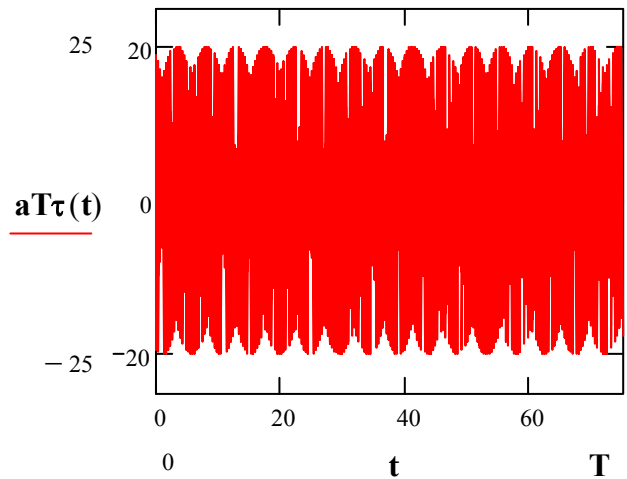

**Pic**.7.9.Target tangent acceleration:$a_T^\tau(t)$.

**Example 8.** $\tau = 0.001, \kappa_1 = 10^{-3}, \kappa_2 = 0.001, \bar{a}_T^r = 20m/\sec^2,$
$\bar{a}_T^\tau = 20m/\sec^2, R(0) = 1000m, V_r(0) = 100m/\sec,$
$z(0) = 600, \dot{z}(0) = 40, a_T^r(t) = \bar{a}_T^r(\sin(\omega \cdot t))^p,$
$a_T^\tau(t) = \bar{a}_T^\tau(\sin(\omega \cdot t))^q, \omega = 50, p = 2, q = 1.$

**Pic**.8.6.Missile acceleration along target-to-missile direction:$a_M^r(t)$

**Pic**.8.7.Missile tangent acceleration:$a_M^\tau(t)$.

**Pic**.8.8.Target acceleration along target-to-missile direction:$a_T^r(t)$.

**Pic**.8.9.Target tangent acceleration:$a_T^\tau(t)$.

# IV.3.Optimal control numerical simulation.
# 2-Persons antagonistic differential game

# $IDG_{2;T}(f, 0, M, 0, \beta)$, with non-linear dynamics and imperfect information."Step-by-step" strategy.

## IV.3.1. Optimal control numerical simulation. 2-Persons antagonistic differential game $IDG_{2;T}(f,0,M,0,\beta)$, with non-linear dynamics and imperfect measurments."Step-by-step" strategy.

**Example 4.3.1.**

Let us consider an 2-persons dissipative differential game $DG_{2;T}^{\#}(f,0,M,0)$, with nonlinear dynamics and imperfect measurments about the nonlinear system:

$$\dot{x}_1(t) = x_2(t),$$

$$\dot{x}_2(t) = -\kappa x_2^3(t) + \alpha_1[t; x_1(t); x_2(t) + \beta(t)] + \alpha_{21}[t; x_1(t); x_2(t) + \beta(t)],$$

$$\kappa > 0. \tag{4.3.1}$$

$$\alpha_1(t) \in [-\rho_1, \rho_1], \alpha_2(t) \in [-\rho_2, \rho_2],$$

$$\mathbf{J}_i = x_1^2(T), i = 1, 2.$$

Thus optimal control problem for the first player:

$$\min_{\alpha_1(t) \in [-\rho_1,\rho_1]} \left( \max_{\alpha_2(t) \in [-\rho_2,\rho_2]} x_1^2(T) \right) \tag{4.3.2}$$

and optimal control problem for the second player:

$$\max_{\alpha_2(t)\in[-\rho_2,\rho_2]} \left( \min_{\alpha_1(t)\in[-\rho_1,\rho_1]} x_1^2(T) \right). \tag{4.3.3}$$

Using the replacement $x_2(t) \to z_2(t) - \beta(t),$ one obtain:

$$\dot{x}_1(t) = z_2(t) - \beta(t),$$

$$\dot{z}_2(t) = -\kappa[z_2(t) - \beta(t)]^3 + \alpha_1[x_1(t); z_2(t)] + \alpha_{21}[x_1(t); z_2(t)] + \dot{\beta}(t),$$

$$\kappa > 0. \tag{4.3.1'}$$

$$\alpha_1(t) \in [-\rho_1, \rho_1], \alpha_2(t) \in [-\rho_2, \rho_2],$$

$$\mathbf{J}_i = x_1^2(T), i = 1, 2.$$

From Eqs.(3.1.15)-(3.1.16) and Eq.(4.3.1') we obtain linear master game for the optimal control problem (4.3.1)-(4.3.3):

$$\dot{u}_1 = u_2 + \lambda_2 + \beta(t),$$

$$\dot{u}_2 = -3\kappa(\lambda_2 - \beta(t))^2 u_2 - \kappa(\lambda_2 - \beta(t))^3 + \check{\alpha}_1(t) + \check{\alpha}_2(t) + \dot{\beta}(t),$$

$$\kappa > 0, \tag{4.3.4}$$

$$\check{\alpha}_1(t) \in [-\rho_1, \rho_1], \check{\alpha}_2(t) \in [-\rho_2, \rho_2],$$

$$\mathbf{J}_i = u_1^2(T), i = 1, 2.$$

Using replacement $\lambda_2 \to \tilde{\lambda}_2 + \beta(t)$ into Eq.(4.3.4) one obtain:

$$\dot{u}_1 = u_2 + \check{\lambda}_2 + 2\beta(t),$$

$$\dot{u}_2 = -3\kappa\check{\lambda}_2^2 u_2 - \kappa\check{\lambda}_2^3 + \check{\alpha}_1(t) + \check{\alpha}_2(t) + \dot{\beta}(t),$$

$$\kappa > 0, \tag{4.3.4'}$$

$$\check{\alpha}_1(t) \in [-\rho_1, \rho_1], \check{\alpha}_2(t) \in [-\rho_2, \rho_2],$$

$$\mathbf{J}_i = u_1^2(T), i = 1, 2.$$

Thus optimal control problem for the first player:

$$\min_{\check{\alpha}_1(t) \in [-\rho_u, \rho_u]} \left( \max_{\check{\alpha}_2(t) \in [-\rho_v, \rho_v]} [u_1^2(T)] \right), \tag{4.3.5}$$

and optimal control problem for the second player:

$$\max_{\check{\alpha}_1(t) \in [-\rho_u, \rho_u]} \left( \min_{\check{\alpha}_2(t) \in [-\rho_v, \rho_v]} [u_1^2(T)] \right). \tag{4.3.6}$$

Suppose that

$$\tau \left| \sup_{t \in [0,T]} \beta(t) - \inf_{t \in [0,T]} \beta(t) \right| \ll 1,$$

$$\tau \left| \sup_{t \in [0,T]} \dot{\beta}(t) - \inf_{t \in [0,T]} \dot{\beta}(t) \right| \ll 1. \tag{4.3.7}$$

Thus from Eq.(A.26) (see Appendix A.) we obtain standard solution for the linear optimal control problem (4.3.4′)-(4.3.6):

$$\check{\alpha}_1(t) = -\rho_1 \text{sign}[x_1(t) + (T-t)x_2(t)]$$

$$\check{\alpha}_2(t) = \rho_2 \text{sign}[x_1(t) + [(T-t)]x_2(t)],$$

(4.3.8)

From Eq.(4.3.8) and Theorem 3.2, one obtain "step-by-step" feedback optimal control for the nonlinear optimal control problem (4.3.1)-(4.3.3). Thus "step-by-step" optimal control for the first player $\alpha_1^*(t)$ one have solving in the next form:

$$\alpha_1^*(t) = -\rho_1 \text{sign}[x_1(t_n) + (t_{n+1} - t)(x_2(t) + \beta(t))],$$

$$t \in [t_n, t_{n+1}],$$

(4.3.9)

$$\tau \triangleq t_{n+1} - t_n = \frac{T}{N}, n = 1, \ldots, N.$$

and "step-by-step" optimal control for the second player $\alpha_2^*(t)$ in the next form:

$$\alpha_2^*(t) = \rho_2 \text{sign}[x_1(t) + (t_{n+1} - t)(x_2(t) + \beta(t))],$$

$$t \in [t_n, t_{n+1}],$$

(4.3.10)

$$\tau \triangleq t_{n+1} - t_n = \frac{T}{N}, n = 1, \ldots, N.$$

Suppose that $\tau \triangleq (t_{n+1} - t_n) \ll 1, t \in [t_n, t_{n+1}], n = 1, \ldots, N$, from Eq.(A.26) we obtain (quasy) optimal control $\alpha_1^*(t)$ for the first player and optimal control $\alpha_2^*(t)$ for the second player in the next form:

$$\alpha_1^*(t) \simeq -\rho_1 \text{sign}[x_1(t) + (t_{n+1} - t)(x_2(t) + \beta(t))],$$

(4.3.11)

$$\alpha_2^*(t) \simeq \rho_2 \text{sign}[x_1(t) + [(t_{n+1} - t)](x_2(t) + \beta(t))].$$

From Eq.(4.3.11) one obtain optimal control $\alpha_1^*(t)$ for the first player and optimal control $\alpha_2^*(t)$ for the second player:

$$\alpha_1^*(t) \simeq -\rho_1 \mathbf{sign}[x_1(t) + \Theta_\tau(t)(x_2(t) + \beta(t))],$$

(4.3.12)

$$\alpha_2^*(t) \simeq \rho_2 \mathbf{sign}[x_1(t) + \Theta_\tau(t)(x_2(t) + \beta(t))].$$

Thus for the numerical simulation we obtain ODE:

$$\dot{x}_1(t) = x_2(t),$$

$$\dot{x}_2(t) = -\kappa x_2^3(t) - \rho_1 \cdot \mathbf{sign}[x_1(t) + \Theta_\tau(t)(x_2(t) + \beta(t))] +$$

(4.3.13)

$$+\rho_2 \cdot \mathbf{sign}[x_1(t) + \Theta_\tau(t)(x_2(t) + \beta(t))],$$

$$\kappa > 0.$$

**Numerical simulation. Example 4.3.2.** $\beta(t) = A\sin(\omega t); \kappa = 1, \rho_1 = 300,$ $\rho_2 = 100, A = 300, \omega = 50, \tau = 10^{-2}.$

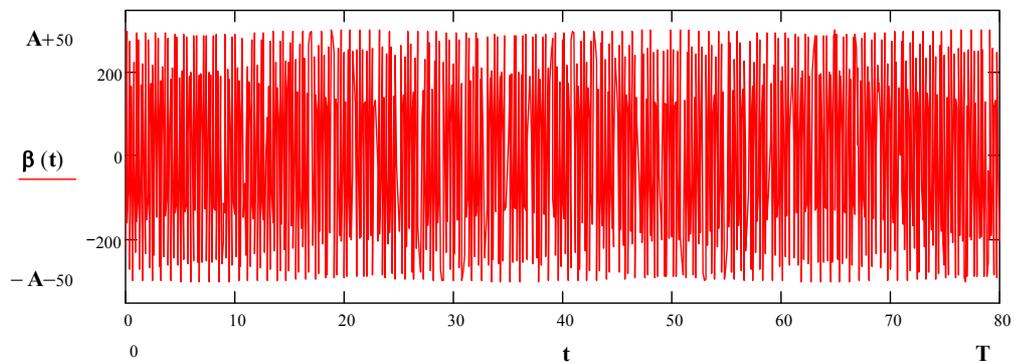

Uncertainty on an output the devices the measured of speed of rapproachement

Pic.4.3.2.1.Ancertainty $\beta(t) = A\sin(\omega t). A = 300, \omega = 50.$

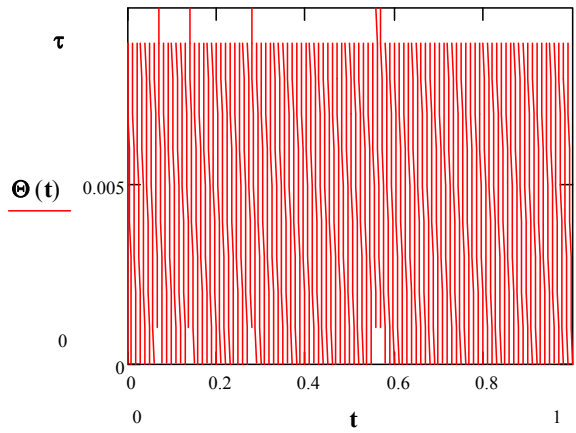

Cutting functuion

Pic.4.3.2.2. Cutting function $\Theta_\tau(t)$. $\tau = 0.01$.

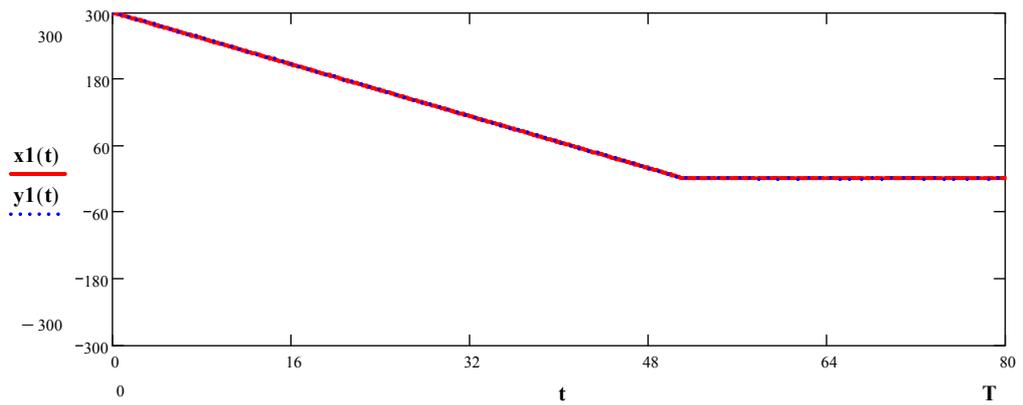

Optimal trajectory x1(t). Game with uncertainty: red curve. Classical game: blue curve.

Pic.4.3.2.3. $\kappa = 1, \rho_1 = 300, \rho_2 = 100, A = 300, \omega = 50, \tau = 0.01$.

$$x_1(T) = 0.345,$$
$$y_1(T) = 8.068 \times 10^{-6}.$$

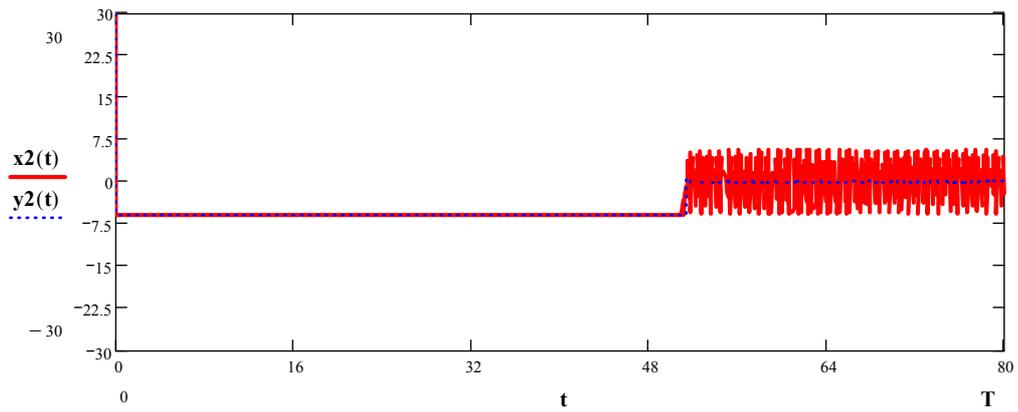

Optimal velocity x2(t). Game with uncertainty: red curve. Classical game: blue curve.

Pic.4.3.2.4. $\kappa = 1, \rho_1 = 300, \rho_2 = 100, A = 300,$
$\omega = 50, \tau = 0.01.$

$$x_2(T) = -2.11,$$
$$y_2(T) = -4.529 \times 10^{-4}.$$

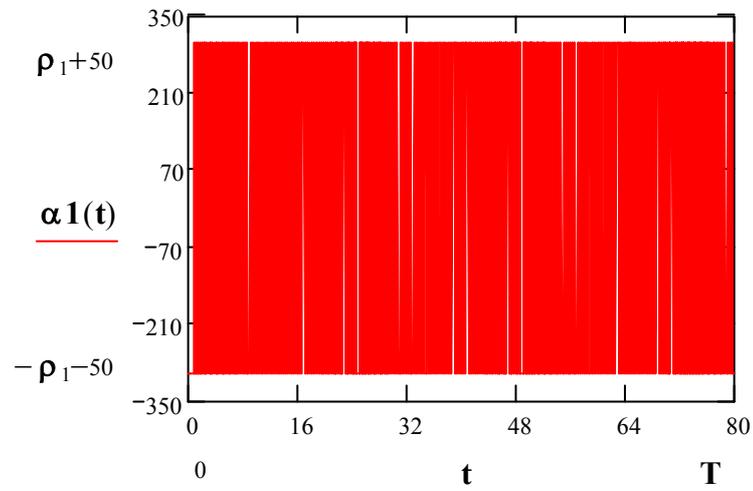

Optimal control of the first player.

Pic.4.3.2.5. Optimal control of the first player.
$\kappa = 1, \rho_1 = 300, \rho_2 = 100, A = 300,$
$\omega = 50, \tau = 0.01.$

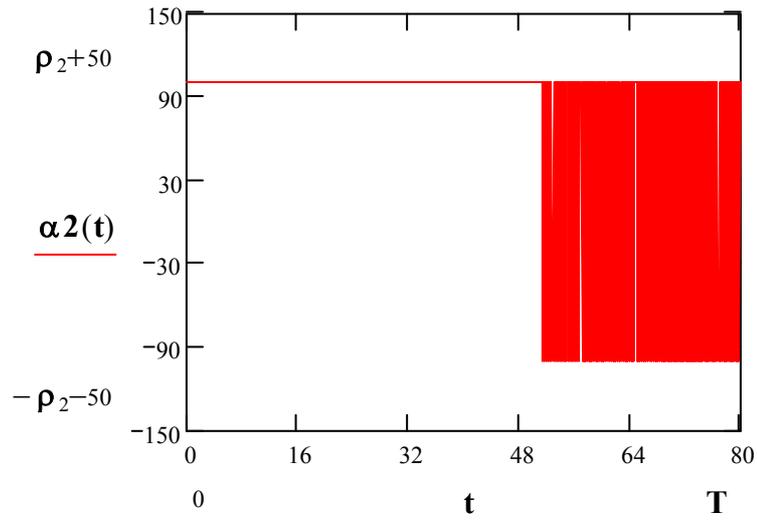

Optimal control of the second player.

Pic.4.3.2.6. $\kappa = 1, \rho_1 = 300, \rho_2 = 100, A = 300, \omega = 50, \tau = 0.01$.

**Numerical simulation. Example 4.3.3.**
$\beta(t) = A\sin(\omega t); \kappa = 1, \rho_1 = 300, \rho_2 = 100, A = 300, \omega = 50, \tau = 10^{-4}$.

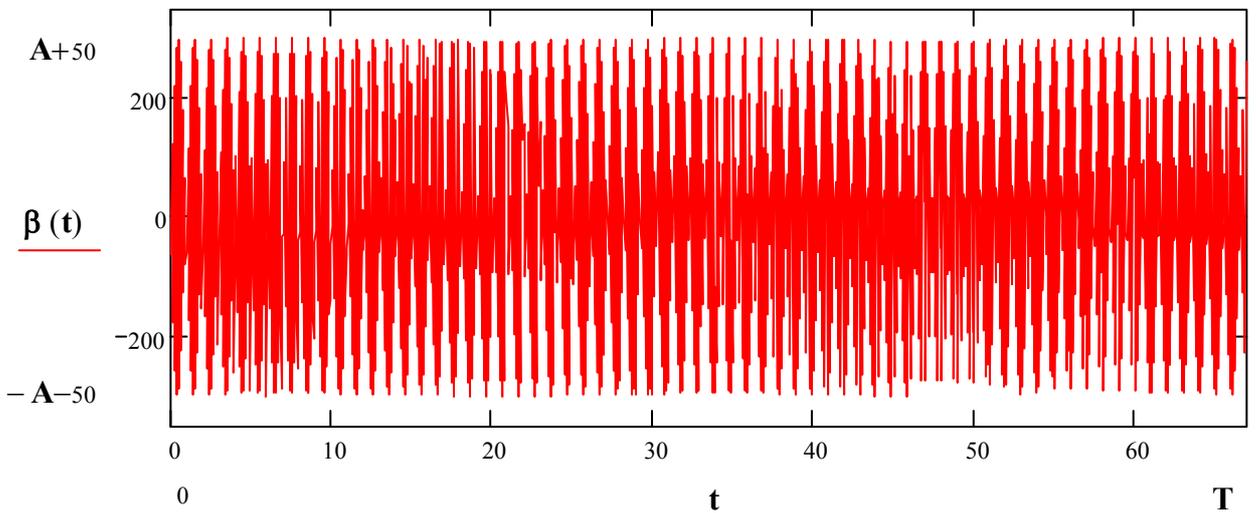

Uncertainty on an output the devices the measured of speed of rapproachement

Pic.4.3.3.1. Ancertainty $\beta(t) = A\sin(\omega t). A = 300, \omega = 50$.

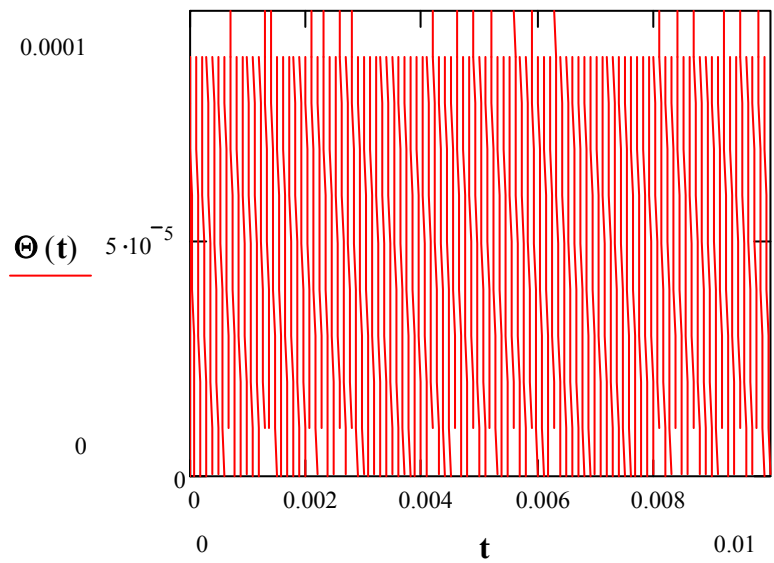

Cutting functuion

Pic.4.3.3.2. Cutting function $\Theta_\tau(t). \tau = 10^{-4}$.

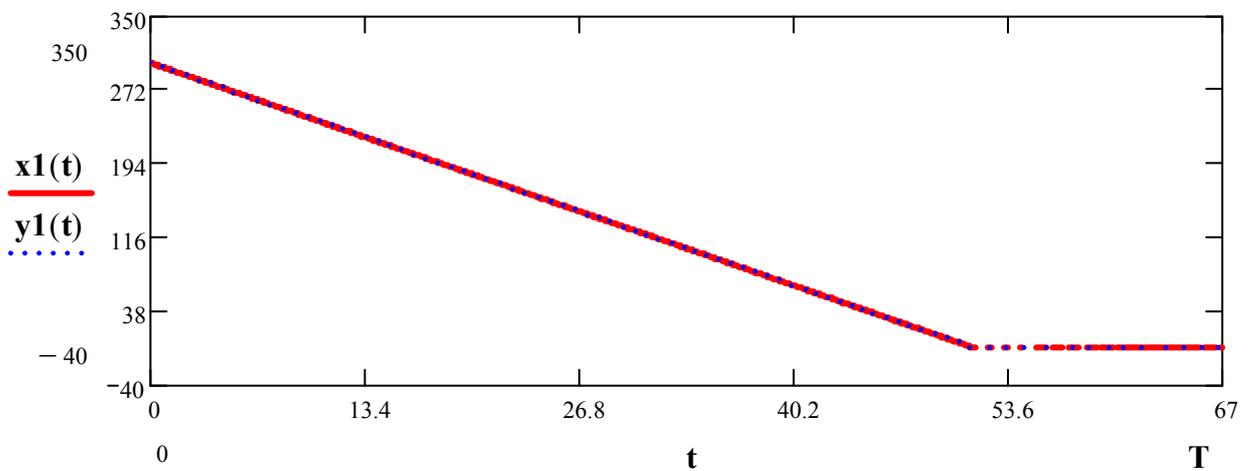

Optimal trajectory. Game with uncertainty: red curve. Classical game: blue curve.

Pic.4.3.3.3. $\kappa = 1, \rho_1 = 300, \rho_2 = 100, A = 300, \omega = 50, \tau = 10^{-4}$.

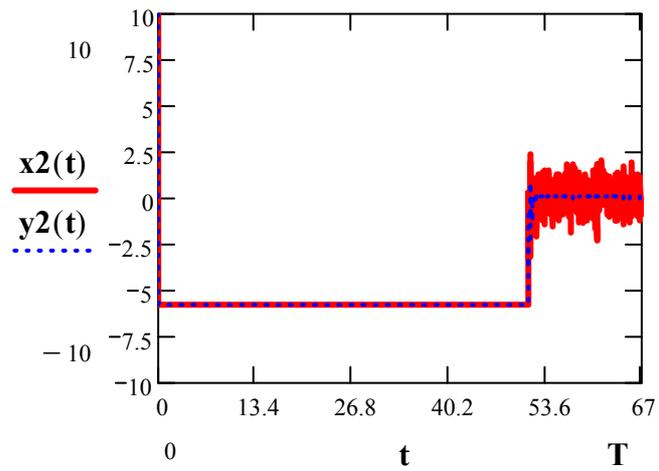

Optimal velocity

Pic.4.3.3.4. $\kappa = 1, \rho_1 = 300, \rho_2 = 100, A = 300,$
$\omega = 50, \tau = 10^{-4}$.

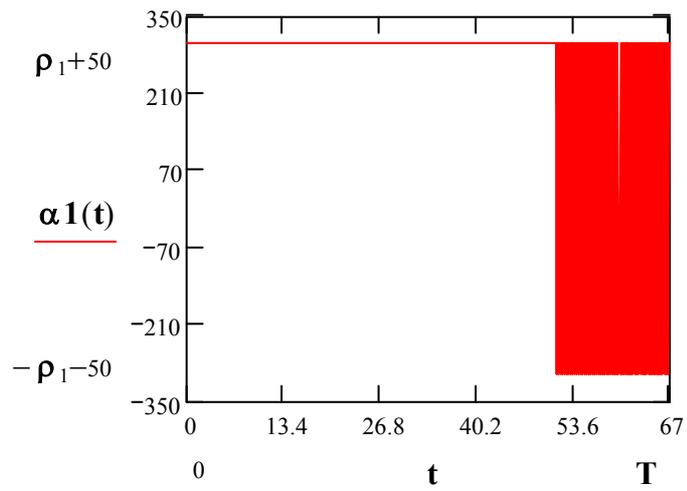

Optimal control for the first player.

Pic.4.3.3.5. $\kappa = 1, \rho_1 = 300, \rho_2 = 100, A = 300, \omega = 50, \tau = 10^{-4}$.

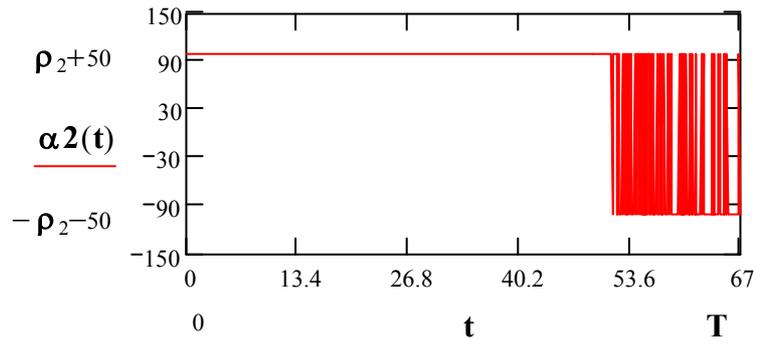

Optimal control of the second player.

Pic.4.3.3.6. $\kappa = 1, \rho_1 = 300, \rho_2 = 100, A = 300, \omega = 50, \tau = 10^{-4}$.

**Numerical simulation. Example 4.3.4.**

$\beta(t) = A\sin(\omega t); \kappa = 1, \rho_1 = 300, \rho_2 = 100, A = 300, \omega = 50, \tau = 10^{-6}$.

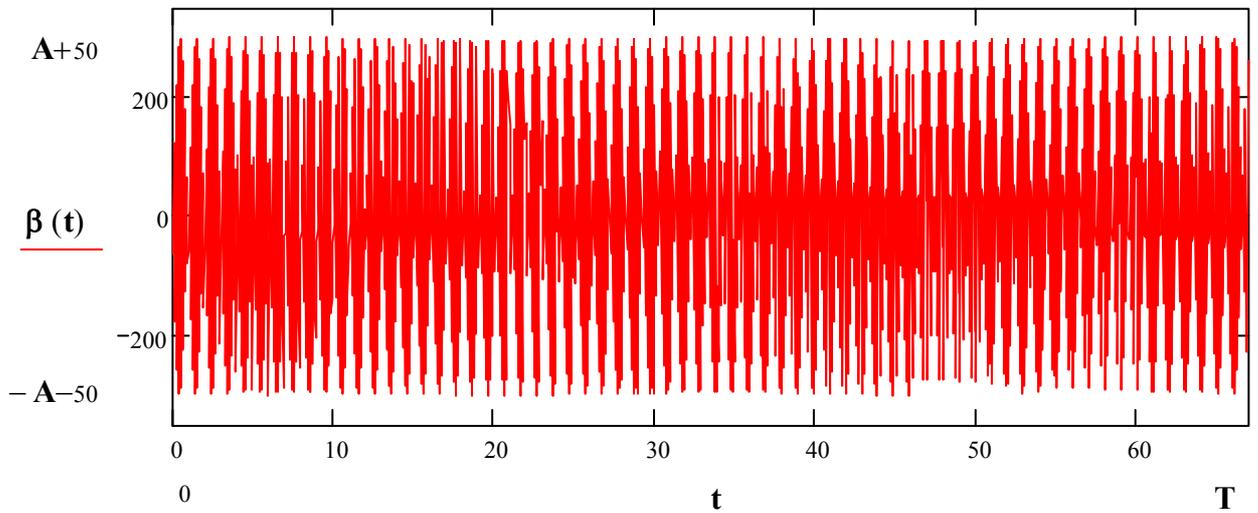

Uncertainty on an output the devices the measured of speed of rapproachement

Pic.4.3.4.1. Ancertainty $\beta(t) = A\sin(\omega t). A = 300, \omega = 50$.

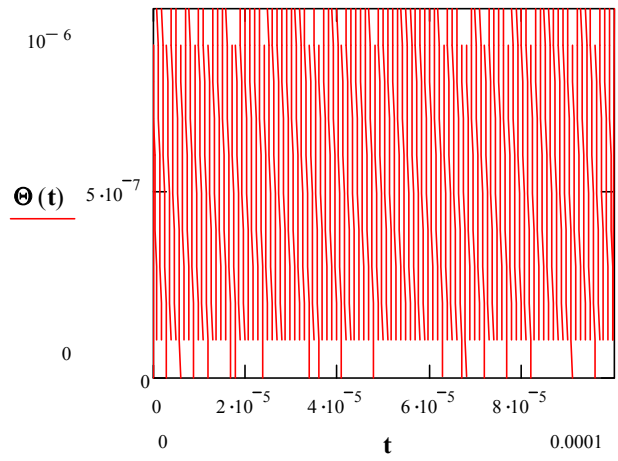

Cutting functuion

Pic.4.3.4.2. Cutting function $\Theta_\tau(t).\tau = 10^{-6}$.

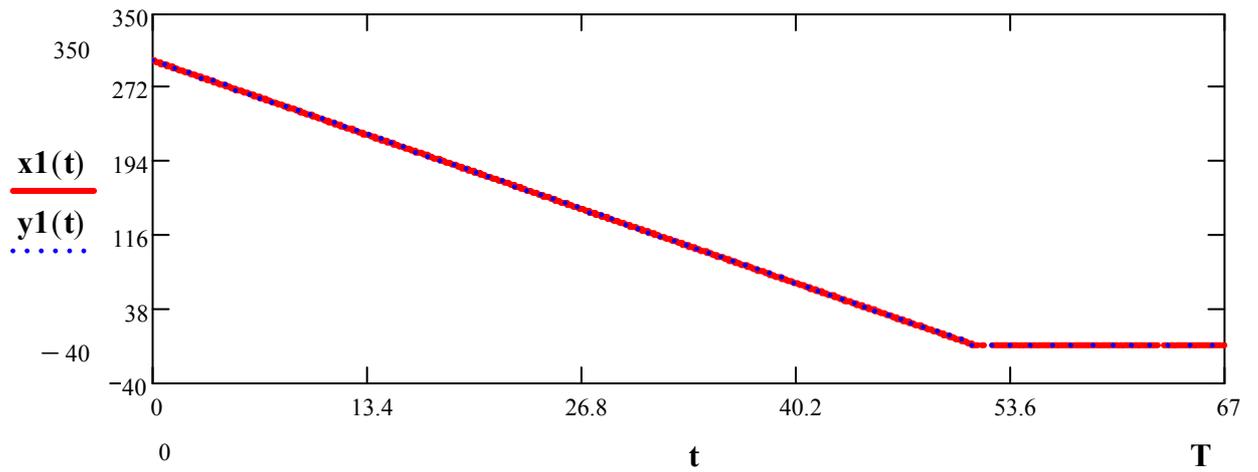

Optimal trajectory. Game with uncertainty: red curve. Classical game: blue curve.

Pic.4.3.4.3. $\kappa = 1, \rho_1 = 300, \rho_2 = 100, A = 300, \omega = 50, \tau = 10^{-6}$.

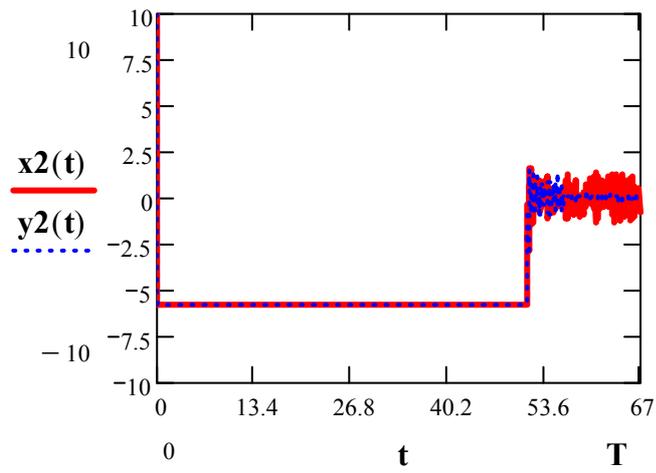

Optimal velocity

Pic.4.3.4.4. $\kappa = 1, \rho_1 = 300, \rho_2 = 100, A = 300, \omega = 50, \tau = 10^{-6}$.

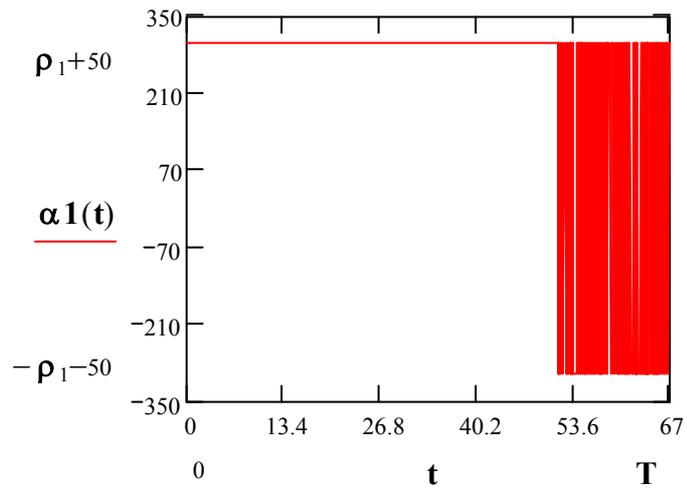

Optimal control for the first player.

Pic.4.3.4.5. $\kappa = 1, \rho_1 = 300, \rho_2 = 100, A = 300, \omega = 50, \tau = 10^{-6}$.

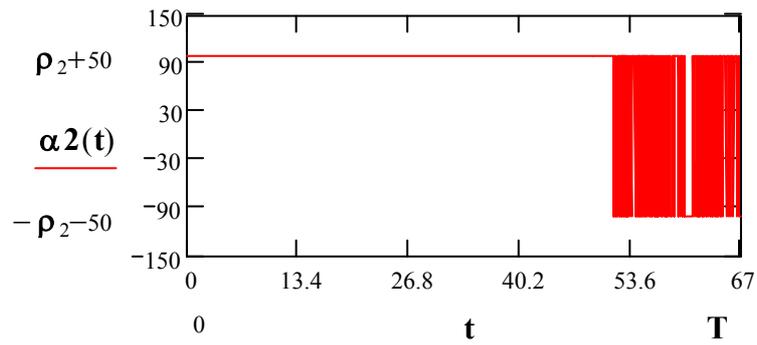

Optimal control of the second player.

Pic.4.3.4.6. $\kappa = 1, \rho_1 = 300, \rho_2 = 100, A = 300, \omega = 50, \tau = 10^{-6}$.

**Numerical simulation. Example 4.3.5.**

$\beta(t) = A\sin(\omega t); \kappa = 1, \rho_1 = 300, \rho_2 = 100, A = 10^4, \omega = 50, \tau = 10^{-8}$.

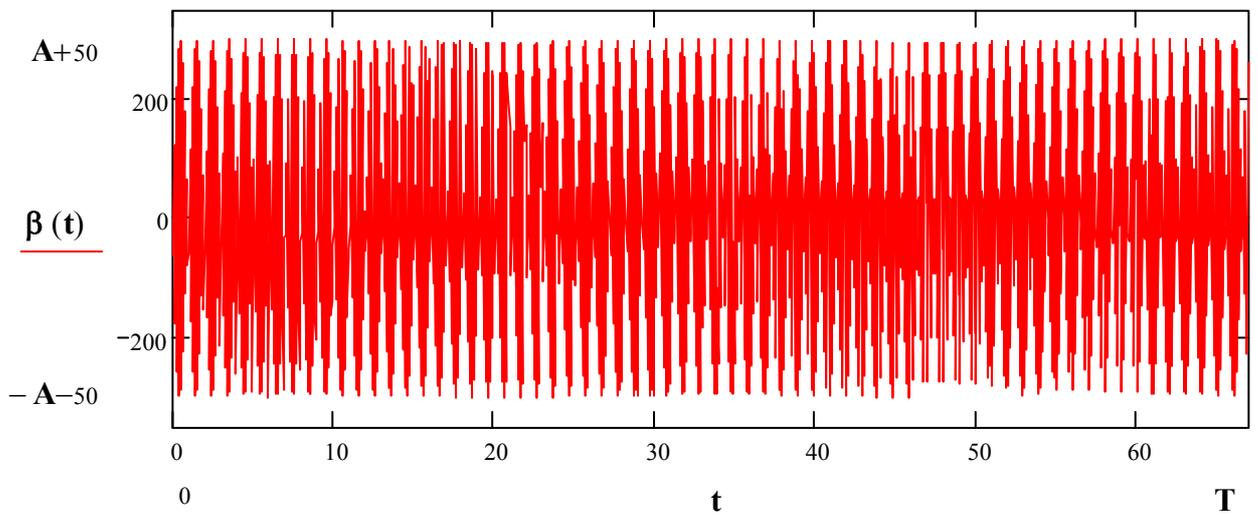

Uncertainty on an output the devices the measured of speed of rapproachement

Pic.4.3.5.1. Ancertainty $\beta(t) = A\sin(\omega t). A = 300, \omega = 50$.

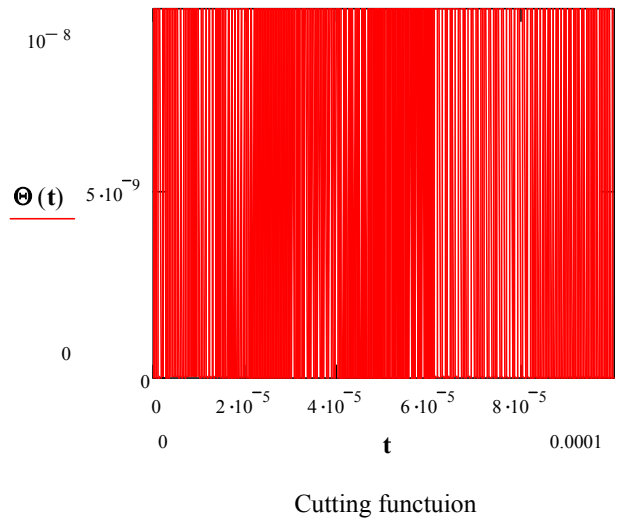

Cutting functuion

Pic.4.3.5.2. Cutting function $\Theta_\tau(t). \tau = 10^{-8}$.

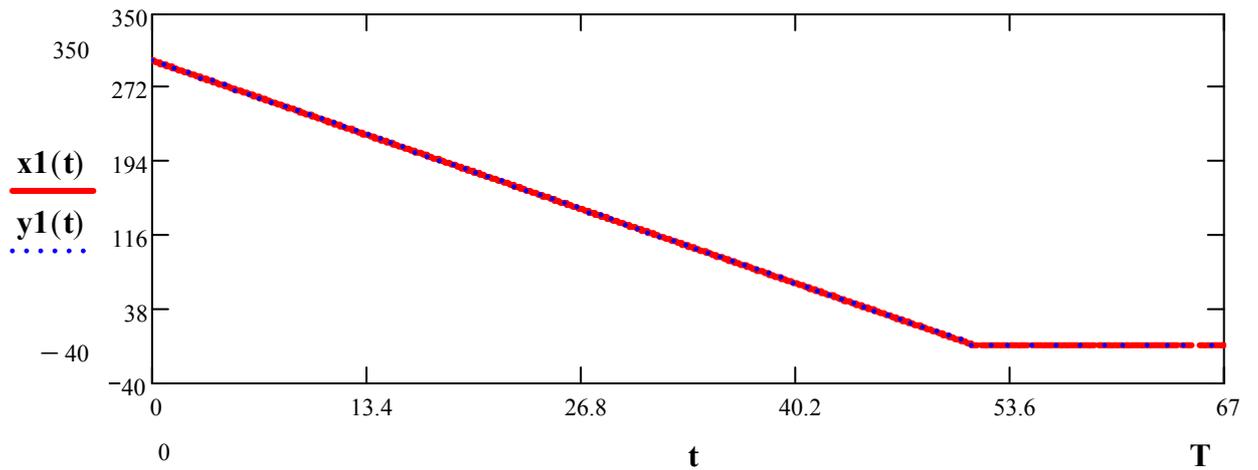

Optimal trajectory. Game with uncertainty: red curve. Classical game: blue curve.

Pic.4.3.5.3. $\kappa = 1, \rho_1 = 300, \rho_2 = 100, A = 300, \omega = 50, \tau = 10^{-8}$.
$x_1(T) = 1.8 \times 10^{-4}, y_1(T) = 7.4 \times 10^{-5}$.

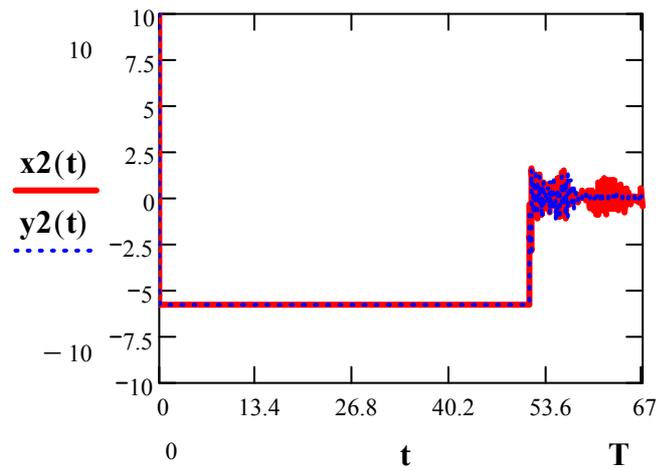

Optimal velocity

Pic.4.3.5.4. $\kappa = 1, \rho_1 = 300, \rho_2 = 100, A = 300, \omega = 50, \tau = 10^{-8}$.

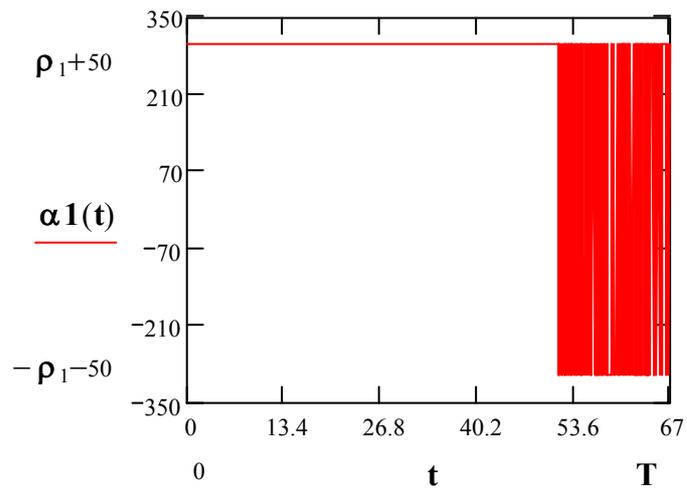

Optimal control for the first player.

Pic.4.3.5.5. $\kappa = 1, \rho_1 = 300, \rho_2 = 100, A = 300, \omega = 50, \tau = 10^{-8}$

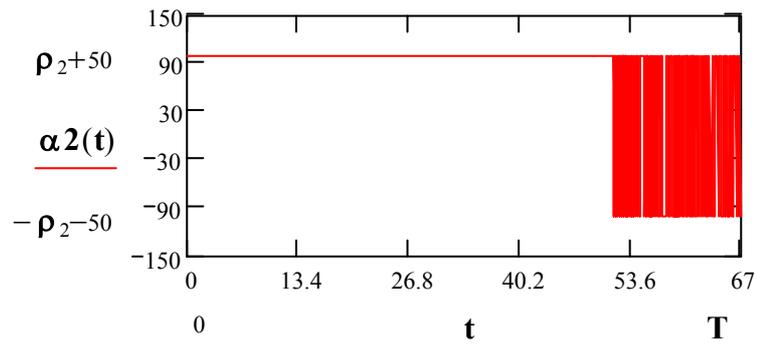

Optimal control of the second player.

Pic.4.3.5.6. $\kappa = 1, \rho_1 = 300, \rho_2 = 100, A = 300, \omega = 50, \tau = 10^{-8}$

**Numerical simulation. Example 4.3.6.**

$\beta(t) = A\sin(\omega t); \kappa = 1, \rho_1 = 300, \rho_2 = 100, A = 300, \omega = 500, \tau = 10^{-8}$.

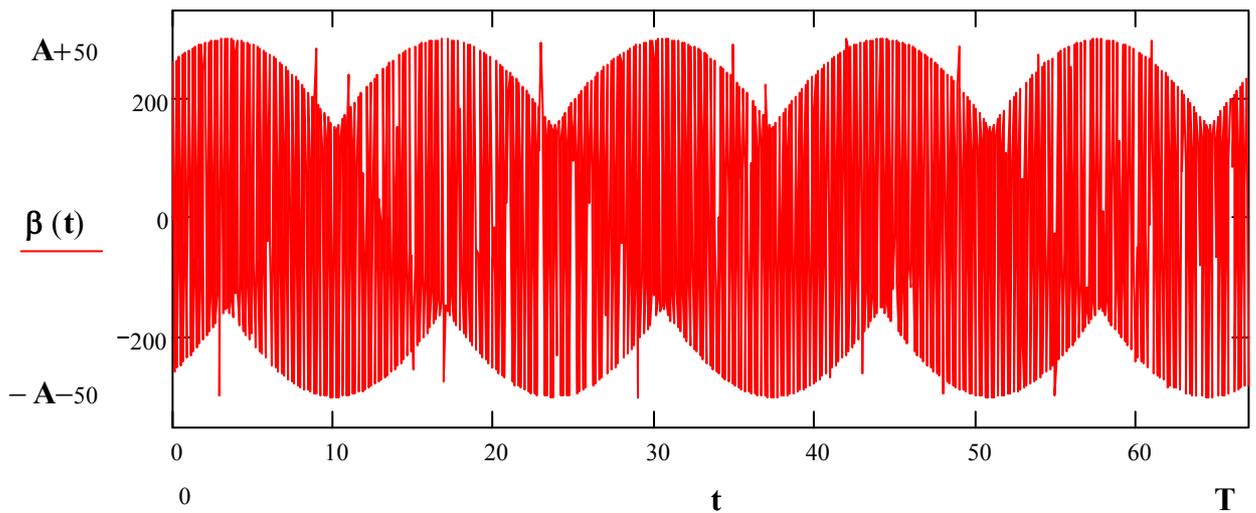

Uncertainty on an output the devices the measured of speed of rapproachement

Pic.4.3.6.1. Ancertainty $\beta(t) = A\sin(\omega t). A = 300, \omega = 500$.

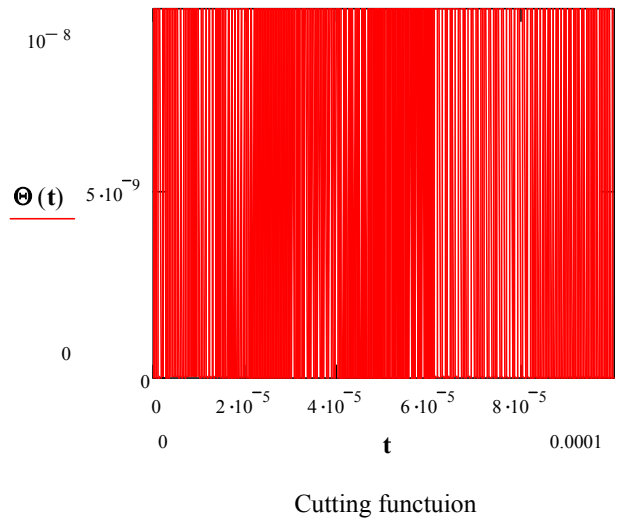

Cutting functuion

Pic.4.3.6.2. Cutting function $\Theta_\tau(t). \tau = 10^{-8}$.

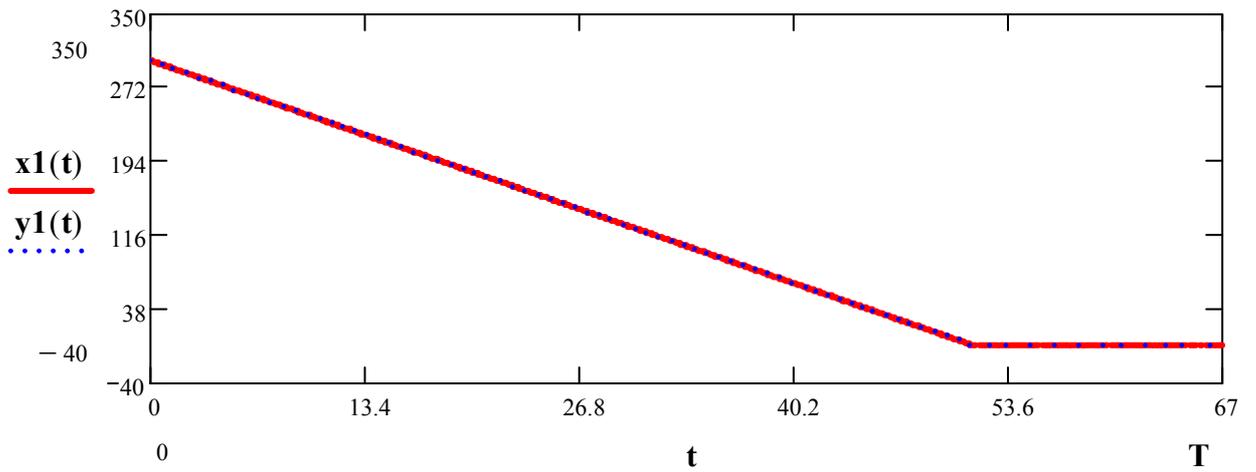

Optimal trajectory. Game with uncertainty: red curve. Classical game: blue curve.

Pic.4.3.6.3. $\kappa = 1, \rho_1 = 300, \rho_2 = 100, A = 300, \omega = 50, \tau = 10^{-8}$.
$x_1(T) = -7 \times 10^{-4}, y_1(T) = 7 \times 10^{-5}$.

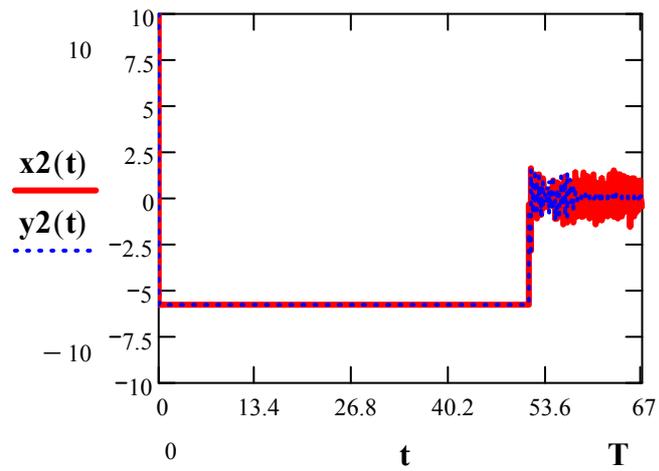

Optimal velocity

Pic.4.3.6.4. $\kappa = 1, \rho_1 = 300, \rho_2 = 100, A = 300, \omega = 50, \tau = 10^{-8}$.

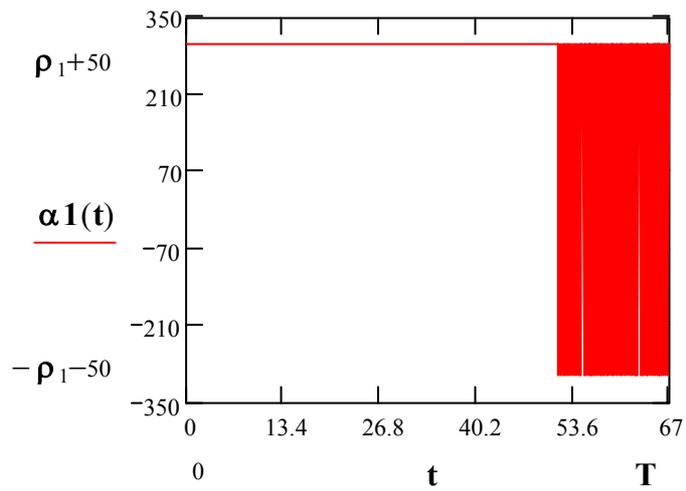

Optimal control for the first player.

Pic.4.3.6.5. $\kappa = 1, \rho_1 = 300, \rho_2 = 100, A = 300, \omega = 50, \tau = 10^{-8}$

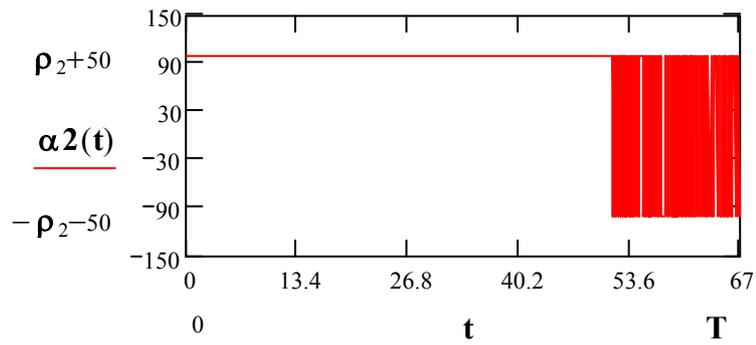

Optimal control of the second player.

Pic.4.3.6.6. $\kappa = 1, \rho_1 = 300, \rho_2 = 100, A = 300, \omega = 50, \tau = 10^{-8}$

**Numerical simulation. Example 4.3.7.**

$\beta(t) = A\sin(\omega t^2); \kappa = 1, \rho_1 = 300, \rho_2 = 100, A = 300, \omega = 500, \tau = 10^{-8}$.

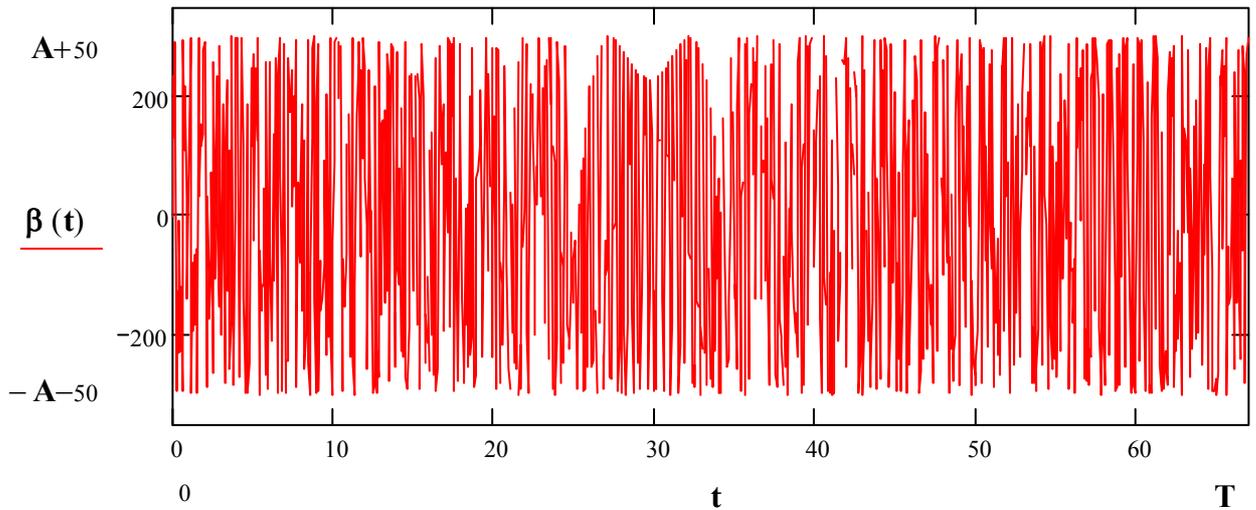

Uncertainty on an output the devices the measured of speed of rapproachement

Pic.4.3.7.1. Ancertainty $\beta(t) = A\sin(\omega t^2). A = 300, \omega = 500$.

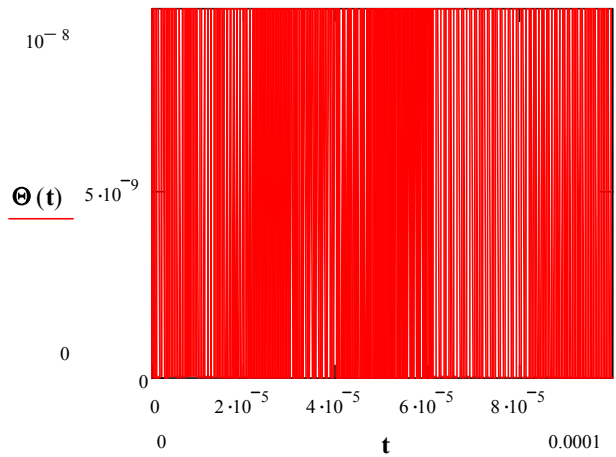

Cutting functuion

Pic.4.3.7.2. Cutting function $\Theta_\tau(t)$. $\tau = 10^{-8}$.

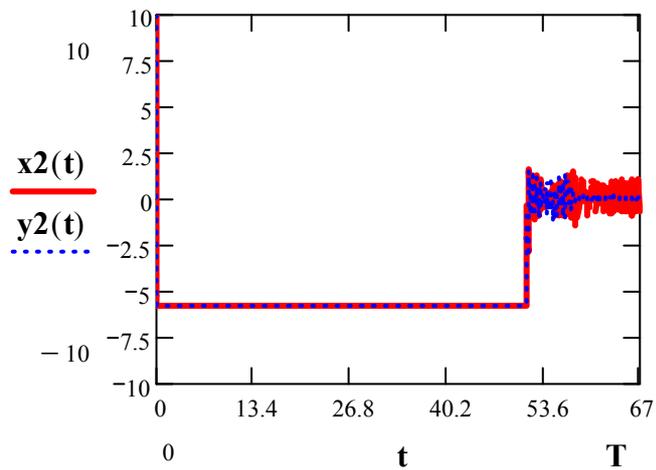

Optimal velocity

Pic.4.3.7.3. $\kappa = 1, \rho_1 = 300, \rho_2 = 100, A = 300, \omega = 50, \tau = 10^{-8}$.

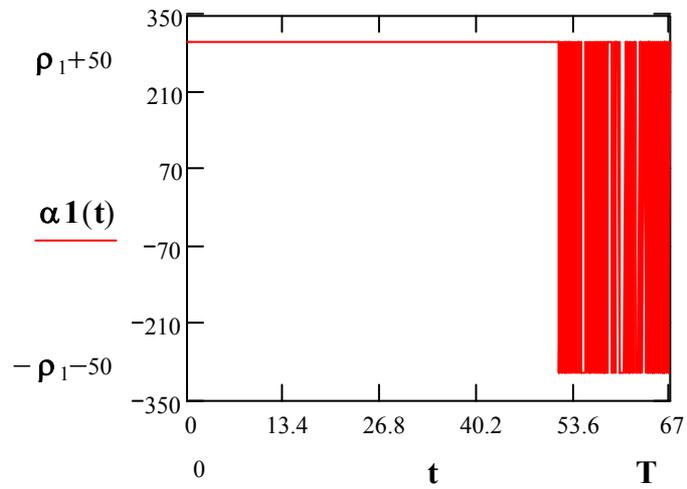

Optimal control for the first player.

Pic.4.3.7.4. $\kappa = 1, \rho_1 = 300, \rho_2 = 100, A = 300, \omega = 50, \tau = 10^{-8}$.

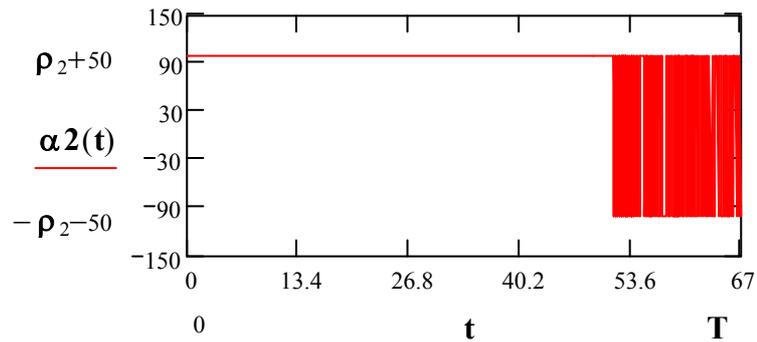

Optimal control of the second player.

Pic.4.3.7.5. $\kappa = 1, \rho_1 = 300, \rho_2 = 100, A = 300, \omega = 50, \tau = 10^{-8}$

**Numerical simulation. Example 4.3.8.**

$\beta(t) = A\sin(\omega t); \kappa = 1, \rho_1 = 300, \rho_2 = 100, A = 10^4, \omega = 50, \tau = 10^{-2}$.

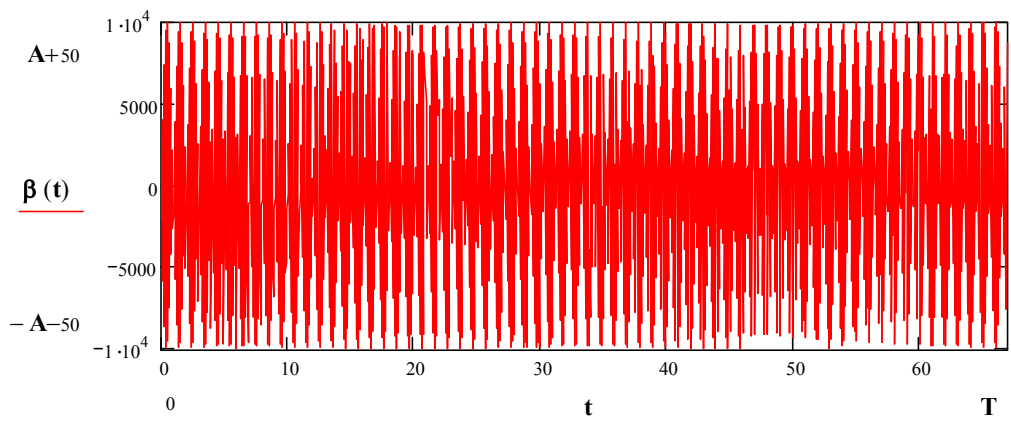

Uncertainty on an output the devices the measured of speed of rapproachement

Pic.4.3.8.1. Ancertainty $\beta(t) = A\sin(\omega t). A = 10^4, \omega = 50.$

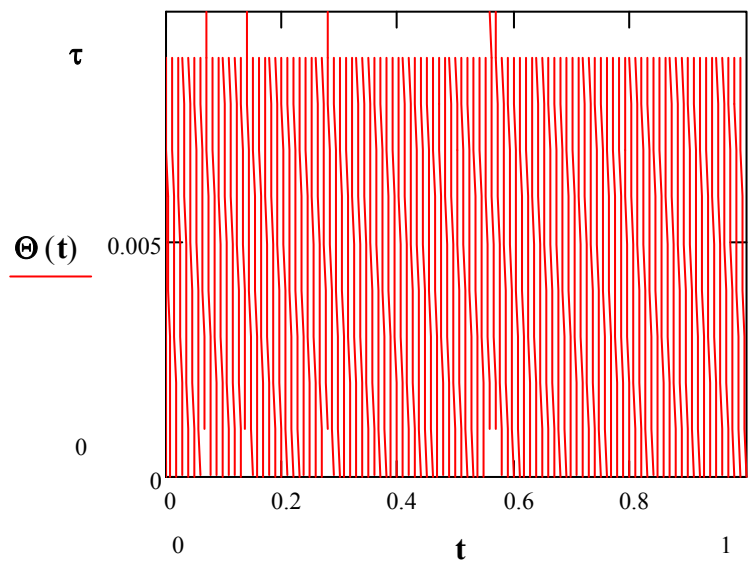

Cutting functuion

Pic.4.3.8.2. Cuttingfunction $\Theta_\tau(t). \tau = 0.01.$

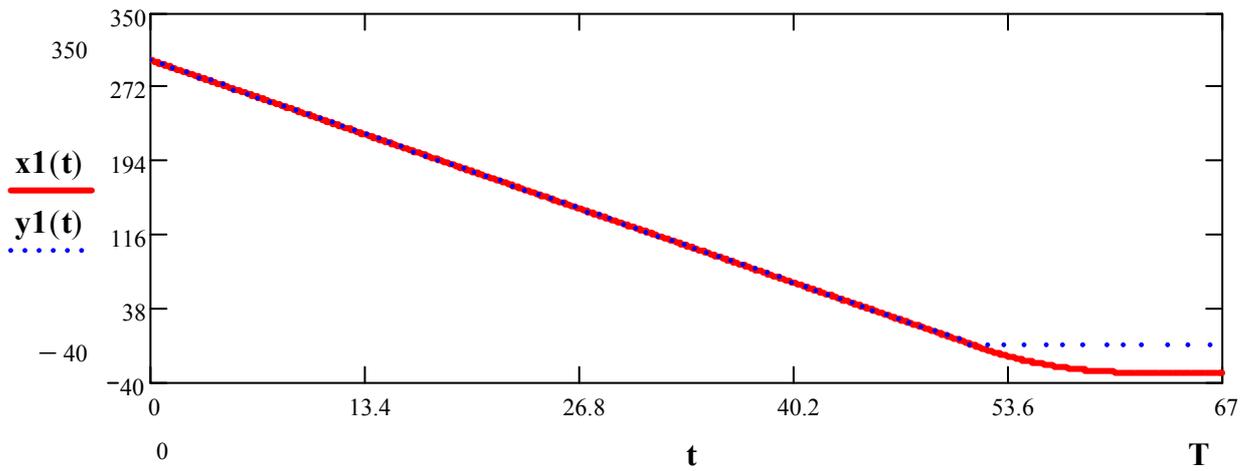

Optimal trajectory. Game with uncertainty: red curve. Classical game: blue curve.

Pic.4.3.8.3. $\kappa = 1, \rho_1 = 300, \rho_2 = 100, A = 10^4, \omega = 50, \tau = 0.01$.

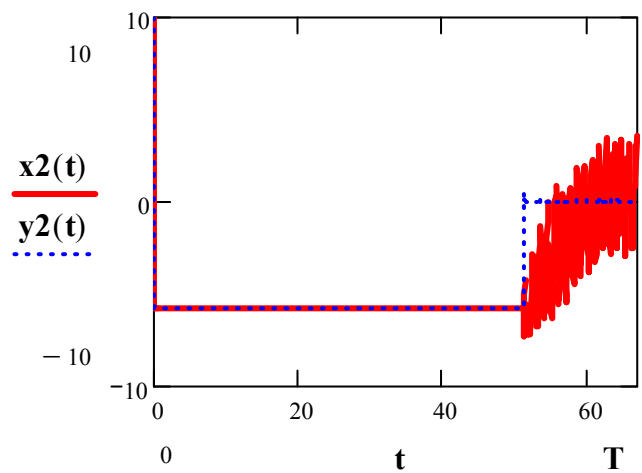

Optimal velocity

Pic.4.3.8.4. $\kappa = 1, \rho_1 = 300, \rho_2 = 100, A = 10^4, \omega = 50$.

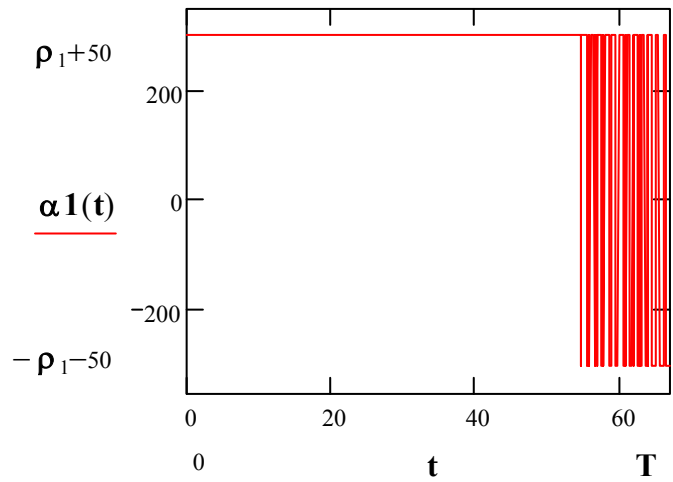

Optimal control for the first player.

Pic.4.3.8.5. $\kappa = 1, \rho_1 = 300, \rho_2 = 100, A = 10^4, \omega = 50$.

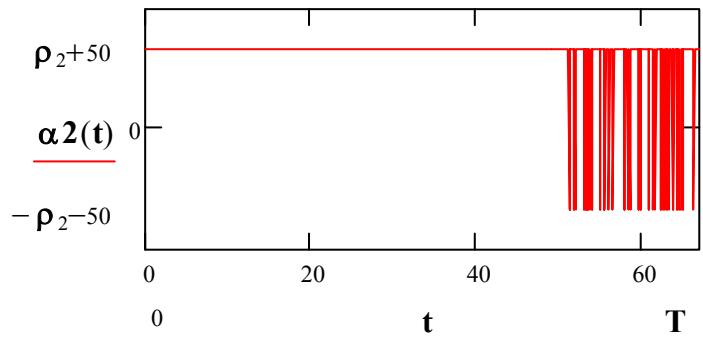

Optimal control of the second player.

Pic.4.3.8.6. $\kappa = 1, \rho_1 = 300, \rho_2 = 100, A = 10^4, \omega = 50$.

**Example 4.3.9.**
$\beta(t) = A\sin(\omega t); \kappa = 1, \rho_1 = 300, \rho_2 = 100, A = 10^4, \omega = 50, \tau = 10^{-3}$.

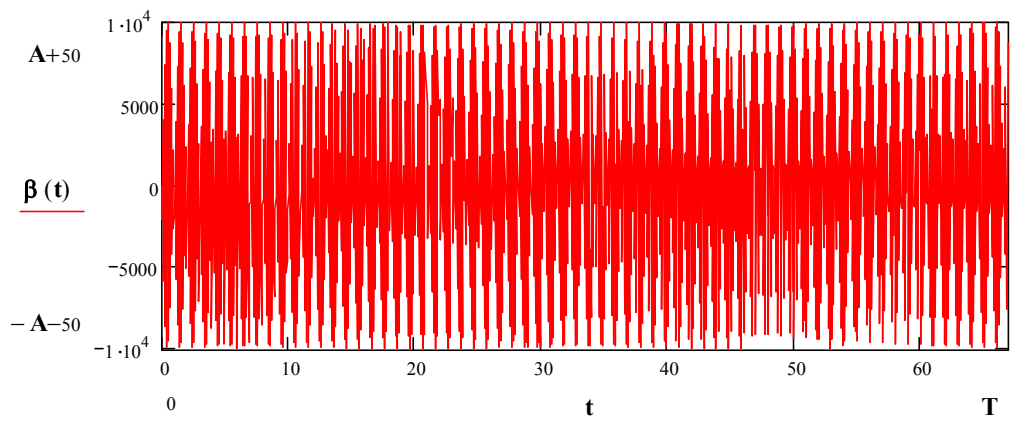

Uncertainty on an output the devices the measured of speed of rapproachement

Pic.4.3.9.1.Ancertainty: $\beta(t) = A\sin(\omega t). A = 10^4, \omega = 50.$

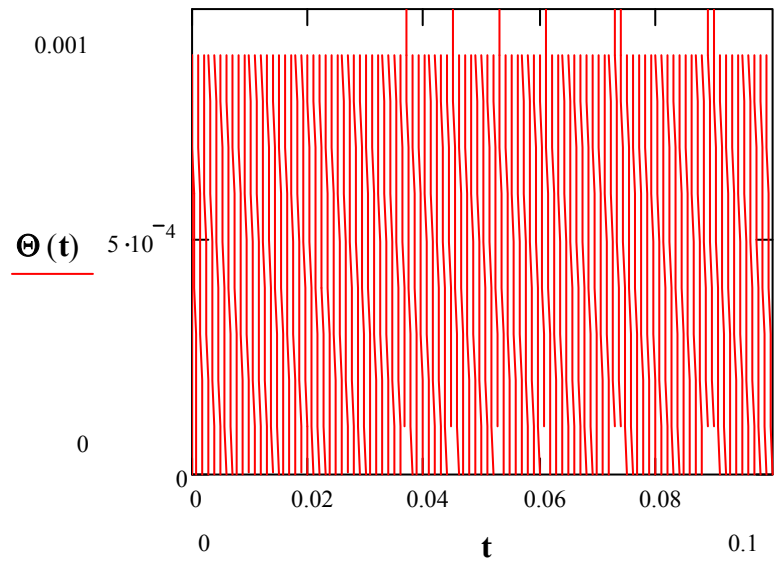

Cutting functuion

Pic.4.3.9.2.Cutting function $\Theta_\tau(t). \tau = 0.001.$

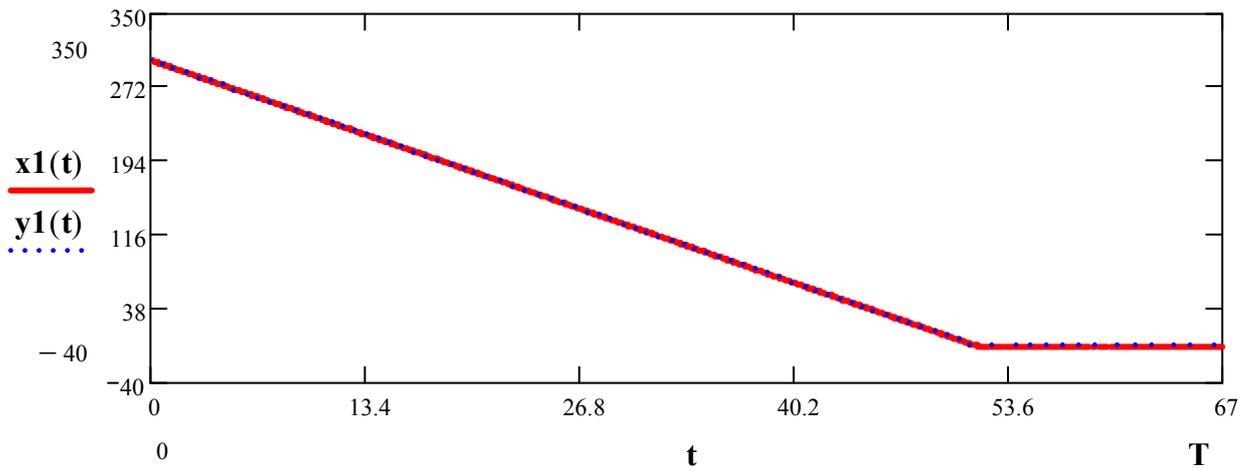

Optimal trajectory. Game with uncertainty: red curve. Classical game: blue curve.

Pic.4.3.9.3. $\kappa = 1, \rho_1 = 300, \rho_2 = 100, A = 10^4, \omega = 50, \tau = 0.001.$

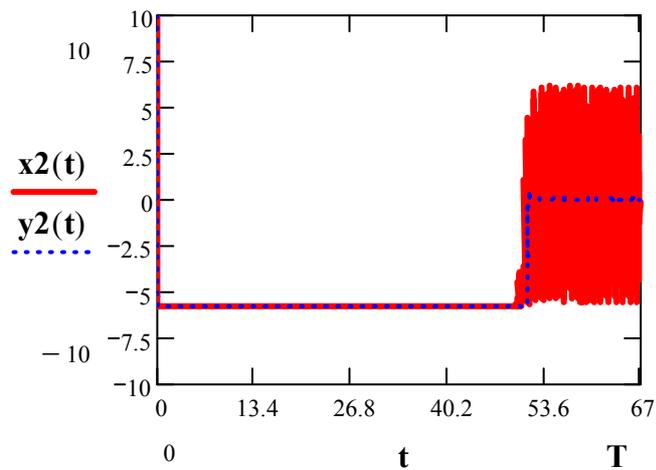

Optimal velocity

Pic.4.3.9.4. $\kappa = 1, \rho_1 = 300, \rho_2 = 100, A = 10^4, \omega = 50,$

$\tau = 0.001.$

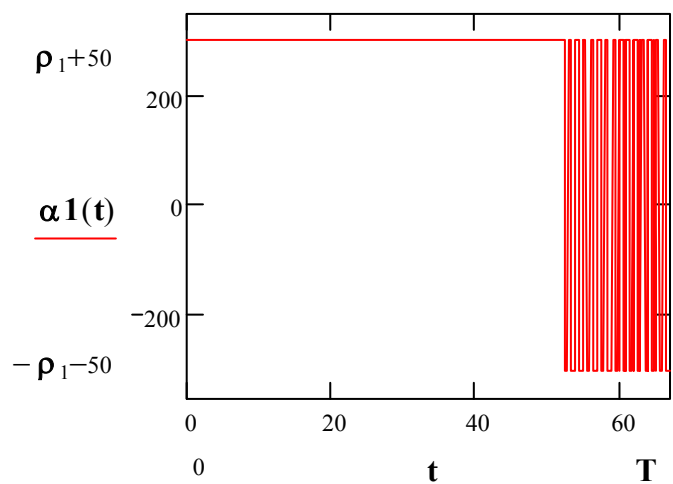

Optimal control for the first player.

Pic.4.3.9.5. $\kappa = 1, \rho_1 = 300, \rho_2 = 100, A = 10^4, \omega = 50,$

$\tau = 0.001.$

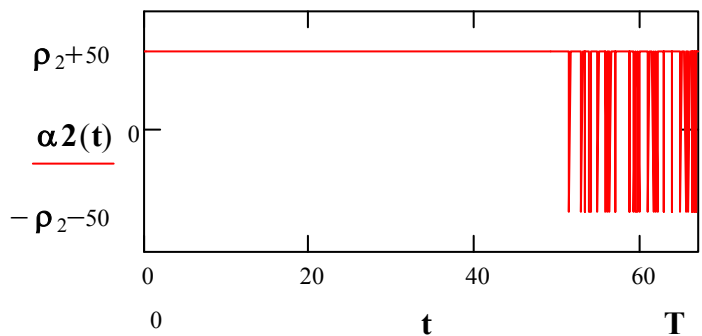

Optimal control of the second player.

Pic.4.3.9.6. $\kappa = 1, \rho_1 = 300, \rho_2 = 100, A = 10^4, \omega = 50,$

$\tau = 0.001.$

**Example 4.3.10.**

$\beta(t) = A\sin(\omega t); \kappa = 1, \rho_1 = 300, \rho_2 = 100, A = 10^4, \omega = 50, \tau = 10^{-9}.$

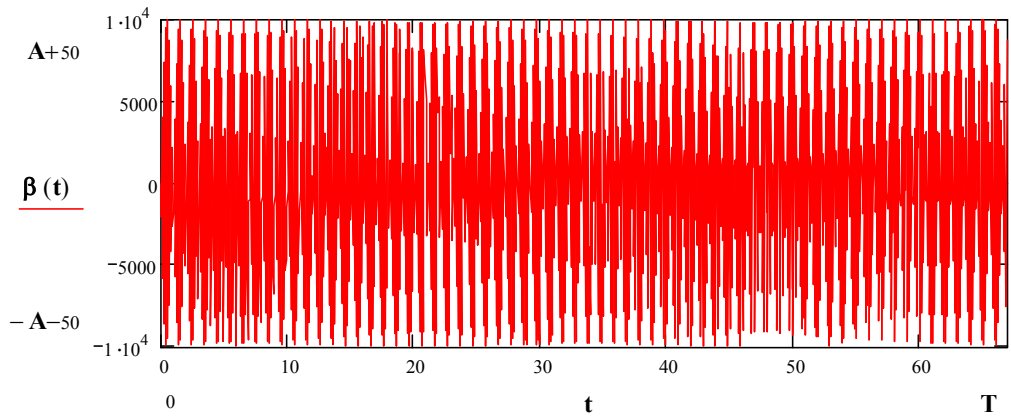

Uncertainty on an output the devices the measured of speed of rapproachement

Pic.4.3.10.1. Ancertainty : $\kappa = 1, \rho_1 = 300, \rho_2 = 100, A = 10^4, \omega = 50, \tau = 10^{-9}$.

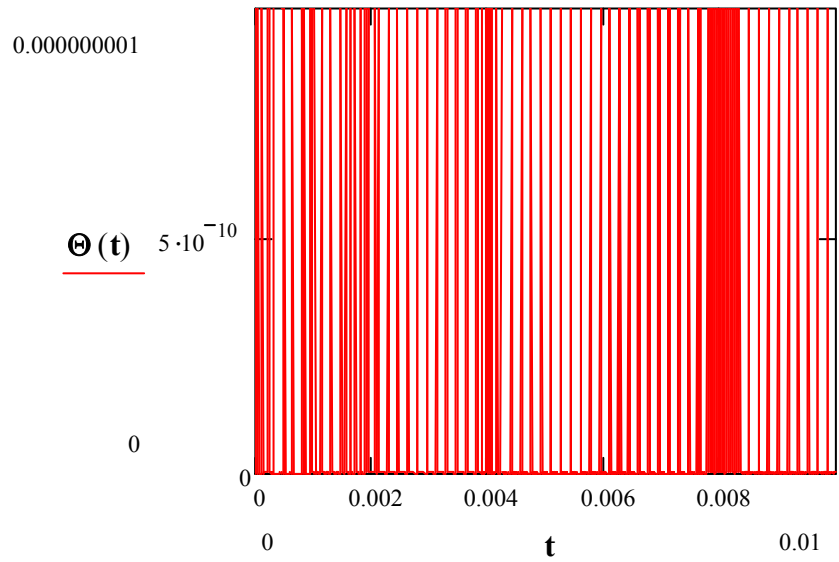

Cutting functuion

Pic.4.3.10.2. Cuttingfunction: $\Theta_\tau(t). \tau = 10^{-9}$.

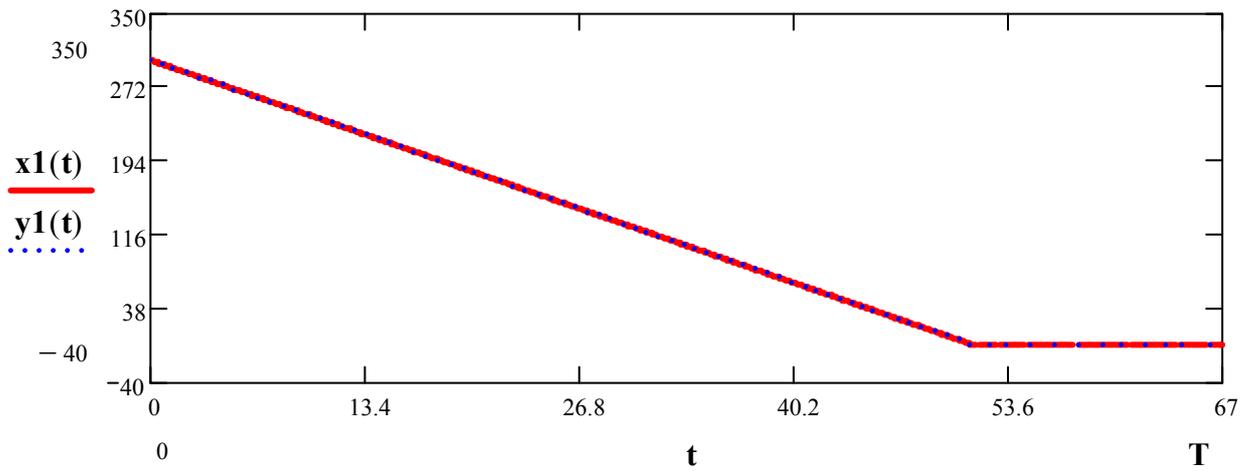

Optimal trajectory. Game with uncertainty: red curve. Classical game: blue curve.

Pic.4.3.10.3. $\kappa = 1, \rho_1 = 300, \rho_2 = 100, A = 10^4, \omega = 50, \tau = 10^{-9}$.
$x_1(T) = -1.369 \times 10^{-3}, y_1(T) = 7.43 \times 10^{-5}$.

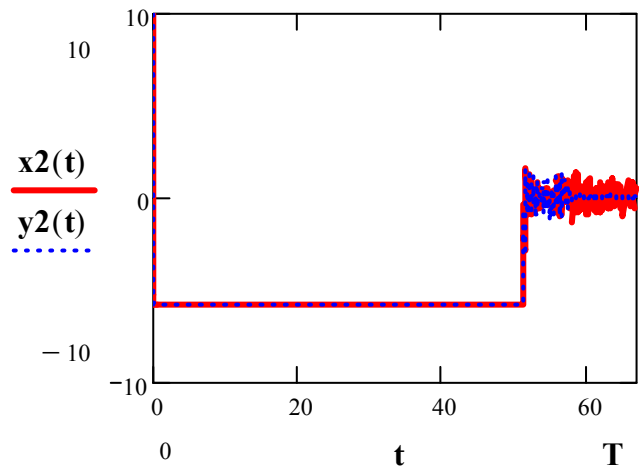

Optimal velocity

Pic.4.3.10.4. $\kappa = 1, \rho_1 = 300, \rho_2 = 100, A = 10^4, \omega = 50,$
$\tau = 10^{-9}$.

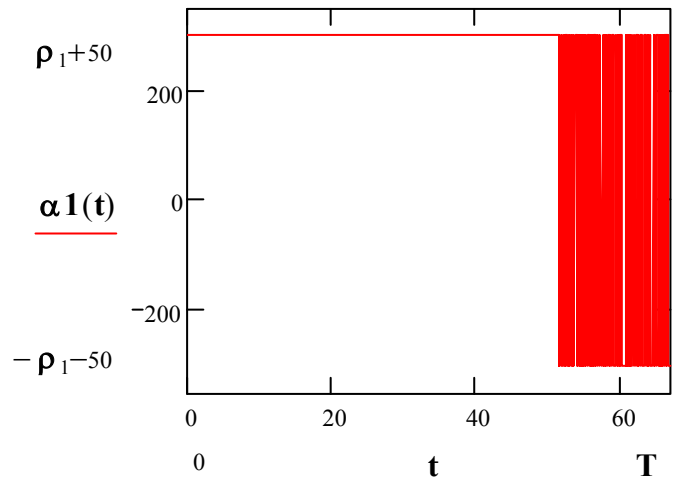

Optimal control for the first player.

Pic.4.3.10.5. $\kappa = 1, \rho_1 = 300, \rho_2 = 100, A = 10^4, \omega = 50,$
$\tau = 10^{-9}.$

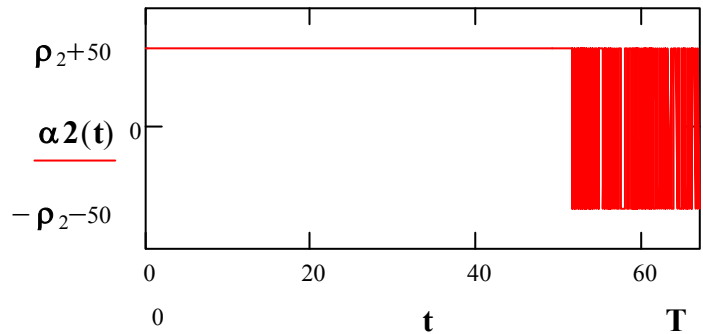

Optimal control of the second player.

Pic.4.10.6. $\kappa = 1, \rho_1 = 300, \rho_2 = 100, A = 10^4, \omega = 50,$
$\tau = 10^{-9}.$

**Numerical simulation. Example 4.3.11.**
$\beta(t) = A\sin(\omega t); \kappa = 1, \rho_1 = 300, \rho_2 = 100, A = 10^4, \omega = 50, \tau = 10^{-11}.$

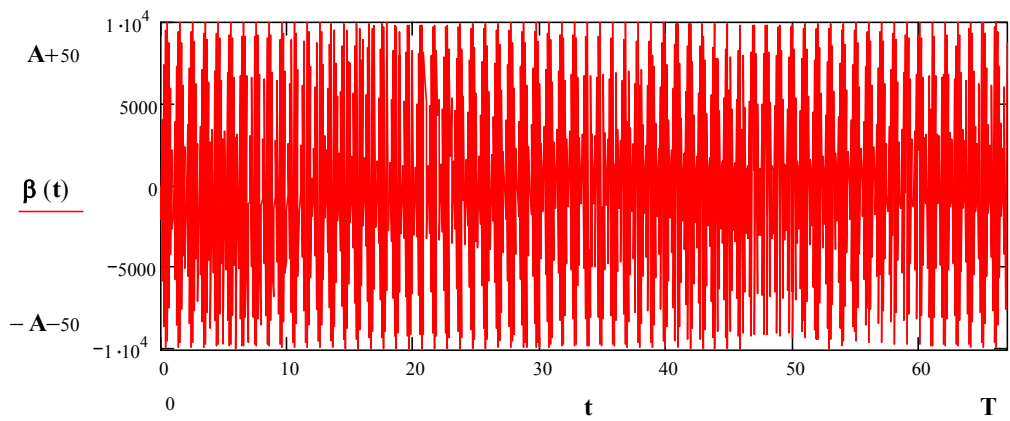

Uncertainty on an output the devices the measured of speed of rapproachement

Pic.4.3.11.1. $\kappa = 1, \rho_1 = 300, \rho_2 = 100, A = 10^4, \omega = 50, \tau = 10^{-11}$.

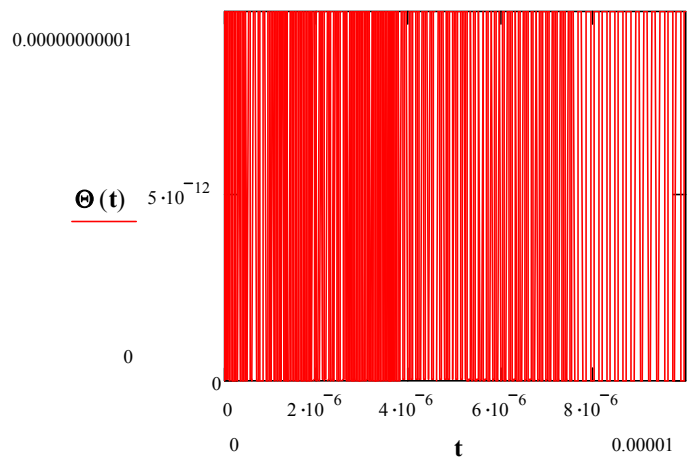

Cutting functuion

Pic.4.3.11.2. Cuttingfunction: $\Theta_\tau(t). \tau = 10^{-11}$.

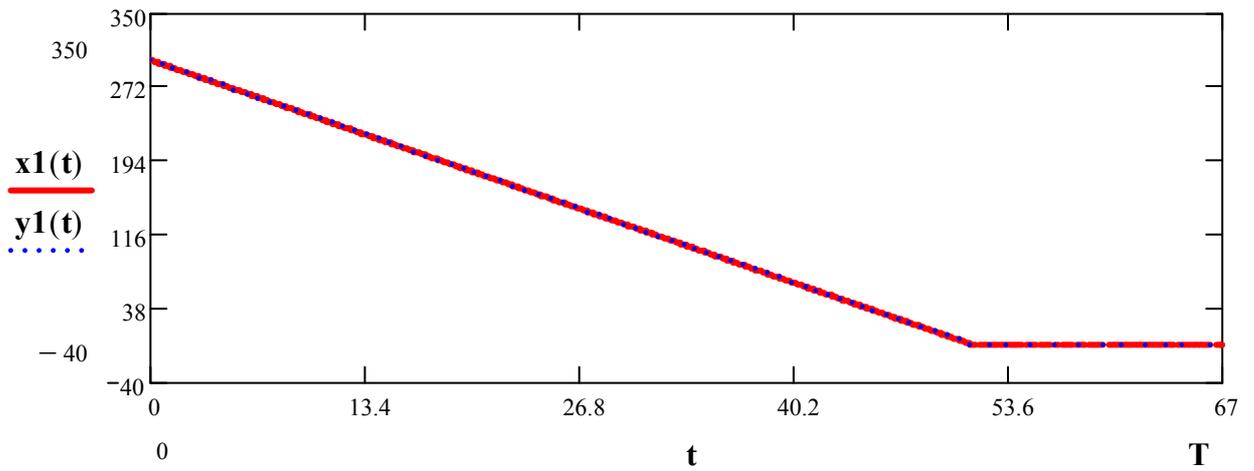

Optimal trajectory. Game with uncertainty: red curve. Classical game: blue curve.

Pic.4.3.11.3. $\kappa = 1, \rho_1 = 300, \rho_2 = 100, A = 10^4, \omega = 50, \tau = 10^{-11}$.

$x_1(T) = -8.2 \times 10^{-4}, y_1(T) = 7.4 \times 10^{-5}$.

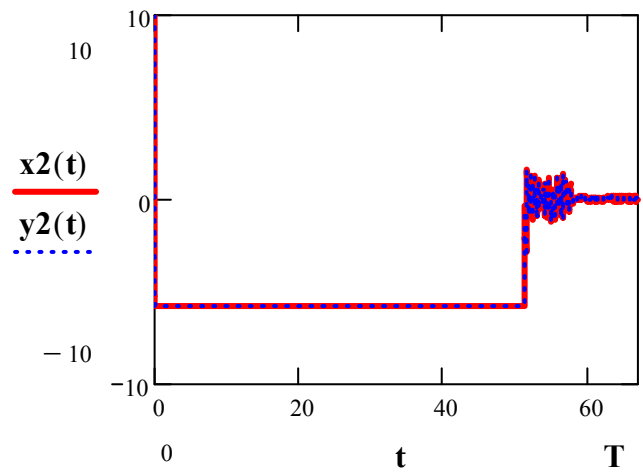

Optimal velocity

Pic.4.3.11.4. $\kappa = 1, \rho_1 = 300, \rho_2 = 100, A = 10^4, \omega = 50,$
$\tau = 10^{-11}$.

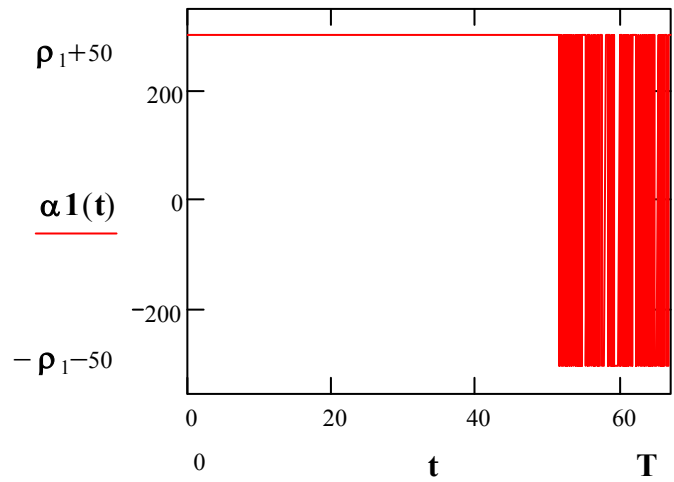

Optimal control for the first player.

Pic.4.3.11.5. $\kappa = 1, \rho_1 = 300, \rho_2 = 100, A = 10^4, \omega = 50,$
$\tau = 10^{-11}.$

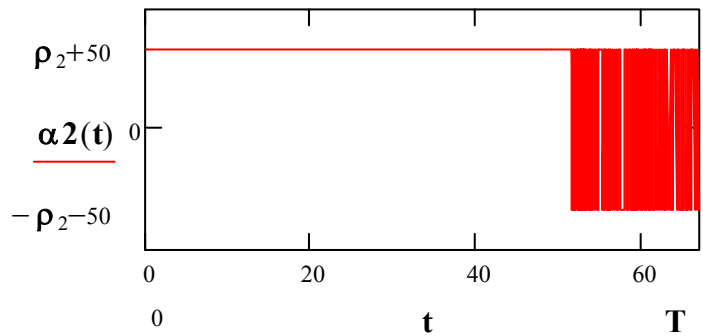

Optimal control of the second player.

Pic.4.3.11.6. $\kappa = 1, \rho_1 = 300, \rho_2 = 100, A = 10^4, \omega = 50,$
$\tau = 10^{-11}.$

**Example 4.3.2.**

Let us consider an 2-persons dissipative differential game $LDG^{\#}_{2;T}(f,0,M,0)$, with nonlinear dynamics and imperfect information about the system:

$$\dot{x}_1(t) = x_2(t),$$

$$\dot{x}_2(t) = -\kappa_1[x_2(t) + \beta(t)]^3 + \kappa_2[x_2(t) + \beta(t)]^2 + \alpha_1(t) + \alpha_2(t),$$

$$\kappa_1 > 0, \kappa_2 \in (-\infty, +\infty), \quad (4.3.13)$$

$$\alpha_1(t) \in [-\rho_1, \rho_1], \alpha_2(t) \in [-\rho_2, \rho_2]$$

$$\mathbf{J}_i = x_1^2(T), i = 1,2.$$

Thus optimal control problem for the first player:

$$\min_{\alpha_1(t)\in[-\rho_1,\rho_1]} \left( \max_{\alpha_2(t)\in[-\rho_2,\rho_2]} [x_1^2(T) + \dot{x}_1^2(T)] \right) \quad (4.3.14)$$

and optimal control problem for the second player:

$$\max_{\alpha_2(t)\in[-\rho_2,\rho_2]} \left( \min_{\alpha_1(t)\in[-\rho_1,\rho_1]} [x_1^2(T) + \dot{x}_1^2(T)] \right). \quad (4.3.15)$$

From Eqs.(3.1.15)-(3.1.16) we obtain linear master game for the optimal control problem (4.3.13)-(4.3.15):

$$\dot{u}_1 = u_2 + \lambda_2 + \beta(t),$$

$$\dot{u}_2 = \left[-3\kappa_1(\lambda_2 + \beta(t))^2 + 2\kappa_2(\lambda_2 + \beta(t))\right]u_2 -$$

$$-\kappa_1(\lambda_2 + \beta(t))^3 + \kappa_2(\lambda_2 + \beta(t))^2 + \check{\alpha}_1(t) + \check{\alpha}_2(t),$$

(4.3.16)

$$\kappa > 0,$$

$$\check{\alpha}_1(t) \in [-\rho_1, \rho_1], \check{\alpha}_2(t) \in [-\rho_2, \rho_2],$$

$$\mathbf{J}_i = u_1^2(T) + \dot{u}_1^2(T), i = 1, 2.$$

Thus optimal control problem for the first player:

$$\min_{\check{\alpha}_1(t) \in [-\rho_u, \rho_u]} \left( \max_{\check{\alpha}_2(t) \in [-\rho_v, \rho_v]} u_1^2(T) \right), \qquad (4.3.17)$$

and optimal control problem for the second player:

$$\max_{\check{\alpha}_2(t) \in [-\rho_v, \rho_v]} \left( \min_{\check{\alpha}_1(t) \in [-\rho_u, \rho_u]} u_1^2(T) \right). \qquad (4.3.18)$$

Thus from Eq.(A.26) (see Appendix A) we obtain standard solution for the linear optimal control problem (4.3.16)-(4.3.18):

$$\check{\alpha}_1(t) = -\rho_1 \mathbf{sign}[x_1(t) + (T-t)x_2(t)],$$

(4.3.19)

$$\check{\alpha}_2(t) = \rho_2 \mathbf{sign}[x_1(t) + (T-t)x_2(t)].$$

From Eq.(4.3.19) and Theorem 3.2, we obtain "step by step" feedback optimal control for the nonlinear optimal control problem (4.3.13)-(4.3.15). Thus "step-by-step" optimal control for the first player $\alpha_1^*(t)$ in the next form:

$$\alpha_1^*(t) = -\rho_1 \mathbf{sign}[x_1(t_n) + (t_{n+1} - t)\check{x}_2(t)],$$

$$\check{x}_2(t_n) = x_2(t_n) + \beta(t_n), \quad (4.3.20)$$

$$t \in [t_n, t_{n+1}], t_{n+1} - t_n = \frac{T}{N}, n = 1, \ldots, N.$$

and "step by step" optimal control for the second player $\alpha_2^*(t)$ in the next form:

$$\alpha_2^*(t) = \rho_2 \mathbf{sign}[x_1(t_n) + [(t_{n+1} - t)]\check{x}_2(t)],$$

$$\check{x}_2(t_n) = x_2(t_n) + \beta(t_n), \quad (4.3.21)$$

Suppose that

$$t \in [t_n, t_{n+1}], t_{n+1} - t_n = \frac{T}{N}, n = 1, \ldots, N.$$

$(t_{n+1} - t_n) \ll 1, t \in [t_n, t_{n+1}], n = 1, \ldots, N$, from Eq.(A.26) we obtain optimal control $\alpha_1^*(t)$ for the first player and optimal control $\alpha_2^*(t)$ for the second player in the next form:

$$\alpha_1^*(t) \simeq -\rho_1 \mathbf{sign}[x_1(t) + (t_{n+1} - t)x_2(t)],$$

$$(4.3.22)$$

$$\alpha_2^*(t) \simeq \rho_2 \mathbf{sign}[x_1(t) + (t_{n+1} - t)x_2(t)].$$

From Eq.(4.3.22) we obtain optimal control $\alpha_1^*(t)$ for the first player and optimal control $\alpha_2^*(t)$ for the second player:

$$\alpha_1^*(t) \simeq -\rho_1 \mathbf{sign}[x_1(t) + \Theta_\tau(t)(x_2(t) + \beta(t))],$$

$$(4.3.23)$$

$$\alpha_2^*(t) \simeq \rho_2 \mathbf{sign}[x_1(t) + \Theta_\tau(t)(x_2(t) + \beta(t))].$$

Thus for the numerical simulation we obtain ODE:

$$\dot{x}_1(t) = x_2(t),$$

$$\dot{x}_2(t) = -\kappa_1 x_2^3(t) + \kappa_2 x_2^2(t) - \rho_1 \cdot \textbf{sign}[x_1(t) + \Theta_\tau(t)(x_2(t) + \beta(t))] +$$

$$+\rho_2 \cdot \textbf{sign}[x_1(t) + \Theta_\tau(t) x_2(t)],$$

(4.3.24)

$$\kappa_1 > 0.$$

**Numerical simulation. Example 4.3.13.**
$\beta(t) = A\sin(\omega t); \kappa_1 = 1, \kappa_2 = 50, \rho_1 = 30000, \rho_2 = 10000, A = 10^4, \omega = 50, \tau = 10^{-11}.$

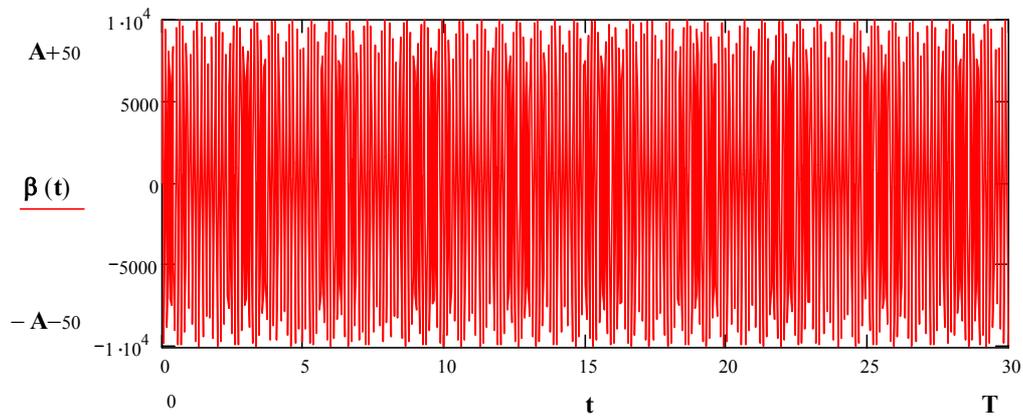

Uncertainty on an output the devices the measured of speed of rapproachement

Pic.4.3.13.1. Uncertainty $\beta(t) = A\sin(\omega t); \kappa = 1, \kappa_2 = 50, \rho_1 = 30000, \rho_2 = 10000,$
$A = 10^4, \omega = 50, \tau = 10^{-11}.$

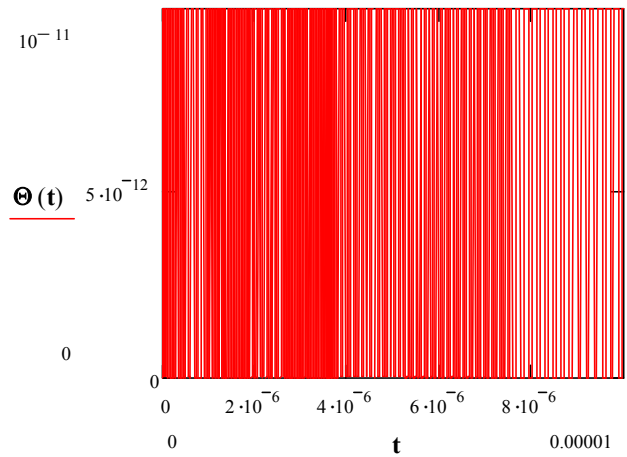

Cutting functuion

Pic.4.3.13.3.Cutting function $\Theta_\tau(t).\tau = 10^{-11}$.

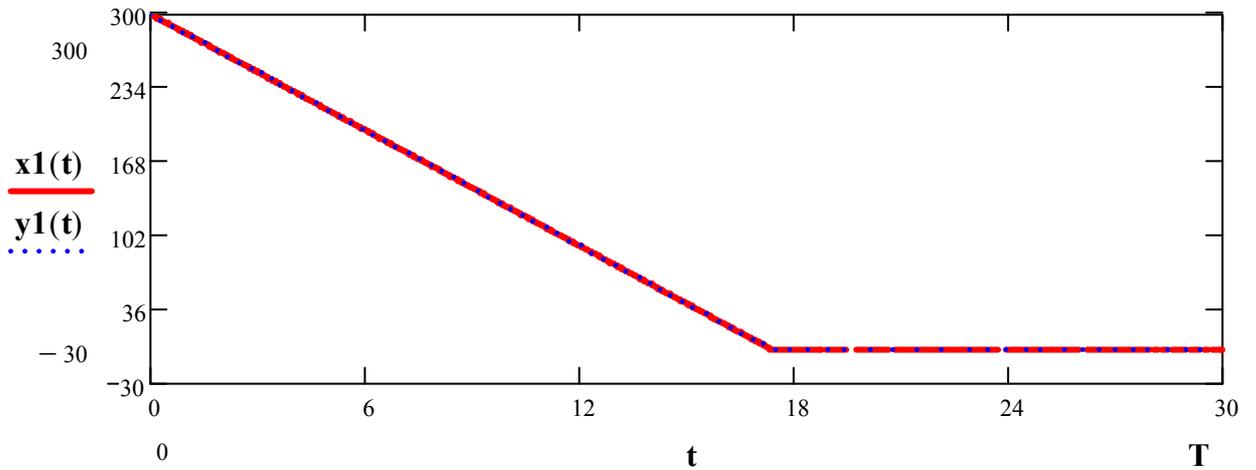

Optimal trajectory. Game with uncertainty: red curve. Classical game: blue curve.

Pic.4.3.13.3.$\kappa_1 = 1, \kappa_2 = 50, \rho_1 = 30000, \rho_2 = 10000, A = 10^4, \omega = 50, \tau = 10^{-11}$.
$x_1(T) = -4.6 \cdot 10^{-4}, y_2(T) = 6.76 \cdot 10^{-4}$.

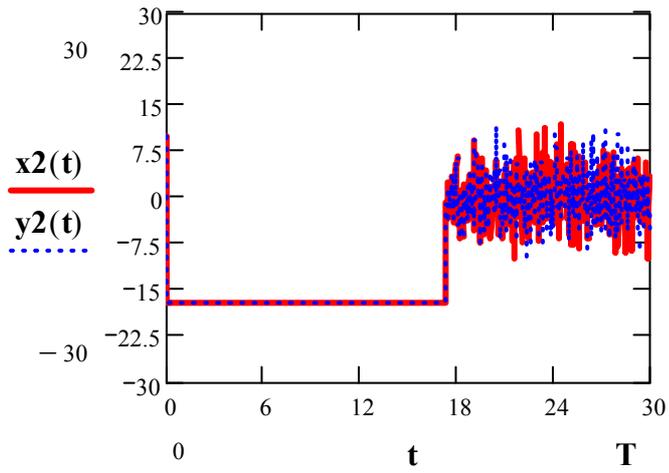

Optimal velocity

Pic.4.3.13.4. $\kappa_1 = 1, \kappa_2 = 50, \rho_1 = 30000, \rho_2 = 10000, A = 10^4, \omega = 50, \tau = 10^{-11}$

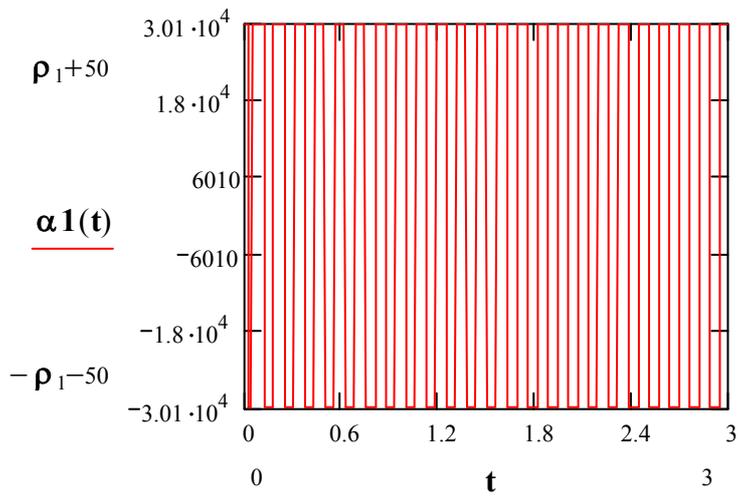

Optimal control for the first player.

Pic.4.3.13.5. $\kappa_1 = 1, \kappa_2 = 50, \rho_1 = 30000, \rho_2 = 10000, A = 10^4, \omega = 50, \tau = 10^{-11}$

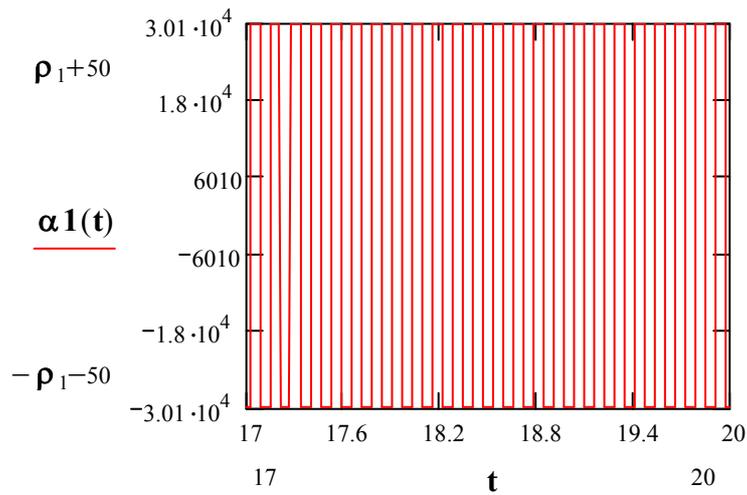

Optimal control for the first player.

Pic.4.3.13.6. $\kappa_1 = 1, \kappa_2 = 50, \rho_1 = 30000, \rho_2 = 10000, A = 10^4, \omega = 50, \tau = 10^{-11}$

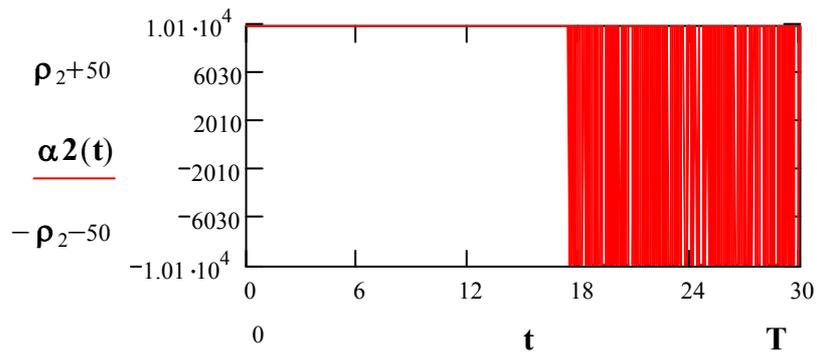

Optimal control of the second player.

Pic.4.3.13.7. $\kappa_1 = 1, \kappa_2 = 50, \rho_1 = 30000, \rho_2 = 10000, A = 10^4, \omega = 50, \tau = 10^{-11}$

**Numerical simulation. Example 4.3.14.**

$\beta(t) = A\sin(\omega t); \kappa_1 = 1, \kappa_2 = 50, \rho_1 = 30000, \rho_2 = 10000, A = 10^4, \omega = 500, \tau = 10^{-11}$

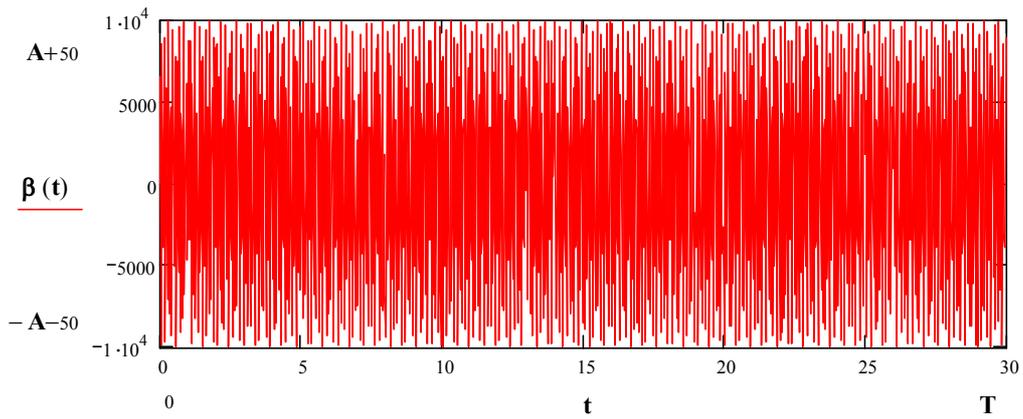

Uncertainty on an output the devices the measured of speed of rapproachement

Pic.4.3.14.1. Uncertainty $\beta(t) = A\sin(\omega t); \kappa_1 = 1, \kappa_2 = 50, \rho_1 = 30000, \rho_2 = 10000,$
$A = 10^4, \omega = 500, \tau = 10^{-11}.$
$x_1(T) = 4.5 \cdot 10^{-4}, y_1(T) = -6.76 \cdot 10^{-4}.$

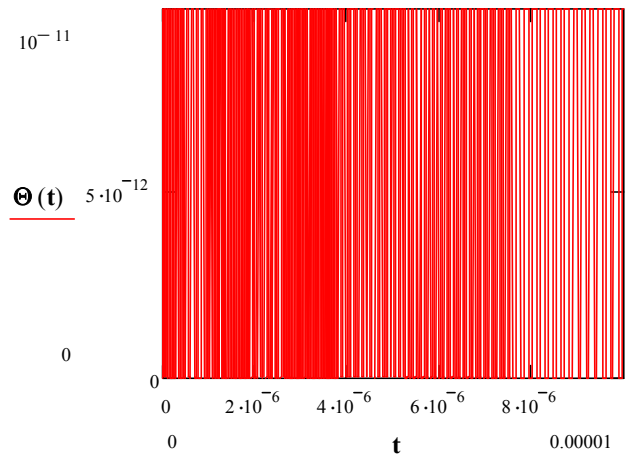

Cutting functuion

Pic.4.3.14.2. Cutting function $\Theta_\tau(t). \tau = 10^{-11}.$

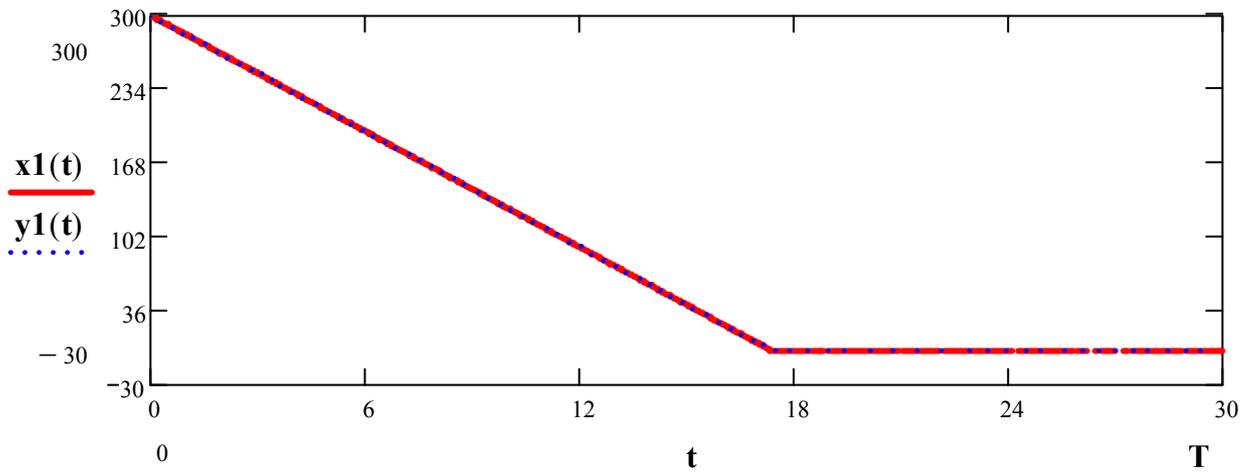

Optimal trajectory. Game with uncertainty: red curve. Classical game: blue curve.

Pic.4.3.14.3. $\kappa_1 = 1, \kappa_2 = 50, \rho_1 = 30000, \rho_2 = 10000,$
$A = 10^4, \omega = 500, \tau = 10^{-11}.$

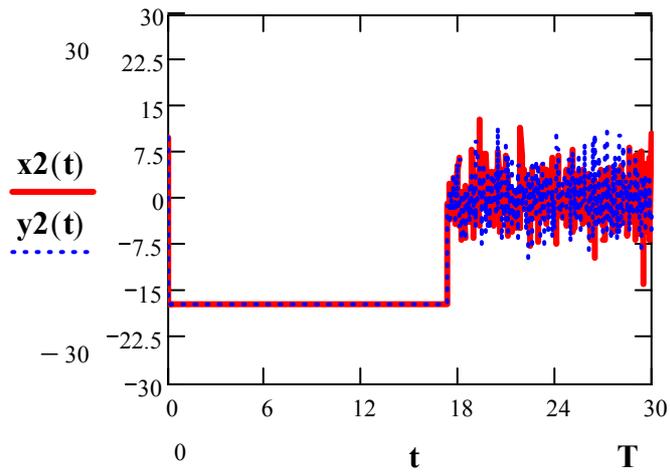

Optimal velocity

Pic.4.3.14.4. $\kappa_1 = 1, \kappa_2 = 50, \rho_1 = 30000, \rho_2 = 10000,$
$A = 10^4, \omega = 500, \tau = 10^{-11}$

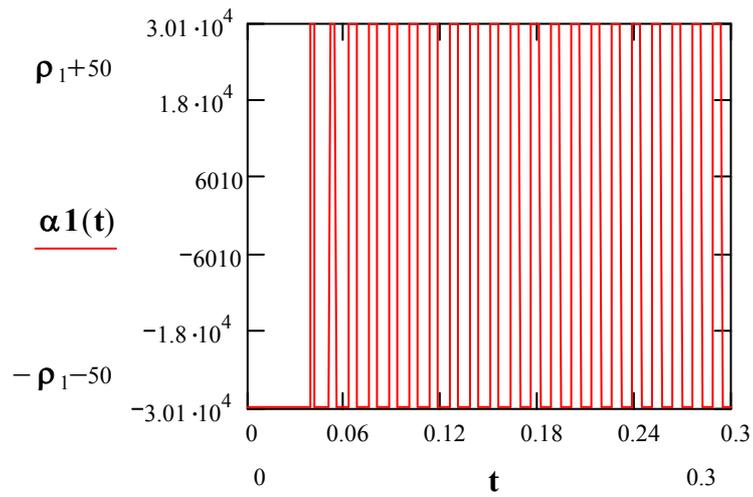

Optimal control for the first player.

Pic.4.3.14.5. $\kappa_1 = 1, \kappa_2 = 50, \rho_1 = 30000, \rho_2 = 10000, A = 10^4, \omega = 500, \tau = 10^{-11}$

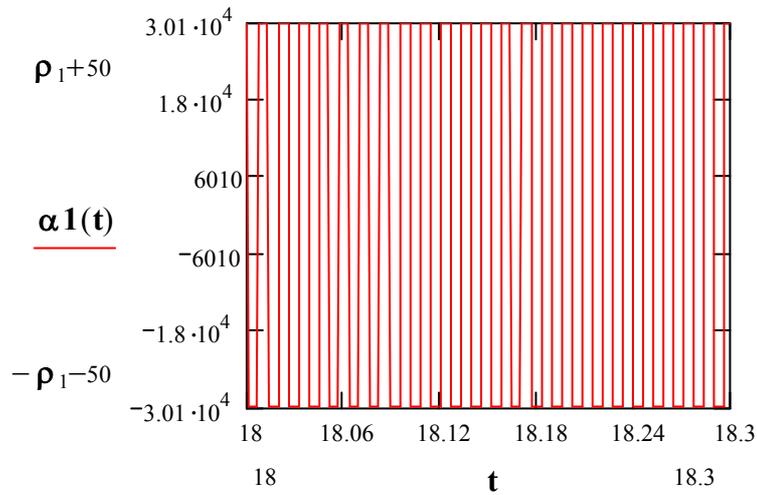

Optimal control for the first player.

Pic.4.3.14.6. $\kappa_1 = 1, \kappa_2 = 50, \rho_1 = 30000, \rho_2 = 10000, A = 10^4, \omega = 500, \tau = 10^{-11}$

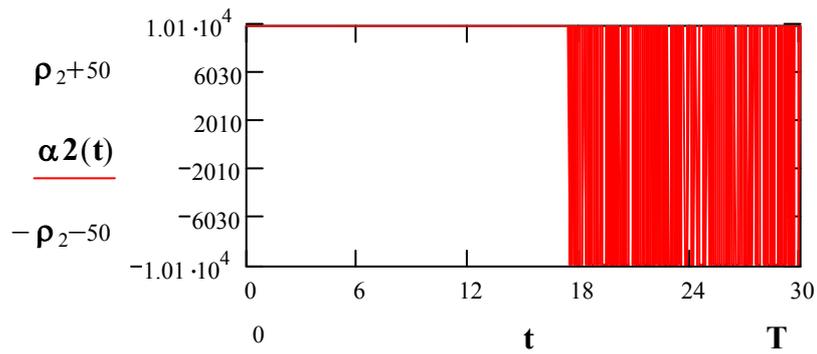

Optimal control of the second player.

Pic.4.3.14.7. $\kappa_1 = 1, \kappa_2 = 50, \rho_1 = 30000, \rho_2 = 10000, A = 10^4, \omega = 500, \tau = 10^{-11}$

**Numerical simulation. Example 4.3.15.**

$\beta(t) = A\sin(\omega t); \kappa_1 = 1, \kappa_2 = 50, \rho_1 = 30000, \rho_2 = 10000, A = 10^4, \omega = 5000, \tau = 10^{-11}$

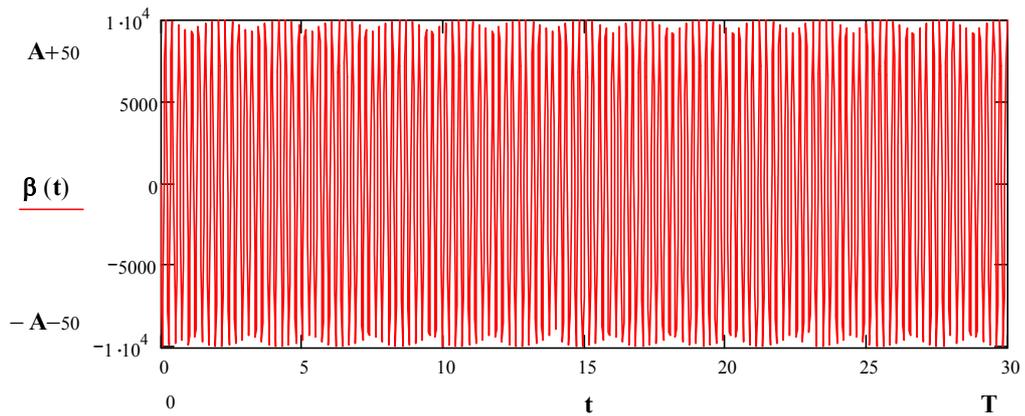

Uncertainty on an output the devices the measured of speed of rapproachement

Pic.4.3.15.1. Uncertainty $\beta(t) = A\sin(\omega t); \kappa_1 = 1, \kappa_2 = 50, \rho_1 = 30000, \rho_2 = 10000$
$A = 10^4, \omega = 5000, \tau = 10^{-11}$.

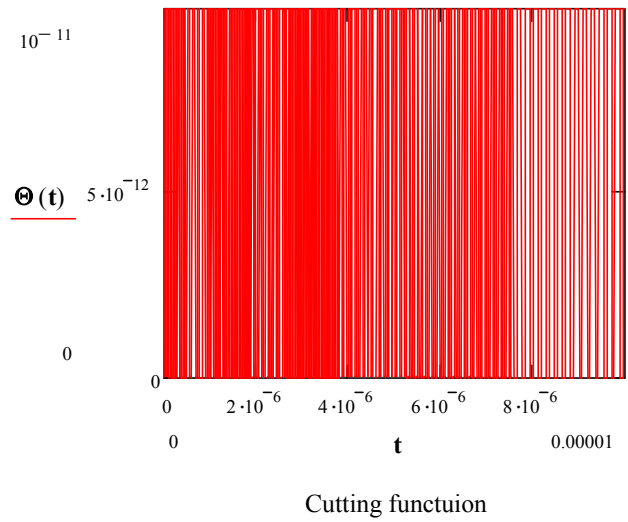

Cutting functuion

Pic.4.3.15.2. Cutting function $\Theta_\tau(t).\tau = 10^{-11}$.

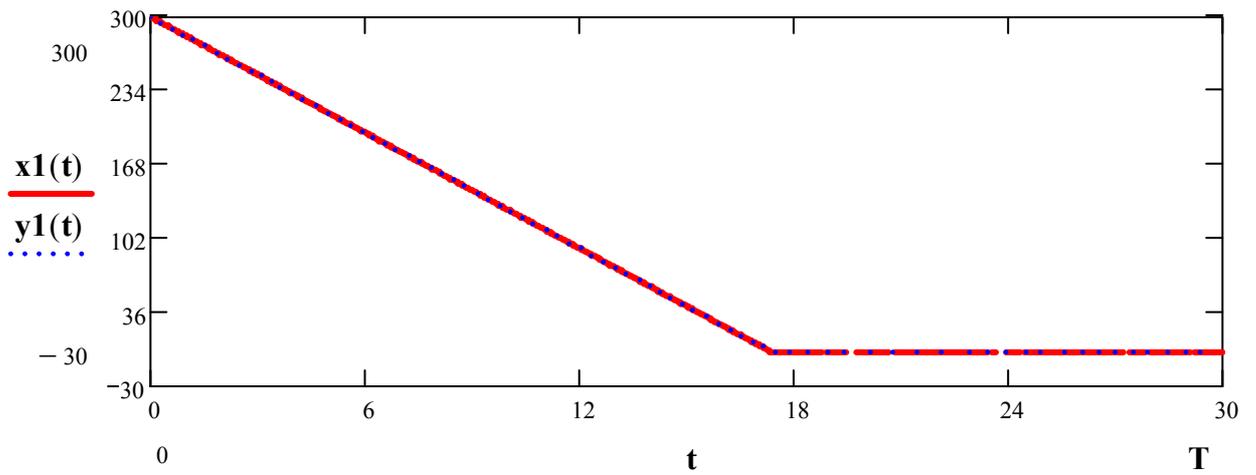

Optimal trajectory. Game with uncertainty: red curve. Classical game: blue curve.

Pic.4.3.15.3. $\kappa_1 = 1, \kappa_2 = 50, \rho_1 = 30000, \rho_2 = 10000, A = 10^4, \omega = 5000, \tau = 10^{-11}$.
$x_1(T) = 6.7 \cdot 10^{-4}, y_1(T) = -6.763 \cdot 10^{-4}$.

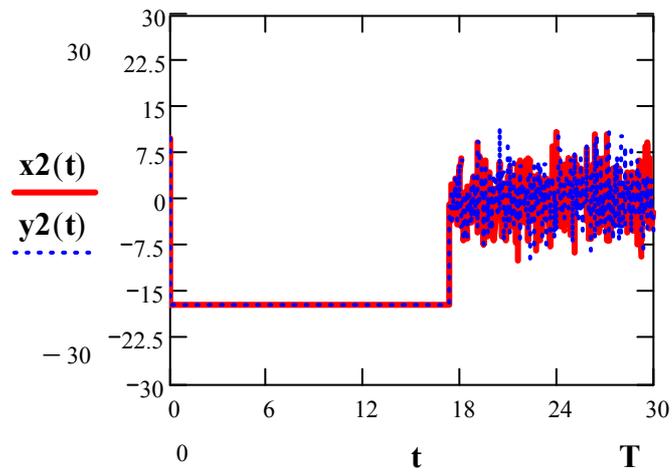

Optimal velocity

Pic.4.3.15.4. $\kappa_1 = 1, \kappa_2 = 50, \rho_1 = 30000, \rho_2 = 10000, A = 10^4, \omega = 5000, \tau = 10^{-11}$

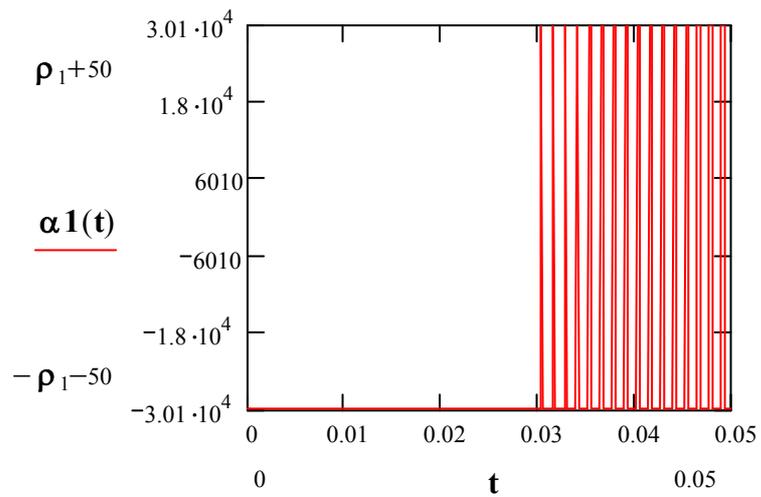

Optimal control for the first player.

Pic.4.3.15.5. $\kappa_1 = 1, \kappa_2 = 50, \rho_1 = 30000, \rho_2 = 10000, A = 10^4, \omega = 5000, \tau = 10^{-11}$

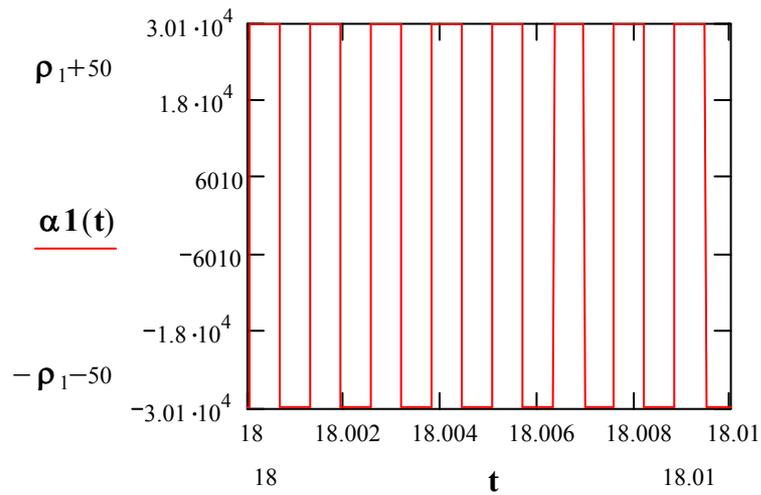

Optimal control for the first player.

Pic.4.3.15.6. $\kappa_1 = 1, \kappa_2 = 50, \rho_1 = 30000, \rho_2 = 10000, A = 10^4, \omega = 5000, \tau = 10^{-11}$

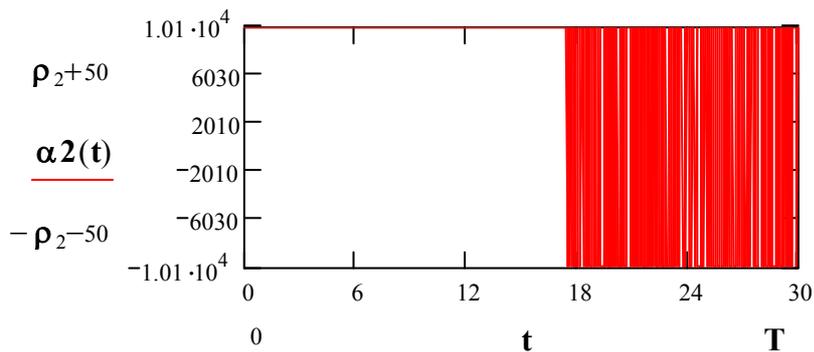

Optimal control of the second player.

Pic.4.3.15.7. $\kappa_1 = 1, \kappa_2 = 50, \rho_1 = 30000, \rho_2 = 10000, A = 10^4, \omega = 5000, \tau = 10^{-11}$

**Numerical simulation. Example 4.3.16.**

$\beta(t) = A\sin(\omega t^2); \kappa_1 = 1, \kappa_2 = 50, \rho_1 = 30000, \rho_2 = 10000, A = 10^4, \omega = 500, \tau = 10^{-11}$

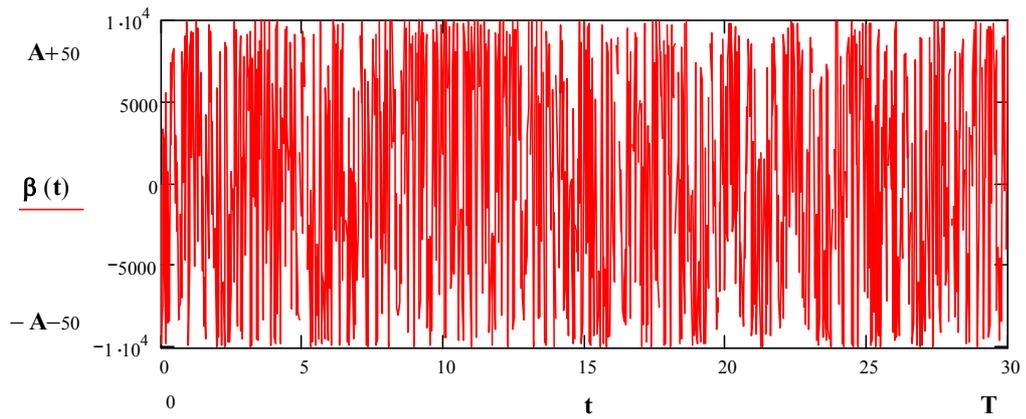

Uncertainty on an output the devices the measured of speed of rapproachement

Pic.4.3.16.1. Uncertainty: $\beta(t) = A\sin(\omega t^2); \kappa_1 = 1, \kappa_2 = 50, \rho_1 = 30000, \rho_2 = 10000,$
$A = 10^4, \omega = 50000, \tau = 10^{-11}.$

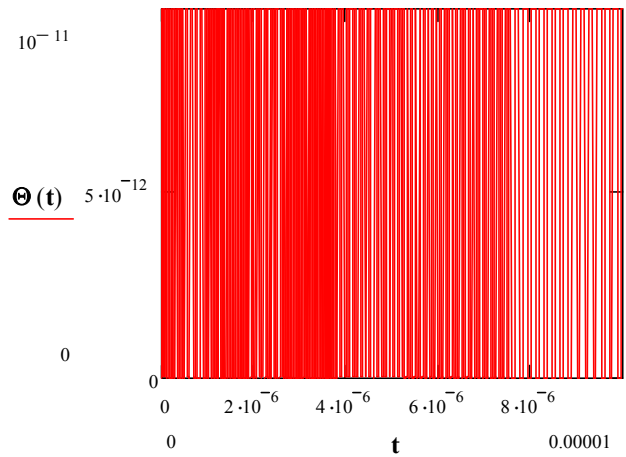

Cutting functuion

Pic.4.3.16.2. Cutting function $\Theta_\tau(t). \tau = 10^{-11}.$

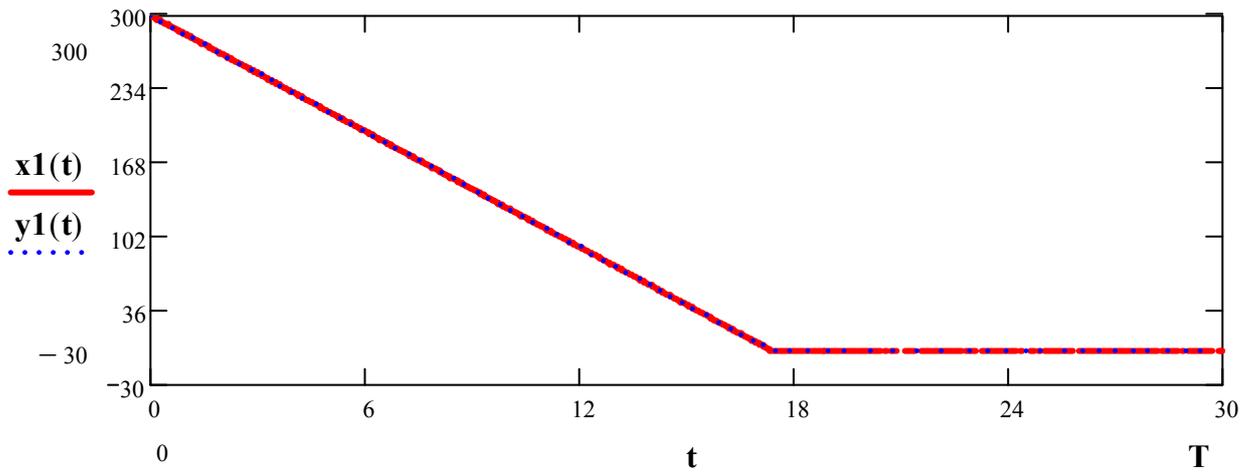

Optimal trajectory. Game with uncertainty: red curve. Classical game: blue curve.

Pic.4.3.16.3. $\kappa_1 = 1, \kappa_2 = 50, \rho_1 = 30000, \rho_2 = 10000, A = 10^4, \omega = 50000, \tau = 10^{-11}$.
$x_1(T) = -4.1 \cdot 10^{-4}, y_1(T) = -6.76 \cdot 10^{-4}$.

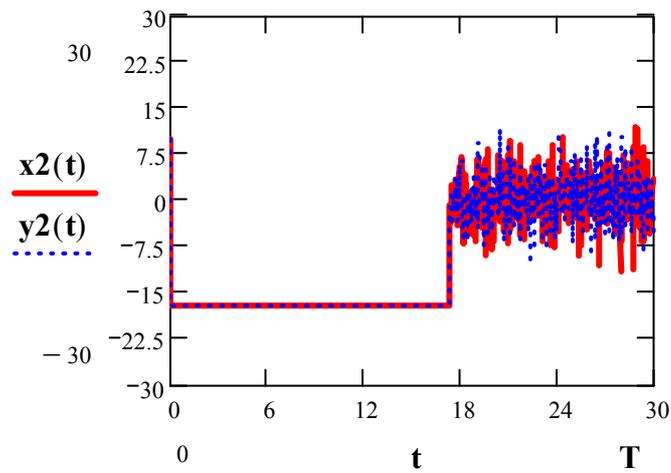

Optimal velocity

Pic.4.3.16.4. $\kappa_1 = 1, \kappa_2 = 50, \rho_1 = 30000, \rho_2 = 10000, A = 10^4, \omega = 50000, \tau = 10^{-11}$

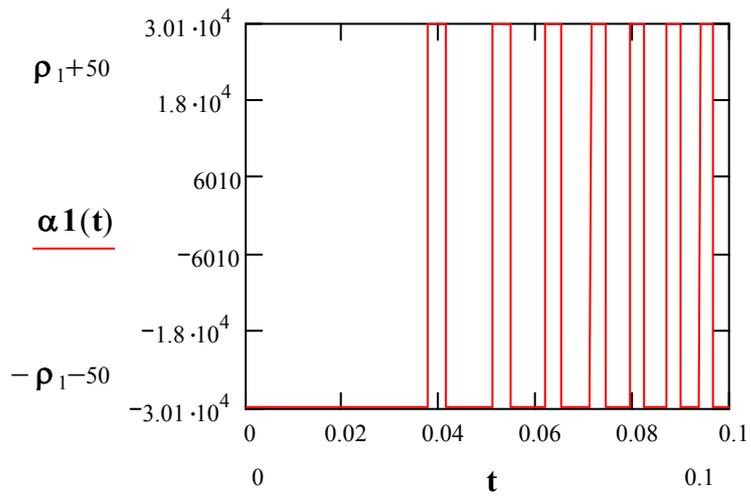

Optimal control for the first player.

Pic.4.3.16.5. $\kappa_1 = 1, \kappa_2 = 50, \rho_1 = 30000, \rho_2 = 10000, A = 10^4, \omega = 50000, \tau = 10^{-11}$

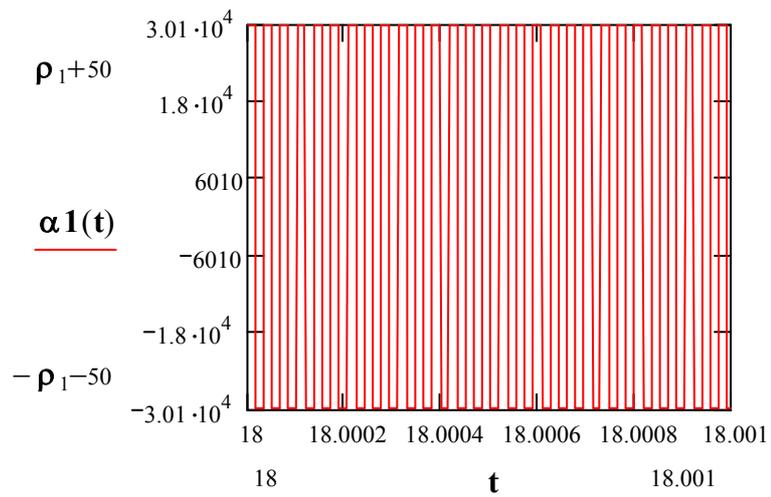

Optimal control for the first player.

Pic.4.3.16.6. $\kappa_1 = 1, \kappa_2 = 50, \rho_1 = 30000, \rho_2 = 10000, A = 10^4, \omega = 500, \tau = 10^{-11}$

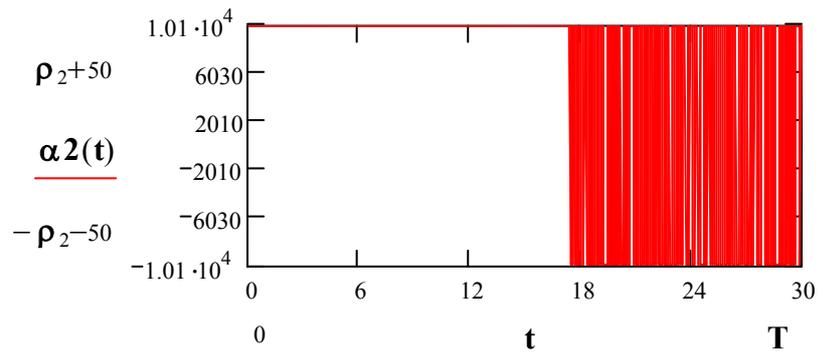

Optimal control of the second player.

Pic.4.3.16.7.$\kappa_1 = 1, \kappa_2 = 50, \rho_1 = 30000, \rho_2 = 10000, A = 10^4, \omega = 50000, \tau = 10^{-11}$

Pic.4.3.16.1.Uncertainty $\beta(t) = A\sin(\omega t^2); \kappa_1 = 1, \kappa_2 = 50, \rho_1 = 30000, \rho_2 = 10000,$
$A = 10^4, \omega = 50000, \tau = 10^{-11}$.

Pic.4.3.16.2.Cutting function $\Theta_\tau(t). \tau = 10^{-11}$.

Pic.4.3.16.3.$\kappa_1 = 1, \kappa_2 = 50, \rho_1 = 30000, \rho_2 = 10000, A = 10^4, \omega = 50000, \tau = 10^{-11}$.
$x_1(T) = -4.1 \cdot 10^{-4}, y_1(T) = -6.76 \cdot 10^{-4}$.

Pic.4.3.16.4.$\kappa_1 = 1, \kappa_2 = 50, \rho_1 = 30000, \rho_2 = 10000, A = 10^4, \omega = 50000, \tau = 10^{-11}$

## IV.3.2.Optimal control numerical simulation. 2-Persons antagonistic differential game $IDG_{2;T}(f, 0, M, 0, \beta, w),$ with non-linear dynamics and imperfect measurements and imperfect information about the system. "Step-by-step" strategy.Homing missile guidance Laws using "Step-by-step" strategy.

Taking the origin of the reference frame to be the instantaneous

position of the missile, the equation of motion in polar form are:

$$\ddot{R} = R\dot{\sigma}^2 + a_M^r\left[t, \widetilde{\mathbf{R}}(t), \dot{\widetilde{\mathbf{R}}}(t)\right] + a_T^r(t),$$

$$\breve{a}_M^r(t) \in [-\bar{a}_M^r, \bar{a}_M^r], a_T^r(t) \in [-\bar{a}_T^r, \bar{a}_T^r].$$
(4.3.24)

$$R\ddot{\sigma} + 2\dot{R}\dot{\sigma} = a_M^\tau\left[t, \widetilde{\sigma}(t), \dot{\widetilde{\sigma}}(t)\right] + a_T^\tau(t),$$

$$\breve{a}_M^\tau(t) \in [-\bar{a}_M^\tau, \bar{a}_M^\tau], a_T^\tau(t) \in [-\bar{a}_T^\tau, \bar{a}_T^\tau]$$

1. The variable $R = R(t)$ denotes the it is *true* target-to-missle range $R_{TM}(t)$.
2. The variable $\widetilde{\mathbf{R}} = \widetilde{\mathbf{R}}(t)$ denotes the it is *real measured* target-to-missile range $R_{TM}(t)$.
3. The variable $\sigma = \sigma(t)$ denotes the it is *true* line-of-sight angle (**LOST**) i.e.,the it is *true* angle between the constant reference direction and target-to-missile direction.
4. The variable $\widetilde{\sigma} = \widetilde{\sigma}(t)$ denotes the it is *real measured* line-of-sight angle (**LOSM**) i.e.,the it is *true* angle between the constant reference direction and target-to-missile direction.
5. The variable $\breve{a}_M^r(t) = a_M^r\left[t, \widetilde{\mathbf{R}}(t), \dot{\widetilde{\mathbf{R}}}(t)\right]$ denotes the missiles tangent acceleration,i.e.missile acceleration along direction which perpendicularly to line-of-sight direction.
6. The variable $\breve{a}_M^\tau(t) = a_M^\tau\left[t, \widetilde{\sigma}(t), \dot{\widetilde{\sigma}}(t)\right]$ denotes the missile acceleration along target-to-missile direction.
7. The variable $a_T^\tau(t)$ denotes the target tangent acceleration.
8. The variable $a_T^r(t)$ denotes the target acceleration along target-to-missile direction.

Using the replacement $z = R\dot{\sigma}$ into Eq.(4.3.24) one obtain:

$$\dot{\sigma} = \frac{z}{R},$$

$$\dot{z} = \dot{R}\dot{\sigma} + R\ddot{\sigma}, R\ddot{\sigma} = \dot{z} - \dot{R}\dot{\sigma}, \tag{4.3.25}$$

$$R\ddot{\sigma} + 2\dot{R}\dot{\sigma} = \dot{z} + \dot{R}\dot{\sigma} = \dot{z} + \frac{\dot{R}z}{R}.$$

Thus

$$\ddot{R} = \frac{z^2}{R} + a_M^r\left[t, \widetilde{\mathbf{R}}(t), \widetilde{\dot{\mathbf{R}}}(t)\right] + a_T^r(t),$$

$$\check{a}_M^r(t) \in [-\bar{a}_M^r, \bar{a}_M^r], a_T^r(t) \in [-\bar{a}_T^r, \bar{a}_T^r].$$

$$\dot{z} = -\frac{\dot{R}z}{R} + a_M^\tau\left[t, \mathbf{Z}(t), \widetilde{\mathbf{Z}}(t)\right] + a_T^\tau(t), \tag{4.3.26}$$

$$\check{a}_M^\tau(t) \in [-\bar{a}_M^\tau, \bar{a}_M^\tau], a_T^\tau(t) \in [-\bar{a}_T^\tau, \bar{a}_T^\tau].$$

$$\widetilde{z}(t) = \widetilde{R}(t)\widetilde{\dot{\sigma}}(t),$$

$$\widetilde{\dot{z}}(t) = \widetilde{\dot{R}}(t)\widetilde{\dot{\sigma}}(t) + \widetilde{R}(t)\widetilde{\ddot{\sigma}}(t).$$

Suppose that:

$$\widetilde{\mathbf{R}}(t) = R(t) + \beta_1(t),$$

$$\widetilde{\sigma}(t) = \sigma(t) + \beta_2(t). \tag{4.3.27}$$

Thus

$$\widetilde{\mathbf{R}}(t) = \dot{R}(t) + \dot{\beta}_1(t) = \dot{R}(t) + \bar{\boldsymbol{\beta}}_1(t), \bar{\boldsymbol{\beta}}_1(t) \triangleq \dot{\beta}_1(t).$$

$$\widetilde{\boldsymbol{\sigma}}(t) = \dot{\sigma}(t) + \dot{\beta}_2(t), \widetilde{\boldsymbol{\sigma}}(t) = \ddot{\sigma}(t) + \ddot{\beta}_2(t).$$

$$\mathbf{Z}(t) = \widetilde{\mathbf{R}}(t)\widetilde{\boldsymbol{\sigma}}(t) = (R(t) + \beta_1(t))(\dot{\sigma}(t) + \dot{\beta}_2(t)) =$$

$$= R(t)\dot{\sigma}(t) + [\beta_1(t)(\dot{\sigma}(t) + \dot{\beta}_2(t)) + R(t)\dot{\beta}_2(t)] \approx$$

$$\approx z(t) + \left[\beta_1(t)\left(\widetilde{\boldsymbol{\sigma}}(t) + \dot{\beta}_2(t)\right) + \widetilde{\mathbf{R}}(t)\dot{\beta}_2(t)\right] =$$

$$= z(t) + \bar{\boldsymbol{\beta}}_2(t),$$

$$\bar{\boldsymbol{\beta}}_2(t) \triangleq \beta_1(t)\left(\widetilde{\boldsymbol{\sigma}}(t) + \dot{\beta}_2(t)\right) + \widetilde{\mathbf{R}}(t)\dot{\beta}_2(t).$$

(4.3.28)

$$\widetilde{\mathbf{Z}}(t) = \dot{\widetilde{\mathbf{R}}}(t)\widetilde{\boldsymbol{\sigma}}(t) + \widetilde{\mathbf{R}}(t)\widetilde{\boldsymbol{\sigma}}(t) =$$

$$(\dot{R}(t) + \dot{\beta}_1(t))(\dot{\sigma}(t) + \dot{\beta}_2(t)) + (R(t) + \beta_1(t))(\ddot{\sigma}(t) + \ddot{\beta}_2(t)) =$$

$$= [\dot{R}(t)\dot{\sigma}(t) + R(t)\ddot{\sigma}(t)] + \dot{\beta}_1(t)(\dot{\sigma}(t) + \dot{\beta}_2(t)) +$$

$$+\dot{R}(t)\dot{\beta}_2(t) + R(t)\ddot{\beta}_2(t) + \beta_1(t)(\ddot{\sigma}(t) + \ddot{\beta}_2(t)) \approx$$

$$\approx \dot{z}(t) + \dot{\widetilde{\mathbf{R}}}(t)\dot{\beta}_2(t) + \widetilde{\mathbf{R}}(t)\ddot{\beta}_2(t) + \beta_1(t)\left(\widetilde{\boldsymbol{\sigma}}(t) + \ddot{\beta}_2(t)\right) =$$

$$\dot{z}(t) + \bar{\boldsymbol{\beta}}_3(t),$$

$$\bar{\boldsymbol{\beta}}_3(t) \triangleq \dot{\widetilde{\mathbf{R}}}(t)\dot{\beta}_2(t) + \widetilde{\mathbf{R}}(t)\ddot{\beta}_2(t) + \beta_1(t)\left(\widetilde{\boldsymbol{\sigma}}(t) + \ddot{\beta}_2(t)\right).$$

Using the replacement $\dot{R} = V_r$ into Eq.(4.3.26) one obtain:

$$\dot{R} = V_r,$$

$$\dot{V}_r = \frac{z^2}{R} + a_M^r\left[t, \widetilde{\mathbf{R}}(t), \widetilde{V}_r(t)\right] + a_T^r(t),$$

$$\widetilde{V}_r(t) = \dot{\widetilde{\mathbf{R}}}(t) = V_r(t) + \bar{\boldsymbol{\beta}}_1(t),$$

$$a_M^r(t) \in [-\bar{a}_M^r, \bar{a}_M^r], a_T^r(t) \in [-\bar{a}_T^r, \bar{a}_T^r].$$

$$\dot{z} = -\frac{V_r z}{R} + a_M^\tau\left[t, \mathbf{Z}(t), \widetilde{\mathbf{Z}}(t)\right] + a_T^\tau(t),$$

$$a_M^\tau(t) \in [-\bar{a}_M^\tau, \bar{a}_M^\tau], a_T^\tau(t) \in [-\bar{a}_T^\tau, \bar{a}_T^\tau].$$

(4.3.29)

Let as considered antagonistic differential game $IDG_{2;T}(f, 0, M, 0, \boldsymbol{\beta}, \mathbf{w})$, $\boldsymbol{\beta} = (\beta_1, \beta_2, \beta_3, \beta_4), \mathbf{w} = (\beta_2, \beta_4)$ with non-linear dynamics and imperfect measurements and imperfect information about the system:

$$\dot{R} = V_r,$$

$$\dot{V}_r = \frac{z^2}{R} + \breve{a}_M^r(t) + a_T^r(t),$$

$$\breve{a}_M^r(t) = a_M^r\left[t, \widetilde{\mathbf{R}}(t), \widetilde{V}_r(t)\right] - \kappa_1(t)\widetilde{V}_r^3(t)$$

$$\widetilde{\mathbf{R}}(t) = R(t) + \beta_1(t); \widetilde{V}_r(t) = V_r(t) + \beta_2(t),$$

$$\breve{a}_M^r(t) \in [-\bar{a}_M^r, \bar{a}_M^r], a_T^r(t) \in [-\bar{a}_T^r, \bar{a}_T^r].$$

(4.3.30)

$$\dot{z} = -\frac{V_r z}{R} + \breve{a}_M^\tau(t) + a_T^\tau(t),$$

$$\breve{a}_M^\tau(t) = a_M^\tau\left[t, \mathbf{Z}(t), \widetilde{\mathbf{Z}}(t)\right] - \kappa_2(t)\left(\widetilde{\mathbf{Z}}(t)\right)^3,$$

$$\mathbf{Z}(t) = z(t) + \beta_3(t), \widetilde{\mathbf{Z}}(t) = \dot{z}(t) + \beta_4(t)$$

$$\breve{a}_M^\tau(t) \in [-\bar{a}_M^\tau, \bar{a}_M^\tau], a_T^\tau(t) \in [-\bar{a}_T^\tau, \bar{a}_T^\tau].$$

$$\mathbf{J}_i = R^2(t_1), i = 1, 2.$$

Optimal control problem for the first player are:

$$\bar{\mathbf{J}}_1 = \min_{\breve{a}_M^r(t) \in [-\bar{a}_M^r, \bar{a}_M^r], \breve{a}_M^\tau(t) \in [-\bar{a}_M^\tau, \bar{a}_M^\tau]} \left\{ \max_{a_T^r(t) \in [-\bar{a}_T^r, \bar{a}_T^r], a_T^\tau(t) \in [-\bar{a}_T^\tau, \bar{a}_T^\tau]} R^2(t_1) \right\} =$$

(4.3.31)

$$= \min_{\substack{a_M^r(t) \in [-\bar{a}_M^r, \bar{a}_M^r], a_M^\tau(t) \in [-\bar{a}_M^\tau, \bar{a}_M^\tau], \\ \kappa_1(t) > 0, \kappa_2(t) > 0}} \left\{ \max_{a_T^r(t) \in [-\bar{a}_T^r, \bar{a}_T^r], a_T^\tau(t) \in [-\bar{a}_T^\tau, \bar{a}_T^\tau]} R^2(t_1) \right\}.$$

Optimal control problem for the second player are:

$$\bar{J}_2 = \max_{a_T^r(t)\in[-\bar{a}_T^r,\bar{a}_T^r],a_T^\tau(t)\in[-\bar{a}_T^\tau,\bar{a}_T^\tau]} \left\{ \min_{\check{a}_M^r(t)\in[-\bar{a}_M^r,\bar{a}_M^r],\check{a}_M^\tau(t)\in[-\bar{a}_M^\tau,\bar{a}_M^\tau]} R^2(t_1) \right\} =$$

$$= \max_{a_T^r(t)\in[-\bar{a}_T^r,\bar{a}_T^r],a_T^\tau(t)\in[-\bar{a}_T^\tau,\bar{a}_T^\tau]} \left\{ \min_{\substack{a_M^r(t)\in[-\bar{a}_M^r,\bar{a}_M^r],a_M^\tau(t)\in[-\bar{a}_M^\tau,\bar{a}_M^\tau], \\ \kappa_1(t)>0,\kappa_2(t)>0}} R^2(t_1) \right\}.$$

(4.3.32)

Suppose that

$$\frac{z^2(0)}{R(0)} \approx 0, \frac{V_r(0)z(0)}{R(0)} \approx 0,$$

$$\delta(t) = \frac{z^2(t)}{R(t)} \approx 0, \eta(t) = \frac{V_r(t)z(t)}{R(t)} \approx 0,$$

(4.3.33)

$$t \in [0, t_1].$$

Let us consider the optimal control problem for the dissipative nonlinear game:

$$\dot{R} = V_r,$$

$$\dot{V}_r = -\kappa_1(t)\widetilde{V}_r^3(t) + a_M^r(t) + a_T^r(t),$$

$$\check{a}_M^r(t) = a_M^r\left[t, \widetilde{\mathbf{R}}(t), \widetilde{V}_r(t)\right] - \kappa_1(t)\widetilde{V}_r^3(t)$$

$$\widetilde{\mathbf{R}}(t) = R(t) + \beta_1(t); \widetilde{V}_r(t) = V_r(t) + \beta_2(t),$$

$$\check{a}_M^r(t) \in [-\bar{a}_M^r, \bar{a}_M^r], a_T^r(t) \in [-\bar{a}_T^r, \bar{a}_T^r].$$

$$\dot{z} = -\kappa_2(t)\left(\widetilde{\mathbf{z}}(t)\right)^3 + a_M^\tau(t) + a_T^\tau(t), \quad (4.3.34)$$

$$\check{a}_M^\tau(t) = a_M^\tau\left[t, \mathbf{z}(t), \widetilde{\mathbf{z}}(t)\right] - \kappa_2(t)\left(\widetilde{\mathbf{z}}(t)\right),$$

$$\mathbf{z}(t) = z(t) + \beta_3(t), \widetilde{\mathbf{z}}(t) = \dot{z}(t) + \beta_4(t)$$

$$\check{a}_M^\tau(t) \in [-\bar{a}_M^\tau, \bar{a}_M^\tau], a_T^\tau(t) \in [-\bar{a}_T^\tau, \bar{a}_T^\tau].$$

$$\delta(t) \approx 0, \eta(t) \approx 0,$$

$$\mathbf{J}_i = R^2(t_1), i = 1, 2.$$

Optimal control problem for the first player are:

$$\bar{\mathbf{J}}_1 = \min_{\substack{(\check{a}_M^r(t) \in [-\bar{a}_M^r, \bar{a}_M^r]) \wedge (\check{a}_M^\tau(t) \in [-\bar{a}_M^\tau, \bar{a}_M^\tau]) \wedge \\ \wedge(\delta(t)\approx 0, \eta(t)\approx 0)}} \left\{ \max_{(a_T^r(t) \in [-\bar{a}_T^r, \bar{a}_T^r]) \wedge (a_T^\tau(t) \in [-\bar{a}_T^\tau, \bar{a}_T^\tau])} R^2(t_1) \right\} =$$

$$= \min_{\substack{(a_M^r(t) \in [-\bar{a}_M^r, \bar{a}_M^r]) \wedge (a_M^\tau(t) \in [-\bar{a}_M^\tau, \bar{a}_M^\tau]) \wedge \\ \wedge(\kappa_1(t)>0, \kappa_2(t)>0) \wedge ((\delta(t)\approx 0, \eta(t)\approx 0))}} \left\{ \max_{(a_T^r(t) \in [-\bar{a}_T^r, \bar{a}_T^r]) \wedge (a_T^\tau(t) \in [-\bar{a}_T^\tau, \bar{a}_T^\tau])} R^2(t_1) \right\}. \quad (4.3.35)$$

Optimal control problem for the second player are:

$$\mathbf{\bar{J}}_2 = \max_{(a_T^r(t)\in[-\bar{a}_T^r,\bar{a}_T^r])\wedge(a_T^\tau(t)\in[-\bar{a}_T^\tau,\bar{a}_T^\tau])} \left\{ \min_{\substack{(\check{a}_M^r(t)\in[-\bar{a}_M^r,\bar{a}_M^r])\wedge(\check{a}_M^\tau(t)\in[-\bar{a}_M^\tau,\bar{a}_M^\tau])\wedge \\ \wedge(\delta(t)\approx 0,\eta(t)\approx 0)}} R^2(t_1) \right\} =$$

$$= \max_{(a_T^r(t)\in[-\bar{a}_T^r,\bar{a}_T^r])\wedge(a_T^\tau(t)\in[-\bar{a}_T^\tau,\bar{a}_T^\tau])} \left\{ \min_{\substack{(a_M^r(t)\in[-\bar{a}_M^r,\bar{a}_M^r])\wedge(a_M^\tau(t)\in[-\bar{a}_M^\tau,\bar{a}_M^\tau])\wedge \\ \wedge(\kappa_1(t)>0,\kappa_2(t)>0)\wedge(\delta(t)\approx 0,\eta(t)\approx 0)}} R^2(t_1) \right\}.$$

(4.3.36)

From Eqs.(4.3.34)-(4.3.35) one obtain corresponding linear master game:

$$\dot{r} = v_r + \lambda_2,$$

$$\dot{v}_r = -3\kappa_1(t)\lambda_2^2 \widetilde{v}_r(t) - \kappa_1(t)\lambda_2^3 + a_M^r(t) + a_T^r(t),$$

$$\check{a}_M^r(t) = a_M^r[t, \widetilde{\mathbf{r}}(t), \widetilde{v}_r(t)] - \kappa_1(t)\widetilde{v}_r^3(t),$$

$$\widetilde{\mathbf{r}}(t) = \lambda_1 + r(t) + \beta_1(t); \widetilde{v}_r(t) = \lambda_2 + v_r(t) + \beta_2(t),$$

$$\check{a}_M^r(t) \in [-\bar{a}_M^r, \bar{a}_M^r], a_T^r(t) \in [-\bar{a}_T^r, \bar{a}_T^r].$$

$$\dot{z}_1 = -3\kappa_2(t)\lambda_4^2 \widetilde{\mathbf{Z}}_1(t) - \kappa_2(t)\lambda_4^3 + a_M^\tau(t) + a_T^\tau(t),$$

$$\check{a}_M^\tau(t) = a_M^\tau\left[t, \widetilde{\mathbf{Z}}(t), \widetilde{\dot{\mathbf{Z}}}(t)\right] - \kappa_2(t)\left(\widetilde{\dot{\mathbf{Z}}}(t)\right)^3,$$

(4.3.37)

$$\widetilde{\mathbf{Z}}(t) = \lambda_3 + z_1(t) + \beta_3(t), \widetilde{\dot{\mathbf{Z}}}(t) = \dot{z}(t) + \beta_4(t)$$

$$\check{a}_M^\tau(t) \in [-\bar{a}_M^\tau, \bar{a}_M^\tau], a_T^\tau(t) \in [-\bar{a}_T^\tau, \bar{a}_T^\tau].$$

$$\bar{\delta}(t) = \frac{\bar{z}^2(t)}{\bar{r}(t)} \approx 0, \bar{\eta}(t) = \frac{\bar{v}_r(t)\bar{z}(t)}{\bar{r}(t)} \approx 0,$$

$$\bar{r}(t) = \lambda_1 + r(t), \bar{v}_r(t) = \lambda_2 + v_r(t), \bar{z}(t) = \lambda_3 + z_1(t),$$

$$\mathbf{J}_i = r^2(t_1), i = 1, 2.$$

Optimal control problem for the first player are:

$$\mathbf{J}_1 = \min_{\substack{(\tilde{a}_M^r(t)\in[-\bar{a}_M^r,\bar{a}_M^r])\wedge(\tilde{a}_M^\tau(t)\in[-\bar{a}_M^\tau,\bar{a}_M^\tau])\wedge \\ \wedge(\tilde{\delta}(t)\approx 0,\bar{\eta}(t)\approx 0)}} \left\{ \max_{(a_T^r(t)\in[-\bar{a}_T^r,\bar{a}_T^r])\wedge(a_T^\tau(t)\in[-\bar{a}_T^\tau,\bar{a}_T^\tau])} r^2(t_1) \right\} =$$

(4.3.38)

$$= \min_{\substack{(a_M^r(t)\in[-\bar{a}_M^r,\bar{a}_M^r])\wedge(a_M^\tau(t)\in[-\bar{a}_M^\tau,\bar{a}_M^\tau]) \\ \wedge(\kappa_1(t)>0,\kappa_2(t)>0)\wedge((\tilde{\delta}(t)\approx 0,\bar{\eta}(t)\approx 0))}} \left\{ \max_{(a_T^r(t)\in[-\bar{a}_T^r,\bar{a}_T^r])\wedge(a_T^\tau(t)\in[-\bar{a}_T^\tau,\bar{a}_T^\tau])} r^2(t_1) \right\}.$$

Optimal control problem for the second player are:

$$\mathbf{J}_2 = \max_{(a_T^r(t)\in[-\bar{a}_T^r,\bar{a}_T^r])\wedge(a_T^\tau(t)\in[-\bar{a}_T^\tau,\bar{a}_T^\tau])} \left\{ \min_{\substack{(\tilde{a}_M^r(t)\in[-\bar{a}_M^r,\bar{a}_M^r])\wedge(\tilde{a}_M^\tau(t)\in[-\bar{a}_M^\tau,\bar{a}_M^\tau]) \\ \wedge(\tilde{\delta}(t)\approx 0,\bar{\eta}(t)\approx 0)}} r^2(t_1) \right\} =$$

(4.3.39)

$$= \max_{(a_T^r(t)\in[-\bar{a}_T^r,\bar{a}_T^r])\wedge(a_T^\tau(t)\in[-\bar{a}_T^\tau,\bar{a}_T^\tau])} \left\{ \min_{\substack{(a_M^r(t)\in[-\bar{a}_M^r,\bar{a}_M^r])\wedge(a_M^\tau(t)\in[-\bar{a}_M^\tau,\bar{a}_M^\tau])\wedge \\ (\tilde{\delta}(t)\approx 0,\bar{\eta}(t)\approx 0)\wedge(\kappa_1(t)>0,\kappa_2(t)>0)}} r^2(t_1) \right\}.$$

By setting for the non principal simplification: $\kappa_1(t) = \kappa_1 = const$, $\kappa_2(t) = \kappa_2 = const$ one obtain the simple linear master game:

$$\dot{r} = v_r + \lambda_2,$$

$$\dot{v}_r = -3\kappa_1\lambda_2^2\widetilde{v}_r(t) - \kappa_1\lambda_2^3 + a_M^r(t) + a_T^r(t),$$

$$\breve{a}_M^r(t) = a_M^r[t, \widetilde{\mathbf{r}}(t), \widetilde{v}_r(t)] - \kappa_1\widetilde{v}_r^3(t),$$

$$\widetilde{\mathbf{r}}(t) = \lambda_1 + r(t) + \beta_1(t); \widetilde{v}_r(t) = \lambda_2 + v_r(t) + \beta_2(t),$$

$$\breve{a}_M^r(t) \in [-\bar{a}_M^r, \bar{a}_M^r], a_T^r(t) \in [-\bar{a}_T^r, \bar{a}_T^r].$$

$$\dot{z}_1 = -3\kappa_2\lambda_4^2\widetilde{\mathbf{Z}}_1(t) - \kappa_2\lambda_4^3 + a_M^\tau(t) + a_T^\tau(t), \tag{4.3.40}$$

$$\breve{a}_M^\tau(t) = a_M^\tau\left[t, \mathbf{Z}(t), \widetilde{\mathbf{Z}}(t)\right] - \kappa_2\left(\widetilde{\mathbf{Z}}(t)\right)^3,$$

$$\mathbf{Z}_1(t) = \lambda_3 + z_1(t) + \beta_3(t), \widetilde{\mathbf{Z}}(t) = \dot{z}(t) + \beta_4(t)$$

$$\breve{a}_M^\tau(t) \in [-\bar{a}_M^\tau, \bar{a}_M^\tau], a_T^\tau(t) \in [-\bar{a}_T^\tau, \bar{a}_T^\tau].$$

$$\bar{\delta}(t) \approx 0, \bar{\eta}(t) \approx 0, \kappa_1 > 0, \kappa_2 > 0,$$

$$\mathbf{J}_i = r^2(t_1), i = 1, 2.$$

Optimal control problem for the first player are:

$$\mathbf{J}_1 = \min_{\substack{(\breve{a}_M^r(t) \in [-\bar{a}_M^r, \bar{a}_M^r]) \wedge (\breve{a}_M^\tau(t) \in [-\bar{a}_M^\tau, \bar{a}_M^\tau]) \wedge \\ \wedge (\bar{\delta}(t) \approx 0, \bar{\eta}(t) \approx 0)}} \left\{ \max_{(a_T^r(t) \in [-\bar{a}_T^r, \bar{a}_T^r]) \wedge (a_T^\tau(t) \in [-\bar{a}_T^\tau, \bar{a}_T^\tau])} r^2(t_1) \right\} =$$

$$= \min_{\substack{(a_M^r(t) \in [-\bar{a}_M^r, \bar{a}_M^r]) \wedge (a_M^\tau(t) \in [-\bar{a}_M^\tau, \bar{a}_M^\tau]) \\ \wedge (\kappa_1 > 0, \kappa_2 > 0) \wedge ((\bar{\delta}(t) \approx 0, \bar{\eta}(t) \approx 0))}} \left\{ \max_{(a_T^r(t) \in [-\bar{a}_T^r, \bar{a}_T^r]) \wedge (a_T^\tau(t) \in [-\bar{a}_T^\tau, \bar{a}_T^\tau])} r^2(t_1) \right\}. \tag{4.3.41}$$

Optimal control problem for the second player are:

$$\bar{J}_2 = \max_{(a_T^r(t)\in[-\bar{a}_T^r,\bar{a}_T^r])\wedge(a_T^\tau(t)\in[-\bar{a}_T^\tau,\bar{a}_T^\tau])} \left\{ \min_{\substack{(\check{a}_M^r(t)\in[-\bar{a}_M^r,\bar{a}_M^r])\wedge(\check{a}_M^\tau(t)\in[-\bar{a}_M^\tau,\bar{a}_M^\tau])\\ \wedge(\tilde{\delta}(t)\approx 0,\bar{\eta}(t)\approx 0)}} r^2(t_1) \right\} =$$

(4.3.42)

$$= \max_{(a_T^r(t)\in[-\bar{a}_T^r,\bar{a}_T^r])\wedge(a_T^\tau(t)\in[-\bar{a}_T^\tau,\bar{a}_T^\tau])} \left\{ \min_{\substack{(a_M^r(t)\in[-\bar{a}_M^r,\bar{a}_M^r])\wedge(a_M^\tau(t)\in[-\bar{a}_M^\tau,\bar{a}_M^\tau])\wedge\\ (\tilde{\delta}(t)\approx 0,\bar{\eta}(t)\approx 0)\wedge(\kappa_1>0,\kappa_2>0)}} r^2(t_1) \right\}.$$

From Eqs.(4.3.40)-(4.3.41) we obtain optimal control $\{\alpha_M^r(t),\alpha_M^\tau(t)\}$ for the first player and optimal control $\{\alpha_T^r(t),\alpha_T^\tau(t)\}$ for the second player in the next form:

$$\check{\alpha}_M^r(t) \simeq -\rho_M^r\mathbf{sign}[[R(t)+\beta_1(t)]+\Theta_\tau(t)[V_r(t)+\beta_2(t)]] - \kappa_1[V_r(t)+\beta_2(t)]^3,$$

$$\check{\alpha}_M^\tau(t) = -\rho_M^\tau\mathbf{sign}[[z(t)+\beta_3(t)]+\Theta_\tau(t)[\dot{z}(t)+\beta_4(t)]] - \kappa_2[z(t)+\beta_4(t)]^3.$$

(4.3.43)

$$\alpha_T^r(t) \simeq \rho_T^r\mathbf{sign}\left[[R(t)+\hat{\beta}_1(t)]+\Theta_\tau(t)\left[V_r(t)+\hat{\beta}_2(t)\right]\right] - \kappa_1\left[V_r(t)+\hat{\beta}_2(t)\right]^3,$$

$$\alpha_T^\tau(t) = \rho_T^\tau\mathbf{sign}\left[[z(t)+\hat{\beta}_3(t)]+\Theta_\tau(t)\left[\dot{z}(t)+\hat{\beta}_4(t)\right]\right] - \kappa_2\left[z(t)+\hat{\beta}_4(t)\right]^3.$$

Thus for the numerical simulation we obtain ODE:

$$\dot{R} = V_r,$$

$$\dot{V}_r = \frac{z^2}{R} - \rho_M^r\mathbf{sign}[[R(t)+\beta_1(t)]+\Theta_\tau(t)[V_r(t)+\beta_2(t)]] -$$
$$-\kappa_1[V_r(t)+\beta_2(t)]^3 + a_T^r(t),$$

(4.3.30)

$$\dot{z} = -\frac{V_r z}{R} - \rho_M^\tau\mathbf{sign}[[z(t)+\beta_3(t)]+\Theta_\tau(t)[\dot{z}(t)+\beta_4(t)]] -$$
$$-\kappa_2[z(t)+\beta_4(t)]^3 + a_T^\tau(t).$$

**Example 4.3.1.** $\tau = 0.001, \kappa_1 = 10^{-3}, \kappa_2 = 0.001, \bar{a}_T^r = 20m/\sec^2,$
$\bar{a}_T^\tau = 20m/\sec^2, R(0) = 200m, V_r(0) = 10m/\sec,$
$z(0) = 60, \dot{z}(0) = 40, a_T^r(t) = \bar{a}_T^r(\sin(\omega \cdot t))^p,$

$$a_T^\tau(t) = \bar{a}_T^\tau(\sin(\omega \cdot t))^q, \omega = 50, w(t) = \beta(t) = \bar{\beta}(\sin(\omega \cdot t))^p,$$
$$\bar{\beta} = 20,\ p = 2, q = 1.$$

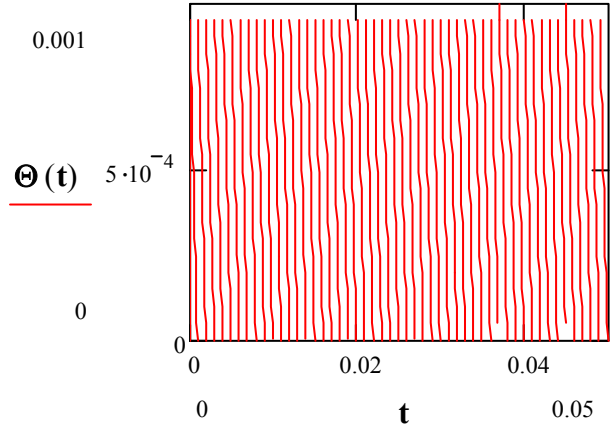

**Pic.1.1**.Cutting function:$\Theta_\tau(t)$.

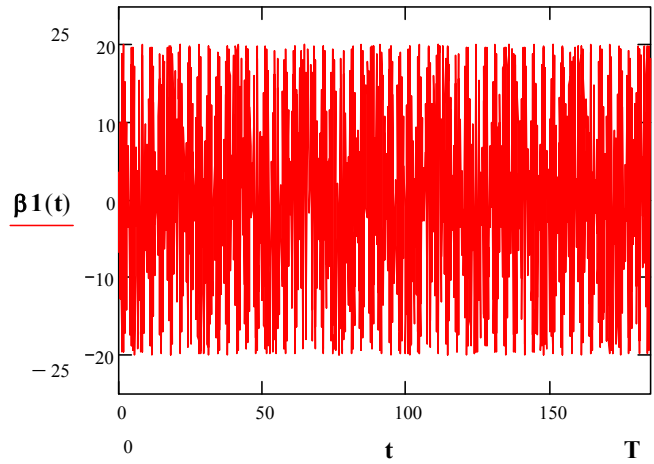

Uncertainty of measurements of a variable dR(t)/dt.

**Pic**.**1**.**2**.Uncertainty of measurements of a variable $\dot{R}(t) : \beta_1(t)$.

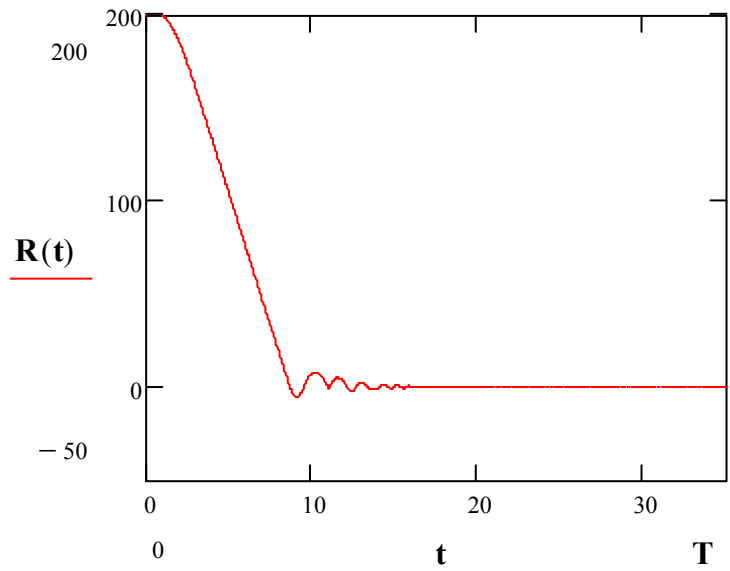

Target-to-missile range R(t)

**Pic**.**1**.3. Target-to-missile range $R(t)$.
$R(T) = 7.2 \times 10^3 m.$

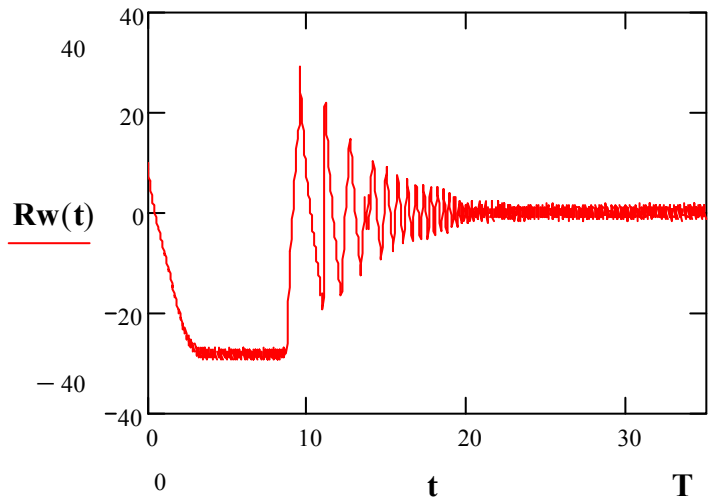

Speed of rapprochement missile-to-target

**Pic**.**1**.4. Speed of rapprochment missile-to-target: $\dot{R}(t)$.

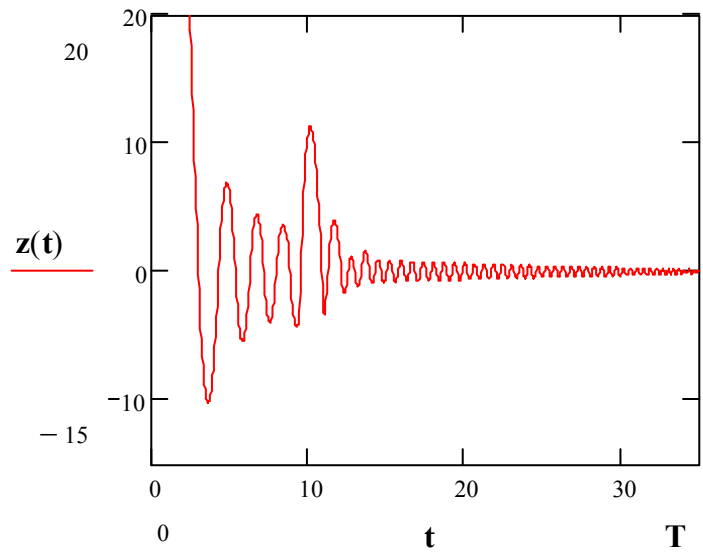

**Pic**.**1**.**5**.Variable $z(t)$.

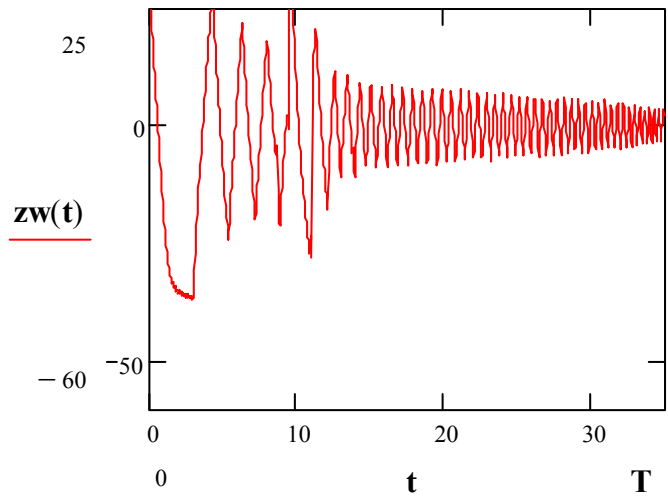

**Pic**.**1**.**6**.Variable $\dot{z}(t)$.
$\dot{z}(T) = 2.172.$

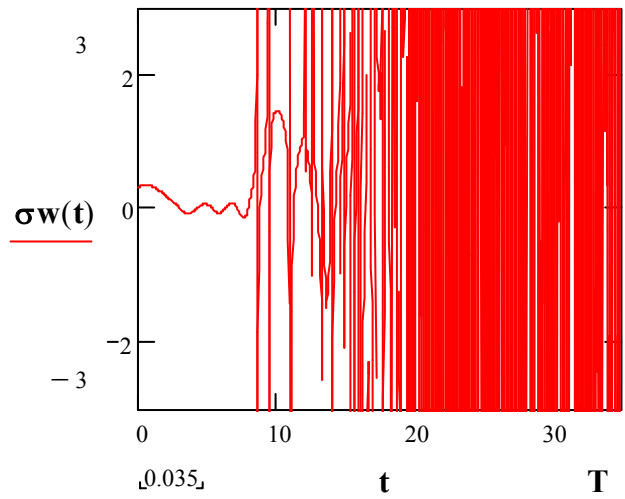

**Pic**.**1**.**7**.Variable $\dot{\sigma}(t)$.

$\dot{\sigma}(0) = 0.3$.

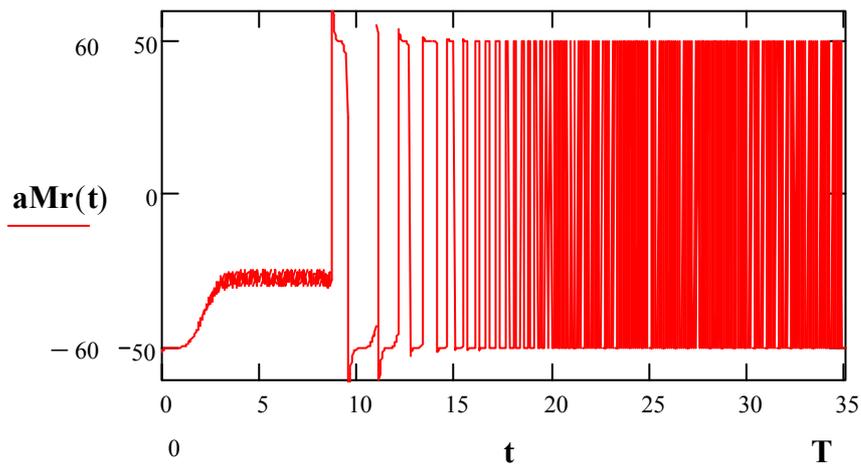

Missile acceleration along target-to-missile direction

**Pic**.**1**.**8**.Missile acceleration along target-to-missile direction:$a_M^r(t)$.

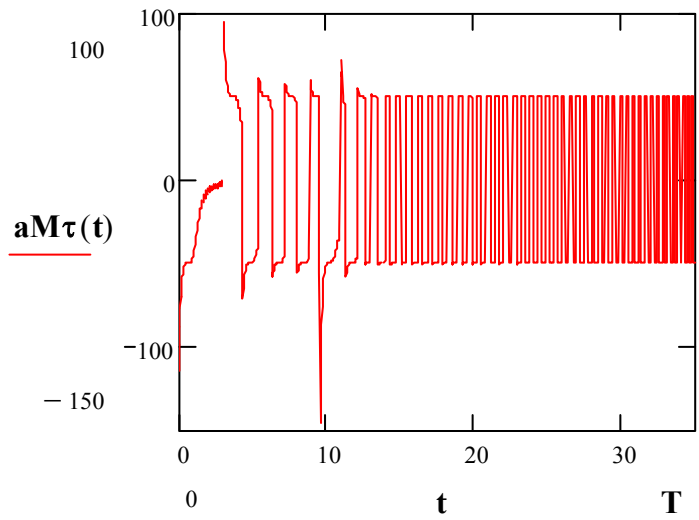

Missile tangent acceleration

**Pic.8.7**.Missile tangent acceleration:$a_M^\tau(t)$.

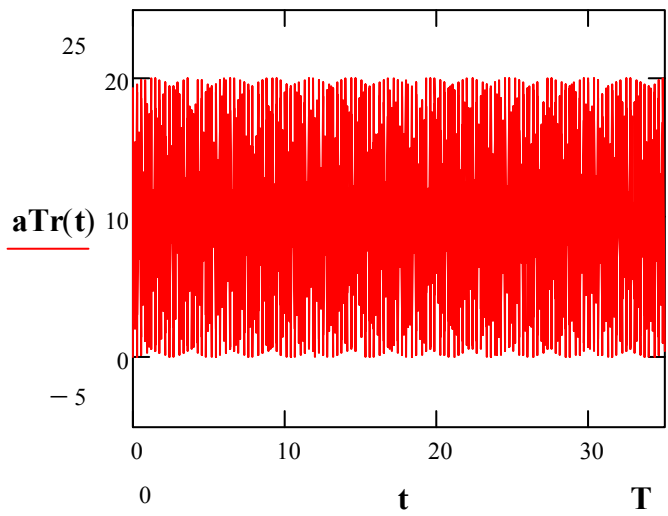

**Pic.1.9**.Target acceleration along target-to-missile direction:$a_T^r(t)$.

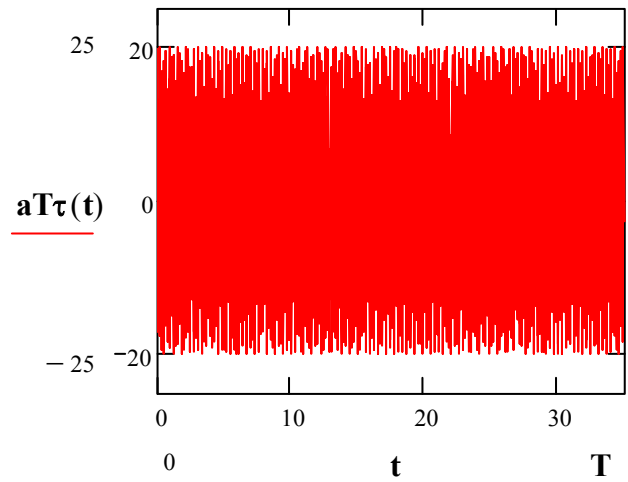

**Pic.1.10.** Target tangent acceleration: $a_T^\tau(t)$.

**Example 4.3.2.** $\tau = 0.001, \kappa_1 = 10^{-4}, \kappa_2 = 10^{-3}, \bar{a}_T^r = 20 m/\sec^2,$
$\bar{a}_T^\tau = 20 m/\sec^2,\ R(0) = 200m,\ V_r(0) = 10 m/\sec,$
$z(0) = 60, \dot{z}(0) = 40, a_T^r(t) = \bar{a}_T^r(\sin(\omega \cdot t))^p,$
$a_T^\tau(t) = \bar{a}_T^\tau(\sin(\omega \cdot t))^q, \omega = 50, w(t) = \beta_1(t) = \bar{\beta}_1(\sin(\omega \cdot t))^p,$
$\bar{\beta}_1 = 200 m/\sec,\ p = 2, q = 1.$

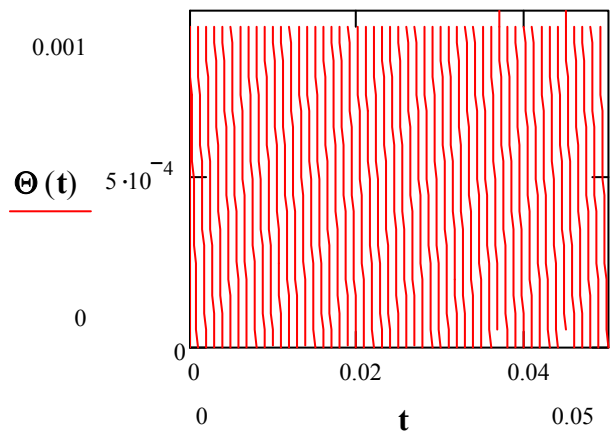

**Pic.2.1.** Cutting function: $\Theta_\tau(t)$.

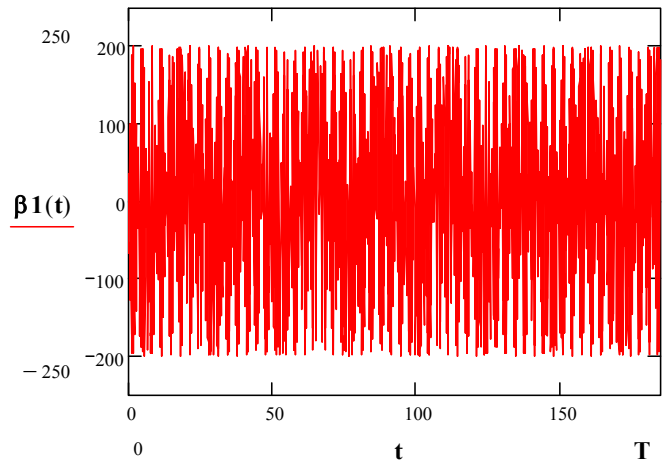

Uncertainty of measurements of a variable dR(t)/dt.

**Pic.2.2.** Uncertainty of measurements of a variable $\dot{R}(t) : \beta_1(t)$.

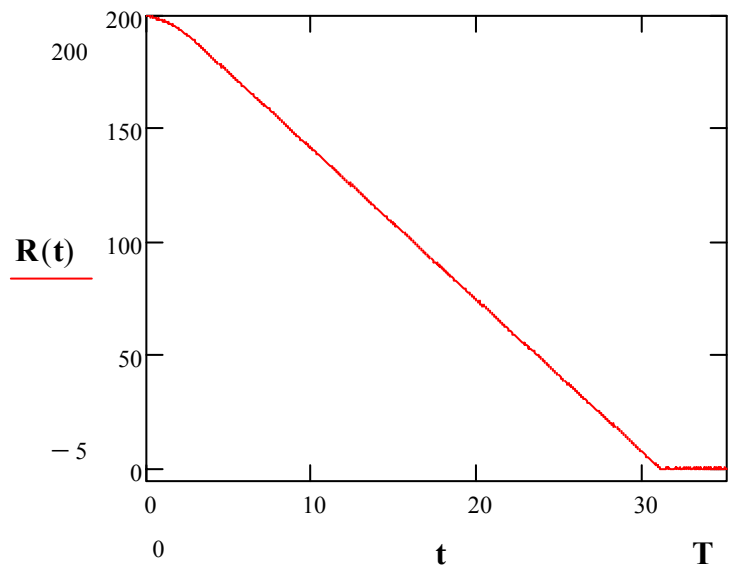

Target-to-missile range R(t)

**Pic.2.3.** Target-to-missile range $R(t)$.

$R(T) = 0.055m.$

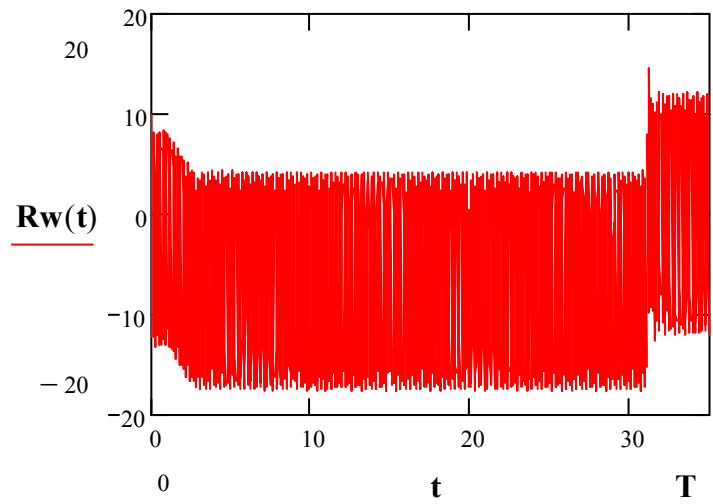

Speed of rapprochement missile-to-target

**Pic**.**2**.**4**.Speed of rapprochment missile-to-target:$\dot{R}(t)$.

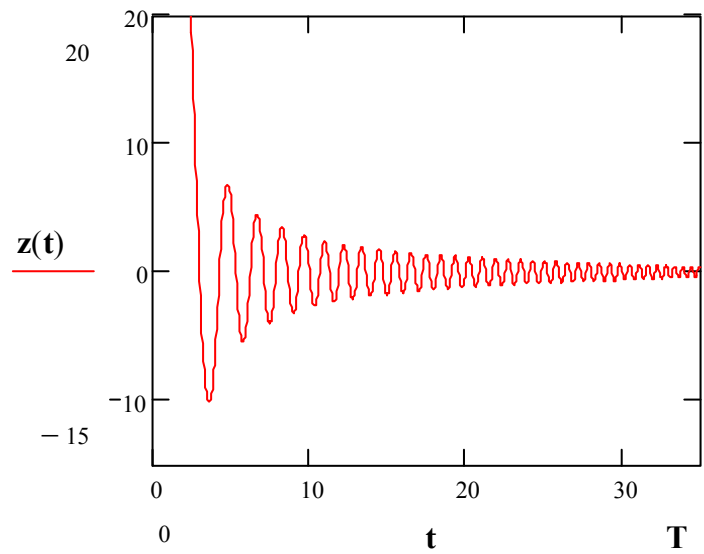

**Pic**.**2**.**5**.Variable $z(t)$.

$z(T) = 0.42.$

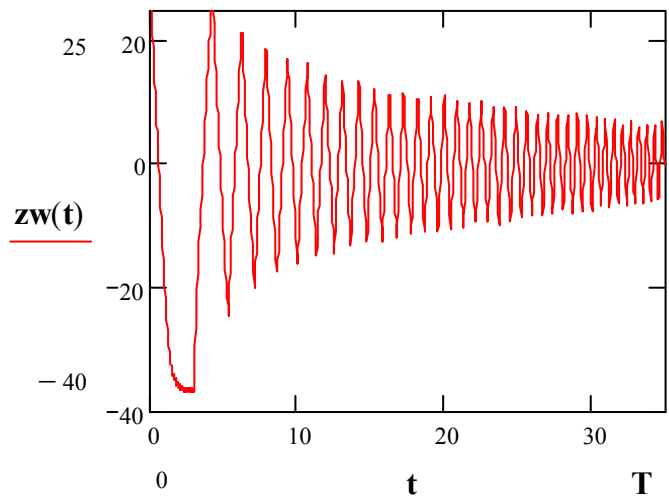

**Pic.2.6.** Variable $\dot{z}(t)$.
$z(T) = -0.149$.

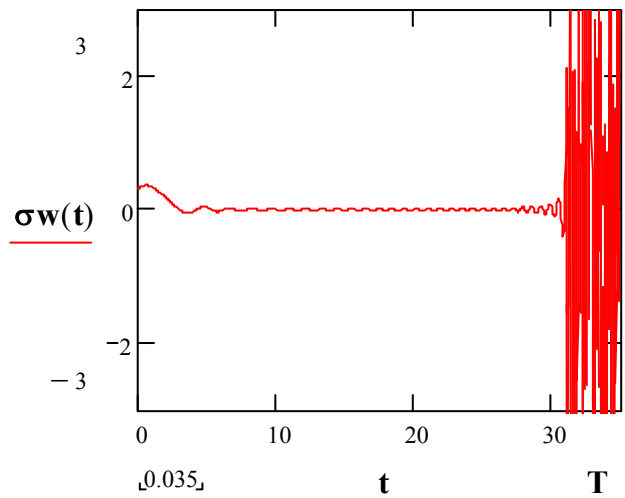

**Pic.2.7.** Variable $\dot{\sigma}(t)$.

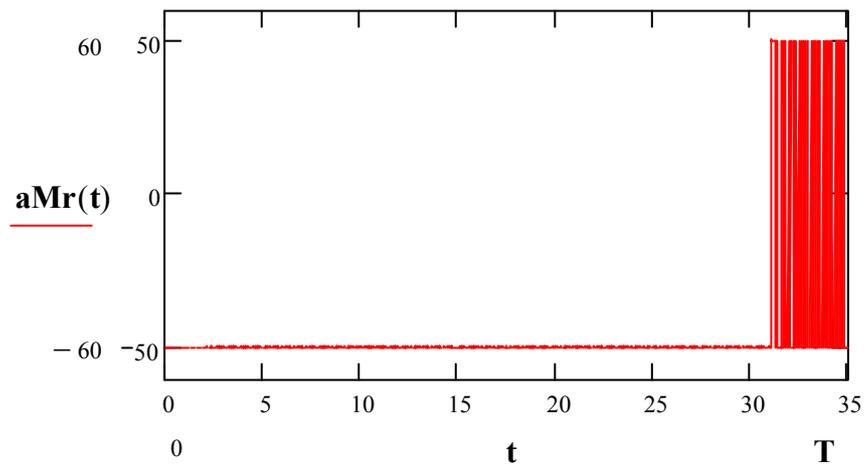

Missile acceleration along target-to-missile direction

**Pic**.2.8.Missile acceleration along target-to-missile direction:$a_M^r(t)$.

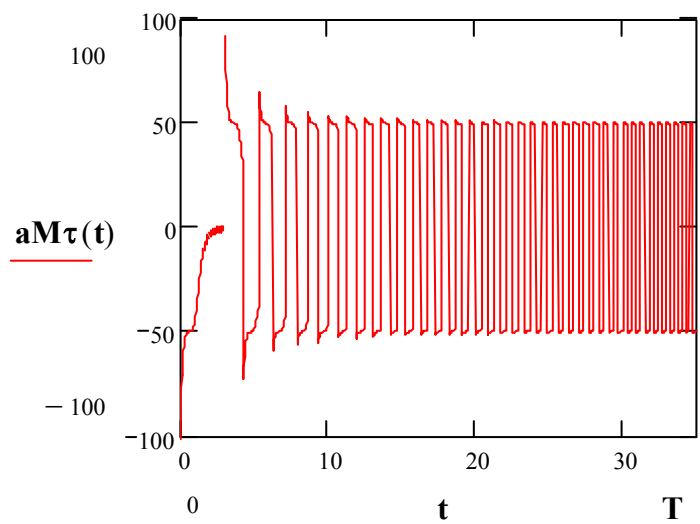

Missile tangent acceleration

**Pic**.2.9.Missile tangent acceleration:$a_M^\tau(t)$.

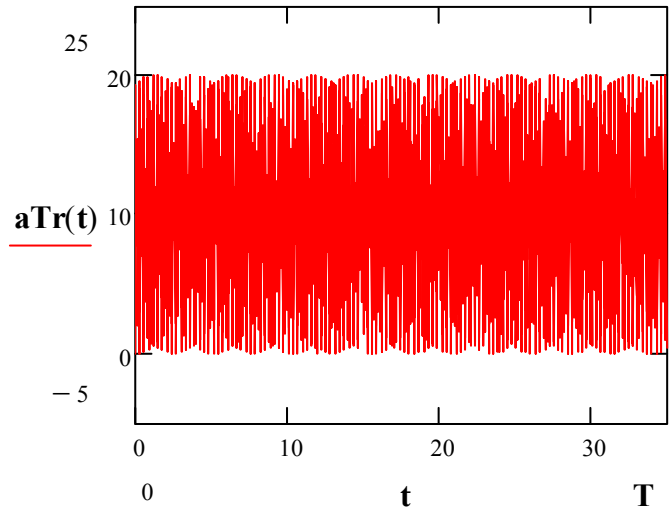

**Pic.2.10.** Target acceleration along target-to-missile direction: $a_T^r(t)$.

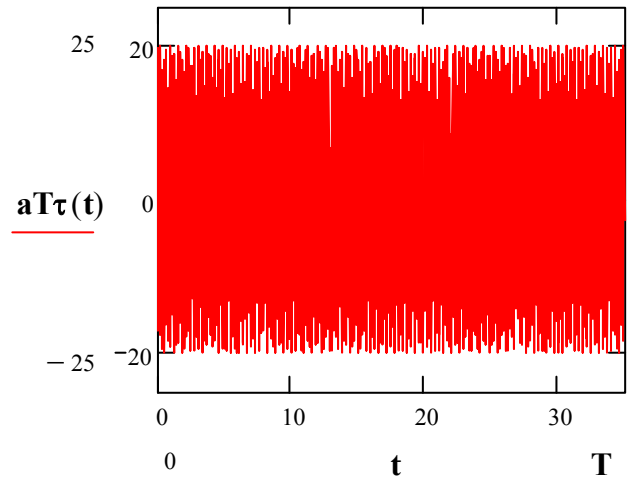

**Pic.2.11.** Target tangent acceleration: $a_T^\tau(t)$.

**Example 4.3.3.** $\tau = 0.00001, \kappa_1 = 10^{-3}, \kappa_2 = 10^{-3}, \bar{a}_T^r = 20m/\sec^2,$
$\bar{a}_T^\tau = 20m/\sec^2, R(0) = 200m, V_r(0) = 10m/\sec,$
$z(0) = 60, \dot{z}(0) = 40, a_T^r(t) = \bar{a}_T^r(\sin(\omega \cdot t))^p,$
$a_T^\tau(t) = \bar{a}_T^\tau(\sin(\omega \cdot t))^q, \omega = 50, w(t) = \beta_1(t) = \bar{\beta}_1(\sin(\omega \cdot t))^q,$
$\bar{\beta}_1 = 200m/\sec, \beta_2(t) = \bar{\beta}_2(\sin(\omega \cdot t))^p, \bar{\beta}_2 = 100m, p = 2, q = 1.$

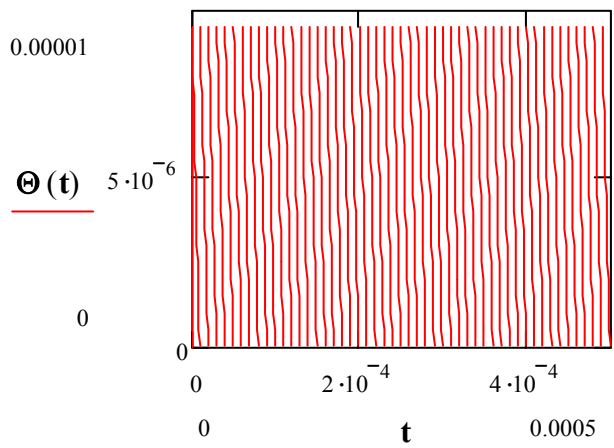

**Pic.3.1**. Cutting function: $\Theta_\tau(t)$.

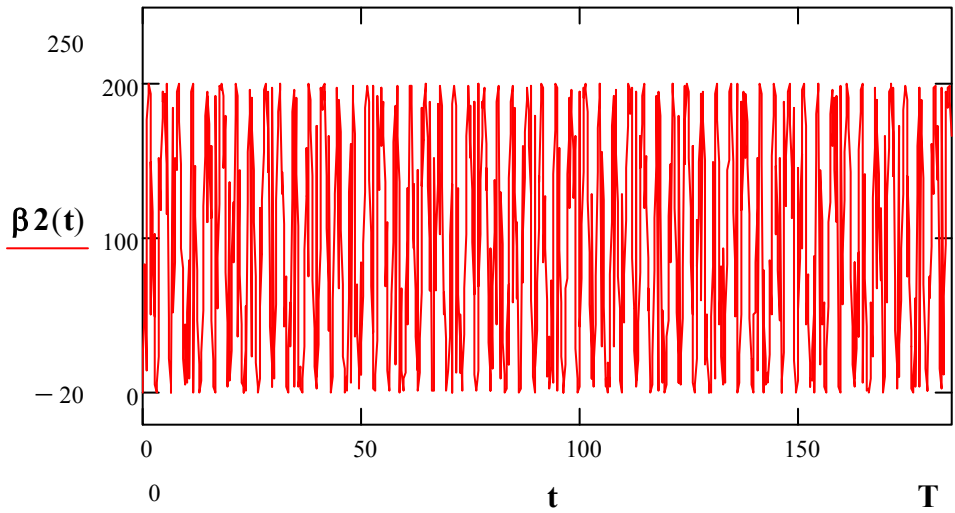

Uncertainty of measurements of a target-to-missile range R(t)

**Pic.3.2**. $\beta_2(t)$ : Uncertainty of measurements of a target-to-missile range $R(t)$.

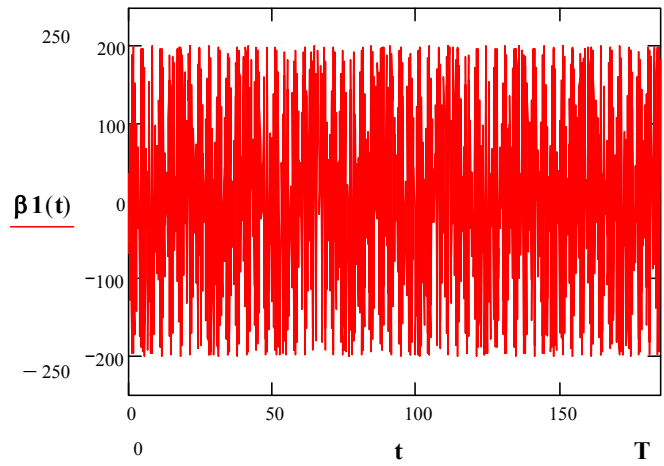

Uncertainty of measurements of a variable dR(t)/dt.

**Pic.3.3.** $\beta_1(t)$ : Uncertainty of measurements of a variable $\dot{R}(t)$.

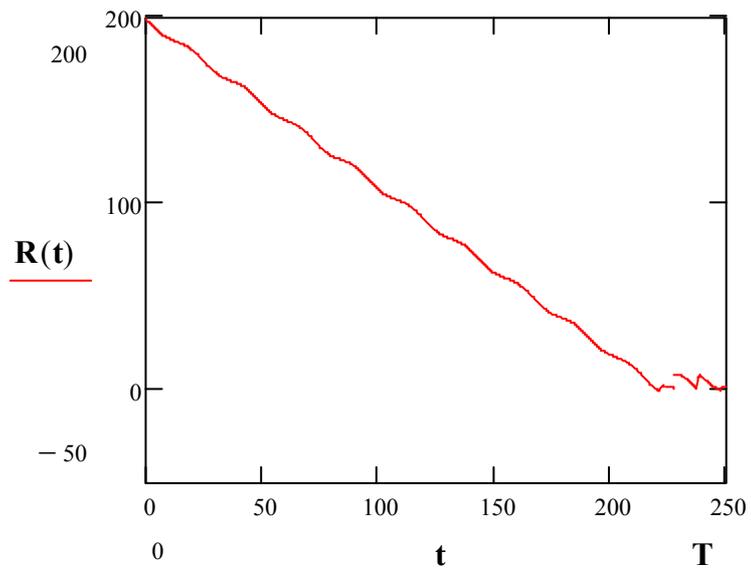

Target-to-missile range R(t)

**Pic.3.4.** Target-to-missile range $R(t)$.

$$R(T) = 1.108 m.$$

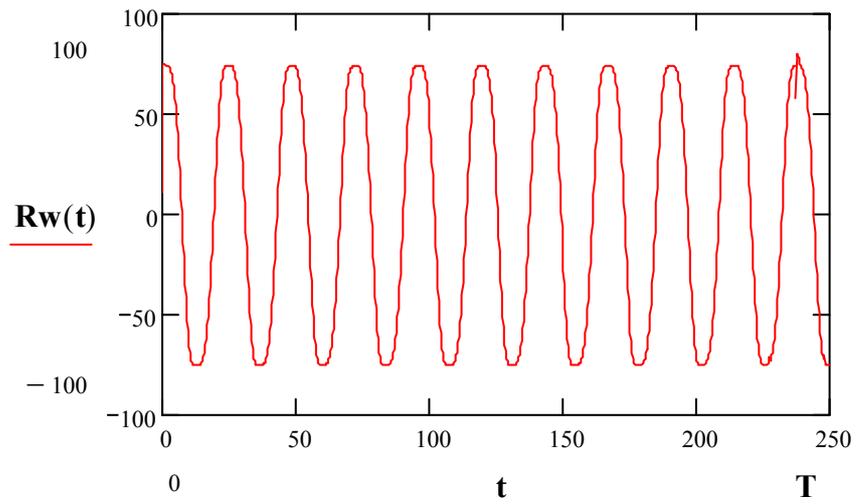

Speed of rapprochement missile-to-target: dR(t)/dt.

**Pic.3.5.** Speed of rapprochment missile-to-target: $\dot{R}(t)$.

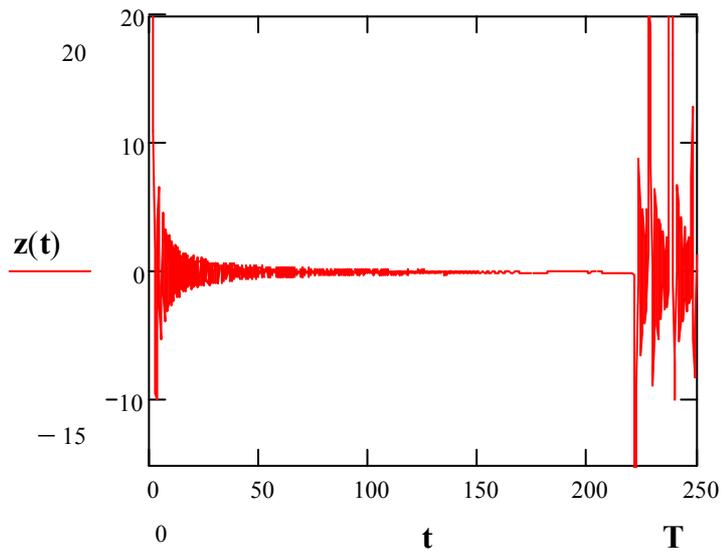

**Pic.3.6.** Variable $z(t)$.

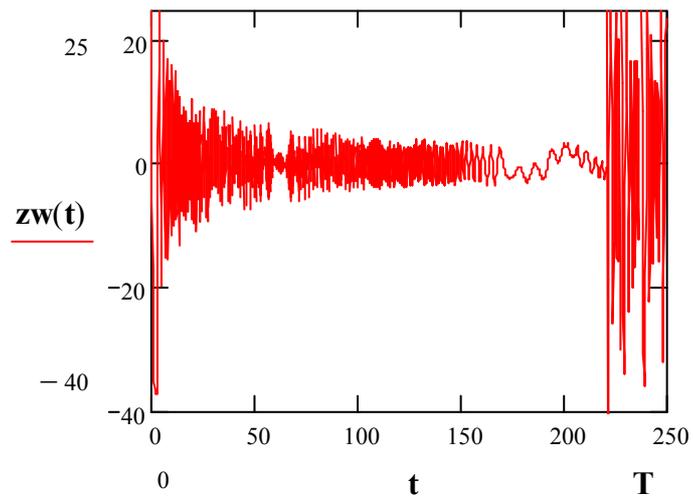

**Pic.3.7.** Variable $\dot{z}(t)$.

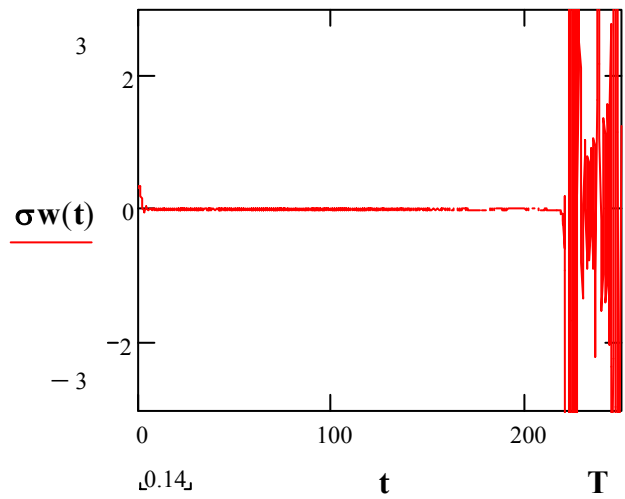

**Pic.3.8.** Variable $\dot{\sigma}(t)$.

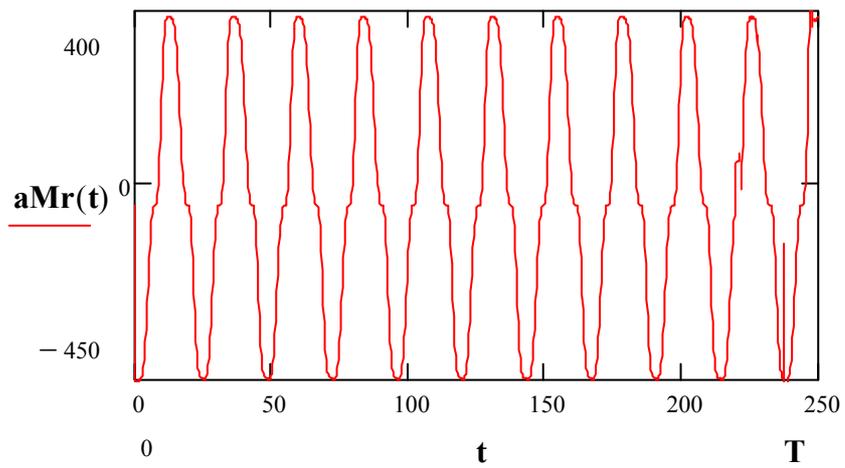

Missile acceleration along target-to-missile direction

**Pic**.**3**.**9**.Missile acceleration along target-to-missile direction: $a_M^r(t)$

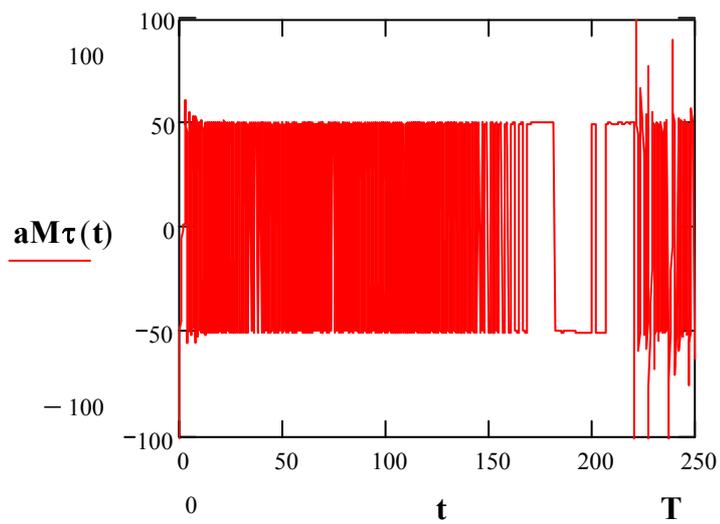

Missile tangent acceleration

**Pic**.**3**.**10**.Missile tangent acceleration: $a_M^\tau(t)$.

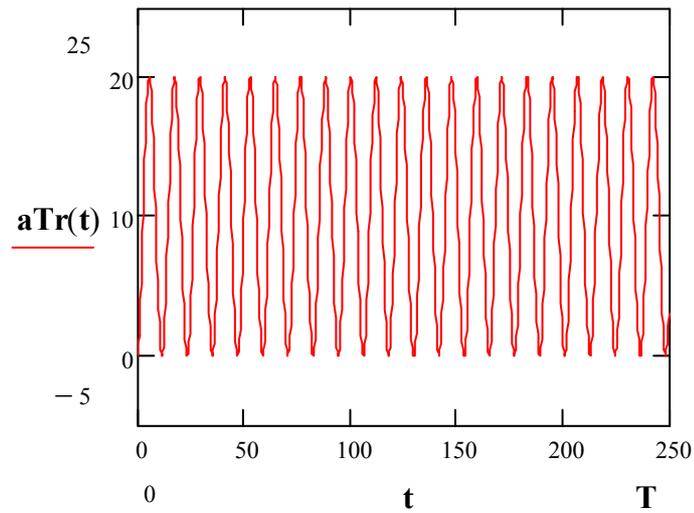

**Pic.3.11.** Target acceleration along target-to-missile direction: $a_T^r(t)$.

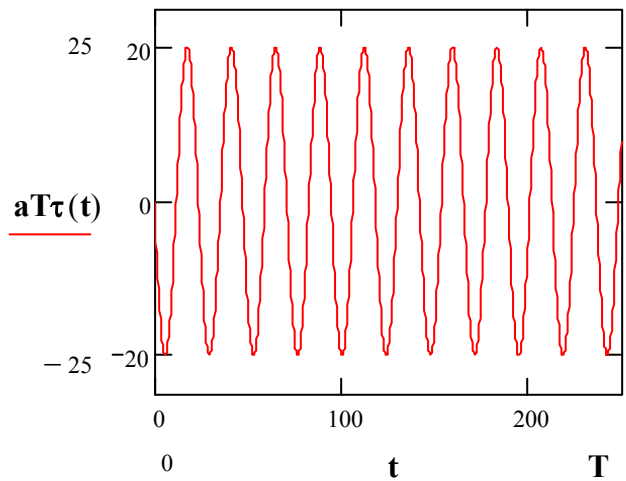

**Pic.3.12.** Target tangent acceleration: $a_T^\tau(t)$.

**Example 4.3.4.** $\tau = 0.00001, \kappa_1 = 8 \times 10^{-3}, \kappa_2 = 10^{-3}, \bar{a}_T^r = 20 m/\sec^2,$
$\bar{a}_T^\tau = 20 m/\sec^2, R(0) = 200m, V_r(0) = 10m/\sec,$
$z(0) = 60, \dot{z}(0) = 40, a_T^r(t) = \bar{a}_T^r(\sin(\omega \cdot t))^p,$
$a_T^\tau(t) = \bar{a}_T^\tau(\sin(\omega \cdot t))^q, \omega = 50, w(t) = \beta_1(t) = \bar{\beta}_1(\sin(\omega \cdot t))^q,$
$\bar{\beta}_1 = 200 m/\sec, \beta_2(t) = \bar{\beta}_2(\sin(\omega \cdot t))^p, \bar{\beta}_2 = 100m, p = 2, q = 1.$

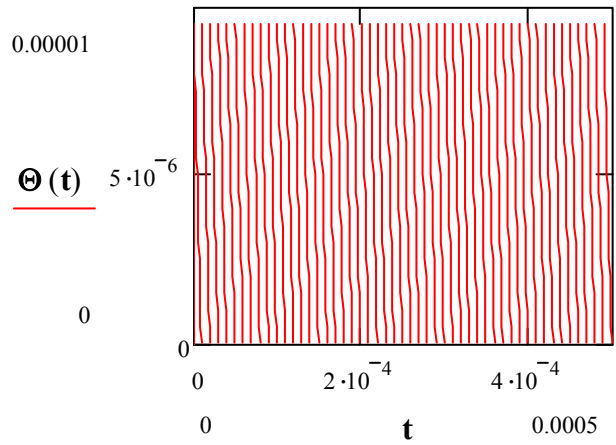

**Pic.4.1.** Cutting function: $\Theta_\tau(t)$.

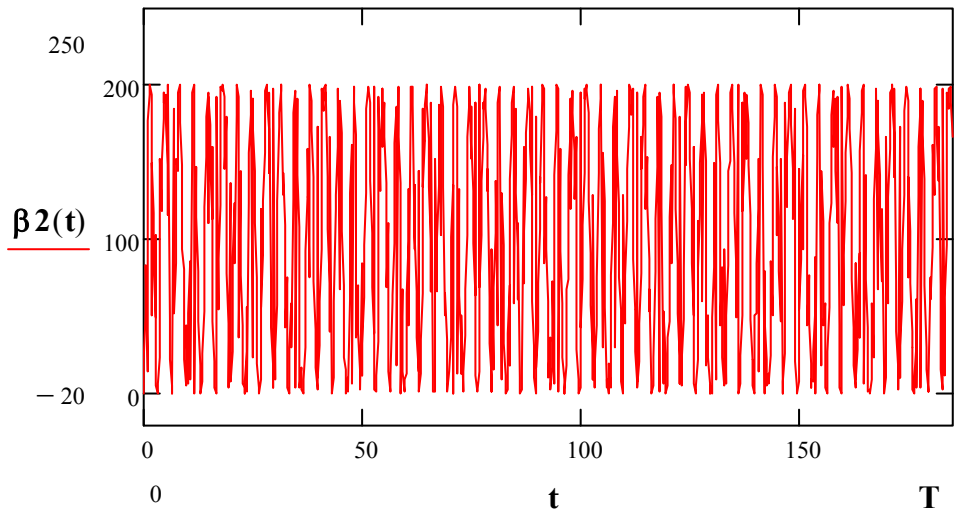

Uncertainty of measurements of a target-to-missile range R(t)

**Pic.4.2.** $\beta_2(t)$ : Uncertainty of measurements of a target-to-missile range $R(t)$.

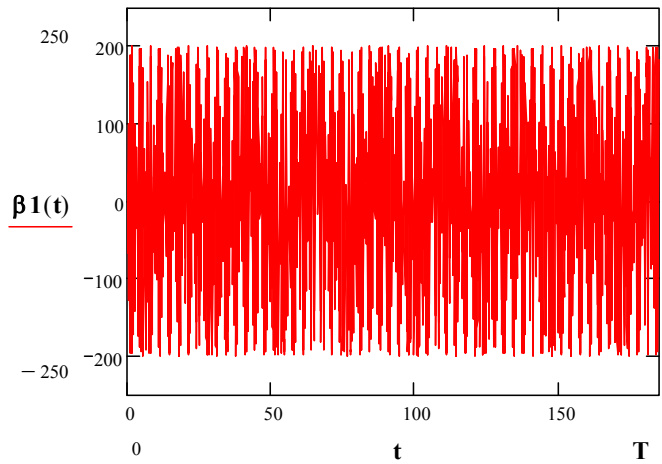

Uncertainty of measurements of a variable dR(t)/dt.

**Pic.4.3.**$\beta_1(t)$ : Uncertainty of measurements of a variable $\dot{R}(t)$.

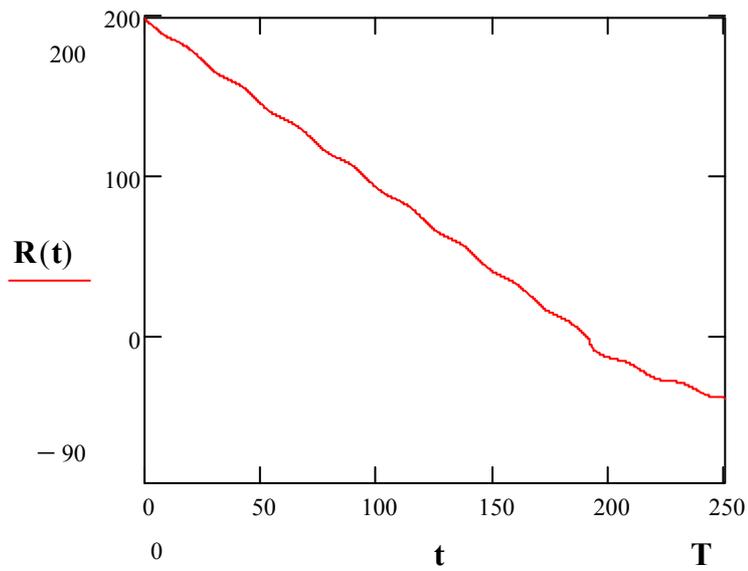

Target-to-missile range R(t)

**Pic.4.4.** Target-to-missile range $R(t)$.

$R(T) = -37.5m.$

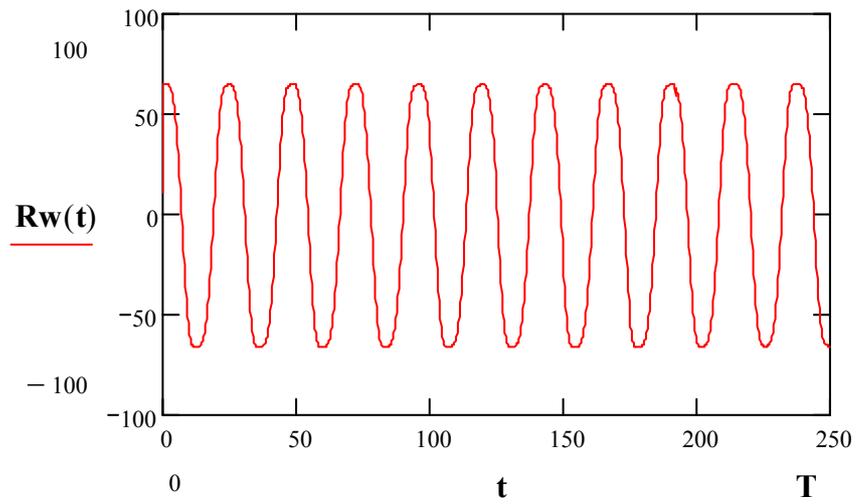

Speed of rapprochement missile-to-target:dR(t)/dt.

**Pic**.**4**.**5**.Speed of rapprochment missile-to-target: $\dot{R}(t)$.

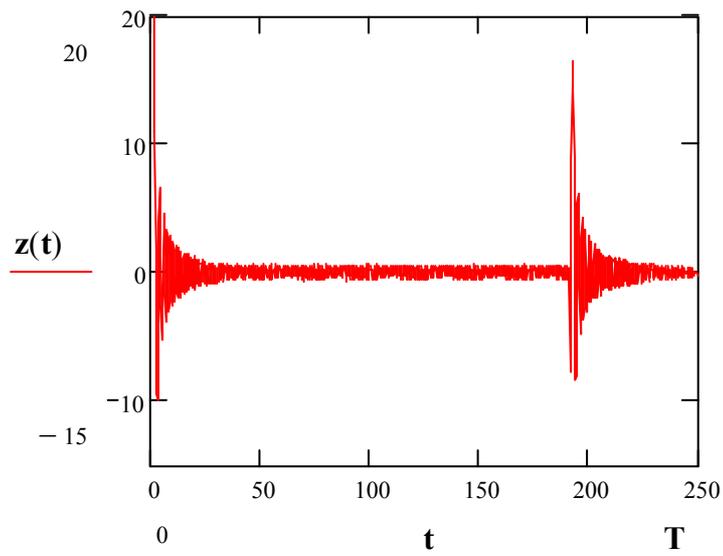

**Pic**.**4**.**6**.Variable $z(t)$.

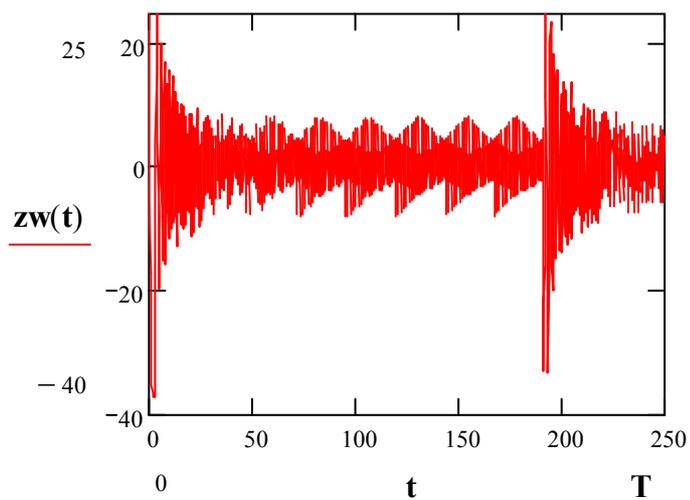

**Pic.4.7.** Variable $\dot{z}(t)$.

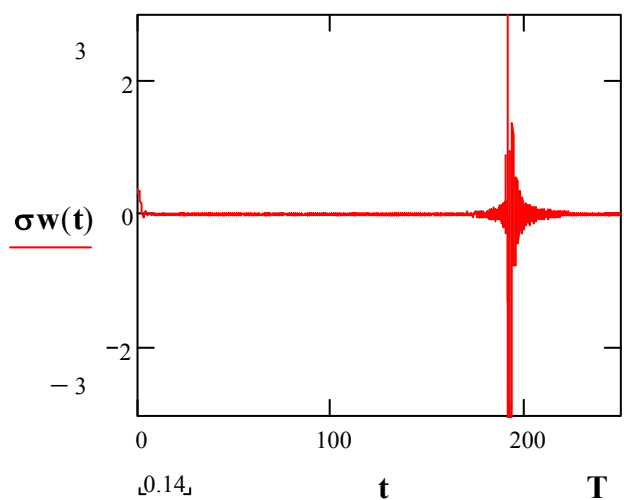

**Pic.4.8.** Variable $\dot{\sigma}(t)$.

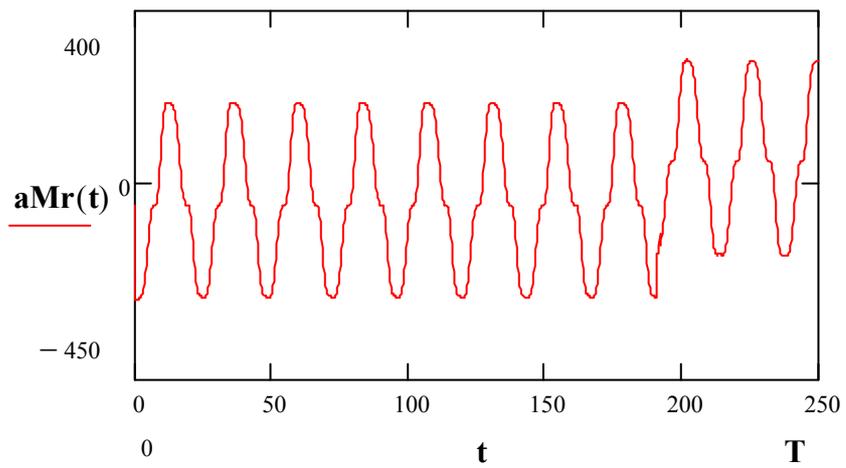

Missile acceleration along target-to-missile direction

**Pic**.**4**.**9**.Missile acceleration along target-to-missile direction: $a_M^r(t)$.

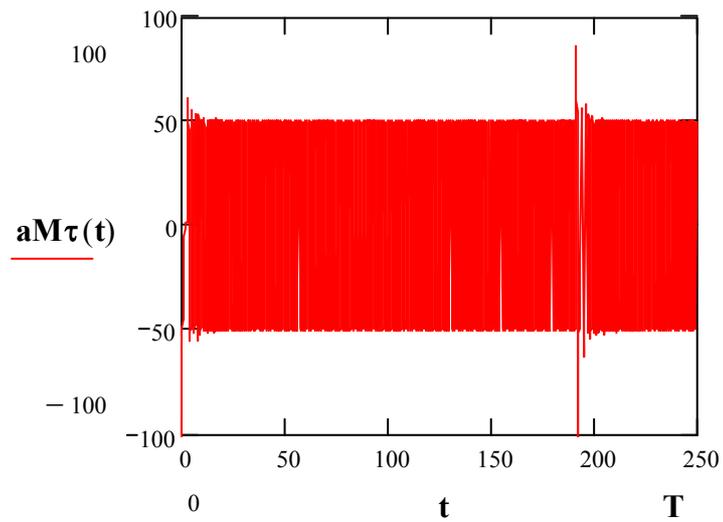

Missile tangent acceleration

**Pic**.**4**.**10**.Missile tangent acceleration: $a_M^\tau(t)$.

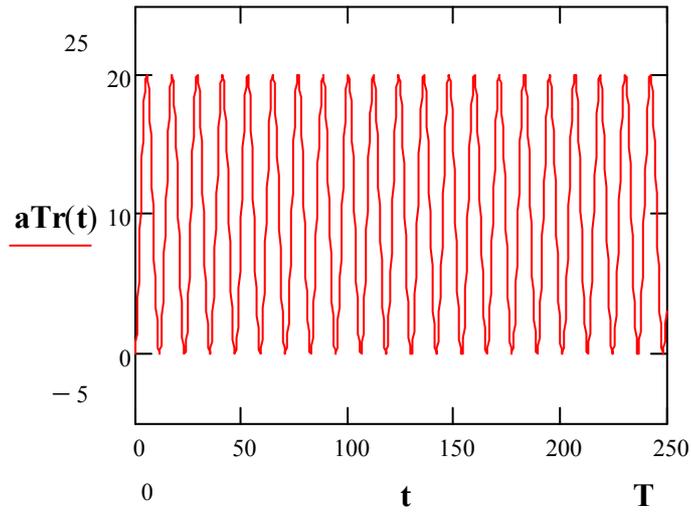

**Pic.4.11.** Target acceleration along target-to-missile direction: $a_T^r(t)$.

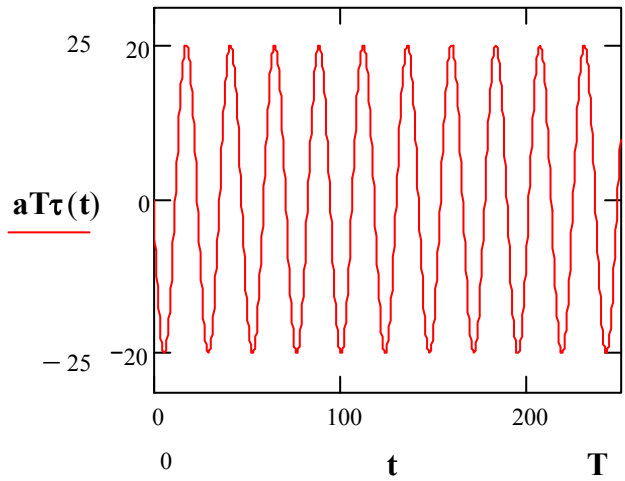

**Pic.4.12.** Target tangent acceleration: $a_T^\tau(t)$.

**Example 4.3.5.** $\tau = 0.00001, \kappa_1 = 5 \times 10^{-4}, \kappa_2 = 10^{-3}, \bar{a}_T^r = 20 m/\sec^2,$
$\bar{a}_T^\tau = 20 m/\sec^2, R(0) = 200 m, V_r(0) = 10 m/\sec,$
$z(0) = 60, \dot{z}(0) = 40, a_T^r(t) = \bar{a}_T^r(\sin(\omega \cdot t))^p,$
$a_T^\tau(t) = \bar{a}_T^\tau(\sin(\omega \cdot t))^q, \omega = 50, w(t) = \beta_1(t) = \bar{\beta}_1(\sin(\omega \cdot t))^q,$
$\bar{\beta}_1 = 200 m/\sec, \beta_2(t) = \bar{\beta}_2(\sin(\omega \cdot t))^p, \bar{\beta}_2 = 100 m, p = 2, q = 1.$

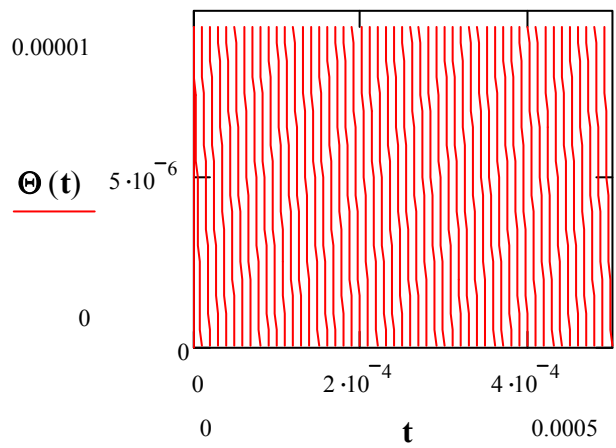

**Pic.5.1.** Cutting function: $\Theta_\tau(t)$.

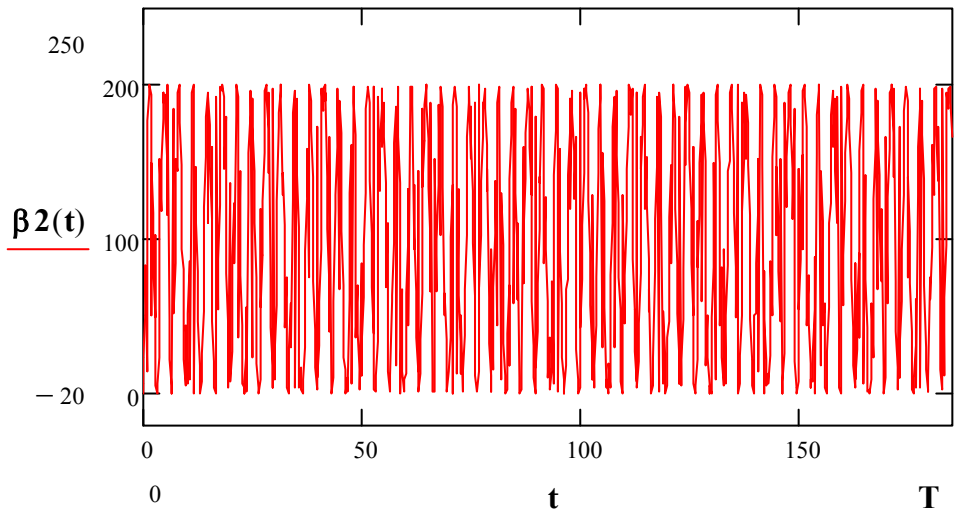

Uncertainty of measurements of a target-to-missile range R(t)

**Pic.5.2.** $\beta_2(t)$ : Uncertainty of measurements of a target-to-missile range $R(t)$.

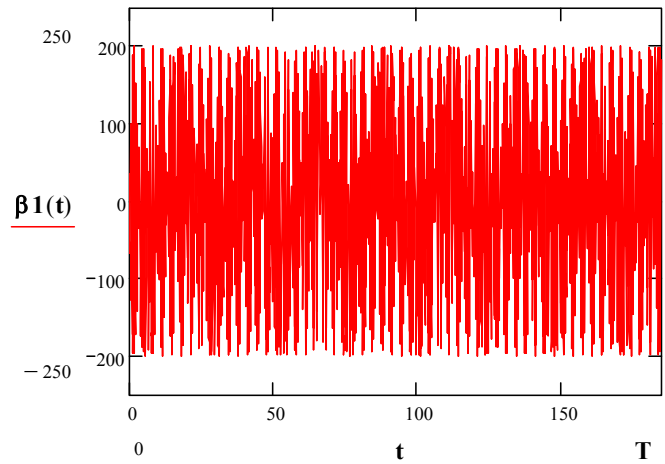

Uncertainty of measurements of a variable dR(t)/dt.

**Pic.5.3.** $\beta_1(t)$ : Uncertainty of measurements of a variable $\dot{R}(t)$.

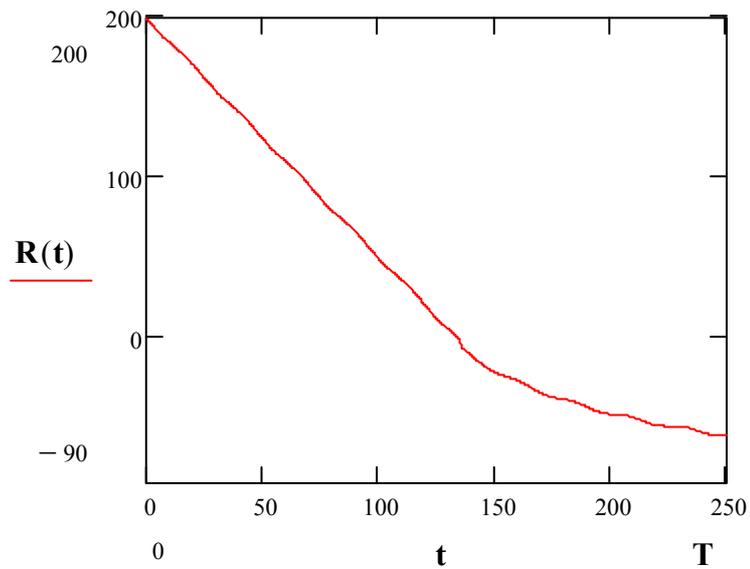

Target-to-missile range R(t)

**Pic.5.4.** Target-to-missile range $R(t)$.

$$R(T) = -60.3m.$$

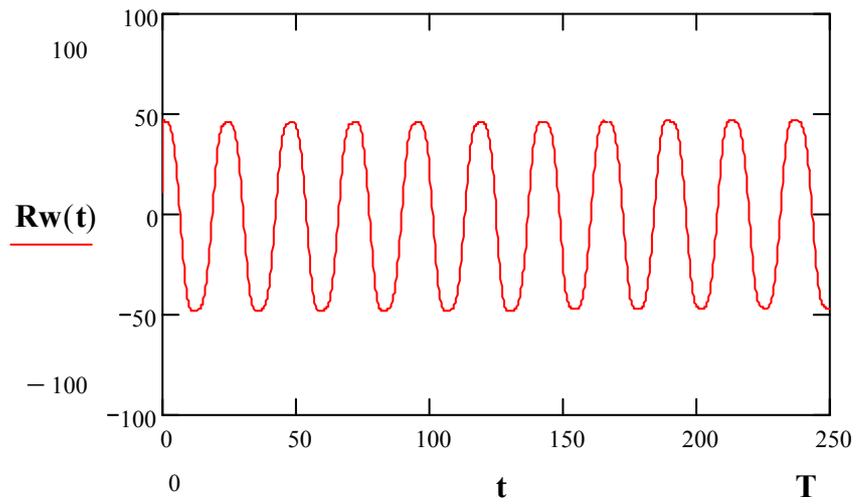

Speed of rapprochement missile-to-target:dR(t)/dt.

**Pic**.**5**.**5**.Speed of rapprochment missile-to-target: $\dot{R}(t)$.

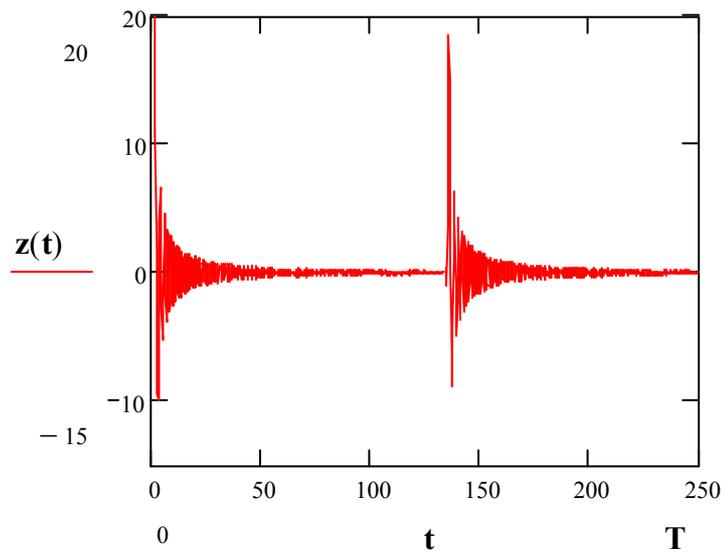

**Pic**.**5**.**6**.Variable $z(t)$.

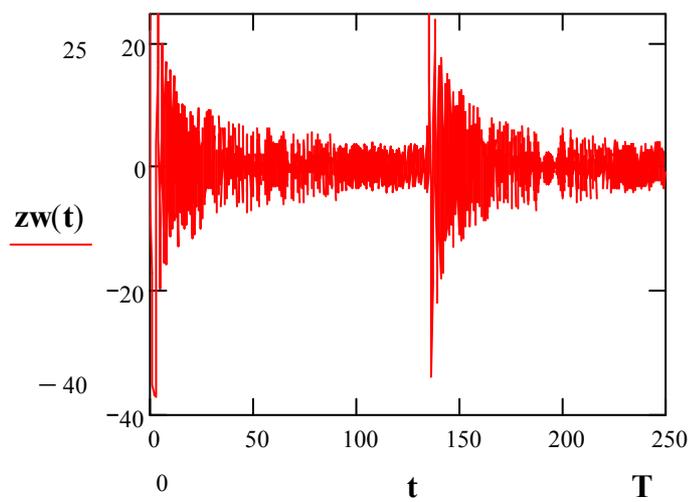

**Pic.5.7**.Variable $\dot{z}(t)$.

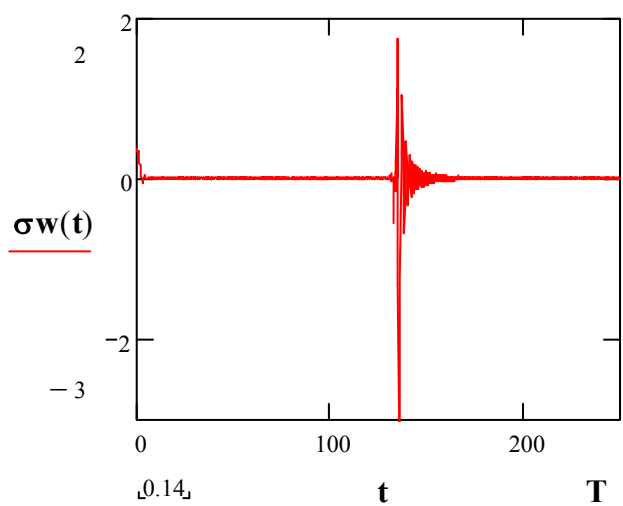

**Pic.5.8**.Variable $\dot{\sigma}(t)$.
$\dot{\sigma}(T) = -1.951 \times 10^{-3}$.

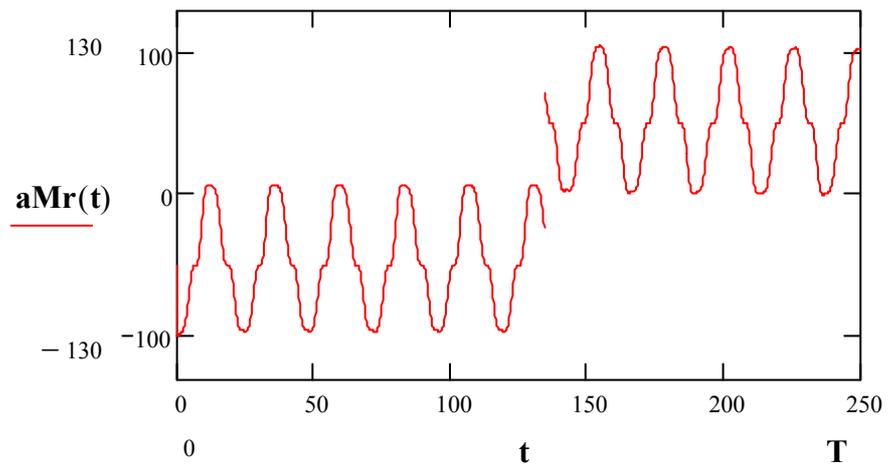

Missile acceleration along target-to-missile direction

**Pic.5.9.** Missile acceleration along target-to-missile direction: $a_M^r(t)$.

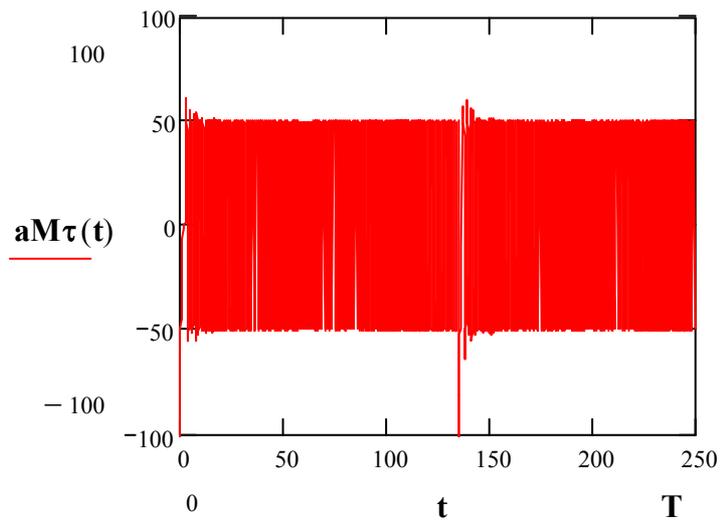

Missile tangent acceleration

**Pic.5.10.** Missile tangent acceleration: $a_M^\tau(t)$.

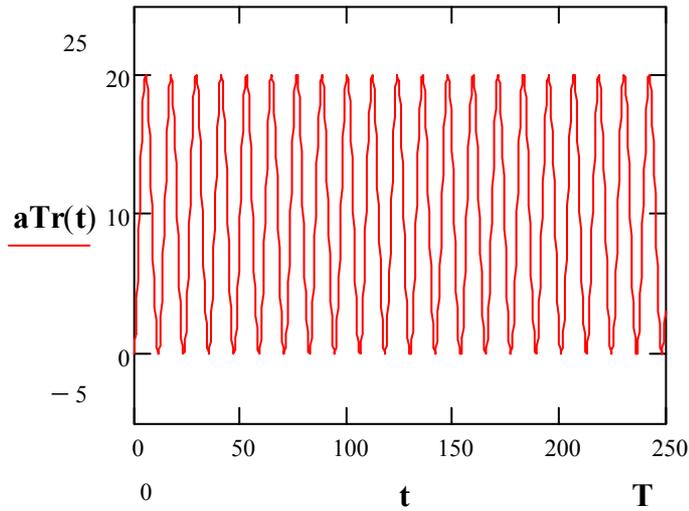

**Pic.5.11.** Target acceleration along target-to-missile direction: $a_T^r(t)$.

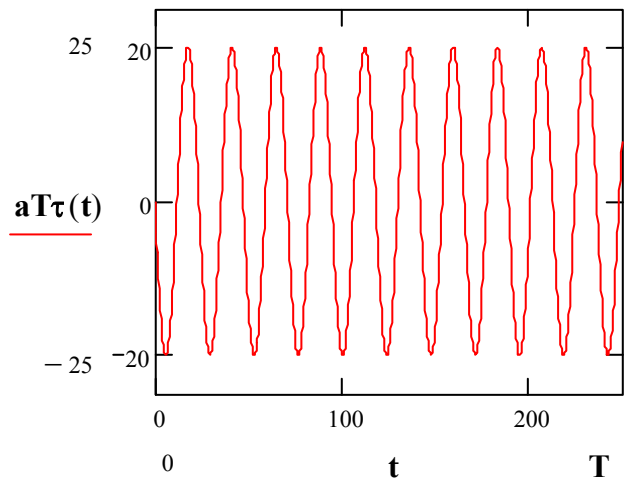

**Pic.5.12.** Target tangent acceleration: $a_T^\tau(t)$.

**Example 4.3.6.** $\tau = 0.00001, \kappa_1 = 5 \times 10^{-4}, \kappa_2 = 10^{-3}, \bar{a}_T^r = 20 m/\sec^2$,
$\bar{a}_T^\tau = 20 m/\sec^2, R(0) = 200m, V_r(0) = 10m/\sec$,
$z(0) = 60, \dot{z}(0) = 40, a_T^r(t) = \bar{a}_T^r(\sin(\omega \cdot t))^p$,
$a_T^\tau(t) = \bar{a}_T^\tau(\sin(\omega \cdot t))^q, \omega = 50, w(t) = \beta_1(t) = \bar{\beta}_1(\sin(\omega \cdot t))^q$,
$\bar{\beta}_1 = 200m/\sec, \beta_2(t) = \bar{\beta}_2(\sin(\omega \cdot t))^p, \bar{\beta}_2 = 100m, p = 2, q = 1$.

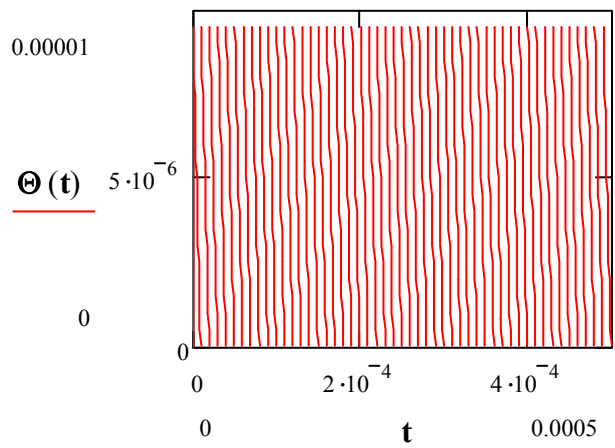

**Pic.6.1.** Cutting function: $\Theta_\tau(t)$.

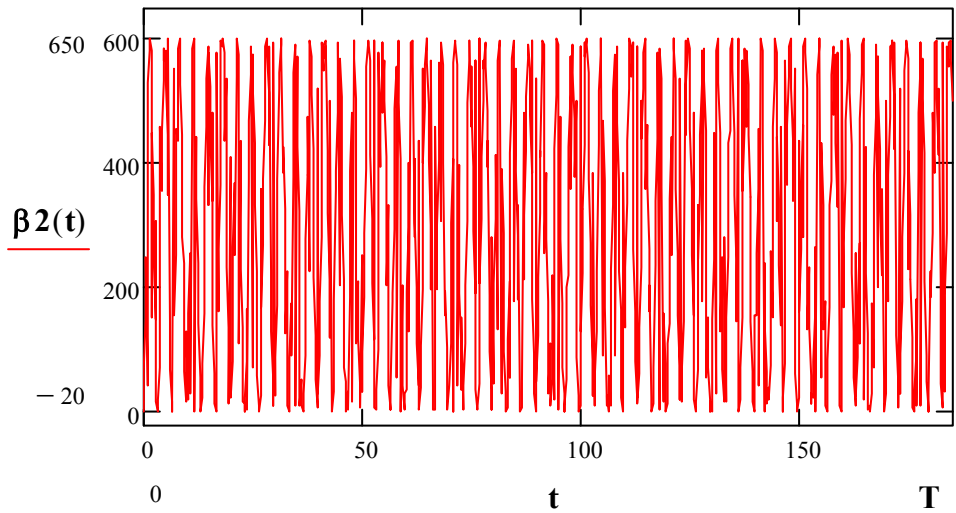

Uncertainty of measurements of a target-to-missile range R(t)

**Pic.6.2.** $\beta_2(t)$ : Uncertainty of measurements of a target-to-missile range $R(t)$.

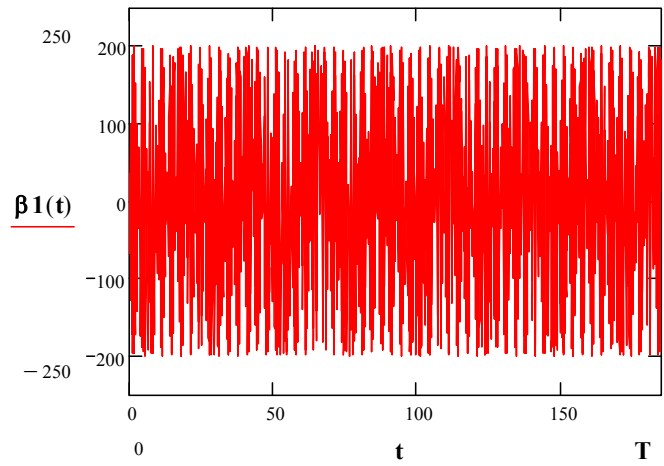

Uncertainty of measurements of a variable dR(t)/dt.

**Pic**.**6**.**3**.$\beta_1(t)$ : Uncertainty of measurements of a variable $\dot{R}(t)$.

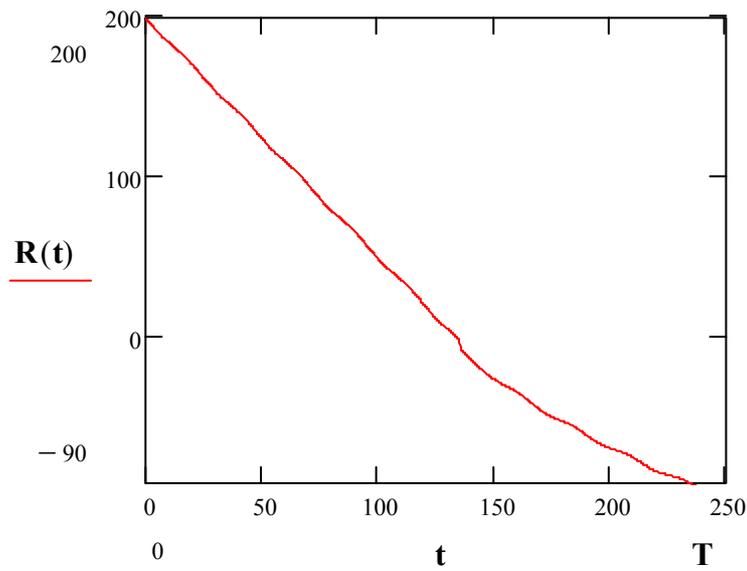

Target-to-missile range R(t)

**Pic**.**6**.**4**.Target-to-missile range $R(t)$.

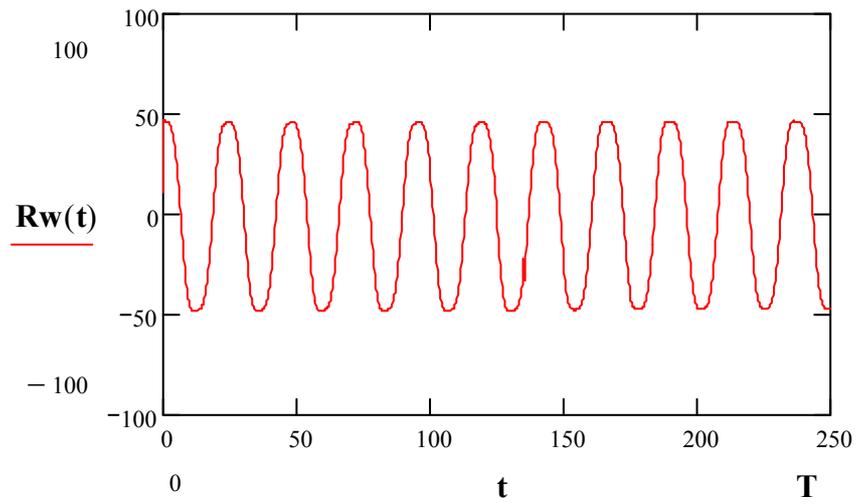

Speed of rapprochement missile-to-target: dR(t)/dt.

**Pic.6.5.** Speed of rapprochment missile-to-target: $\dot{R}(t)$.

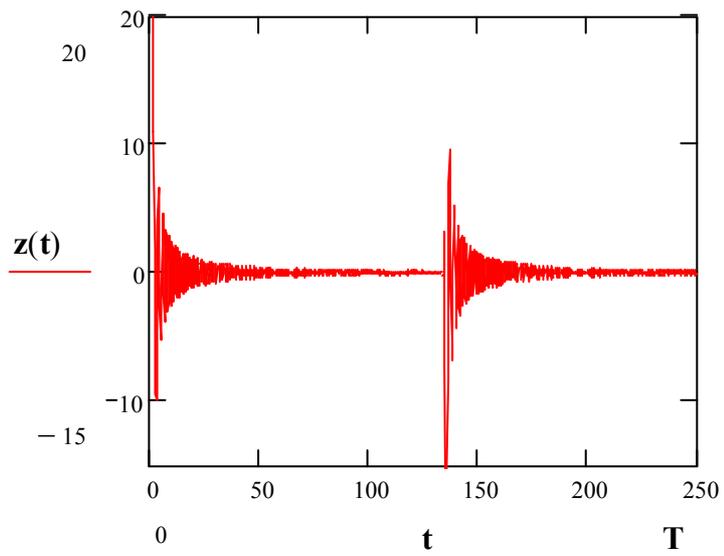

**Pic.6.6.** Variable $z(t)$.

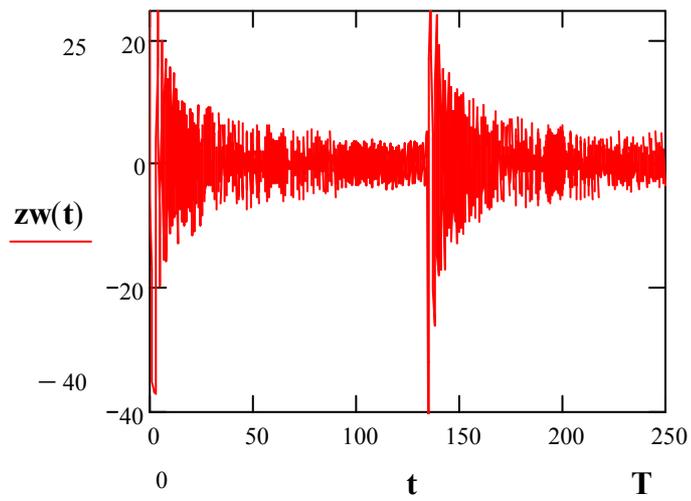

**Pic.6.7.** Variable $\dot{z}(t)$.

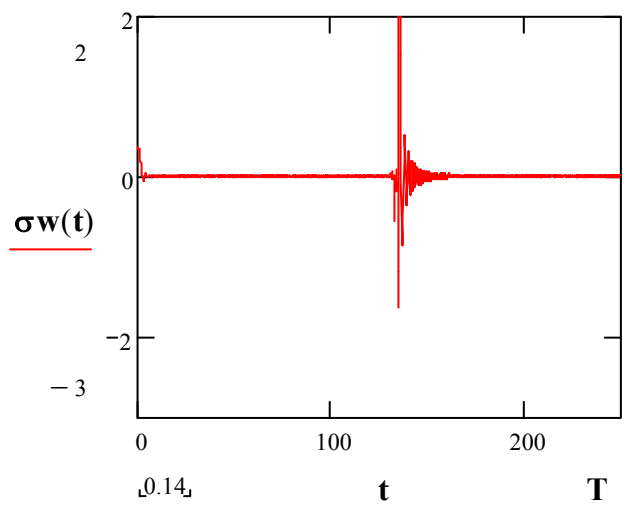

**Pic.6.8.** Variable $\dot{\sigma}(t)$.

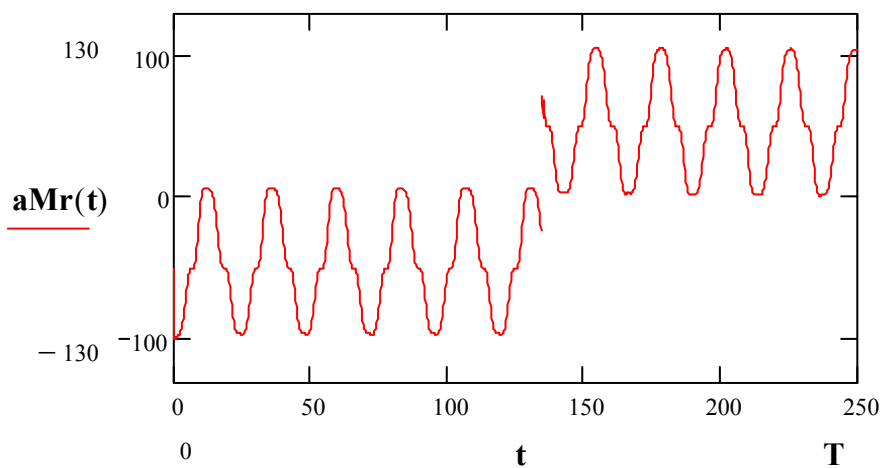

Missile acceleration along target-to-missile direction

**Pic**.**6**.**9**.Missile acceleration along target-to-missile direction: $a_M^r(t)$.

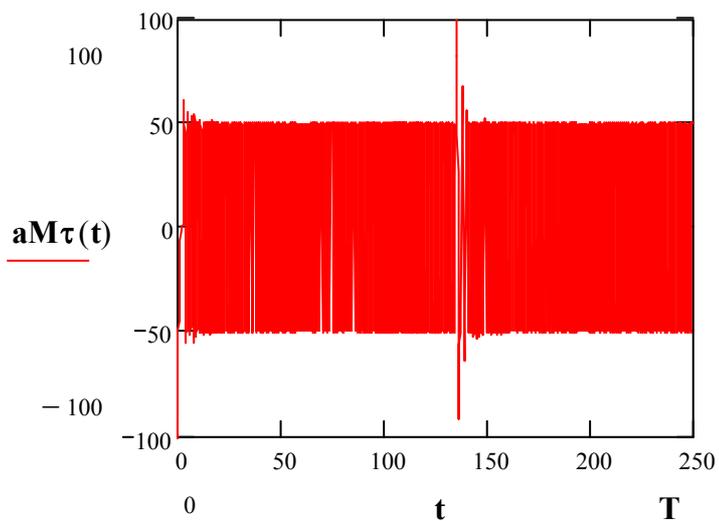

Missile tangent acceleration

**Pic**.**6**.**10**.Missile tangent acceleration: $a_M^\tau(t)$.

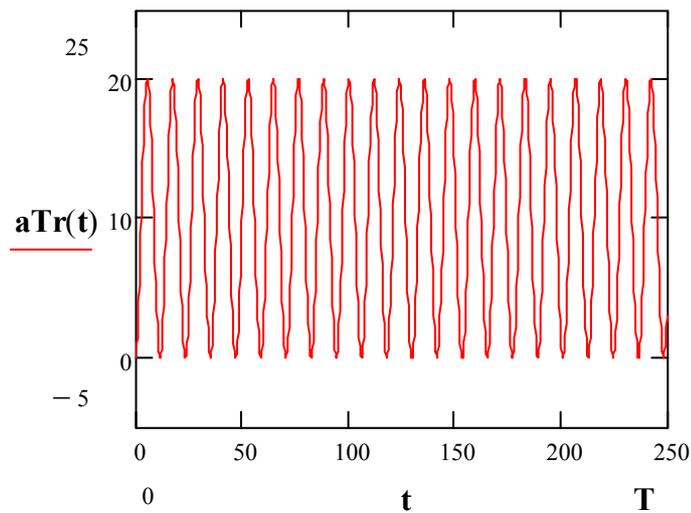

**Pic**.**6**.**11**.Target acceleration along target-to-missile direction: $a_T^r(t)$.

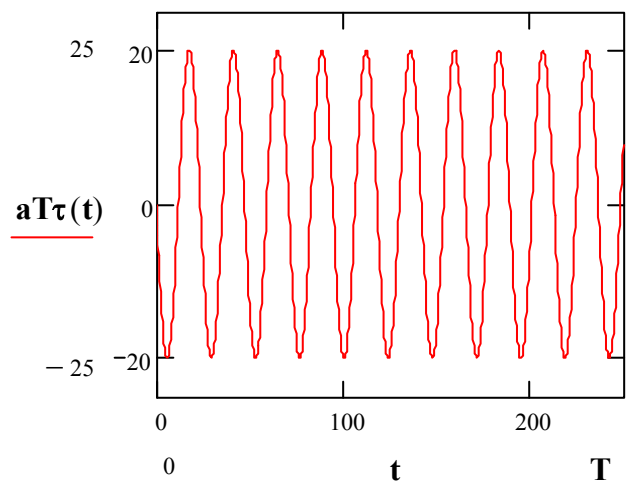

**Pic**.**6**.**12**.Target tangent acceleration: $a_T^\tau(t)$.

# Appendix. A. Optimal control for $2$-persons antagonistic differential game with linear dynamics.

In more details the qustions concerning differential games with linear dynamics look a papers [1],[3],[7].

Let us consider an $2$-persons antagonistic differential games $LDG_{2;T}(\mathbf{f},\mathbf{0},\mathbf{M},\mathbf{0})$, with linear dynamics:

$$\dot{\mathbf{x}} = \mathbf{A}(t)\mathbf{x} + \mathbf{B}(t)\mathbf{u} + \mathbf{C}(t)\mathbf{v}$$

$$\mathbf{u} \in U \subsetneq \mathbb{R}^n, \mathbf{v} \in V \subsetneq \mathbb{R}^n. \tag{A.1}$$

$$\mathbf{J} = \|\mathbf{M}\mathbf{x}(T)\|.$$

Optimal control problem for the first player:

$$\min_{\alpha_1(t) \in V} \left( \max_{\alpha_2(t) \in U} \|\mathbf{M}\mathbf{x}(T)\| \right) \tag{A.1'}$$

and optimal control problem for the second player:

$$\max_{\alpha_2(t) \in V} \left( \min_{\alpha_1(t) \in U} \|\mathbf{M}\mathbf{x}(T)\| \right). \tag{A.1''}$$

For the solutions of ordinary differential equation

$$\dot{\mathbf{x}} = \mathbf{A}(t)\mathbf{x} + \mathbf{f}(t), t \geq t_0, \tag{A.2}$$

the Cauchy formula

$$\mathbf{x}(t) = \mathbf{X}(t)\mathbf{x}(t_0) + \int_{t_0}^{t} \Phi(t,s) \cdot \mathbf{f}(s)ds, \tag{A.3}$$

where $\mathbf{X}(t) \triangleq n \times n$ matrix whose columns constitute $n$ lynearly independet solutions to ordinary differential equation

$$\dot{\mathbf{x}}(t) = \mathbf{A}(t)\mathbf{x}(t), \tag{A.4}$$

is the fundamental matrix.

$$\dot{\mathbf{X}}(t) = \mathbf{A}(t)\mathbf{X}(t),$$

$$\Phi(t,s) = \mathbf{X}(t)\mathbf{X}^{-1}(s). \tag{A.5}$$

**Notation** A.1. $\Phi(T,t) \triangleq$ *Cauchy matrix for the solutions of ordinary differential equation (A.4).*

Thus

$$\dot{\Phi}(T,t) = \frac{\partial \Phi(T,t)}{\partial t} = -\Phi(T,t)\mathbf{A}(t). \tag{A.6}$$

**Notation** A.2.

$$\mathbf{y} \triangleq \mathbf{M}\Phi(T,t)\mathbf{x},$$

$$\mathbf{X}(T,t) \triangleq \mathbf{M}\Phi(T,t)\mathbf{B}(t), \tag{A.7}$$

$$\mathbf{Y}(T,t) \triangleq \mathbf{M}\Phi(T,t)\mathbf{C}(t).$$

*Thus*

$$\dot{\mathbf{y}}(T,t) = \mathbf{M}\dot{\Phi}(T,t)x(t) + \mathbf{M}\Phi(T,t)\dot{x}(t). \tag{A.8}$$

*Substitution Eqs.(A.1) and (A.7) into Eq.(A.8) gives:*

$$\dot{\mathbf{y}}(T,t) = \mathbf{M}\dot{\Phi}(T,t)\mathbf{x}(t) + \mathbf{M}\Phi(T,t)[\mathbf{A}(t)\mathbf{x}(t) + \mathbf{B}(t)\mathbf{u}(t) + \mathbf{C}(t)\mathbf{v}(t)] =$$

$$= [\mathbf{M}\dot{\Phi}(T,t) + \mathbf{M}\Phi(T,t)\mathbf{A}(t)]\mathbf{x}(t) + \mathbf{M}\Phi(T,t)\mathbf{B}(t)\mathbf{u}(t) + \mathbf{M}\Phi(T,t)\mathbf{C}(t)\mathbf{v}(t) =$$

$$= \mathbf{X}(T,t)\mathbf{u}(t) + \mathbf{Y}(T,t)\mathbf{v}(t), \qquad (A.9)$$

i.e.

$$\dot{\mathbf{y}}(T,t) = \mathbf{X}(T,t)\mathbf{u}(t) + \mathbf{Y}(T,t)\mathbf{v}(t).$$

$$\mathbf{J} = \|\mathbf{y}(T)\|.$$

Thus from 2-persons differential game $LDG_{2;T}(\mathbf{f}, 0, \mathbf{M}, \mathbf{0})$ (A.1) we obtain 2-persons differential game $LDG_{2;T}(\tilde{\mathbf{f}}, 0, \mathbf{M}, \mathbf{0})$, with simple linear dynamics:

$$\dot{\mathbf{y}}(T,t) = \mathbf{X}(T,t)\mathbf{u}(t) + \mathbf{Y}(T,t)\mathbf{v}(t).$$
$$\mathbf{u} \in U, \mathbf{v} \in V. \qquad (A.10)$$
$$\mathbf{J} = \|\mathbf{y}(T)\|.$$

Thus optimal control problem for the first player:

$$\min_{\alpha_1(t) \in U} \left( \max_{\alpha_2(t) \in V} \|\mathbf{y}(T)\|^2 \right) \qquad (A.10')$$

and optimal control problem for the second player:

$$\max_{\alpha_2(t) \in V} \left( \min_{\alpha_1(t) \in U} \|\mathbf{y}(T)\|^2 \right). \qquad (A.10'')$$

Let us consider an 2-persons differential game $LDG_{2;T}(\mathbf{f}, 0, \mathbf{M}, 0)$, with linear dynamics

$$\dot{x}_1 = x_2,$$

$$\dot{x}_2 = u(t) + v(t),$$

$$u \in [-\rho_u, \rho_u], v \in [-\rho_v, \rho_v]. \quad (A.11)$$

$$\mathbf{J} = \|\mathbf{M}\mathbf{x}(T)\|.$$

$$\mathbf{M} = \begin{pmatrix} 1 & 0 \\ 0 & 0 \end{pmatrix}.$$

Thus optimal control problem for the first player:

$$\min_{\alpha_1(t) \in [-\rho_u, \rho_u]} \left( \max_{\alpha_2(t) \in [-\rho_v, \rho_v]} \|\mathbf{x}(T)\|^2 \right) =$$

$$= \min_{\alpha_1(t) \in [-\rho_u, \rho_u]} \left( \max_{\alpha_2(t) \in [-\rho_v, \rho_v]} [x_1^2(T) + x_2^2(T)] \right) \quad (A.11')$$

and optimal control problem for the second player:

$$\max_{\alpha_2(t) \in [-\rho_v, \rho_v]} \left( \min_{\alpha_1(t) \in [-\rho_u, \rho_u]} \|x(T)\|^2 \right)$$

$$\max_{\alpha_2(t) \in [-\rho_v, \rho_v]} \left( \min_{\alpha_1(t) \in [-\rho_u, \rho_u]} [x_1^2(T) + x_2^2(T)] \right). \quad (A.11'')$$

For the Cauchy matrix of the game (A.11), (i.e. for the Cauchy matrix of the ordinary differential equation Eq.(A.4)) we obtain:

$$\Phi(T,t) = \exp\left[\begin{pmatrix} 0 & 1 \\ 0 & 0 \end{pmatrix}(T-t)\right] =$$

$$= \begin{pmatrix} 1 & T-t \\ 0 & 0 \end{pmatrix}.$$

(A.12)

Thus

$$y(T,t) = \mathbf{M}\Phi(T,t)x(t) =$$

$$= \begin{pmatrix} 1 & 0 \\ 0 & 0 \end{pmatrix} \times \begin{pmatrix} 1 & T-t \\ 0 & 0 \end{pmatrix} \times \begin{pmatrix} x_1(t) \\ x_2(t) \end{pmatrix} =$$

$$= \begin{pmatrix} 1 & 0 \\ 0 & 0 \end{pmatrix} \times \begin{pmatrix} x_1(t) + (T-t)x_2(t) \\ 0 \end{pmatrix} =$$

(A.13)

$$= \begin{pmatrix} x_1(t) + (T-t)x_2(t) \\ 0 \end{pmatrix}.$$

$$y(T,t) = \begin{pmatrix} x_1(t) + (T-t)x_2(t) \\ 0 \end{pmatrix}.$$

Using the results of the paper [7] (Theorem.1) we obtain optimal control $u^*(t)$ of the first player and optimal control $v^*(t)$ of the second player:

$$u^*(t) = -\rho_u \mathbf{sign}[x_1(t) + (T-t)x_2(t)],$$

(A.14)

$$v^*(t) = \rho_v \mathbf{sign}[x_1(t) + (T-t)x_2(t)].$$

Let us consider now an 2-persons antagonistic differential game $LDG_{2;T}(\mathbf{f}, 0, \mathbf{M}, 0)$, with linear dynamics:

$$\dot{\mathbf{x}} = \mathbf{A}(t)x + \mathbf{B}(t)u + \mathbf{C}(t)v + \mathbf{f}(t),$$

$$\mathbf{u} \in U, \mathbf{v} \in V.$$

$$\mathbf{J} = \|\mathbf{M}x(T, 0)\|. \tag{A.15}$$

$$\mathbf{M} = \begin{pmatrix} 1 & 0 \\ 0 & 0 \end{pmatrix}.$$

Thus optimal control problem for the first player:

$$\min_{\alpha_1(t) \in U} \left( \max_{\alpha_2(t) \in V} \|\mathbf{y}(T)\| \right) \tag{A.15'}$$

and optimal control problem for the second player:

$$\max_{\alpha_2(t) \in V} \left( \min_{\alpha_1(t) \in U} \|\mathbf{y}(T)\| \right). \tag{A.15''}$$

**Notation**   A.3.

$$1. \mathbf{y}(T, t) \triangleq \mathbf{M}\Phi(T, t)\mathbf{x}(t),$$

$$2. \mathbf{X}(T, t) \triangleq \mathbf{M}\Phi(T, t)\mathbf{B}(t),$$

$$3. \mathbf{Y}(T, t) \triangleq \mathbf{M}\Phi(T, t)\mathbf{C}(t), \tag{A.16}$$

$$4. \mathbf{Z}(T, t) \triangleq \mathbf{M}\Phi(T, t)\mathbf{f}(t)$$

Substitution Eqs.(A.15) and (A.16) into Eq(A.8) gives:

$$\dot{\mathbf{y}}(T,t) = \mathbf{M}\dot{\boldsymbol{\Phi}}(T,t)\mathbf{x}(t) + \mathbf{M}\boldsymbol{\Phi}(T,t)[\mathbf{A}(t)\mathbf{x}(t) + \mathbf{B}(t)\mathbf{u}(t) + \mathbf{C}(t)\mathbf{v}(t) + \mathbf{f}(t)] =$$

$$= [\mathbf{M}\dot{\boldsymbol{\Phi}}(T,t) + \mathbf{M}\boldsymbol{\Phi}(T,t)A(t)]\mathbf{x}(t) + \mathbf{M}\boldsymbol{\Phi}(T,t)\mathbf{B}(t)\mathbf{u}(t) + \mathbf{M}\boldsymbol{\Phi}(T,t)\mathbf{C}(t)\mathbf{v}(t) +$$

$$+\mathbf{M}\boldsymbol{\Phi}(T,t)\mathbf{f}(t) =$$

$$= \mathbf{X}(T,t)\mathbf{u}(t) + \mathbf{Y}(T,t)\mathbf{v}(t) + \mathbf{Z}(T,t), \qquad (A.17)$$

i.e.

$$\dot{\mathbf{y}}(T,t) = \mathbf{X}(T,t)\mathbf{u}(t) + \mathbf{Y}(T,t)\mathbf{v}(t) + \mathbf{Z}(T,t).$$

$$\mathbf{J} = \|\mathbf{y}(T,0)\|.$$

Thus from 2-persons differential game $LDG_{2;T}(\mathbf{f},0,\mathbf{M},\mathbf{0})$ (A.15) we obtain simple 2-persons differential game $LDG_{2;T}(\tilde{\mathbf{f}},0,\mathbf{M},\mathbf{0})$, with linear dynamics:

$$\dot{\mathbf{y}}(T,t) = \mathbf{X}(T,t)\mathbf{u}(t) + \mathbf{Y}(T,t)\mathbf{v}(t) + \mathbf{Z}(T,t),$$

$$\mathbf{u} \in U, \mathbf{v} \in V, \qquad (A.18)$$

$$\mathbf{J} = \|\mathbf{y}(T,0)\|.$$

Substitution Eq.(A.12) into Eq.(A.16.4) gives:

$$\mathbf{Z}(T,t) \triangleq \mathbf{M}\mathbf{\Phi}(T,t)\mathbf{f}(t) =$$

$$= \begin{pmatrix} 1 & 0 \\ 0 & 0 \end{pmatrix} \times \begin{pmatrix} 1 & T-t \\ 0 & 0 \end{pmatrix} \times \begin{pmatrix} \mathbf{f}_1(t) \\ \mathbf{f}_2(t) \end{pmatrix} =$$

$$= \begin{pmatrix} 1 & 0 \\ 0 & 0 \end{pmatrix} \times \begin{pmatrix} \mathbf{f}_1(t) + (T-t)\mathbf{f}_2(t) \\ 0 \end{pmatrix} = \qquad (A.19)$$

$$= \begin{matrix} \mathbf{f}_1(t) + (T-t)\mathbf{f}_2(t) \\ 0 \end{matrix}.$$

$$\mathbf{Z}(T,t) = \begin{pmatrix} \mathbf{f}_1(t) + (T-t)\mathbf{f}_2(t) \\ 0 \end{pmatrix}.$$

**Notation**   A.3.

$$\mathbf{z}(T,t) \triangleq \mathbf{y}(T,t) - \int_0^t \mathbf{Z}(T,t)dt. \qquad (A.20)$$

*Substitution (A.20) into (A.18) gives:*

$$\dot{\mathbf{z}}(T,t) = \mathbf{X}(T,t)\mathbf{u}(t) + \mathbf{Y}(T,t)\mathbf{v}(t)$$

$$\mathbf{u} \in U, \mathbf{v} \in V. \qquad (A.21)$$

$$J = \|\mathbf{z}(T,0)\|.$$

Let us consider an 2-persons differential game $DG_{2;T}(\mathbf{f},0,\mathbf{M},\mathbf{0})$, with linear dynamics

$$\dot{x}_1 = x_2,$$

$$\dot{x}_2 = \kappa x_2 + u(t) + v(t),$$

$$u \in U, v \in V. \tag{A.22}$$

$$J = \|\mathbf{M}x(T)\|.$$

$$\mathbf{M} = \begin{pmatrix} 1 & 0 \\ 0 & 0 \end{pmatrix}.$$

For the Cauchy matrix of the game (A.22), (i.e. for the Cauchy matrix of the ordinary differential equation Eq.(A.4)) we obtain:

$$\Phi(T,t) = \exp\left[\begin{pmatrix} 0 & 1 \\ 0 & \kappa \end{pmatrix}(T-t)\right] = \begin{pmatrix} 1 & T-t \\ 0 & \xi(T-t,\kappa) \end{pmatrix}, \tag{A.23}$$

where

$$\xi(T-t,\kappa) = \exp(\kappa(T-t)). \tag{A.24}$$

Thus

$$y(T,t) = \mathbf{M}\Phi(T,t)x(t) =$$

$$= \begin{pmatrix} 1 & 1 \\ 0 & 0 \end{pmatrix} \times \begin{pmatrix} 1 & T-t \\ 0 & \xi(T-t,\kappa) \end{pmatrix} \times \begin{pmatrix} x_1(t) \\ x_2(t) \end{pmatrix} =$$

$$= \begin{pmatrix} 1 & 1 \\ 0 & 0 \end{pmatrix} \times \begin{pmatrix} x_1(t) + (T-t)x_2(t) \\ \xi(T-t,\kappa)x_2(t) \end{pmatrix} = \qquad (A.25)$$

$$= \begin{pmatrix} x_1(t) + (T-t)x_2(t) + \xi(T-t,\kappa)x_2(t) \\ 0 \end{pmatrix}.$$

$$y(T,t) = \begin{pmatrix} x_1(t) + (T-t)x_2(t) + \xi(T-t,\kappa)x_2(t) \\ 0 \end{pmatrix}.$$

From the results of the paper [7] (Theorem.1) by simple calculation we obtain optimal control $u^*(t)$ for the first player and optimal control $v^*(t)$ for the second player:

$$u^*(t) = -\rho_u \mathbf{sign}[x_1(t) + (T-t)x_2(t)],$$

$$\qquad (A.26)$$

$$v^*(t) = \rho_v \mathbf{sign}[x_1(t) + (T-t)x_2(t)].$$

Let us consider an 2-persons differential game $DG_{2;T}(\mathbf{f},0,\mathbf{M},\mathbf{0})$, with linear dynamics

$$\dot{x}_1 = x_2,$$

$$\dot{x}_2 = \kappa x_2 + u(t) + v(t),$$

$$u \in U, v \in V.$$

(A.27)

$$J = \|\mathbf{M} x(T)\|.$$

$$\mathbf{M} = \begin{pmatrix} 1 & 1 \\ 0 & 0 \end{pmatrix}.$$

For the Cauchy matrix of the game (A.27), (i.e. for the Cauchy matrix of the ordinary differential equation Eq.(A.4)) we obtain:

$$\Phi(T,t) = \exp\left[\begin{pmatrix} 0 & 1 \\ 0 & \kappa \end{pmatrix}(T-t)\right] = \begin{pmatrix} 1 & T-t \\ 0 & \xi(T-t,\kappa) \end{pmatrix},$$

(A.28)

$$\xi(T-t,\kappa) = \exp(\kappa(T-t)).$$

Thus

$$y(T,t) = \mathbf{M}\Phi(T,t)x(t) =$$

$$= \begin{pmatrix} 1 & 1 \\ 0 & 0 \end{pmatrix} \times \begin{pmatrix} 1 & T-t \\ 0 & \xi(T-t,\kappa) \end{pmatrix} \times \begin{pmatrix} x_1(t) \\ x_2(t) \end{pmatrix} =$$

$$= \begin{pmatrix} 1 & 1 \\ 0 & 0 \end{pmatrix} \times \begin{pmatrix} x_1(t) + (T-t)x_2(t) \\ \xi(T-t,\kappa)x_2(t) \end{pmatrix} = \qquad (A.29)$$

$$= \begin{pmatrix} x_1(t) + (T-t)x_2(t) + \xi(T-t,\kappa)x_2(t) \\ 0 \end{pmatrix}.$$

$$y(T,t) = \begin{pmatrix} x_1(t) + (T-t)x_2(t) + \xi(T-t,\kappa)x_2(t) \\ 0 \end{pmatrix}.$$

From the results of the paper [7] (Theorem.1) by simple calculation we obtain optimal control $u^*(t)$ for the first player and optimal control $v^*(t)$ for the second player:

$$u^*(t) = -\rho_u \mathbf{sign}[x_1(t) + [(T-t) + \xi(T-t,\kappa)]x_2(t)],$$

$$\qquad (A.30)$$

$$v^*(t) = \rho_v \mathbf{sign}[x_1(t) + [(T-t) + \xi(T-t,\kappa)]x_2(t)].$$

# Appendix.B.
# Canonical criteria for linear system about existence of controls to get from any given state to any other.

Consider the linear controllable dynamical system [41],[42] with state-space $\mathbb{R}^d$ and action-space $\mathbb{R}^m$, given by

$$f(x,a) = \mathbf{A}x + \mathbf{B}a,$$
$$x \in \mathbb{R}^d, a \in \mathbb{R}^m. \tag{B.1}$$

Here $\mathbf{A}$ is a $d \times d$ matrix and $\mathbf{B}$ is a $d \times m$ matrix.

**Definition** B.1. We say that $f$ is fully controllable in $n$ steps if, for all $x_0, x \in \mathbb{R}^d$, there is a control $(u_0, \ldots, u_{n-1})$ such that $x_n = x$. Here, $(x_0, \ldots, x_n)$ is the controlled sequence, given by $x_{k+1} = f(x_k, u_k)$ for $0 \le k \le n-1$.

We then seek to minimize the energy $E_n(x_0, x)$:

$$E_n(x_0, x) = \sum_{k=0}^{n-1} |u_2|^2 \tag{B.2}$$

over the set of such controls.

**Proposition** B.1. [42]. The system $f$ is fully controllable in $n$ steps if and only if $\mathbf{rank}(\mathbf{M}_n) = d$, where $\mathbf{M}_n$ is the $d \times n \cdot m$ matrix $[\mathbf{A}^{n-1}\mathbf{B}, \ldots, \mathbf{AB}, \mathbf{B}]$. Set $y = x - \mathbf{A}^n x_0$ and $\mathbf{G}_n = \mathbf{M}_n \mathbf{M}_n^T$. Then the minimal energy from $x_0$ to $x$ in $n$ steps is $y^T \mathbf{G}_n^{-1} y$ and this is achieved uniquely by the control:

$$u^T = y^T \mathbf{G}_n^{-1} \mathbf{A}^{n-k-1},$$
$$0 \le k \le n-1. \tag{B.3}$$

**Remark** B.1. Note that $\mathrm{rank}(\mathbf{M}_n)$ is non-decreasing in $n$ and, by Cayley-Hamilton, is constant for $n > d$.

Let us consider now the continuous-time linear controllable dynamical system

$$\mathbf{b}(x,u) = \mathbf{A}x + \mathbf{B}u,$$
$$x \in \mathbb{R}^d, a \in \mathbb{R}^m. \tag{B.4}$$

**Definition** B.2. Given a starting point $x_0$, the controlled process for control $(u_t)_{t>0}$ is given by the solution of $\dot{x}_t = b(x_t, u_t)$ for $t > 0$. We say that $\mathbf{b}$ is fully controllable in time $t$ if, for all $x_0, x \in \mathbb{R}^d$, there exists a control $(u_s)_{0 \le s \le t}$ such that $x_t = x$. We then seek to minimize the energy

$$E(x_0,x;t) = \int_0^t \|u_s\|^2 ds \qquad (B.5)$$

subject to $x_t = x$.

Note that

$$\frac{d}{dt}(\exp(-\mathbf{A}t)x_t) = \exp(-\mathbf{A}t)(\dot{x}_t - \mathbf{A}x_t) = \exp(-\mathbf{A}t)\mathbf{B}u_t.$$
$$\frac{d}{dt}(\exp(-\mathbf{A}t)x_t) = \exp(-\mathbf{A}t)\mathbf{B}u_t. \qquad (B.6)$$

Thus

$$x_t = \exp(-\mathbf{A}t)x_0 + \int_0^t \exp[\mathbf{A}(t-s)]\mathbf{B}u_s ds. \qquad (B.7)$$

Consider for $t > 0$ the $d \times d$ matrix

$$\mathbf{G}(t) = \int_0^t \exp(\mathbf{A}s)\mathbf{B}\mathbf{B}^T(\exp(\mathbf{A}s))^T ds. \qquad (B.8)$$

Note that for all $t > 0$, $\mathbf{G}(t)$ is invertible if and only if $\mathbf{rank}(\mathbf{M}_d) = d$.

**Proposition** B.2.[42].*The linear system $\mathbf{b}(x,u)$ is fully controllable in time $t$ if and only if $\mathbf{G}(t)$ is invertible. The minimal energy for a control from $x_0$ to $x$ in time $t$ is $y^T G^{-1}(t)y$, where*

$$y = x - \exp(-\mathbf{A}t)x_0, \qquad (B.9)$$

*and is achieved uniquely by the control:*

$$y_s^T = y^T \mathbf{G}^{-1}(t) \exp[\mathbf{A}(t-s)]\mathbf{B}. \qquad (B.10)$$